\documentclass[11pt,dvips
]{scrbook}

\usepackage{etex}

\usepackage[utf8]{inputenc}
\usepackage[english]{babel}

\usepackage[makeindex,useindex]{splitidx}
\makeindex
\usepackage{showidx}

\usepackage{setspace}
\usepackage{tikz}
\usetikzlibrary{cd,automata,arrows,positioning,decorations.pathreplacing,calc}

\tikzcdset{
  shift left/.default=+0.7ex,
  shift right/.default=+0.7ex}
\newenvironment{tikzar}[1][]{{}\kern-4pt\tikzset{nothing/.tip={},cart/.tip={Glyph[glyph math command = bullet]}}
\begin{tikzcd}[ampersand replacement=\&,#1]}%
{\end{tikzcd}\kern-4pt{}}
\tikzstyle{edge label
}=[font=\scriptsize]
\tikzcdset{every label/.append style = {font = \scriptsize}}
\tikzcdset{every node/.append style = {font = \scriptsize}}
\tikzcdset{arrow style = tikz,diagrams={>={Straight Barb[scale=0.8]}}
}

\usepackage{graphicx}
\usepackage{pgf}
\usepackage{color}

\usepackage{multicol}
\usepackage{multirow}
\usepackage{hhline}

\usepackage{rotating}

\usepackage{pst-plot}
\usepackage{pstricks}

\usepackage{array}
\usepackage{multirow}
\usepackage{mathtools}

\newcolumntype{L}[1]{>{\raggedright\let\newline\\\arraybackslash\hspace{0pt}}m{#1}}
\newcolumntype{C}[1]{>{\centering\let\newline\\\arraybackslash\hspace{0pt}}m{#1}}
\newcolumntype{R}[1]{>{\raggedleft\let\newline\\\arraybackslash\hspace{0pt}}m{#1}}

\usepackage{amsfonts,amssymb,amsmath,amsthm}
\usepackage{stmaryrd}
\usepackage{mathtools}
\usepackage{txfonts}
\usepackage{yfonts}
\newfont{\gothic}{ygoth at 12pt}
\usepackage{bbm}
\usepackage{alltt}

\theoremstyle{plain}

\theoremstyle{definition}

\theoremstyle{remark}

\usepackage{fullpage}
\usepackage{pgf}
\usepackage{enumerate}

\usepackage{color}
\usepackage{pstricks}
\usepackage{extdash}

\makeatletter
\DeclareOldFontCommand{\rm}{\normalfont\rmfamily}{\mathrm}
\DeclareOldFontCommand{\sf}{\normalfont\sffamily}{\mathsf}
\DeclareOldFontCommand{\tt}{\normalfont\ttfamily}{\mathtt}
\DeclareOldFontCommand{\bf}{\normalfont\bfseries}{\mathbf}
\DeclareOldFontCommand{\it}{\normalfont\itshape}{\mathit}
\DeclareOldFontCommand{\sl}{\normalfont\slshape}{\@nomath\sl}
\DeclareOldFontCommand{\sc}{\normalfont\scshape}{\@nomath\sc}
\makeatother

\providecommand*{\Ifstr}{\ifstr}
\usepackage{blindtext}
\usepackage{xcolor}

\usepackage{scrwfile}
\usepackage{xpatch}

\colorlet{partcolor}{blue}
\addtokomafont{partprefix}{\color{partcolor}}
\addtokomafont{part}{\normalcolor}
\RedeclareSectionCommand[
  tocdynnumwidth=true,%
  tocbeforeskip=1em
]{part}
\RedeclareSectionCommand[
  beforeskip=0pt,
  afterindent=false,
  afterskip=2\baselineskip,
  tocdynnumwidth,
  tocbeforeskip=1em plus 1pt minus 1pt,
  toclinefill=\TOCLineLeaderFill,
    tocindent=1.5em
]{chapter}
\RedeclareSectionCommand[
  beforeskip=\baselineskip,
  afterindent=false,
  afterskip=.5\baselineskip,
  tocindent=1.5em,
  tocnumwidth=3.5em
]{section}
\RedeclareSectionCommand[
  beforeskip=.75\baselineskip,
  afterindent=false,
  afterskip=.5\baselineskip,
  tocindent=3.8em,
  tocnumwidth=4em
]{subsection}
\RedeclareSectionCommand[
  beforeskip=.5\baselineskip,
  afterindent=false,
  afterskip=.25\baselineskip,
  tocindent=7em,
  tocnumwidth=4.1em
]{subsubsection}
\RedeclareSectionCommand[
  tocindent=10em,
  tocnumwidth=5em
]{paragraph}
\RedeclareSectionCommand[
  tocindent=12em,
  tocnumwidth=6em
]{subparagraph}

\RedeclareSectionCommands
  [tocpagenumberwidth=3em]
  {part,chapter,section,subsection,paragraph,subparagraph}

\makeatletter
\newif\ifuseparttoc
\newcommand*{\parttoc}[1][\thepart]{
  \useparttoctrue
  \edef\ext@parttoc{tcp#1}
  \DeclareNewTOC[
    listname=Contents,
  ]{\ext@parttoc}
  \begingroup
    \value{tocdepth}=\chaptertocdepth
    \listoftoc{\ext@parttoc}
  \endgroup
}
\xapptocmd\addtocentrydefault{
  \ifuseparttoc
    \expandafter\tocbasic@addxcontentsline\expandafter{\ext@parttoc}{#1}{#2}{#3}
  \fi
}{}{}
\xpretocmd\part{\useparttocfalse}{}{}
\newif\ifusechaptertoc
\newcommand*{\chaptertoc}[2][\thechapter]{
  \usechaptertoctrue
  \edef\ext@chaptoc{tcc#1}
  \DeclareNewTOC{\ext@chaptoc}
  \setchapterpreamble{%
    \begin{minipage}{\linewidth}
      \hrulefill\par
      \value{tocdepth}=\subsectiontocdepth
      \listoftoc*{\ext@chaptoc}
    \end{minipage}%
    \par\bigskip\nobreak\noindent\hrulefill\par
    \bigskip\noindent\ignorespaces
  }%
}
\xapptocmd\addtocentrydefault{
  \ifusechaptertoc
    \Ifstr{#1}{chapter}{}
      {\expandafter\tocbasic@addxcontentsline\expandafter{\ext@chaptoc}{#1}{#2}{#3}}
  \fi
}{}{}
\xpretocmd\chapter{\usechaptertocfalse}{}{}
\xpretocmd\part{\usechaptertocfalse}{}{}
\makeatother

\newcommand\setchaptertoc[1][]{%
  \Ifstr{#1}{}
    {\AddtoOneTimeDoHook{heading/preinit/chapter}{\chaptertoc}}
    {\AddtoOneTimeDoHook{heading/preinit/chapter}{\chaptertoc[#1]}}%
}

\usepackage[dvips]{hyperref} 

\newcommand{\EQLS}{$=$}
\newcommand{\DOTT}{\Huge$\bullet$}
\newcommand{\WDOTT}{\Huge$\mathbf{\circ}$}
\newcommand{\Dottt}{\LARGE$\bullet$}

\mathcode`\<="4268 
\mathcode`\>="5269 
\mathchardef\gt="313E 
\mathchardef\lt="313C 

\newcommand{\pleq}{\preceq}

\newcommand{\plt}{\prec}
\newcommand{\pgeq}{\succeq}

\newcommand{\pgt}{\succ}
\newcommand{\pwedge}{\curlywedge}
\newcommand{\wo}{\textstyle\bigcurlywedge}
\newcommand{\wob}[1]{\wo\!\left[{#1}\right]}

\renewcommand{\to}{\begin{tikzar}[column sep=1.3em]\hspace{-.3em}\arrow{r}\&\hspace{-.2em}\end{tikzar}}

\newcommand{\tto}[1]{\begin{tikzar}[column sep=2.2em]\hspace{-.3em}\arrow{r}{#1}\&\hspace{-.2em}\end{tikzar}}
\newcommand{\oot}[1]{\begin{tikzar}[column sep=2.2em]\hspace{-.3em}\&\hspace{-.2em}\arrow{l}[swap]{#1}\end{tikzar}}

\newcommand{\epi}{\begin{tikzar}[column sep=1.3em]\hspace{-.3em}\arrow[two heads]{r}\&\hspace{-.2em}\end{tikzar}}
\newcommand{\eepi}[1]{\begin{tikzar}[column sep=2.2em]\hspace{-.3em}\arrow[two heads]{r}{#1}\&\hspace{-.2em}\end{tikzar}}

\renewcommand{\mapsfrom}{\mathrel{\reflectbox{\ensuremath{\mapsto}}}}

\newcommand{\stricto}{\begin{tikzar}[column sep=1.8em]
\hspace{-.3em}%
\arrow[nothing-cart]{r}\&%
\hspace{-.3em}\end{tikzar}}

\newcommand{\strictto}[1]{\begin{tikzar}[column sep=2em]
\hspace{-.3em}%
\arrow[nothing-cart]{r}{#1}\&%
\hspace{-.3em}\end{tikzar}}

\newcommand{\strictoot}[1]{\begin{tikzar}[column sep=2em]
\hspace{-.3em}%
\&\arrow[nothing-cart]{l}[swap]{#1}
\hspace{-.3em}\end{tikzar}}

\newcommand{\sstrictto}[1]{\begin{tikzar}[column sep=2em]
\hspace{-.3em}%
\arrow[nothing-cart]{r}{#1}\&%
\hspace{-.3em}\end{tikzar}}

\newcommand{\strictinto}{\begin{tikzar}[column sep=1.8em]
\hspace{-.3em}%
\arrow[nothing-cart,tail]{r}\&%
\hspace{-.3em}\end{tikzar}}

\newcommand{\strictintto}[1]{\begin{tikzar}[column sep=2.4em]
\hspace{-.3em}%
\arrow[nothing-cart,tail]{r}{#1}\&%
\hspace{-.3em}\end{tikzar}}

\newcommand{\enco}[1]{\left\ulcorner{#1}\right\urcorner}
\newcommand{\enc}{\enco-}

\newcommand{\id}{{\rm id}}

\newcommand{\CCC}{{\cal C}}

\newcommand{\EEE}{{\cal E}}
\newcommand{\FFF}{{\cal F}}

\newcommand{\ZZZ}{{\cal Z}}

\newcommand{\BBb}{{\mathbb B}}
\newcommand{\CCc}{{\mathbb C}}

\newcommand{\HHh}{{\mathbb H}}

\newcommand{\LLl}{{\mathbb L}}

\newcommand{\NNn}{{\mathbb N}}

\newcommand{\PPp}{{\mathbb P}}

\newcommand{\SSs}{{\mathbb S}}

\newcommand{\ZZz}{{\mathbb Z}}

\newcommand{\QqQ}{{\mathfrak Q}}

\newcommand{\bbB}{{\mathfrak b}}

\newcommand{\ddD}{{\mathfrak d}}

\newcommand{\hhH}{{\mathfrak h}}

\newcommand{\qqQ}{{\mathfrak q}}

\newcommand{\iii}{\mathbf{i}}

\newcommand{\ppp}{\mathbf{p}}
\newcommand{\qqq}{\mathbf{q}}
\newcommand{\rrr}{\mathbf{r}}
\newcommand{\sss}{\mathbf{s}}

\newcommand{\www}{\mathbf{w}}

\newcommand{\Eee}{{\mathbold E}}

\def\pushright#1{{
   \parfillskip=0pt            
   \widowpenalty=10000         
   \displaywidowpenalty=10000  
   \finalhyphendemerits=0      
   \leavevmode                 
   \unskip                     
   \nobreak                    
   \hfil                       
   \penalty50                  
   \hskip.2em                  
   \null                       
   \hfill                      
   {#1}                        
   \par}}                      

\def\qed{\pushright{$\boxempty$}\penalty-700 \smallskip}

\newenvironment{prf}[1]{\begin{trivlist} \item[{\bf ~Proof}#1.]}%
{\qed\end{trivlist}}

\newcommand{\beq}{\begin{equation}}
\newcommand{\eeq}{\end{equation}}
\newcommand{\bea}{\begin{eqnarray}}
\newcommand{\eea}{\end{eqnarray}}
\newcommand{\bear}{\begin{eqnarray*}}
\newcommand{\eear}{\end{eqnarray*}}
\newcommand{\bpr}{\begin{prf}{}}
\newcommand{\epr}{\end{prf}}
\newcommand{\bprf}[1]{\begin{prf}{#1}}
\newcommand{\eprf}{\end{prf}}

{\end{itemize}}

\newdimen\proofrulebreadth \proofrulebreadth=.05em
\newdimen\proofdotseparation \proofdotseparation=1.25ex
\newdimen\proofrulebaseline \proofrulebaseline=2ex
\newcount\proofdotnumber \proofdotnumber=3
\let\then\relax
\def\hfi{\hskip0pt plus.0001fil}
\mathchardef\squigto="3A3B
\newif\ifinsideprooftree\insideprooftreefalse
\newif\ifonleftofproofrule\onleftofproofrulefalse
\newif\ifproofdots\proofdotsfalse
\newif\ifdoubleproof\doubleprooffalse
\let\wereinproofbit\relax
\newdimen\shortenproofleft
\newdimen\shortenproofright
\newdimen\proofbelowshift
\newbox\proofabove
\newbox\proofbelow
\newbox\proofrulename
\def\shiftproofbelow{\let\next\relax\afterassignment\setshiftproofbelow\dimen0 }
\def\shiftproofbelowneg{\def\next{\multiply\dimen0 by-1 }%
\afterassignment\setshiftproofbelow\dimen0 }
\def\setshiftproofbelow{\next\proofbelowshift=\dimen0 }
\def\setproofrulebreadth{\proofrulebreadth}

\def\prooftree{
\ifnum  \lastpenalty=1
\then   \unpenalty
\else   \onleftofproofrulefalse
\fi
\ifonleftofproofrule
\else   \ifinsideprooftree
        \then   \hskip.5em plus1fil
        \fi
\fi
\bgroup
\setbox\proofbelow=\hbox{}\setbox\proofrulename=\hbox{}%
\let\justifies\proofover\let\leadsto\proofoverdots\let\Justifies\proofoverdbl
\let\using\proofusing\let\[\prooftree
\ifinsideprooftree\let\]\endprooftree\fi
\proofdotsfalse\doubleprooffalse
\let\thickness\setproofrulebreadth
\let\shiftright\shiftproofbelow \let\shift\shiftproofbelow
\let\shiftleft\shiftproofbelowneg
\let\ifwasinsideprooftree\ifinsideprooftree
\insideprooftreetrue
\setbox\proofabove=\hbox\bgroup$\displaystyle 
\let\wereinproofbit\prooftree
\shortenproofleft=0pt \shortenproofright=0pt \proofbelowshift=0pt
\onleftofproofruletrue\penalty1
}

\def\eproofbit{
\ifx    \wereinproofbit\prooftree
\then   \ifcase \lastpenalty
        \then   \shortenproofright=0pt  
        \or     \unpenalty\hfil         
        \or     \unpenalty\unskip       
        \else   \shortenproofright=0pt  
        \fi
\fi
\global\dimen0=\shortenproofleft
\global\dimen1=\shortenproofright
\global\dimen2=\proofrulebreadth
\global\dimen3=\proofbelowshift
\global\dimen4=\proofdotseparation
\global\count255=\proofdotnumber
$\egroup  
\shortenproofleft=\dimen0
\shortenproofright=\dimen1
\proofrulebreadth=\dimen2
\proofbelowshift=\dimen3
\proofdotseparation=\dimen4
\proofdotnumber=\count255
}

\def\proofover{
\eproofbit 
\setbox\proofbelow=\hbox\bgroup 
\let\wereinproofbit\proofover
$\displaystyle
}%
\def\proofoverdbl{
\eproofbit 
\doubleprooftrue
\setbox\proofbelow=\hbox\bgroup 
\let\wereinproofbit\proofoverdbl
$\displaystyle
}%
\def\proofoverdots{
\eproofbit 
\proofdotstrue
\setbox\proofbelow=\hbox\bgroup 
\let\wereinproofbit\proofoverdots
$\displaystyle
}%
\def\proofusing{
\eproofbit 
\setbox\proofrulename=\hbox\bgroup 
\let\wereinproofbit\proofusing
\kern0.3em$
}

\def\endprooftree{
\eproofbit 
  \dimen5 =0pt
\dimen0=\wd\proofabove \advance\dimen0-\shortenproofleft
\advance\dimen0-\shortenproofright
\dimen1=.5\dimen0 \advance\dimen1-.5\wd\proofbelow
\dimen4=\dimen1
\advance\dimen1\proofbelowshift \advance\dimen4-\proofbelowshift
\ifdim  \dimen1<0pt
\then   \advance\shortenproofleft\dimen1
        \advance\dimen0-\dimen1
        \dimen1=0pt
        \ifdim  \shortenproofleft<0pt
        \then   \setbox\proofabove=\hbox{%
                        \kern-\shortenproofleft\unhbox\proofabove}%
                \shortenproofleft=0pt
        \fi
\fi
\ifdim  \dimen4<0pt
\then   \advance\shortenproofright\dimen4
        \advance\dimen0-\dimen4
        \dimen4=0pt
\fi
\ifdim  \shortenproofright<\wd\proofrulename
\then   \shortenproofright=\wd\proofrulename
\fi
\dimen2=\shortenproofleft \advance\dimen2 by\dimen1
\dimen3=\shortenproofright\advance\dimen3 by\dimen4
\ifproofdots
\then
        \dimen6=\shortenproofleft \advance\dimen6 .5\dimen0
        \setbox1=\vbox to\proofdotseparation{\vss\hbox{$\cdot$}\vss}%
        \setbox0=\hbox{%
                \advance\dimen6-.5\wd1
                \kern\dimen6
                $\vcenter to\proofdotnumber\proofdotseparation
                        {\leaders\box1\vfill}$%
                \unhbox\proofrulename}%
\else   \dimen6=\fontdimen22\the\textfont2 
        \dimen7=\dimen6
        \advance\dimen6by.5\proofrulebreadth
        \advance\dimen7by-.5\proofrulebreadth
        \setbox0=\hbox{%
                \kern\shortenproofleft
                \ifdoubleproof
                \then   \hbox to\dimen0{%
                        $\mathsurround0pt\mathord=\mkern-6mu%
                        \cleaders\hbox{$\mkern-2mu=\mkern-2mu$}\hfill
                        \mkern-6mu\mathord=$}%
                \else   \vrule height\dimen6 depth-\dimen7 width\dimen0
                \fi
                \unhbox\proofrulename}%
        \ht0=\dimen6 \dp0=-\dimen7
\fi
\let\doll\relax
\ifwasinsideprooftree
\then   \let\VBOX\vbox
\else   \ifmmode\else$\let\doll=$\fi
        \let\VBOX\vcenter
\fi
\VBOX   {\baselineskip\proofrulebaseline \lineskip.2ex
        \expandafter\lineskiplimit\ifproofdots0ex\else-0.6ex\fi
        \hbox   spread\dimen5   {\hfi\unhbox\proofabove\hfi}%
        \hbox{\box0}%
        \hbox   {\kern\dimen2 \box\proofbelow}}\doll%
\global\dimen2=\dimen2
\global\dimen3=\dimen3
\egroup 
\ifonleftofproofrule
\then   \shortenproofleft=\dimen2
\fi
\shortenproofright=\dimen3
\onleftofproofrulefalse
\ifinsideprooftree
\then   \hskip.5em plus 1fil \penalty2
\fi
}

\usepackage{array}
\usepackage{multirow}

\newcolumntype{L}[1]{>{\raggedright\let\newline\\\arraybackslash\hspace{0pt}}m{#1}}
\newcolumntype{C}[1]{>{\centering\let\newline\\\arraybackslash\hspace{0pt}}m{#1}}
\newcolumntype{R}[1]{>{\raggedleft\let\newline\\\arraybackslash\hspace{0pt}}m{#1}}

\newcommand{\sta}[1]{{#1}^{\triangleright}}
\newcommand{\out}[1]{{#1}^{\odot}}

\newcommand{\cata}[1]{\llparenthesis {#1} \rrparenthesis}

\newcommand{\bbag}[1]{\lbag {#1} \rbag}

\newcommand{\uev}[1]{\left\{{#1}\right\}}
\newcommand{\pev}[1]{\left[{#1}\right]}

\newcommand{\puev}[1]{\big\{\!\big\{{#1}\big\}\!\big\}}
\newcommand{\universal}{\big\{\big\}}
\newcommand{\puniversal}{\puev{}}

\newcommand{\Universal}{\Big\{\Big\}}
\newcommand{\pUniversal}{\Big\{\!\Big\{\Big\}\!\Big\}}
\newcommand{\prtial}{\big[\big]}

\newcommand{\Run}[1]{\left\{\!\lvert {#1} \rvert\!\right\}}

\newcommand{\Runn}{\left\{\!\lvert \ \rvert\!\right\}}

\newcommand{\runn}{{\tt run}}

\providecommand{\dotdiv}{
  \mathbin{
    \vphantom{+}
    \text{
      \mathsurround=0pt 
      \ooalign{
        \noalign{\kern-.35ex}
        \hidewidth$\smash{\cdot}$\hidewidth\cr 
        \noalign{\kern.35ex}
        $-$\cr 
      }%
    }%
  }%
}

\newcommand{\veryhigh}[3]
{   \begin{tikzpicture}
        \node (tempnode-0) at (0,0) {$#1$};
        \foreach \mytext [count=\c] in {#2}
        { \pgfmathtruncatemacro{\b}{\c-1}
            \node[above right,font=\tiny,inner sep=3pt] (tempnode-\c) at (tempnode-\b) {$\mytext$};
            \xdef\maxexp{\c}
        }
		
        \draw [decoration={brace,amplitude=4pt,mirror,raise=2pt},decorate] ($(tempnode-1.south east)+(-0.13,0.13)$) -- node[below right=1mm,font=\tiny] {#3} ($(tempnode-\maxexp.south east)+(-0.13,0.13)$);
    \end{tikzpicture}
}

\newcommand{\blank}{\square}

\newcommand{\iif}{\mathit{ifte}}

\newcommand{\IIF}{\mathit{IFTE}}

\newcommand{\pairr}[2]{\pairrr {{#1}, {#2}}}
\newcommand{\unpairr}[1]{\lfloor {#1}\rfloor}
\newcommand{\fstt}[1]{\unpairr{#1}_{0}}
\newcommand{\sndd}[1]{\unpairr{#1}_{1}}

\newcommand{\suce}{\sss}
\newcommand{\addd}{\aaa}

\newcommand{\pred}{\rrr}

\newcommand{\bpredc}{\bbB}
\newcommand{\dpredc}{\ddD}
\newcommand{\hpredc}{\hhH}
\newcommand{\qpredc}{\qqQ}

\newcommand{\iszero}{0^?}

\newcommand{\true}{\top}
\newcommand{\flse}{\bot}
\newcommand{\MU}{MU}
\newcommand{\WH}{WH}

\newcommand{\iseq}{\stackrel{?}=}

\newcommand{\Upsa}{\dot\Upsilon}
\newcommand{\Upsb}{\ddot\Upsilon}
\newcommand{\Upsc}{\overset{\ldots}\Upsilon}
\newcommand{\upsa}{\dot\upsilon}
\newcommand{\upsb}{\ddot\upsilon}
\newcommand{\upsc}{\overset{\ldots}\upsilon}
\newcommand{\psia}{\dot\psi}
\newcommand{\psib}{\ddot\psi}
\newcommand{\psic}{\overset{\ldots}\psi}
\newcommand{\Psia}{\dot\Psi}
\newcommand{\Psib}{\ddot\Psi}
\newcommand{\Psic}{\overset{\ldots}\Psi}

\newcommand{\Fixp}{\Omega}

\newcommand{\fixp}{\rotatebox[origin=c]{90}{$\rightsquigarrow$}}

\newcommand{\divg}{\rotatebox[origin=c]{90}{$\multimap$}}
\newcommand{\convg}{\mbox{$\rotatebox[origin=c]{90}{$\multimap$}^{\raisebox{-.3ex}{\hspace{-1ex}\footnotesize$\bullet$}}$}}
\newcommand{\sconvg}{\mbox{$\rotatebox[origin=c]{90}{$\multimap$}^{\raisebox{-.33ex}{\hspace{-1ex}$\scriptstyle\bullet$}}$}}

\newcommand{\congv}{\mbox{$\rotatebox[origin=c]{-90}{$\multimap$}_{\raisebox{0.15ex}{\hspace{-0.9ex}\footnotesize$\bullet$}}$}}
\newcommand{\scongv}{\mbox{$\rotatebox[origin=c]{-90}{$\multimap$}_{\raisebox{0.15ex}{\hspace{-0.9ex}$\scriptstyle\bullet$}}$}}

\newcommand{\unt}{\congv}
\newcommand{\sunt}{\scongv}
\newcommand{\cmn}{\Delta}
\newcommand{\cun}{\convg}
\newcommand{\scun}{\sconvg}
\newcommand{\halts}{\!\downarrow}

\newcommand{\inup}{\mathbin{\rotatebox[origin=c]{90}{$\in$}}}

\newcommand{\DP}{\PPp}

\newcommand{\SP}{\SSs}
\newcommand{\Bool}{\BBb}

\newcommand{\incl}{\iii}
\newcommand{\retr}{\qqq}

\newcommand{\comp}[2]{{#2}\circ {#1}}
\newcommand{\eext}[2]{\overset{#2}{\underset{#1}{\equiv}}}

\newcommand{\textbe}[1]{\textbf{\emph{#1}}}

\newcommand{\para}[1]{\noindent\textbf{\textsf{#1}}}

\renewcommand{\paragraph}[1]{\noindent\textbf{#1}}

\newcommand{\Janos}{J\'anos}

\newcommand{\ttimes}{\times}

\newcommand{\fuind}[2]{{#1}^{(#2)}}

\newcommand{\funnn}{function}
\newcommand{\Funnn}{Function}
\newcommand{\procs}{process}

\newcommand{\seq}[1]{\Big(\, #1\, \Big)}
\newcommand{\sseq}[1]{\left(\, #1\, \right)}

\newcommand{\phhi}{\phi}
\newcommand{\beeta}{q}
\newcommand{\seeta}{\sigma}

\newcommand{\prd}{{\sf p}}
\newcommand{\Prd}{\widehat{\sf p}}
\newcommand{\nill}{<>}

\newcommand{\imfun}[1]{\iota f}
\newcommand{\numidem}{\NNn}
\newcommand{\numidemm}{\nu}

\newcommand{\strict}{{cartesian}}
\newcommand{\Strict}{{Cartesian}}

\newcommand{\ottao}{\overline\vartheta_{00}}
\newcommand{\ottai}{\overline\vartheta_{11}}
\newcommand{\ottaoi}{\overline\vartheta_{01}}
\newcommand{\ottaio}{\overline\vartheta_{10}}

\newcommand{\tttao}{\underline\vartheta_{00}}
\newcommand{\tttai}{\underline\vartheta_{11}}
\newcommand{\tttaoi}{\underline\vartheta_{01}}
\newcommand{\tttaio}{\underline\vartheta_{10}}

\newcommand{\tot}[1]{{#1}^\bullet}

\begin{document}

\frontmatter
\pagenumbering{roman}
\pagestyle{empty}

\begin{titlepage}
\title{{\Huge Programs as Diagrams}
\\[3ex]
{\huge From Categorical Computability}
\\[.5ex] 
{\huge to Computable Categories}
} 

\author{Dusko Pavlovic}
\date{
\vspace{3.5cm}
\begin{center}
\includegraphics[height=10cm
]{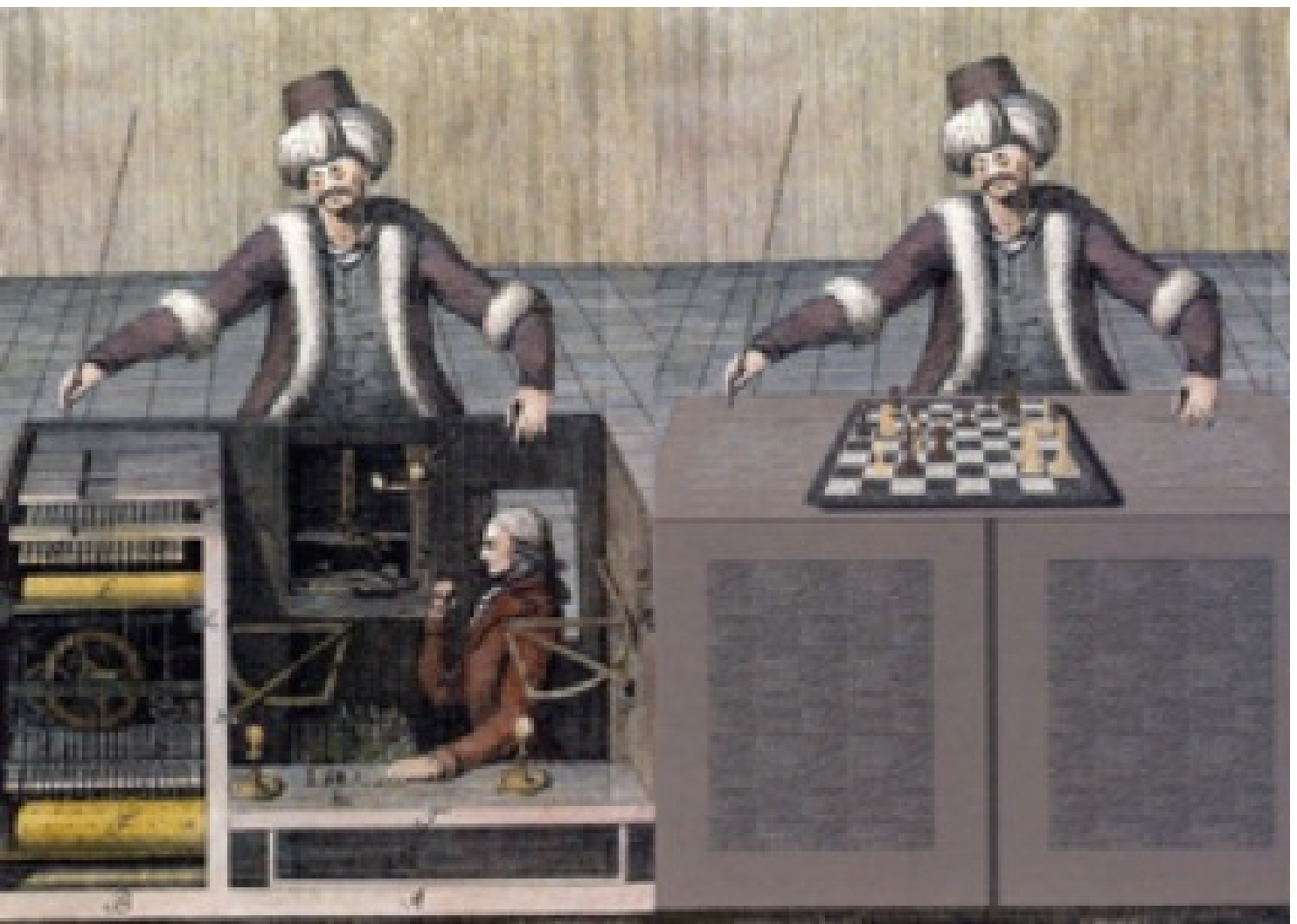}
\end{center}
}

\end{titlepage}

\maketitle

%
%
%
%
\tableofcontents

\chapter{Preface}\label{Preface}

Pythagoreans discovered that the world was built of numbers. Their rituals with numbers, allowing you to measure some sides of objects and to compute the size of other sides, were called \emph{mathemata}. Systematized into \emph{mathematics}, those rituals demonstrated that the world was regular and predictable.

The world spanned by computers and networks is definitely built of numbers. Mathematics, in the meantime, evolved beyond and away from the Pythagorean computing rituals. While programmers program computers in a great variety of languages, mathematicians study mathematics in a great variety of completely different languages. A great variety of languages is a good thing, provided that the languages do not isolate but diversify the views of the world and that there are multi-lingual communities to connect them. Many mathematicians speak programming languages; many programmers are conversant with mathematics. The language of categories seems convenient for both, and to some extent connects the communities.

In this book, we use a categorical language based on string diagrams that is so basic that it must be contained in every mathematical model of computation, and so compact that every programming language must project onto it. It evolved through many years of teaching mathematics of computation. The diversity of the mathematical and the programming languages makes such teaching and learning tasks into lengthy affairs, loaded with standard prerequisites. By peeling off what got standardized by accident and following the shortcuts that emerged, and then peeling off what became unnecessary after that, we arrived at a mathematical model of computation generated by a single diagrammatic element, which is also a single-instruction programming language. I know it sounds crazy, but read on.  

The presentation is self-contained, assuming enough confidence and curiosity. The early chapters have been taught as undergraduate courses, the middle chapters as graduate courses. The final chapters contain the results that make a point of it all. Without a teacher, a reader familiar with categories is hoped to get easy access to the basic concepts of computability, a reader familiar with computability to basic category theory, and a reader familiar with both is hoped to make use of the many opportunities to improve the approach.

\para{The \emph{"workouts"}\/ and \emph{"stories"} should probably be skipped the first time around.} They used to be called \emph{Exercises}\/ and \emph{Historic Background}\/ but that was misleading. The workouts concern the issues that the reader might like to consider on their own first. Some are presented as guided exercises, but some are much more than exercises, and some are just asides. Most are worked out in Appendix~\ref{Sec:Help}. The stories are sort of like philosophical gossips, precisely what the math is meant to defend us from, so you are welcome to skip them altogether.

\setchaptertoc
\chapter{What?}\label{What}

\section{What are we talking about?}
We spell out the basic math behind computer science. Everyone knows what is a computer. Most people also know what is science, almost as confidently. Computers run programs. Programmers program programs.  Why is there a computer science?

\section{What is computer science?} 
Science inputs processes that exist in nature and outputs their descriptions.  
\begin{figure}[!ht]
\begin{center}
\begin{tikzpicture}[shorten >=1pt,node distance=3cm]
  \node[state,minimum size=35pt,draw=black,very thick]   
  (nocarry)                {\begin{minipage}[c]{1.75cm}\begin{center} World \end{center}\end{minipage}};
   \node[state,minimum size=0pt,draw=black,very thick] (carry)    [right=of nocarry]              
   {\begin{minipage}[c]{1.75cm}\begin{center} Mind \end{center}\end{minipage}};
  \path[->] 
  (nocarry) 
  edge [bend left] node [above] {science} 
  (carry)
  (carry) 
  edge [bend left] node [below] {engineering}  
  (nocarry);
   \end{tikzpicture}
   \caption{Civilisation as a conversation} 
\label{Fig:sci-eng}
\end{center}
\end{figure}
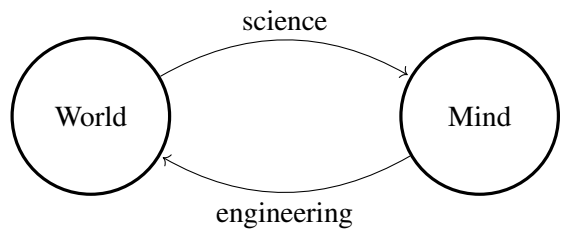
Engineering inputs descriptions of some desirable processes that do not exist in nature and outputs their realizations. Fig.~\ref{Fig:sci-eng} shows the resulting conversation staged by scientists and engineers. Computers and programs were originally built by  engineers. After lots of computers were built and lots of programs programmed, they formed networks, and the Big Bang of the Internet created cyberspace. The cyberspace clumped into galaxies, life emerged in it, botnets and viruses, influencers and celebrities, genomic databases, online yoga studios. We live in cyberspace. Our children grow there. Like our cities, freeways, and space stations, it was built by engineers, but then life took over. Life in cyberspace is not programmable and it cannot be engineered anymore. Economy, security,  epidemics are natural processes, whether they spread through physical space or cyberspace, or through a mixture. Cyberspace is the space of computations, spanned by the networks of people and computers. They act together, evolve together, and it is hard to tell who is who. They are the subject of the same science. 

\section{What is computer science about?}
Biology is about life, and life comes about by evolution. Chemistry is about molecules, and they come about by binding atoms. And so on. Every science is about some basic principle (or two, in the case of modern physics) and when you come in to study it, on day one they tell you what it is about. You study biology and the first lesson goes: "Biology is about evolution". Not so when you study computer science. What is the basic principle of computation? What is $X$ in the following equations?
\beq\label{eq:X}
\frac{\mbox{\ evolution\ }}{\mbox{biology}} \ \  = \ \ \frac{\mbox{chemical bonds}}{\mbox{chemistry}} \ \  =\  \cdots\ = \ \  \frac X {\mbox{computer science}} 
\eeq
The answer $X$ = \emph{computability} is usually taught in a ``theory" course in the final year of computer science. Many students avoid this course. Computer science and computer engineering are often taught together, which makes many courses avoidable. Those who do take a ``theory" course learn that a function is computable if it can be implemented using a Turing machine\sindex{Turing machine}, or the $\lambda$-calculus\sindex{lambda-calculus@$\lambda$-calculus}, or recursion\sindex{recursion} and minimization schemas\sindex{minimization}, or a cellular automaton, or a uniform boolean circuit, or one of many other theoretical models of computation. The Turing machine and the $\lambda$-calculus stories go back to the 1930s. The other models go back to the surrounding decades. Each of them is a story about a different aspect of computation, as seen by different people in different contexts. Each makes the same class of functions computable, but in each case they are encoded differently. The \emph{Church-Turing Thesis}\/ \sindex{Church-Turing!Thesis} postulates that each of the defined notions of computation is the same, modulo the encodings. What is their common denominator? 

\section{What is computation?} 
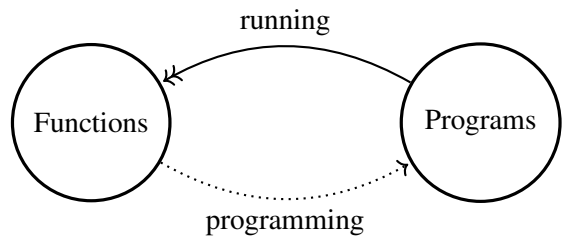
\begin{figure}[!ht]
\begin{center}
\begin{tikzpicture}[shorten >=1pt,node distance=3cm]
  \node[state,minimum size=35pt,draw=black,very thick]   
  (nocarry)                {\begin{minipage}[c]{1.75cm}\begin{center} Functions \end{center}\end{minipage}};
   \node[state,minimum size=0pt,draw=black,very thick] (carry)    [right=of nocarry]              
   {\begin{minipage}[c]{1.75cm}\begin{center} Programs \end{center}\end{minipage}};
  \path[->] 
  (nocarry) 
  edge [bend right,dotted,thick] node [below] {programming} 
  (carry);
   \path[->>] 
  (carry) 
  edge [bend right,thick] node [above] {running}  
  (nocarry);
   \end{tikzpicture}
   \caption{Computation as a conversation} 
\label{Fig:prog-run}
\end{center}
\end{figure}

The process of computation can also be viewed as a conversation. Fig.~\ref{Fig:prog-run} shows the idea.   
Programs are the mind of a computer. Functions\sindex{function} are its world. Programmers call anything that takes inputs and produces outputs a function. They analyze functions and synthesize programs. Computers run the programs and compute the functions. While running a given program is a routine operation, programming requires creativity and luck. That is why the two arrows in Fig.~\ref{Fig:prog-run} are different. The difference impacts the sciences concerned with complexity, information, evolution, language \cite{BennettC:depth}. The computational origin of this difference is described in Ch.~\ref{Chap:Undec}. A version with the symmetry of running and abstracting restored  is displayed in Fig.~\ref{Fig:CTA}. 

\section{What is a computer?}
We interact with many different computers, many different models of computation, many different programming languages. But they are avatars of the same character in the same story. We speak different languages in different societies, but they are manifestations of the same capability of speech and social interaction. What makes a language a language? What makes a sentence different from the bark of a dog? What makes a program different from a sentence? What makes a computer a computer? 

\begin{figure}[!ht]
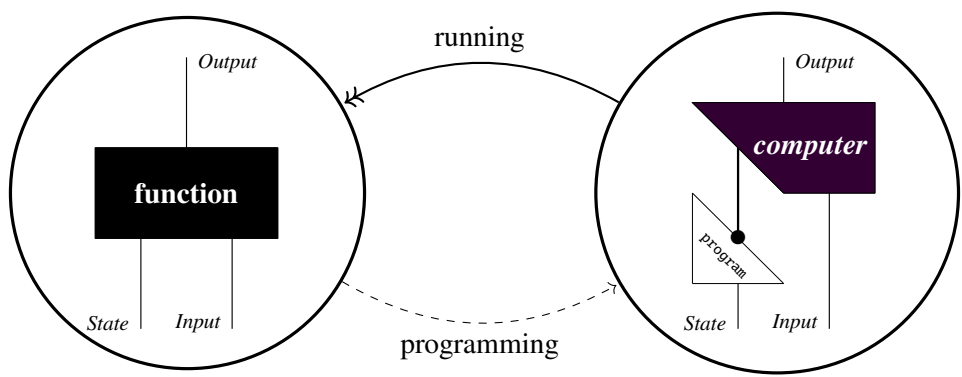

\begin{center}
\begin{tikzpicture}[shorten >=1pt,node distance=3cm]
  \node[state,minimum size=35pt,draw=black,very thick]   
  (nocarry)                {\newcommand{\fee}{\textbf{\color{white}  function}}
\newcommand{\Aee}{\mbox{\scriptsize \it State}}
\newcommand{\Bee}{\mbox{\scriptsize \it Input}}
\newcommand{\Cee}{\mbox{\scriptsize \it Output}}
\newcommand{\Code}{}
\def\JPicScale{.3}
\input{PIC/black-func.tex}};
   \node[state,minimum size=0pt,draw=black,very thick] (carry)    [right=of nocarry]              
   {\newcommand{\Fee}{\mbox{\ \ \ \tiny \texttt{program}}}
\newcommand{\Aee}{\mbox{\scriptsize \it State}}
\newcommand{\Bee}{\mbox{\scriptsize \it Input}}
\newcommand{\Cee}{\mbox{\scriptsize \it Output}}
\newcommand{\Code}{}
\newcommand{\Univ}{\mbox{\textbf{\textit{\color{white}computer}}}\ }
\newcommand{\Dott}{\mbox{\Large$\bullet$}}
\def\JPicScale{.3}
\input{PIC/black-prog.tex}};
  \path[->] 
  (nocarry) 
  edge [bend right,dashed] node [below] {programming} 
  (carry);
   \path[->>] 
   (carry) 
  edge [bend right,thick] node [above] {running}  
  (nocarry);
   \end{tikzpicture}
   \caption{Any computer can be programmed to compute any computable function} 
\label{Fig:head-tail}
\end{center}
\end{figure}

Computers and programs are the two sides of the same coin. Programs are programs because there are computers to run them. A computer is a computer because there are programs to compute. While Fig.~\ref{Fig:prog-run} displays the two sides of the coin, Fig.~\ref{Fig:head-tail} distinguishes the head and the tail. The equation identifying them is the general schema of computation, overarching all different models of computation and underlying all different computers. 
This is their common denominator. This is what makes a computer a computer. It is also the main structural component of the monoidal computer,  described in  Ch.~\ref{Chap:Comput}, formalized in Sec.~\ref{Sec:uev}. The ``running" surjection \sindex{running programs} is spelled out in Sec.~\ref{Sec:surj}. All capabilities and properties of real computers can be derived from it. An abstract computer can thus also be viewed as a single-instruction programming language, called Run, with \runn\ as its only instruction\sindex{instruction!run}\sindex{Run language|see {Run, programming language}}.  The main computational constructs are derived in chapters~\ref{Chap:Prog}--\ref{Chap:Metaprog}. The conceptual insights and the technical stepping stones into further research are in chapters~\ref{Chap:State}--\ref{Chap:Effop}. The string diagrams, convenient for building programs from \runn s and for composing computations are introduced in Ch.~\ref{Chap:Wires}.

\mainmatter

\parttoc
\def\thechapter{1}
\setchaptertoc
\chapter{Drawing types and \funnn s}\label{Chap:Wires}
\newpage

\section[Types as strings]{Types as strings: Data passing without variables}\label{Sec:type}

Whatever else there might be in a computer, there are always lots of wires. In modern computers, most wires are printed on a chip. In higher animals, the wires grow as nerves. In \sindex{string diagram} string diagrams, the wires are drawn as strings.

\subsection{Types}\label{Sec:typset}
\begin{figure}[!ht]
\begin{minipage}{.45\linewidth}
\begin{center}
\includegraphics[width=1\linewidth]{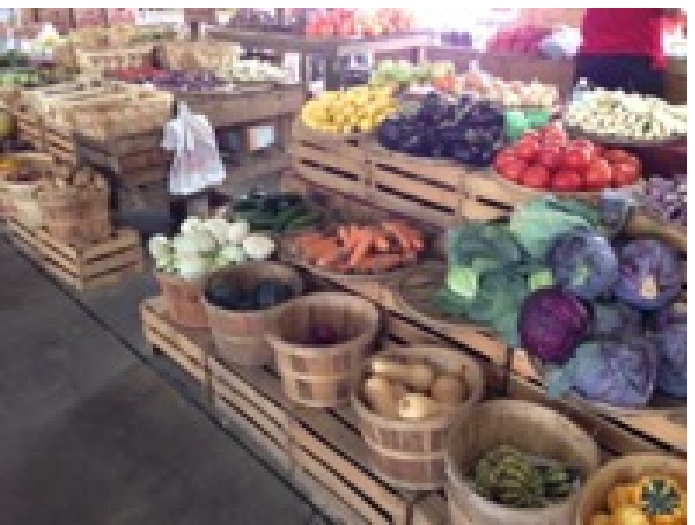}
\end{center}
\end{minipage}
\hspace{.05\linewidth}
\begin{minipage}{.45\linewidth}
\begin{center}
\includegraphics[width=1\linewidth]{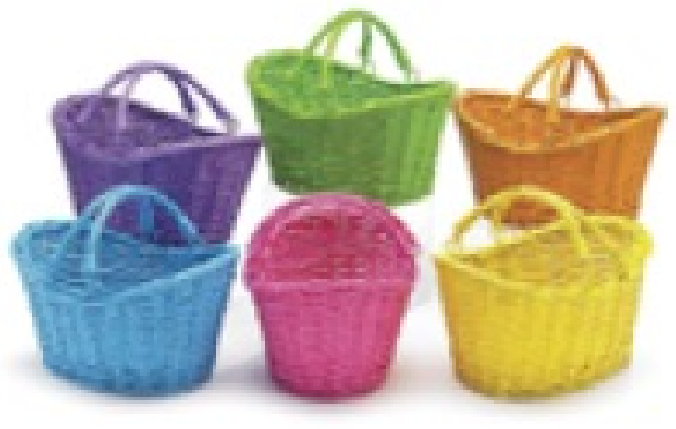}
\end{center}
\end{minipage}
\caption{Sets are containers with contents. Types are just the containers.}
\label{Fig:set-type}
\end{figure}

\begin{figure}[!ht]
\begin{minipage}{.45\linewidth}
\begin{center}
\includegraphics[width=.75\linewidth]{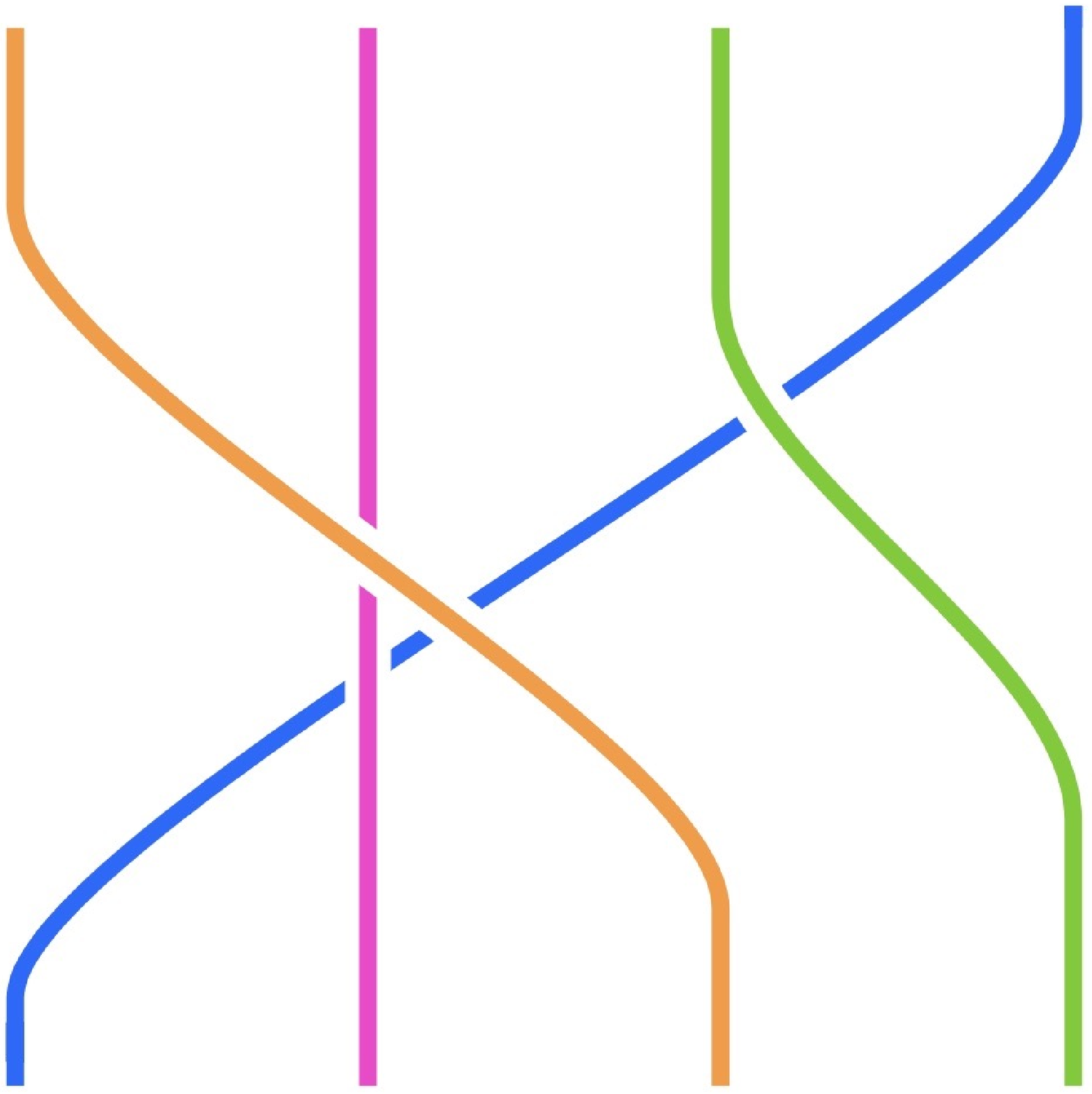}
\end{center}
\end{minipage}
\hspace{.05\linewidth}
\begin{minipage}{.45\linewidth}
\begin{center}
\includegraphics[width=.75\linewidth]{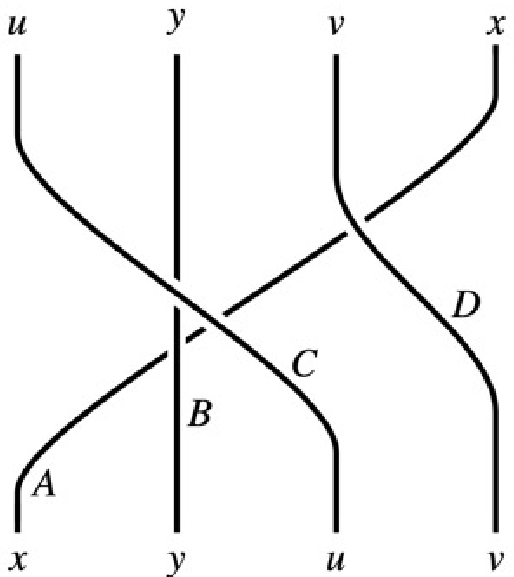}
\end{center}
\end{minipage}
\caption{Types are wires drawn as strings.}
\label{Fig:strings}
\end{figure}

\sindex{type}
Sets are containers with contents. Types are just the containers. When you add or remove an element from a set, you get another set, but the type remains the same. The difference is illustrated in Fig.~\ref{Fig:set-type}, sets on the left, types on the right. \sindex{type} When the elements keep coming and going a lot, programmers use types to keep them apart. A set is completely determined by its elements, whereas a type is determined by a property of that characterize its elements, no matter how many of them there might be. E.g., there is only one empty set, but there are as many empty types as there are colors of unicorns: the type of blue unicorns is completely different from the type of green unicorns, although they are both empty.  
In Fig.~\ref{Fig:strings} on the left, the types are drawn as strings. The same types are drawn on the right but distinguished by names and not by colors. The strings separate different data items. Their names (or colors) tell which ones are of the same type.  

\subsection{Strings instead of variables} 
When we write a program, we usually plan how to process some as\-yet\-unknown data, that will become known when the program is run. In algebra and in most programming languages, such \emph{indeterminate}\/ data values are manipulated using \sindex{variable} \emph{variables}. In computers, they are moved around through wires, depicted as strings. They carry the data and thus play the role of variables. The type annotations constrain the data that can flow through each string. In Fig.~\ref{Fig:strings} on the right, the strings are typed $A, B, C$ and $D$, and the variables $x:A$, $y:B$, $u:C$ and $v:D$, corresponding to each particular wire, are also displayed (just this time). Different variables of the same type are represented by different strings with the same type annotation.

\para{Intuition: types and sets as waves and particles.} While sets provide a foundation of mathematics, types provide a foundation of computation. If sets are collections of particles of mathematical objects, types can be viewed as collections of data waves processed in computations. Drawing types as strings in diagrams supports this view.

\subsection{Tupling and switching}\label{Sec:prodtyp}
\begin{figure}[!ht]
\begin{center}
\newcommand{\Aeex}{A_1}
\newcommand{\Awhy}{A_2}
\newcommand{\Avea}{A_n}
\newcommand{\theemes}{\times}
\newcommand{\theeemes}{\cdots \times \cdots}
\newcommand{\eex}{\scriptstyle x_1}
\newcommand{\why}{\scriptstyle x_2}
\newcommand{\vea}{\scriptstyle x_n}
\newcommand{\Ah}{\scriptstyle A_1}
\newcommand{\Bh}{\scriptstyle A_2}
\newcommand{\Ce}{\Large \ldots}
\newcommand{\De}{\scriptstyle A_n}
\def\JPicScale{1} 
\input{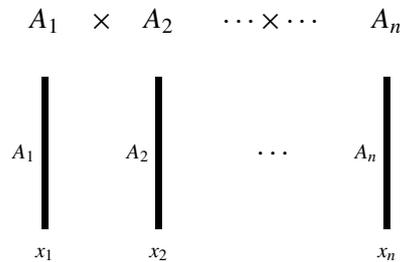}
\caption{Parallel wires correspond to product types}
\label{Fig:prodtype}
\end{center}
\end{figure}

\para{Product types.} \sindex{product}\sindex{type!product} An $n$-tuple of data is carried by of an $n$-tuple of variables through an $n$-tuple of parallel wires. The wires of types $A_{1}, A_{2},\ldots, A_{n}$ run in parallel correspond to the product type $A_{1}\times A_{2}\times\cdots \times A_{n}$, as displayed in Fig.~\ref{Fig:prodtype}.
An $n$-tuple product is the type of an $n$-tuple of values or variables $<x_{1}, x_{2}, \ldots, x_{n}> : A_{1}\times A_{2}\times \cdots \times A_{n}$. Some of the $A_i$s may be the occurrences of the same types. If $A_1=A_2=\cdots = A_n$ are the same type $A$, then the product of $n$ copies of this type is written $A^n$.

\para{The unit type}  \sindex{type!unit} $I$ satisfies $I\times A = A = A\times I$ for any type $A$. The types $A\times I$, $I \times A \times I \times I$, $I^m\times A\times I^{n}$ all boil down the type $A$. Weaving any number of strings representing the type $I$ around the string representing any type $A$ boil down to the string $A$ alone. The string representing the unit type $I$ is thus \emph{invisible}\footnote{Having invisible graphic elements will turn out to be quite convenient. Having invisible points in geometry and empty instructions in programs is not just convenient, but necessary \cite{DijkstraE:guarded}.}. Since the type $A^{n}$ is presented as $n$ copies of the string $A$, the unit type $I$ can be thought of as the product of $0$ copies of  $A$, i.e. $I = A^0$. If an element of $A^{n}$ is in the form $<x_{1},\ldots,x_{n}>$ then an element of $I$ should be  $<>$.

\begin{figure}[!ht]
\begin{center}
\newcommand{\eex}{x}
\newcommand{\why}{y}
\newcommand{\Ah}{A}
\newcommand{\Uh}{B}
\newcommand{\swap}{\varsigma}
\def\JPicScale{1.4} \input{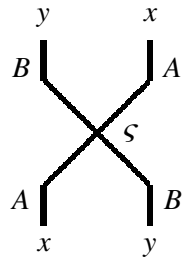}
\caption{The wires are twisted to get the data where needed}
\label{Fig:swap}
\end{center}
\end{figure}

\para{Twisting the wires} like in Fig.~\ref{Fig:swap}, corresponds to changing the order of the variables in an algebraic expression. In type theory, this operation is a \funnn\ $\varsigma\colon A\times B\to B\times A$ where $\varsigma(x,y) = <y,x>$. 

%

\section[Data services]{Data services: copying and deleting, cartesianness}\label{Sec:service}

\sindex{data services}\sindex{copying|see {data services}}\sindex{deleting|see {data services}}\sindex{data services!copying}\sindex{data services!deleting} When the same value, determined or unknown, is needed in several places in a program, or in an algebraic expression, that value can be denoted by a variable, and the variable can be \emph{copied}\/ wherever needed. Since it may be unknown whether a value will be needed, a variable denoting that value may be declared but unused and \emph{deleted}. The \textbf{\emph{data services}}\/ boil down to the operations of
\begin{itemize}
\item \textbf{copying} $\cmn\colon A \to A\times A$, and 
\item \textbf{deleting} $\cun \colon A\to I$.
\end{itemize}
The intuition is that $\cmn(x)=<x,x>$ and $\cun(x) = <>$. In string diagrams, these \funnn s are depicted as
\beq\label{eq:dataserv}
\begin{split}
\newcommand{\AAh}{\scriptstyle A}
\newcommand{\ccopy}{\cmn}
\newcommand{\delete}{\cun}
\def\JPicScale{.35} 
\ifx\JPicScale\undefined\def\JPicScale{1}\fi
\psset{unit=\JPicScale mm}
\psset{linewidth=0.3,dotsep=1,hatchwidth=0.3,hatchsep=1.5,shadowsize=1,dimen=middle}
\psset{dotsize=0.7 2.5,dotscale=1 1,fillcolor=black}
\psset{arrowsize=1 2,arrowlength=1,arrowinset=0.25,tbarsize=0.7 5,bracketlength=0.15,rbracketlength=0.15}
\begin{pspicture}(0,0)(77.5,44.07)
\psline[linewidth=1.5](70,40)(70,0)
\newrgbcolor{userFillColour}{0.2 0 0.2}
\rput{0}(70,40){\psellipse[linewidth=1.5,fillcolor=userFillColour,fillstyle=solid](0,0)(4.07,-4.07)}
\psline[linewidth=1.5](30,30)(15,15)
\psline[linewidth=1.5](15,15)(15,0)
\newrgbcolor{userFillColour}{0.2 0 0.2}
\rput{0}(15,15){\psellipse[linewidth=1.5,fillcolor=userFillColour,fillstyle=solid](0,0)(3.79,-3.79)}
\psline[linewidth=1.5](0,30)(15,15)
\rput[r](12.5,0){$\AAh$}
\rput[r](67.5,0){$\AAh$}
\psline[linewidth=1.5](0,40)(0,30)
\psline[linewidth=1.5](30,40)(30,30)
\rput[l](32.5,40){$\AAh$}
\rput[l](2.5,40){$\AAh$}
\rput[l](77.5,40){$\delete$}
\rput[l](21.25,15){$\ccopy$}
\end{pspicture}

\end{split}
\eeq
To be able to write $<x,x,x>\colon A^{3}$ we need  $<<x,x>,x>=<x,<x,x>>$ and to assure $I\times A =A = A\times I$ we need $<<>,x>=x=<x,<>>$, so the following equations are imposed
\begin{alignat}{7}
\comp {\cmn}{(\cmn \ttimes \id)}  &\ \ =\ \   \comp{\cmn}{(\id \ttimes \cmn)} &\quad&\quad&\quad&\quad&
\comp{\cmn}{(\cun \ttimes\, \id)}  &\ \  =\ \  &\   \id  &\ \   =\ \  &\ \  \comp{\cmn}{(\id \, \ttimes \cun)}\notag
\\[1ex]
\def\JPicScale{.85} 
\ifx\JPicScale\undefined\def\JPicScale{1}\fi
\psset{unit=\JPicScale mm}
\psset{linewidth=0.3,dotsep=1,hatchwidth=0.3,hatchsep=1.5,shadowsize=1,dimen=middle}
\psset{dotsize=0.7 2.5,dotscale=1 1,fillcolor=black}
\psset{arrowsize=1 2,arrowlength=1,arrowinset=0.25,tbarsize=0.7 5,bracketlength=0.15,rbracketlength=0.15}
\begin{pspicture}(0,0)(15,7.5)
\psline[linewidth=0.55](5.62,-0.62)
(10.62,-5.62)
(9.38,-5.62)(10,-5.62)
\rput{0}(4.38,0.62){\psellipse[linewidth=0.55,fillstyle=solid](0,0)(1.56,-1.57)}
\rput{0}(10.32,-5.31){\psellipse[linewidth=0.55,fillstyle=solid](0,0)(1.56,-1.57)}
\psline[linewidth=0.55](15,7.5)
(15,5)
(15,-0.63)(11.25,-4.38)
\psline[linewidth=0.55](9.37,7.5)
(9.37,6.87)
(9.38,5)(5.63,1.25)
\psline[linewidth=0.55](-0.63,7.5)
(-0.63,6.87)
(-0.62,5)(3.13,1.25)
\psline[linewidth=0.55](10,-6.88)
(10,-10)
(10,-10.62)(10,-10)
\end{pspicture}
\   &\ \ =\ \   \def\JPicScale{.85} 
\ifx\JPicScale\undefined\def\JPicScale{1}\fi
\psset{unit=\JPicScale mm}
\psset{linewidth=0.3,dotsep=1,hatchwidth=0.3,hatchsep=1.5,shadowsize=1,dimen=middle}
\psset{dotsize=0.7 2.5,dotscale=1 1,fillcolor=black}
\psset{arrowsize=1 2,arrowlength=1,arrowinset=0.25,tbarsize=0.7 5,bracketlength=0.15,rbracketlength=0.15}
\begin{pspicture}(0,0)(15.63,8.12)
\rput{0}(10.62,0.62){\psellipse[linewidth=0.55,fillstyle=solid](0,0)(1.57,-1.56)}
\psline[linewidth=0.55](15.62,8.12)
(15.62,7.5)
(15.63,5.62)(11.88,1.87)
\psline[linewidth=0.55](5.62,8.12)
(5.62,7.5)
(5.63,5.62)(9.38,1.87)
\rput{0}(4.38,-5.62){\psellipse[linewidth=0.55,fillstyle=solid](0,0)(1.57,-1.57)}
\psline[linewidth=0.55](-0.63,7.5)
(-0.62,1.25)
(-0.62,-0.63)(3.13,-4.38)
\pscustom[linewidth=0.55]{\psline(9.38,-0.62)(4.38,-5.62)
\psbezier(4.38,-5.62)(4.38,-5.62)(4.38,-5.62)
\psbezier(4.38,-5.62)(4.38,-5.62)(4.38,-5.62)
}
\psline[linewidth=0.55](4.38,-6.88)
(4.38,-10)
(4.38,-10.62)(4.38,-10)
\end{pspicture}
 &&&&& 
\def\JPicScale{.85} 
\ifx\JPicScale\undefined\def\JPicScale{1}\fi
\psset{unit=\JPicScale mm}
\psset{linewidth=0.3,dotsep=1,hatchwidth=0.3,hatchsep=1.5,shadowsize=1,dimen=middle}
\psset{dotsize=0.7 2.5,dotscale=1 1,fillcolor=black}
\psset{arrowsize=1 2,arrowlength=1,arrowinset=0.25,tbarsize=0.7 5,bracketlength=0.15,rbracketlength=0.15}
\begin{pspicture}(0,0)(10,8.12)
\psline[linewidth=0.55](0.62,0)
(5.62,-5)
(4.38,-5)(5,-5)
\rput{0}(-0.62,1.25){\psellipse[linewidth=0.55,fillstyle=solid](0,0)(1.57,-1.56)}
\rput{0}(5.31,-4.69){\psellipse[linewidth=0.55,fillstyle=solid](0,0)(1.57,-1.56)}
\psline[linewidth=0.55](10,8.12)
(10,5.62)
(10,0)(6.25,-3.75)
\psline[linewidth=0.55](5,-6.25)
(5,-10)
(5,-10.62)(5,-10)
\end{pspicture}
 \hspace{1em}&\ \ =\ \ & \def\JPicScale{.85} 
\ifx\JPicScale\undefined\def\JPicScale{1}\fi
\psset{unit=\JPicScale mm}
\psset{linewidth=0.3,dotsep=1,hatchwidth=0.3,hatchsep=1.5,shadowsize=1,dimen=middle}
\psset{dotsize=0.7 2.5,dotscale=1 1,fillcolor=black}
\psset{arrowsize=1 2,arrowlength=1,arrowinset=0.25,tbarsize=0.7 5,bracketlength=0.15,rbracketlength=0.15}
\begin{pspicture}(0,0)(0,8.74)
\psline[linewidth=0.55](0,8.74)
(0,-9.38)
(0,-10)(0,-9.38)
\end{pspicture}
\ \   &\ \ = &  \def\JPicScale{.85} 
\ifx\JPicScale\undefined\def\JPicScale{1}\fi
\psset{unit=\JPicScale mm}
\psset{linewidth=0.3,dotsep=1,hatchwidth=0.3,hatchsep=1.5,shadowsize=1,dimen=middle}
\psset{dotsize=0.7 2.5,dotscale=1 1,fillcolor=black}
\psset{arrowsize=1 2,arrowlength=1,arrowinset=0.25,tbarsize=0.7 5,bracketlength=0.15,rbracketlength=0.15}
\begin{pspicture}(0,0)(12.19,8.12)
\rput{0}(10.62,1.25){\psellipse[linewidth=0.55,fillstyle=solid](0,0)(1.56,-1.56)}
\rput{0}(4.38,-5){\psellipse[linewidth=0.55,fillstyle=solid](0,0)(1.57,-1.56)}
\psline[linewidth=0.55](4.38,-6.25)
(4.38,-10)
(4.38,-10.62)(4.38,-10)
\psline[linewidth=0.55](-0.63,8.12)
(-0.62,1.87)
(-0.62,0)(3.13,-3.75)
\pscustom[linewidth=0.55]{\psline(9.38,0)(4.38,-5)
\psbezier(4.38,-5)(4.38,-5)(4.38,-5)
\psbezier(4.38,-5)(4.38,-5)(4.38,-5)
}
\end{pspicture}
\hspace{1em}\label{eq:comonoid}\\[3ex]\notag
\end{alignat}
\bear
\cmn\ \  &\ \ =\ \  \varsigma \circ \cmn
\\[1.5ex]
\def\JPicScale{.85} 
\ifx\JPicScale\undefined\def\JPicScale{1}\fi
\psset{unit=\JPicScale mm}
\psset{linewidth=0.3,dotsep=1,hatchwidth=0.3,hatchsep=1.5,shadowsize=1,dimen=middle}
\psset{dotsize=0.7 2.5,dotscale=1 1,fillcolor=black}
\psset{arrowsize=1 2,arrowlength=1,arrowinset=0.25,tbarsize=0.7 5,bracketlength=0.15,rbracketlength=0.15}
\begin{pspicture}(0,0)(5,10)
\rput{0}(2.49,-2.5){\psellipse[linewidth=0.55,fillstyle=solid](0,0)(1.56,-1.56)}
\psline[linewidth=0.55](2.5,-2.5)(2.5,-7.5)
\psline[linewidth=0.55](5,0)(2.5,-2.5)
\psline[linewidth=0.55](0,0)(2.5,-2.5)
\psline[linewidth=0.55](5,10)(5,0)
\psline[linewidth=0.55](0,10)(0,0)
\end{pspicture}
 &\ \  = \ \  \def\JPicScale{.85} 
\ifx\JPicScale\undefined\def\JPicScale{1}\fi
\psset{unit=\JPicScale mm}
\psset{linewidth=0.3,dotsep=1,hatchwidth=0.3,hatchsep=1.5,shadowsize=1,dimen=middle}
\psset{dotsize=0.7 2.5,dotscale=1 1,fillcolor=black}
\psset{arrowsize=1 2,arrowlength=1,arrowinset=0.25,tbarsize=0.7 5,bracketlength=0.15,rbracketlength=0.15}
\begin{pspicture}(0,0)(5,10)
\rput{0}(2.49,-2.5){\psellipse[linewidth=0.55,fillstyle=solid](0,0)(1.56,-1.56)}
\psline[linewidth=0.55](2.5,-2.5)(2.5,-7.5)
\psline[linewidth=0.55](5,0)(2.5,-2.5)
\psline[linewidth=0.55](0,0)(2.5,-2.5)
\psline[linewidth=0.55](5,2.5)(5,0)
\psline[linewidth=0.55](0,2.5)(0,0)
\psline[linewidth=0.55](0,7.5)(5,2.5)
\psline[linewidth=0.55](0,2.5)(5,7.5)
\psline[linewidth=0.55](5,10)(5,7.5)
\psline[linewidth=0.55](0,10)(0,7.5)
\end{pspicture}

\eear
 
 \medskip
The last equation prevents asymmetric tuplings. While \funnn s in general will be drawn as boxes on strings, the identity \funnn\ $\id:A\to A$, mapping $x$ to $x$, is drawn as an invisible box, just like the unit type $I$ is drawn as an invisible string.  An example of use of data services is in Fig.~\ref{Fig:type-string}.
\begin{figure}[!ht]
\begin{center}
\newcommand{\eex}{x}
\newcommand{\why}{y}
\newcommand{\northeast}{}
\newcommand{\northwest}{}
\def\JPicScale{.5}
\input{PIC/strings-1}
\caption{Rearranging $<x,y>$ into $<y,x,y,x>$ by data services}
\label{Fig:type-string}
\end{center}
\end{figure}

\para{Copying and deleting the empty tuple.} The data services on the unit type $I$ boil down to the identity map. Intuitively, since $\nill$ is the empty tuple, deleting leaves it unchanged and $\cun_I = \left(I\tto \id I\right)$. Since $I \times I = I$, copying the empty tuple gives $<\nill, \nill> = \nill$ and $\cmn_I =\left( I\tto\id I\right)$. 

\para{Cartesianness.}\sindex{product!cartesian} If types are thought of as sets, then the product types are the cartesian products of the corresponding sets, the unit type is the singleton set, and the data services are the \funnn\ s copying and deleting the set-elements. In categories, cartesian products are defined by requiring that the morphisms into $A\times B$ correspond to the pairs of the morphisms to $A$ and to $B$, as briefly explained in  Sec.~\ref{Sec:cat-logic}, and in detail in most books on categories, such as \cite{Lambek-Scott:book}. We shall see in Sec.~\ref{Sec:map} that data services make product types into cartesian products with respect to precisely those \funnn s that preserve the data services, which will be called \emph{\strict}. \sindex{function!cartesian} But we are here to study computable \funnn s, and many computations do not preserve data services. The \funnn s that may not preserve data services will be called \emph{monoidal}, \sindex{function!cartesian} since the product types for them are monoidal. \sindex{product!monoidal}String diagrams conveniently capture all of them.

%
%
%
%
%

\section{Monoidal {\funnn}s as boxes}\label{Sec:fun}

\para{The 2-dimensional text of string diagrams.}
A {\funnn} $f:A\to B$ maps inputs of type $A$ to outputs of type $B$. A process $g:X\times A\to B$ is a {\funnn} that maps causes of type $A$ to effects of type $B$, possibly also depending on states of type $X$. 

The set-theoretic concept of \funnn\ is more restrictive.\sindex{function} It requires that the possible inputs and outputs are collected in sets and that there is a unique output for every input. Programmers deviated from these requirements early on, and have been using the more relaxed concept of function at least since the 1960s. Theoretical computer scientists deviated even earlier when they developed the $\lambda$-calculus as a formal theory of \funnn s that can be fed to themselves as inputs \cite{ChurchA:calculi}. This seemed like an inexorable property of computable functions, that take their programs as inputs. The task of finding set-theoretic models of such \funnn s was tackled in the denotational semantics of computation \cite{Abramsky-Jung:domains,compendium:2003}.

In many programming languages, in type theory, and in category theory, \funnn s are specified as black boxes. The \emph{black}\/ness does not mean that the boxes are painted black, but that no light shines in or out: the interior of the \funnn\ box is hidden from the user. A functionality implemented in a black box is accessed through its input and output interfaces, which are drawn as strings.  A {\funnn} $f:A\to B$ is drawn as a box $f$ hanging from a string $B$ above it, and with a string $A$ hanging below it: 
\[\newcommand{\machine}{$f$}
\newcommand{\inputt}{\scriptstyle A}
\newcommand{\nameslang}{\scriptstyle B} 
\def\JPicScale{.55}
\ifx\JPicScale\undefined\def\JPicScale{1}\fi
\psset{unit=\JPicScale mm}
\psset{linewidth=0.3,dotsep=1,hatchwidth=0.3,hatchsep=1.5,shadowsize=1,dimen=middle}
\psset{dotsize=0.7 2.5,dotscale=1 1,fillcolor=black}
\psset{arrowsize=1 2,arrowlength=1,arrowinset=0.25,tbarsize=0.7 5,bracketlength=0.15,rbracketlength=0.15}
\begin{pspicture}(0,0)(20,40)
\psline[linewidth=0.75](0,30)(20,30)
\psline[linewidth=0.75](0,30)(0,10)
\psline[linewidth=0.75](20,10)(20,30)
\psline[linewidth=0.75](10,40)(10,30)
\rput[l](12.5,37.5){$\nameslang$}
\rput[r](7.5,2.5){$\inputt$}
\psline[linewidth=0.75](0,10)(20,10)
\psline[linewidth=0.75](10,10)(10,0)
\rput(10,20){\machine}
\end{pspicture}

\]
The inputs flow in from the bottom, and the outputs flow out at the top. This bottom-up processing is a drawing convention. At least four other drawing conventions are equally justified. The idea is that the processed data or resources flow through the strings in one direction, which is the direction of time, and that there are no flows in other directions, or between the strings. The two dimensions of string diagrams thus display where the data flow and where they do not flow. The flow through the strings bottom-up, and they do not flow between the strings.

The two dimensions also correspond to the two ways to compose \funnn s.

\subsection{Composing \funnn s}\label{Sec:compos} 
\begin{figure}[!ht]
\begin{center}
\newcommand{\machine}{$f$}
\newcommand{\gee}{$g$}
\newcommand{\kee}{$s$}
\newcommand{\hee}{t}
\newcommand{\nameslang}{\scriptstyle B}
\newcommand{\seqcompp}{{\color{red}g\circ f}}
\newcommand{\parcompp}{{\color{blue}f\ttimes t}}
\newcommand{\inputt}{\scriptstyle A} 
\newcommand{\outpt}{$\scriptstyle C$}
\newcommand{\otherinputt}{\scriptstyle U}
\newcommand{\otheroutpt}{\scriptstyle V} 
\newcommand{\outpttt}{$\scriptstyle W$}
\def\JPicScale{.5}
\input{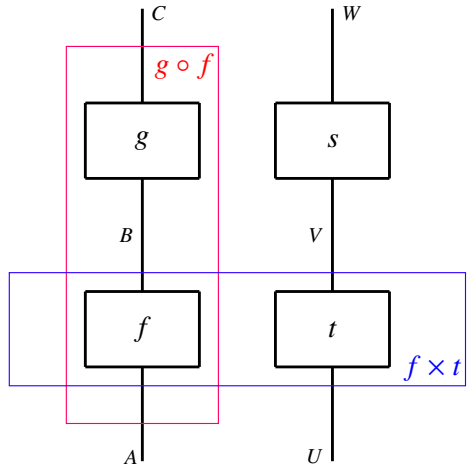}
\vspace{.5\baselineskip}
\caption{$(g\circ f)\times(s\circ t) \ \  = \ \  (g\times s)\circ(f\times t)$}
\label{Fig:godement}
\end{center}
\end{figure}

\sindex{composition}
The functions $f$ and $g$ are composed \emph{sequentially} when the outputs of $f$ flow as the inputs to $g$, provided that output type of $f$ is the input type of $g$. The functions $f$ and $t$ are composed \emph{in parallel} when there are no flows between them. In string diagrams, the sequential compositions are thus drawn by hanging the \funnn s $f$ and $g$ on one another vertically, whereas the parallel compositions are drawn by placing the \funnn s $f$ and $t$ next to one another horizontally. The typing constraints are:
\[\prooftree
A\tto f B\qquad B\tto g C
\justifies A\tto{\color{red}g\circ f} C
\endprooftree
\qquad\qquad \qquad
\prooftree
A\tto f B\qquad U\tto t V
\justifies
A\ttimes U\tto{\color{blue}f\ttimes t} B\ttimes V
\endprooftree\]
The string diagrams for the sequential composite $g\circ f$ and for the parallel composite $f\times t$ are displayed in  Fig.~\ref{Fig:godement}. If a \funnn\ $h:C\to D$ is attached above to the composite $g\circ f:A\to C$, the three boxes hanging on four stings will look the same as if we attached $f:A\to B$ below $h\circ g:B\to D$. String diagrams thus impose the associativity $h\circ (g\circ f) = (h\circ g)\circ f$ as a geometric property of the notation. Similary, if $h:C\to D$ is adjoined to the right of the composite $f\times t:A\times U\to B\times V$, resulting triple composite will forget that $f$ was adjoined on the left of $t$ before $h$ was adjoined on the right. The string diagram notation thus also imposes the associativity $(f\times t)\times h = f\times (t\times h)$ as a geometric property. 

The units for the composition operations are the identity \funnn\ $\id_{A}$ and $\id_{B}$ for the sequential composition, and the identity $\id_{I}$ on unit type $I$ for the parallel composition, in the sense of the equations
\beq\label{eq:unitary-laws}
\id_{B}\circ f\circ \id_{A}\ \  =\ \  f \ \ =\ \  \id_{I}\times f \times \id_{I} 
\eeq
If the the identity \funnn s $\id_{A}$ and $\id_{B}$ are thought of as the invisible boxes on the strings $A$ and $B$, and the unit type $I$ is thought of as the invisible string, present in any diagram wherever convenient, then the unitary laws \eqref{eq:unitary-laws} are also imposed in the string-and-box diagrams tacitly, as the geometric properties of the invisible strings and boxes.

A string diagram is read as 2-dimensional text because the strings assure that it can either be read first vertically then horizontally, or first horizontally and then vertically, but not diagonally. The two ways of reading a string diagram impose the algebraic law that constrains the two composition operations:

\subsection{The middle-two-interchange law}\label{Sec:godement}\sindex{middle-two-interchange}
A string diagram, as 2-dimensional text, can be read from left to right, or from the bottom to the top. In Fig.~\ref{Fig:godement},  we can thus choose to first go vertically and compose $f$ with $g$ and $s$ with $t$ sequentially, and after that to go horizontally and compose the sequential composites $g\circ f$ and $t\circ g$ in parallel, to get $(g\circ f)\ttimes(t\circ g)$. Alternatively, we can also choose to go horizontally first, and compose in parallel $f$ with $t$, and $g$ with $s$, to get $f\ttimes t$ and $g\ttimes s$, and then we can go vertically, to produce the sequential composite $(g\ttimes s)\circ(f\ttimes t)$. These two ways of reading the string diagram in Fig.~\ref{Fig:godement} thus correspond to two different algebraic expressions, but they describe the same composite function:
\bea\label{eq:godement}
(g\circ f)\ttimes(s\circ t) & = & (g\ttimes s)\circ(f\times t)
\eea
This is the \emph{middle-two-interchange}\/ law of the function composition algebra. This algebraic law is hardwired in the geometry of string diagrams. More precisely, the assumption that a string diagram describes a unique \funnn\  implies the middle-two-interchange law, which identifies the two ways to read a string diagram: first  horizontally or first vertically.

\subsection{Sliding boxes}\label{Seq:sliding}
\begin{figure}[!ht]
\begin{center}
\newcommand{\machine}{$f$}
\newcommand{\gee}{$g$}
\newcommand{\hee}{t}
\newcommand{\nameslang}{}
\newcommand{\seqcompp}{\scriptstyle (f;g)}
\newcommand{\parcompp}{\scriptstyle f\ttimes h}
\newcommand{\inputt}{}
\newcommand{\outpt}{}
\newcommand{\otherinputt}{}
\newcommand{\otheroutpt}{}
\def\JPicScale{.35}
\input{PIC/seq-4.tex}
\caption{$(g\ttimes \id)\circ(f\ttimes t) =(g\ttimes\id)\circ(\id\ttimes t)\circ(f\ttimes \id) = (g\ttimes t)\circ(f\ttimes \id)$}
\label{Fig:slidingfuns}
\end{center}
\end{figure}
The net effect of encodings of algebra by geometry is that some algebraic correspondences are encoded as geometric properties. E.g., the equations in the caption of Fig.~\ref{Fig:slidingfuns} can be derived algebraically from the middle-two-interchange law. On the other hand, they can also be derived by sliding boxes along the strings, as displayed in Fig.~\ref{Fig:slidingfuns}. They are thus a geometric property of the string diagram displayed there, as three of its topological deformations.  In general, any pair of {\funnn} compositions, expressed algebraically in terms of $\circ$ and $\ttimes$ and the same function boxes will denote the same composite function if and only if they correspond to two string diagrams that can be deformed into each other by some geometric transformations that may twist or extend strings, and slide boxes along them, but without tearing or disconnecting anything. 

\para{Untwisting.} The boxes also slide through the twists, and the twists can be untwisted.
\begin{align*}
\newcommand{\Ah}{}
\newcommand{\Bh}{}
\newcommand{\Uh}{}
\newcommand{\Vh}{}
\newcommand{\fee}{\scriptstyle f}
\newcommand{\tee}{\scriptstyle t}
\newcommand{\symmetry}{\scriptstyle \varsigma}
\def\JPicScale{.85} 
\ifx\JPicScale\undefined\def\JPicScale{1}\fi
\psset{unit=\JPicScale mm}
\psset{linewidth=0.3,dotsep=1,hatchwidth=0.3,hatchsep=1.5,shadowsize=1,dimen=middle}
\psset{dotsize=0.7 2.5,dotscale=1 1,fillcolor=black}
\psset{arrowsize=1 2,arrowlength=1,arrowinset=0.25,tbarsize=0.7 5,bracketlength=0.15,rbracketlength=0.15}
\begin{pspicture}(0,0)(12.5,20)
\psline(0,-10)(0,-20)
\psline(10,-10)(10,-20)
\psline(12.5,-5)(7.5,-5)
\psline(12.5,-10)(7.5,-10)
\psline(0,5)(0,-5)
\psline(10,5)(10,-5)
\psline(10,15)(0,5)
\psline(10,5)(0,15)
\psline(0,20)(0,15)
\psline(10,20)(10,15)
\rput[r](-1.25,-20){$\Ah$}
\rput[l](11.25,20){$\Bh$}
\rput[l](7.5,10){$\symmetry$}
\psline(12.5,-5)(12.5,-10)
\psline(7.5,-5)(7.5,-10)
\psline(2.5,-5)(-2.5,-5)
\psline(2.5,-10)(-2.5,-10)
\psline(2.5,-5)(2.5,-10)
\psline(-2.5,-5)(-2.5,-10)
\rput(0,-7.5){$\fee$}
\rput(10,-7.5){$\tee$}
\rput[r](-1.25,20){$\Vh$}
\rput[l](11.25,-20){$\Uh$}
\rput[r](-1.25,0){$\Bh$}
\rput[l](11.25,0){$\Vh$}
\end{pspicture}
\ \ \ \ & =\ \ \ \   \ \ \newcommand{\Ah}{}
\newcommand{\Bh}{}
\newcommand{\Uh}{}
\newcommand{\Vh}{}
\newcommand{\fee}{\scriptstyle f}
\newcommand{\tee}{\scriptstyle t}
\newcommand{\symmetry}{\scriptstyle \varsigma}\def\JPicScale{.85} 
\ifx\JPicScale\undefined\def\JPicScale{1}\fi
\psset{unit=\JPicScale mm}
\psset{linewidth=0.3,dotsep=1,hatchwidth=0.3,hatchsep=1.5,shadowsize=1,dimen=middle}
\psset{dotsize=0.7 2.5,dotscale=1 1,fillcolor=black}
\psset{arrowsize=1 2,arrowlength=1,arrowinset=0.25,tbarsize=0.7 5,bracketlength=0.15,rbracketlength=0.15}
\begin{pspicture}(0,0)(12.5,20)
\psline(10,5)(10,-5)
\psline(0,5)(0,-5)
\psline(2.5,10)(-2.5,10)
\psline(2.5,5)(-2.5,5)
\psline(10,20)(10,10)
\psline(0,20)(0,10)
\psline(10,-5)(0,-15)
\psline(10,-15)(0,-5)
\psline(0,-15)(0,-20)
\psline(10,-15)(10,-20)
\rput[r](-1.25,-20){$\Ah$}
\rput[l](11.25,20){$\Bh$}
\rput[l](7.5,-10){$\symmetry$}
\psline(2.5,10)(2.5,5)
\psline(-2.5,10)(-2.5,5)
\psline(12.5,10)(7.5,10)
\psline(12.5,5)(7.5,5)
\psline(12.5,10)(12.5,5)
\psline(7.5,10)(7.5,5)
\rput(10,7.5){$\fee$}
\rput(0,7.5){$\tee$}
\rput[r](-1.25,20){$\Vh$}
\rput[l](11.25,-20){$\Uh$}
\rput[l](11.25,0){$\Ah$}
\rput[r](-1.25,0){$\Uh$}
\end{pspicture}

&
\newcommand{\Ah}{\scriptstyle A}
\newcommand{\Bh}{\scriptstyle B}
\newcommand{\symmetry}{\scriptstyle \varsigma}
\def\JPicScale{.85} 
\ifx\JPicScale\undefined\def\JPicScale{1}\fi
\psset{unit=\JPicScale mm}
\psset{linewidth=0.3,dotsep=1,hatchwidth=0.3,hatchsep=1.5,shadowsize=1,dimen=middle}
\psset{dotsize=0.7 2.5,dotscale=1 1,fillcolor=black}
\psset{arrowsize=1 2,arrowlength=1,arrowinset=0.25,tbarsize=0.7 5,bracketlength=0.15,rbracketlength=0.15}
\begin{pspicture}(0,0)(11.25,20)
\psline(0,-15)(0,-18.75)
\psline(10,-15)(10,-18.75)
\psline(10,-5)(0,-15)
\psline(10,-15)(0,-5)
\psline(0,5)(0,-5)
\psline(10,5)(10,-5)
\psline(10,15)(0,5)
\psline(10,5)(0,15)
\psline(0,20)(0,15)
\psline(10,20)(10,15)
\rput[r](-1.25,-20){$\Ah$}
\rput[l](11.25,-20){$\Bh$}
\rput[r](-1.25,20){$\Ah$}
\rput[l](11.25,20){$\Bh$}
\rput[l](11.25,0){$\Ah$}
\rput[r](-1.25,0){$\Bh$}
\rput[l](7.5,-10){$\symmetry$}
\rput[l](7.5,10){$\symmetry$}
\end{pspicture}
\ \ \ \ \  & = \ \ \ \  
\newcommand{\Ah}{\scriptstyle A}
\newcommand{\Bh}{\scriptstyle B}
\def\JPicScale{.85} 
\ifx\JPicScale\undefined\def\JPicScale{1}\fi
\psset{unit=\JPicScale mm}
\psset{linewidth=0.3,dotsep=1,hatchwidth=0.3,hatchsep=1.5,shadowsize=1,dimen=middle}
\psset{dotsize=0.7 2.5,dotscale=1 1,fillcolor=black}
\psset{arrowsize=1 2,arrowlength=1,arrowinset=0.25,tbarsize=0.7 5,bracketlength=0.15,rbracketlength=0.15}
\begin{pspicture}(0,0)(11.25,20)
\psline(0,20)(0,-18.75)
\psline(10,20)(10,-18.75)
\rput[r](-1.25,-20){$\Ah$}
\rput[l](11.25,-20){$\Bh$}
\end{pspicture}

\end{align*}

\bigskip
\bigskip
\bigskip

\section{{\Strict} \funnn s as boxes with a dot}\label{Sec:map}

A \funnn\ is said to be 
\begin{itemize}
\item \textbe{total}\/ \sindex{function!total} if it produces at least one output for each input, and
\item \textbe{single-valued}\/ \sindex{function!single-valued} if it produces at most one output for each input.
\end{itemize}
The diagrammatic depictions are in \eqref{eq:map}: $s:A\to B$ is total if it satisfies the left-hand equation, single-valued if it satisfies the right-hand equation.
\begin{alignat}{5}\label{eq:map}
\newcommand{\aah}{s}
\newcommand{\AAh}{\scriptstyle A}
\newcommand{\BBh}{\scriptstyle B}\def\JPicScale{.35} 
\ifx\JPicScale\undefined\def\JPicScale{1}\fi
\psset{unit=\JPicScale mm}
\psset{linewidth=0.3,dotsep=1,hatchwidth=0.3,hatchsep=1.5,shadowsize=1,dimen=middle}
\psset{dotsize=0.7 2.5,dotscale=1 1,fillcolor=black}
\psset{arrowsize=1 2,arrowlength=1,arrowinset=0.25,tbarsize=0.7 5,bracketlength=0.15,rbracketlength=0.15}
\begin{pspicture}(0,0)(20,38.79)
\rput[l](12.5,25){$\BBh$}
\psline[linewidth=0.75](10,-5)(10,-30)
\newrgbcolor{userFillColour}{0.2 0 0.2}
\rput{0}(10,35){\psellipse[fillcolor=userFillColour,fillstyle=solid](0,0)(3.79,-3.79)}
\rput[r](7.5,-30){$\AAh$}
\pspolygon[linewidth=0.7,fillcolor=white,fillstyle=solid](0,15)(20,15)(20,-5)(0,-5)
\rput(10,5){$\aah$}
\psline[linewidth=0.75](10,33.75)(10,15)
\newrgbcolor{userFillColour}{0.2 0 0.2}
\rput{0}(10,15){\psellipse[fillcolor=userFillColour,fillstyle=solid](0,0)(3.79,-3.79)}
\end{pspicture}
\   &\ \ =\ \   \def\JPicScale{.35} \newcommand{\aah}{s}
\newcommand{\AAh}{\scriptstyle A}
\newcommand{\BBh}{\scriptstyle B}
\ifx\JPicScale\undefined\def\JPicScale{1}\fi
\psset{unit=\JPicScale mm}
\psset{linewidth=0.3,dotsep=1,hatchwidth=0.3,hatchsep=1.5,shadowsize=1,dimen=middle}
\psset{dotsize=0.7 2.5,dotscale=1 1,fillcolor=black}
\psset{arrowsize=1 2,arrowlength=1,arrowinset=0.25,tbarsize=0.7 5,bracketlength=0.15,rbracketlength=0.15}
\begin{pspicture}(0,0)(9.07,39.07)
\psline[linewidth=0.75](5,35)(5,-30)
\newrgbcolor{userFillColour}{0.2 0 0.2}
\rput{0}(5,35){\psellipse[fillcolor=userFillColour,fillstyle=solid](0,0)(4.07,-4.07)}
\rput[r](2.5,-30){$\AAh$}
\end{pspicture}
 &&&&& 
\newcommand{\aah}{s}
\newcommand{\AAh}{\scriptstyle A}
\newcommand{\BBh}{\scriptstyle B}\def\JPicScale{.35} 
\ifx\JPicScale\undefined\def\JPicScale{1}\fi
\psset{unit=\JPicScale mm}
\psset{linewidth=0.3,dotsep=1,hatchwidth=0.3,hatchsep=1.5,shadowsize=1,dimen=middle}
\psset{dotsize=0.7 2.5,dotscale=1 1,fillcolor=black}
\psset{arrowsize=1 2,arrowlength=1,arrowinset=0.25,tbarsize=0.7 5,bracketlength=0.15,rbracketlength=0.15}
\begin{pspicture}(0,0)(22.5,38.75)
\rput[r](2.5,-35){$\AAh$}
\psline[linewidth=0.75](5,-25)(5,-35)
\pspolygon[linewidth=0.7,fillcolor=white,fillstyle=solid](-5,-5)(15,-5)(15,-25)(-5,-25)
\rput(5,-15){$\aah$}
\psline[linewidth=0.75](20,30)(5,15)
\psline[linewidth=0.75](5,15)(5,-5)
\newrgbcolor{userFillColour}{0.2 0 0.2}
\rput{0}(5,15){\psellipse[fillcolor=userFillColour,fillstyle=solid](0,0)(3.79,-3.79)}
\psline[linewidth=0.6](-10,38.75)(-10,30)
\psline[linewidth=0.75](-10,30)(5,15)
\psline[linewidth=0.6](20,38.75)(20,30)
\rput[l](22.5,37.5){$\BBh$}
\rput[l](-7.5,37.5){$\BBh$}
\newrgbcolor{userFillColour}{0.2 0 0.2}
\rput{0}(5,-5){\psellipse[fillcolor=userFillColour,fillstyle=solid](0,0)(3.79,-3.79)}
\end{pspicture}
& \ \ \ \ = &
\newcommand{\aah}{s}
\newcommand{\AAh}{\scriptstyle A}
\newcommand{\BBh}{\scriptstyle B}\def\JPicScale{.35} 
\ifx\JPicScale\undefined\def\JPicScale{1}\fi
\psset{unit=\JPicScale mm}
\psset{linewidth=0.3,dotsep=1,hatchwidth=0.3,hatchsep=1.5,shadowsize=1,dimen=middle}
\psset{dotsize=0.7 2.5,dotscale=1 1,fillcolor=black}
\psset{arrowsize=1 2,arrowlength=1,arrowinset=0.25,tbarsize=0.7 5,bracketlength=0.15,rbracketlength=0.15}
\begin{pspicture}(0,0)(50,40)
\psline[linewidth=0.75](40,0)(25,-15)
\psline[linewidth=0.75](25,-15)(25,-35)
\newrgbcolor{userFillColour}{0.2 0 0.2}
\rput{0}(25,-15){\psellipse[fillcolor=userFillColour,fillstyle=solid](0,0)(3.79,-3.79)}
\psline[linewidth=0.6](10,10)(10,0)
\psline[linewidth=0.75](10,0)(25,-15)
\psline[linewidth=0.6](40,10)(40,0)
\rput[r](22.5,-35){$\AAh$}
\rput[l](42.5,38.75){$\BBh$}
\rput[l](12.5,38.75){$\BBh$}
\psline[linewidth=0.75](10,40)(10,30)
\pspolygon[linewidth=0.7,fillcolor=white,fillstyle=solid](0,30)(20,30)(20,10)(0,10)
\rput(10,20){$\aah$}
\pspolygon[linewidth=0.7,fillcolor=white,fillstyle=solid](30,30)(50,30)(50,10)(30,10)
\rput(40,20){$\aah$}
\psline[linewidth=0.75](40,40)(40,30)
\newrgbcolor{userFillColour}{0.2 0 0.2}
\rput{0}(10,30){\psellipse[fillcolor=userFillColour,fillstyle=solid](0,0)(3.79,-3.79)}
\newrgbcolor{userFillColour}{0.2 0 0.2}
\rput{0}(40,30){\psellipse[fillcolor=userFillColour,fillstyle=solid](0,0)(3.79,-3.79)}
\end{pspicture}
\\[8ex]
\comp {s}{\cun_B} \ \  &\ \ =\ \   \cun_A &\qquad&\qquad&\qquad&\qquad\qquad&
\comp{s}{\cmn_B}\ \  \   & \ \ \ \  =\ \   &\ \  \   \comp{\cmn_A}{(s\ttimes s)}\notag
\end{alignat}
The left-hand equation says that $s$ preserves the deletions: whenever an input $a$ can be deleted by $\cun_A\Big(a\Big) = \nill$ then the output $s(a)$ that can be deleted as  $\cun_B\Big(s(a)\Big) = \nill$. The right-hand equation says that $s$ preserves copying: when an input $a$ is cloned by $\cmn_A(a) = <a,a>$ and two copies of $s$ output $\big<s(a), s(a) \big>$, the output is the same as when the output $b=s(a)$ is cloned by $\cmn_B(b) = <b,b>$.

\para{\Strict\ \funnn s} \sindex{function!cartesian}\sindex{cartesian function|see {function}} are the \funnn s $s$ that satisfy \eqref{eq:map} and preserve data services. With respect to \strict\ \funnn s, the data services provide the \strict\ structure, i.e. make the products \strict. The black bead $\bullet$ on the output port of $s:A\stricto B$ indicates that $s$ is a {\strict} \funnn. The usual \funnn\ notation $f:A\to B$ is left for \emph{monoidal} \funnn s, \sindex{function!monoidal} \sindex{monoidal!function} that do not necessarily preserve the data services. 

\begin{figure}[!ht]
\begin{center}
\def\JPicScale{.5} \newcommand{\pizero}{\pi_A} \newcommand{\pione}{\pi_B} \newcommand{\aah}{\scriptstyle a} \newcommand{\bbh}{\scriptstyle b} \renewcommand{\pairr}{<a,b>} 
\newcommand{\AAh}{\scriptstyle A} \newcommand{\BBh}{\scriptstyle B}\newcommand{\XXh}{\scriptstyle X}\input{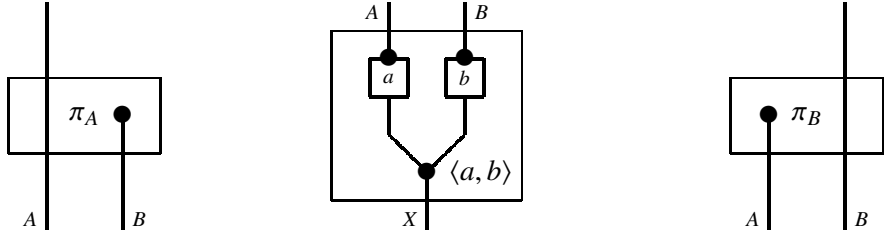}
\caption{Pairing and projections}
\label{Fig:projs}
\end{center}
\end{figure}

\para{Cartesian structure} \sindex{cartesian structure} consists of two components, derivable for all $A$ and $B$: 
\begin{itemize}
\item the projections \sindex{projection}
\beq 
\pi_{A}\ =\ \left(A\times B\sstrictto{A\times\scun} A\right) \qquad \qquad\qquad \pi_{B}\ =\ \left(A\times B\sstrictto{\scun\times B} B\right)
\eeq
\item the pairing
\sindex{pairing}
\beq\label{eq:cartpair}
\begin{split}
\prooftree
a\colon X\stricto A\qquad \qquad b\colon X\stricto B
\justifies
<a,b>\ =\ \left(X\strictto \cmn X\times X\sstrictto{a\times b} A\times B\right)
\endprooftree
\end{split}
\eeq
\end{itemize}
They are displayed in Fig.~\ref{Fig:projs}.
Together, the pairing and the projections provide a bijective correspondence between the pairs of {\strict} \funnn s to $A$ and $B$ and the {\strict} \funnn s to $A\times B$. One direction of the correspondence is given by the pairing in \eqref{eq:cartpair}, the other by the projections in
\beq\label{eq:cartproj}
\begin{split}
\prooftree
h\colon X\stricto A\times B
\justifies
\pi_A h\ =\   \left(X\strictto h A\times B\sstrictto{A\times\scun} A\right)\qquad \qquad \pi_B h\ =\   \left(X\strictto h A\times B\sstrictto{\scun\times B} B\right)
\endprooftree
\end{split}
\eeq
The correspondence is summarized by the equations
\begin{gather}\label{eq:projections}
a = \pi_A<a,b> \qquad \qquad\qquad \quad h\  =\ <\pi_A h, \pi_B h> \qquad \qquad\qquad \quad  b = \pi_B<a,b> 
\end{gather}
It is not necessary that $a$, $b$, and $h$ are {\strict} for the definitions of $<a,b>$, $\pi_A h$, and $\pi_B h$ in (\ref{eq:cartpair}--\ref{eq:cartproj}), but it is necessary for the validity of the equations in  \eqref{eq:projections}. The first equation holds if and only if $b$ is total, the third if and only if $a$ is total, and the second equation holds whenever $h$ is single-valued, but the converse is not always valid. Checking these claims is an instructive exercise. See e.g., \cite{Lambek-Scott:book} for more details.

\para{Elements.} \sindex{element} An element $a\colon A$ is a \emph{single-valued} \funnn\ from $I$ to $A$. The single-valuedness requirement assures that $<a,a> = a\times a$. A \textbe{total}\/  element is a \strict\ \funnn\ $a: I \stricto A$. A \textbf{\emph{partial}}\/ element is an element that is not total.\sindex{element!total, partial} A general \funnn\ $a\colon I\to A$ is not viewed as an element of $A$ as it may produce different outputs in $A$ on different calls, cause different side-effects, and lead to problems in substitution.

\para{Functions with side-effects are not \strict.} \sindex{side-effect} In set-theoretic frameworks, it is usually required that a function $f$ from $A$ to $B$ assigns to every single input $a:A$ a single output $f(a):B$, and all functions are assumed to be \strict. When a function $f$ needs to be computed, this cannot be guaranteed. If a computation of $f$ on an input $a$ gets stuck in a loop, it may not produce any output.  The left-hand equation in  \eqref{eq:map} then does not hold, because the input can be deleted, but there is no output to delete. A computation of $f$ may also  depend on hidden variables or covert parameters of the environment, and the same overt input may lead to one output today and a different output tomorrow, if the environment changes. Besides the outputs, computations thus often produce some \emph{side-effects}, such as \emph{non-termination}\/ when there is no output, or \emph{non-determinism}\/ when there are multiple possible outputs. That is why \textbe{computable \funnn s\ are generally not \strict, but monoidal}. \sindex{function!monoidal} That is why the computer is monoidal. The next section spells out what that means. In Sec.~\ref{Sec:divergence} we shall see that it cannot be made \strict\ as long as all of its \funnn s are programmable in it. 

\para{Terminology: partial, many-valued, monoidal \funnn s.} A function that is not total is called \emph{partial}. A function that is not single-valued is called\sindex{function!many-valued} \emph{many-valued}. The functions that may be partial, many-valued, or cause other side-effects, are called \emph{monoidal}. They form \emph{monoidal categories}, described next.\sindex{function!partial}\sindex{function!monoidal}\sindex{monoidal function|see {function}}

\section[Categories]{Categories: Universes of types and \funnn s}\label{Sec:cat}

A \emph{\textbf{category}} $\CCC$\sindex{category} is comprised of 
\begin{itemize}
\item a \emph{class of objects} 
\bear
\lvert \CCC \rvert & = & \left\{\ \newcommand{\Ah}{A}
\def\JPicScale{.5} 
\ifx\JPicScale\undefined\def\JPicScale{1}\fi
\unitlength \JPicScale mm
\begin{picture}(5,10)(0,0)
\put(3.75,2.5){\makebox(0,0)[cr]{$\Ah$}}

\linethickness{0.25mm}
\put(5,-5){\line(0,1){15}}
\end{picture}
\, ,\  \renewcommand{\Ah}{B}
\ifx\JPicScale\undefined\def\JPicScale{1}\fi
\unitlength \JPicScale mm
\begin{picture}(5,10)(0,0)
\put(3.75,2.5){\makebox(0,0)[cr]{$\Ah$}}

\linethickness{0.25mm}
\put(5,-5){\line(0,1){15}}
\end{picture}
\, ,\ldots,\  \renewcommand{\Ah}{\DP}
\ifx\JPicScale\undefined\def\JPicScale{1}\fi
\unitlength \JPicScale mm
\begin{picture}(5,10)(0,0)
\put(3.75,2.5){\makebox(0,0)[cr]{$\Ah$}}

\linethickness{0.4mm}
\put(5,-5){\line(0,1){15}}
\end{picture}
\, ,\ldots,\ \renewcommand{\Ah}{X}
\ifx\JPicScale\undefined\def\JPicScale{1}\fi
\unitlength \JPicScale mm
\begin{picture}(5,10)(0,0)
\put(3.75,2.5){\makebox(0,0)[cr]{$\Ah$}}

\linethickness{0.25mm}
\put(5,-5){\line(0,1){15}}
\end{picture}
\, ,\  \renewcommand{\Ah}{Y} 
\ifx\JPicScale\undefined\def\JPicScale{1}\fi
\unitlength \JPicScale mm
\begin{picture}(5,10)(0,0)
\put(3.75,2.5){\makebox(0,0)[cr]{$\Ah$}}

\linethickness{0.25mm}
\put(5,-5){\line(0,1){15}}
\end{picture}
\, ,\ldots\right\}
\eear
\item for any pair $X,Y\in \lvert \CCC\rvert$ a \emph{hom-set} 
\bear
\CCC(X,Y) & = & \left\{\ \newcommand{\machine}{$f$}
\newcommand{\inputt}{\scriptstyle X}
\newcommand{\nameslang}{\scriptstyle Y} 
\def\JPicScale{.5}
\ifx\JPicScale\undefined\def\JPicScale{1}\fi
\unitlength \JPicScale mm
\begin{picture}(10,17.5)(0,0)
\linethickness{0.25mm}
\put(0,7.5){\line(1,0){10}}
\linethickness{0.25mm}
\put(0,-2.5){\line(0,1){10}}
\linethickness{0.25mm}
\put(10,-2.5){\line(0,1){10}}
\linethickness{0.25mm}
\put(5,7.5){\line(0,1){10}}
\put(6.25,16.25){\makebox(0,0)[cl]{$\nameslang$}}

\put(3.75,-11.25){\makebox(0,0)[cr]{$\inputt$}}

\linethickness{0.25mm}
\put(0,-2.5){\line(1,0){10}}
\linethickness{0.25mm}
\put(5,-12.5){\line(0,1){10}}
\put(5,2.5){\makebox(0,0)[cc]{\machine}}

\end{picture}
\ ,\ \  \renewcommand{\machine}{$g$}
\ ,\ \  \renewcommand{\machine}{$h$}\ ,\ldots\  \right\}
\eear
\end{itemize}
together with the \emph{sequential composition}\/ \sindex{composition!sequential ($\circ$)} operation $\circ$ from Sec.~\ref{Sec:compos}, presented by linking the boxes vertically along the strings of matching types. If needed, see Appendix~\ref{appendix:cat-def} for more. 

\para{Monoidal categories.} \sindex{monoidal!category} A category is\sindex{category!monoidal} \emph{\textbf{monoidal}}\/ when it also comes with \sindex{composition!parallel ($\times$)} the \emph{parallel composition}\/ operation $\times$, presented in Sec.~\ref{Sec:compos} as the horizontal juxtaposition of strings and boxes next to each other. The monoidal structures are often written in the form $(\CCC,\times, I)$, echoing the notation for monoids. 

Since drawing the strings $A$, $B$, and $C$ in parallel does not show the difference between $(A\times B) \times C$ and $A\times (B \times C)$, and the string $A$ does not show how many invisible strings $I$ are running in parallel with it, the string diagrams usually impose the assumption that \emph{\textbf{the monoidal structure is \emph{strictly}\/ associative and unitary.}} This causes no loss of generality since every monoidal category is equivalent to a strict one. The types in monoidal categories presented by string diagrams thus satisfy the monoid equations $(A\times B) \times C= A\times (B \times C)$ and $A\times I=A=I\times A$ on the nose. In general, monoidal categories only satisfy these equations up to coherent isomorphisms. Even with string diagrams, the types $A\times B$ and $B\times A$ are not identical but only coheretnly isomorphic along the twistings like in Fig.~\ref{Fig:swap}. Since the twistings are available for all pairs of types, \textbf{the monoidal structure is in the present narratives always assumed to be \emph{symmetric}}.

\para{Data services and \strict\ \funnn s form a cartesian category.} The data services in a monoidal category $\CCC$ distinguish a family of \strict\ \funnn s as those that satisfy \eqref{eq:map}. Keeping all of the objects of $\CCC$ but restricting to the \strict\ \funnn s yields the category $\CCC^\bullet$, with
\bear
\lvert \CCC^\bullet \rvert & = & \lvert \CCC\rvert\\
\CCC^\bullet (X,Y) & = & \left\{\ \newcommand{\dttt}{\bullet}\newcommand{\machine}{$r$}
\newcommand{\inputt}{\scriptstyle X}
\newcommand{\nameslang}{\scriptstyle Y} 
\def\JPicScale{.5}
\ifx\JPicScale\undefined\def\JPicScale{1}\fi
\unitlength \JPicScale mm
\begin{picture}(10,17.5)(0,0)
\linethickness{0.25mm}
\put(0,7.5){\line(1,0){10}}
\linethickness{0.25mm}
\put(0,-2.5){\line(0,1){10}}
\linethickness{0.25mm}
\put(10,-2.5){\line(0,1){10}}
\linethickness{0.25mm}
\put(5,7.5){\line(0,1){10}}
\put(6.25,16.25){\makebox(0,0)[cl]{$\nameslang$}}

\put(3.75,-11.25){\makebox(0,0)[cr]{$\inputt$}}

\linethickness{0.25mm}
\put(0,-2.5){\line(1,0){10}}
\linethickness{0.25mm}
\put(5,-12.5){\line(0,1){10}}
\put(5,2.5){\makebox(0,0)[cc]{\machine}}

\put(5,7.5){\makebox(0,0)[cc]{$\dttt$}}

\end{picture}
\ ,\ \  \renewcommand{\machine}{$s$}
\ ,\ \  \renewcommand{\machine}{$t$}\ ,\ldots\  \right\}
\eear
The pairing and the projections derived from the data services in Sec.~\ref{Sec:map} assure that the products inherited from $\CCC$ become the cartesian products in $\CCC^\bullet$. The details are spelled out in the exercises below. 

\para{Cartesian categories}\sindex{category!cartesian}\sindex{projection} \sindex{pairing} are the categories where all \funnn s are {\strict}. The category $\tot \CCC$ defined above is cartesian by definition. It is the largest cartesian subcategory of the monoidal category $\CCC$. The bijective correspondence
\beq\label{eq:cartesian}
\prooftree
a=\pi_Ah \colon X\stricto A\qquad \qquad b=\pi_Bh\colon X\stricto B
\Justifies
h = <a,b>\ \colon X\stricto A\times B 
\endprooftree
\eeq 
induced by (\ref{eq:cartpair}--\ref{eq:cartproj}), shows the universal property of the cartesian products with respect to the \strict\ \funnn s, which implies that they are unique up to isomorphism \cite[III.4]{MacLaneS:CWM}. The correspondence in \eqref{eq:cartesian} ceases to be bijective when the 
\funnn s are not \strict. Nevertheless, the cartesian structure continues to play an important role in the monoidal computer, and at the entire gamut of monoidal frameworks  \cite{PavlovicD:CQStruct}.

\para{Notation.} When the category $\CCC$ is clear from the context, we write $f:A\to B$ instead of $f\in \CCC(A,B)$, and $s:A\stricto B$ instead of $s\in \CCC^{\bullet}(A,B)$.

\para{More about categories} is disseminated through the rest of the book. There are many ways to learn even more, and many popular textbooks and handbooks: \cite{AdamekJ:CT,AwodeyS:CT,BorceuxF:handbook,MacLaneS:CWM,RiehlE:CT,SimmonsH:CT}. Workouts~\ref{Sec:work-wire} on the following pages may convey the flavor.  Stories~\ref{Sec:story-wire} provide fragments of the conceptual background. An outline of the basic ideas about the morphisms between categories (called \emph{functors}), the morphisms between their morphisms (called \emph{natural transformations}), as well as  some concrete examples, are available in  Appendix~\ref{Appendix:cat}.

%
%
%
%
%

\section{Workout}\label{Sec:work-wire}


\begin{enumerate}[a.]
\item Given the type $\NNn$ of natural numbers (non-negative integers), with the data service as in Sec.~\ref{Sec:service} and the addition and the multiplication operations presented as boxes
\begin{center}
\newcommand{\gee}{$(+)$}
\newcommand{\kee}{$(\cdot)$}
\newcommand{\mee}{\scriptstyle \NNn}
\newcommand{\nee}{\scriptstyle \NNn} 
\newcommand{\outplus}{\scriptstyle \NNn}
\newcommand{\outtimes}{\scriptstyle \NNn} \def\JPicScale{.5}
\ifx\JPicScale\undefined\def\JPicScale{1}\fi
\unitlength \JPicScale mm
\begin{picture}(105,32.5)(0,0)
\linethickness{0.35mm}
\put(5,-2.5){\line(0,1){12.5}}
\linethickness{0.35mm}
\put(0,20){\line(1,0){30}}
\linethickness{0.35mm}
\put(0,10){\line(0,1){10}}
\linethickness{0.35mm}
\put(30,10){\line(0,1){10}}
\linethickness{0.35mm}
\put(15,20){\line(0,1){12.5}}
\linethickness{0.35mm}
\put(0,10){\line(1,0){30}}
\put(15,15){\makebox(0,0)[cc]{\gee}}

\linethickness{0.35mm}
\put(80,-2.5){\line(0,1){12.5}}
\linethickness{0.35mm}
\put(75,10){\line(0,1){10}}
\linethickness{0.35mm}
\put(105,10){\line(0,1){10}}
\linethickness{0.35mm}
\put(90,20){\line(0,1){12.5}}
\linethickness{0.35mm}
\put(75,10){\line(1,0){30}}
\linethickness{0.35mm}
\put(75,20){\line(1,0){30}}
\put(90,15){\makebox(0,0)[cc]{\kee}}

\linethickness{0.35mm}
\put(25,-2.5){\line(0,1){12.5}}
\linethickness{0.35mm}
\put(100,-2.5){\line(0,1){12.5}}
\put(3.75,0){\makebox(0,0)[cr]{$\mee$}}

\put(23.75,0){\makebox(0,0)[cr]{$\nee$}}

\put(78.75,0){\makebox(0,0)[cr]{$\mee$}}

\put(98.75,0){\makebox(0,0)[cr]{$\nee$}}

\put(17.5,30){\makebox(0,0)[cl]{$\outplus$}}

\put(92.5,30){\makebox(0,0)[cl]{$\outtimes$}}

\end{picture}

\vspace{.5\baselineskip}
\end{center}
and  
draw the string diagrams for the following \funnn s:
\begin{enumerate}[i)]
\item $\Big<(+), (\cdot)\Big> {\colon} \NNn \ttimes \NNn \to \NNn\ttimes \NNn \colon
<m,n>  \mapsto  <m+n, m\cdot n>$
\item $f {\colon} \NNn \ttimes \NNn \to \NNn \colon
<m,n>  \mapsto  m(m+n)$
\item $g {\colon} \NNn \ttimes \NNn \to \NNn \colon
<m,n>  \mapsto  2mn$
\item $h {\colon} \NNn \ttimes \NNn \to \NNn \colon
<m,n>  \mapsto  n^3$
\end{enumerate}

\item Translate the following string diagram into the algebraic notation. 
\beq\label{eq:string-function}
\begin{split}
\newcommand{\product}{A\times B}
\newcommand{\ductpro}{B\times A}
\newcommand{\Cee}{\scriptstyle C}
\newcommand{\Dee}{\scriptstyle D}
\renewcommand{\Eee}{\scriptstyle E}
\newcommand{\fun}{\scriptstyle f}
\newcommand{\gee}{\scriptstyle g}
\newcommand{\eich}{\scriptstyle h}
\newcommand{\zero}{\scriptstyle 0}
\newcommand{\Ah}{\scriptstyle A}
\newcommand{\Bee}{\scriptstyle B}
\newcommand{\eex}{\scriptstyle x}
\newcommand{\why}{\scriptstyle y}
\newcommand{\northeast}{}
\newcommand{\northwest}{}
\newcommand{\keh}{\mbox{\Large $k$}}
\newcommand{\eeh}{\mbox{\Large $e$}}
\def\JPicScale{.7}
\ifx\JPicScale\undefined\def\JPicScale{1}\fi
\unitlength \JPicScale mm
\begin{picture}(57.5,90)(0,0)
\linethickness{0.3mm}
\put(7.5,55){\line(1,0){10}}
\multiput(12.5,50)(0.12,0.12){42}{\line(1,0){0.12}}
\multiput(7.5,55)(0.12,-0.12){42}{\line(1,0){0.12}}
\linethickness{0.3mm}
\put(22.5,75){\line(0,1){6.25}}
\linethickness{0.3mm}
\put(7.5,75){\line(1,0){45}}
\put(7.5,70){\line(0,1){5}}
\put(52.5,70){\line(0,1){5}}
\put(7.5,70){\line(1,0){45}}
\linethickness{0.3mm}
\put(12.5,65){\line(0,1){5}}
\linethickness{0.3mm}
\put(43.75,65){\line(1,0){7.5}}
\put(43.75,60){\line(0,1){5}}
\put(51.25,60){\line(0,1){5}}
\put(43.75,60){\line(1,0){7.5}}
\linethickness{0.3mm}
\put(32.5,50){\line(0,1){20}}
\linethickness{0.3mm}
\put(37.5,35){\line(0,1){35}}
\linethickness{0.3mm}
\put(47.5,65){\line(0,1){5}}
\linethickness{0.3mm}
\put(47.5,45){\line(0,1){15}}
\linethickness{0.3mm}
\multiput(37.5,35)(0.12,0.12){83}{\line(1,0){0.12}}
\linethickness{0.3mm}
\multiput(17.5,35)(0.12,0.12){125}{\line(1,0){0.12}}
\linethickness{0.3mm}
\multiput(22.5,50)(0.12,-0.12){125}{\line(1,0){0.12}}
\linethickness{0.3mm}
\put(0,65){\line(1,0){25}}
\put(0,60){\line(0,1){5}}
\put(25,60){\line(0,1){5}}
\put(0,60){\line(1,0){25}}
\linethickness{0.3mm}
\multiput(2.5,50)(0.12,-0.12){125}{\line(1,0){0.12}}
\linethickness{0.3mm}
\put(2.5,50){\line(0,1){10}}
\linethickness{0.3mm}
\put(22.5,50){\line(0,1){10}}
\linethickness{0.3mm}
\put(37.5,75){\line(0,1){15}}
\linethickness{0.3mm}
\put(17.5,25){\line(0,1){10}}
\linethickness{0.3mm}
\multiput(17.5,5)(0.12,0.12){167}{\line(1,0){0.12}}
\linethickness{0.3mm}
\put(37.5,25){\line(0,1){10}}
\linethickness{0.3mm}
\multiput(17.5,25)(0.12,-0.12){167}{\line(1,0){0.12}}
\linethickness{0.3mm}
\put(37.5,0){\line(0,1){5}}
\linethickness{0.3mm}
\put(17.5,0){\line(0,1){5}}
\linethickness{0.3mm}
\put(12.5,55){\line(0,1){5}}
\put(47.5,62.5){\makebox(0,0)[cc]{$\gee$}}

\put(12.5,62.5){\makebox(0,0)[cc]{$\fun$}}

\put(30,72.5){\makebox(0,0)[cc]{$\eich$}}

\put(12.5,53.12){\makebox(0,0)[cc]{$\zero$}}

\put(16.88,5.62){\makebox(0,0)[cr]{$\Ah$}}

\put(38.12,5.62){\makebox(0,0)[cl]{$\Bee$}}

\put(16.88,24.38){\makebox(0,0)[cr]{$\Bee$}}

\put(38.75,25){\makebox(0,0)[cl]{$\Ah$}}

\put(17.5,-1.88){\makebox(0,0)[tc]{$\eex$}}

\put(37.5,-1.25){\makebox(0,0)[tc]{$\why$}}

\put(36.88,78.75){\makebox(0,0)[cr]{$\Eee$}}

\put(37.5,35){\makebox(0,0)[cc]{\DOTT}}

\put(17.5,35){\makebox(0,0)[cc]{\DOTT}}

\put(22.5,81.25){\makebox(0,0)[cc]{\DOTT}}

\linethickness{0.15mm}
\put(-2.5,30){\line(1,0){57.5}}
\linethickness{0.15mm}
\put(-2.5,67.5){\line(1,0){57.5}}
\linethickness{0.15mm}
\put(55,30){\line(0,1){37.5}}
\linethickness{0.15mm}
\put(-2.5,30){\line(0,1){37.5}}
\linethickness{0.15mm}
\put(-5,2.5){\line(0,1){82.5}}
\linethickness{0.15mm}
\put(57.5,2.5){\line(0,1){82.5}}
\linethickness{0.15mm}
\put(-5,2.5){\line(1,0){62.5}}
\linethickness{0.15mm}
\put(-5,85){\line(1,0){62.5}}
\put(2.5,35){\makebox(0,0)[cc]{$\eeh$}}

\put(0,7.5){\makebox(0,0)[cc]{$\keh$}}

\end{picture}

\end{split}
\eeq
Write the function expression in terms of the function names $f, g,h,\pi_E$, the constant 0, and the variables  $x,y$. 

\item\label{work:pairing} Prove that the projections and the pairing displayed in Fig.~\ref{Fig:projs} satisfy equations \eqref{eq:projections} for all \strict\ \funnn s $a,b,h$. More precisely show that 
\begin{enumerate}[i)]
\item the first equation in \eqref{eq:projections} holds if and only if $b$ is total, 
\item the second if $h$ is single-valued, 
\item the third if and only if $a$ is total.
Note that the two equations establish a bijection between the \strict\ \funnn s $X\stricto A\times B$ and the pairs of \strict \funnn s $X\stricto A$ and $X\stricto B$.
\end{enumerate}

\item Assuming that the unit type\sindex{type!unit} satisfies $I\times I = I$, as explained in Sec.~\ref{Sec:prodtyp}, show that the data service $I\tto{\scun} I \oot{\cmn} I\times I$ is the identity, i.e. $\cun\  = \id_I = \cmn$. 

\item Given the data services $I\tto{\scun} A \oot{\cmn} A\times A$ and $I\tto{\scun} B\oot{\cmn} B\times B$, define a data service on $A\times B$.

\item Prove that for any type $A$
\begin{enumerate}[i)]
\item the copying $\cmn:A\to A\times A$ is \strict 
\item the deletion $\cun:A\to I$ is the only \strict\ fun to $I$.
\end{enumerate}

\item Show that the composites $g\circ f\colon A\to C$ and $f\times t\colon A\times U \to B\times V$ of {\strict} \funnn s $f\colon A\stricto B$, $g\colon B\stricto C$ and $t\colon U \stricto V$ are also {\strict}.

\item\label{ex:cart-unique} Prove that the cartesian structure in a category is unique up to isomorphism. Towards this goal, suppose that
\begin{itemize}
\item $I$ and $\widetilde I$ are objects such that for every object $X$ there is a unique morphism $X\to I$ and a unique morphism $X\to \widetilde I$;
\item $A\times B$ and $A\widetilde\times B$ are objects such that each of them supports the surjective pairing in the sense of the bijective correspondence in \eqref{eq:cartesian}.
\end{itemize}
Prove that this implies that there are isomorphisms $I\cong \widetilde I$ and $A\times B\cong A \widetilde\times B$.

\item\label{ex:exponent} Prove that a type $B^A$ can be determined uniquely up to isomorphism by a family of bijections
\bear
 \CCC(X\times A, B) & \stackrel{\varphi_X}\cong & \CCC(X,B^A)
\eear
indexed by $X$ which is \emph{natural}\/ in the sense that for every $s\in \CCC(Y,X)$ the following diagram commutes:
\[\begin{tikzar}[row sep=3cm,column sep=3cm]
\CCC(X \times A, B) \arrow{d}[description]{\CCC(s\times A, B)}
\&
\CCC(X,B^A)  \arrow{d}[description]{\CCC(s, B^A)} 
\arrow[leftarrow,bend right=13]
{l}[description]{\varphi_X}
\\
 \CCC(Y\times A, B)
 \& \CCC(Y,B^A)  \arrow[leftarrow,bend right=13]
 {l}[description]{\varphi_Y} 
\end{tikzar}\]
In other words, every $g\in \CCC(X\times A,B)$ satisfies
\bear
\varphi_Y \left(Y\times A\tto{s\times A}X\times A\tto{g} B\right) &=& \left(Y\tto{s}X\tto{\varphi_X(g)} B^A\right)
\eear
\end{enumerate}

%
%
%
%
%

\section{Stories}\label{Sec:story-wire}

\subsection{Prehistory of types}
\label{Sec:story-type}
The concept of type is usually attributed to Bertrand Russell. He introduced a type hierarchy as means of preventing set-theoretic paradoxes in \cite{RussellB:types}. Lord Russell discovered the paradox that carries his name, here presented in Appendix~\ref{Appendix:numnot}, by applying Cantor's diagonal argument to the element relation used in Frege's arithmetic. Russell's type hierarchy, like von Neumann's later distinction of sets and classes, was ostensibly based on Cantor's distinction of ``consistent" and ``inconsistent" sets \cite{CantorG:beitraege,CantorG:dauben,Cantor-Dedekind:Briefwechsel}. Even Russell's use of the word ``type" was probably inspired by Cantor's \emph{``order types"}, which would nowadays be called the \sindex{type!as isomorphism class} \emph{isomorphism classes of ordered sets}.\sindex{type!order types} We will get to ``order'' part at the end of Ch.~\ref{Chap:PCC}. What did he mean by ``types''?

The central idea of Cantor's work, which made possible his breakthrough into the theory of infinities \cite{CantorG:dauben,HilbertD:unendliche}, was to consider sets modulo bijections, to compare them by specifying injections \cite{CantorG:beitraege}, and to compare their orders by specifying order-preserving injections \cite[III.9]{CantorG:collected}. Order types are the equivalence classes of orders modulo the order-preserving bijections. E.g., identifying the orders $\{a\lt b\lt c\}$, $\{x\lt y\lt z\}$, and all other 3-element linear orders, whatever their elements might be, was the order type of the ordinal 3, whereas identifying the sets $\{a,b,c\}$, $\{x,y,z\}$, and all other 3-element sets, made them into the cardinal 3. Cantor's view of cardinals as the \emph{``virtual patterns''}\/ of equivalent sets  \cite[p.~283]{CantorG:collected} is a precursor of the general concept of type. The notion that the equivalence class of all 3-element linear orders and the equivalence class of all 3-element sets are not sets but proper classes did not yet exist, since the notion of a proper class did not exist, but Cantor gleaned the diagonal argument over the element relation that became Russell's paradox, and isolated the ``inconsistent" sets. Some 35 years and many paradoxes later, Zermelo, as the editor of Cantor's collected works,  \cite{CantorG:collected} systematically erred on the side of declaring as Cantor's error everything that he did not understand, as illustrated, for instance, by his informal argument that Cantor's informal argument could not be salvaged by Cantor's well-ordering assumptions, but only by Zermelo's own Axiom of Choice, ``which Cantor uses unconsciously and instinctively everywhere, but does not formulate explicitly anywhere" \cite[p.~451]{CantorG:collected}\footnote{This English translation is quoted from \cite[p.~129]{vanHeijenoortJ}. For the broader context of the Cantor-Dedekind correspondence, see \cite{Cantor-Dedekind:Briefwechsel}, prepared by Emmy Noether. Both Cantor's method of well-ordering and the induced choice functions are used as  programming tools in Ch.~\ref{Chap:State}.} By the time of Cantor's death in 1918, his foundational quest for ``virtual patterns" modulo bijections got overshadowed by the research opportunities in the areas that it created, including transfinite arithmetic, point-set topology, measure theory. In the 1940s, the type formalism re-emerged in Church's work as the framework of function abstraction \cite{ChurchA:types}. We return to this in Sec.~\ref{Sec:CTT}. By the 1960s, type declarations became a standard preamble in the program templates of high-level programming languages. N.G. de Bruijn merged the theory and the practice of typing in his\sindex{Automath Project} Automath Project \cite{Automath}, which gave a lasting impetus to type-based programming 
\cite{BarendregtH:types,GirardJY:F,HudakP:history,Martin-LoefP:inttt,Martin-Loef:programming}.

\begin{figure}[ht]
\begin{minipage}[b]{0.5\linewidth}
\centering
\includegraphics[height=5cm
]{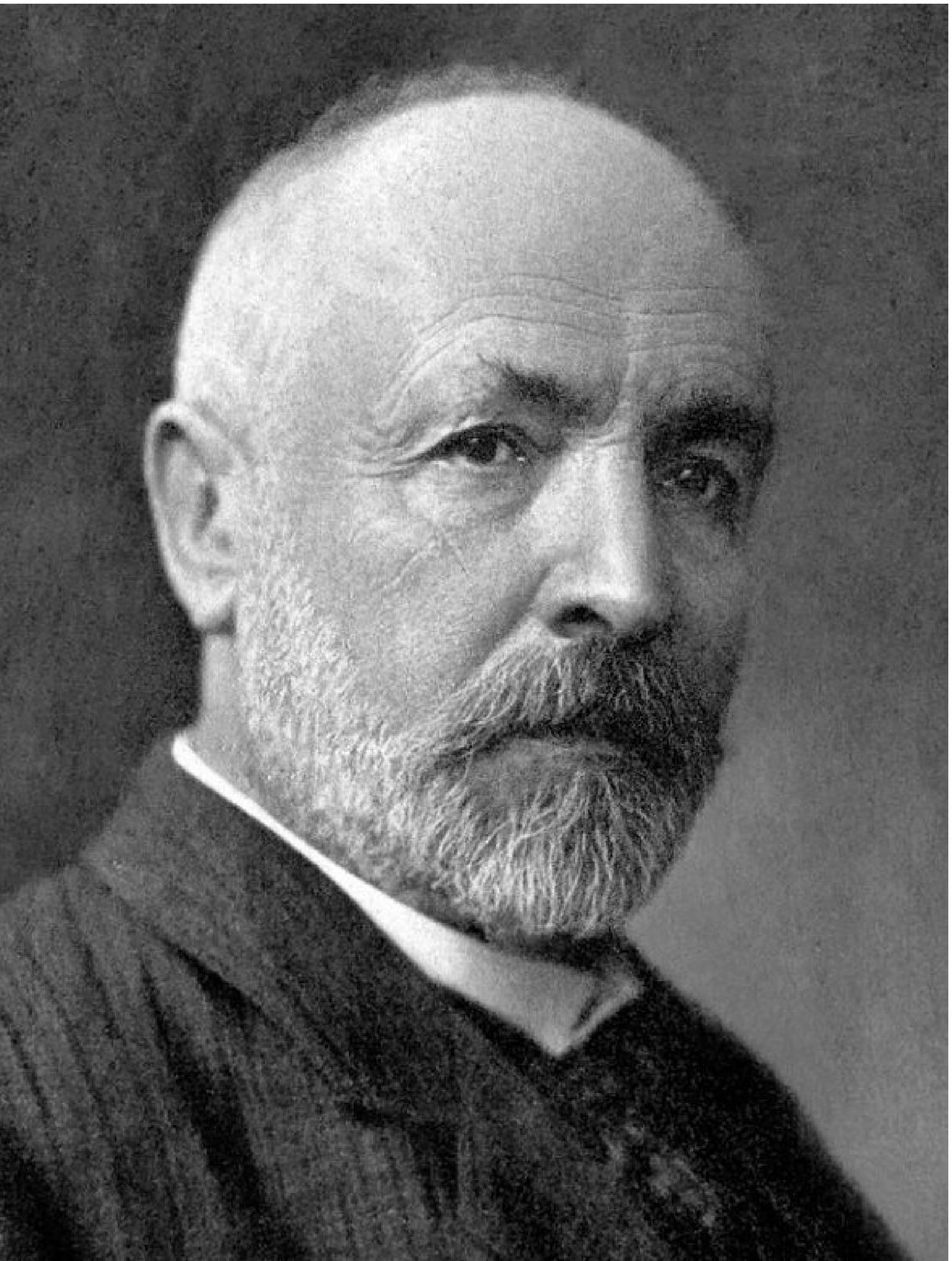}
\caption{Georg Cantor}
\label{Fig:cantor}
\end{minipage}
\begin{minipage}[b]{0.5\linewidth}
\centering
\includegraphics[height=5cm
]{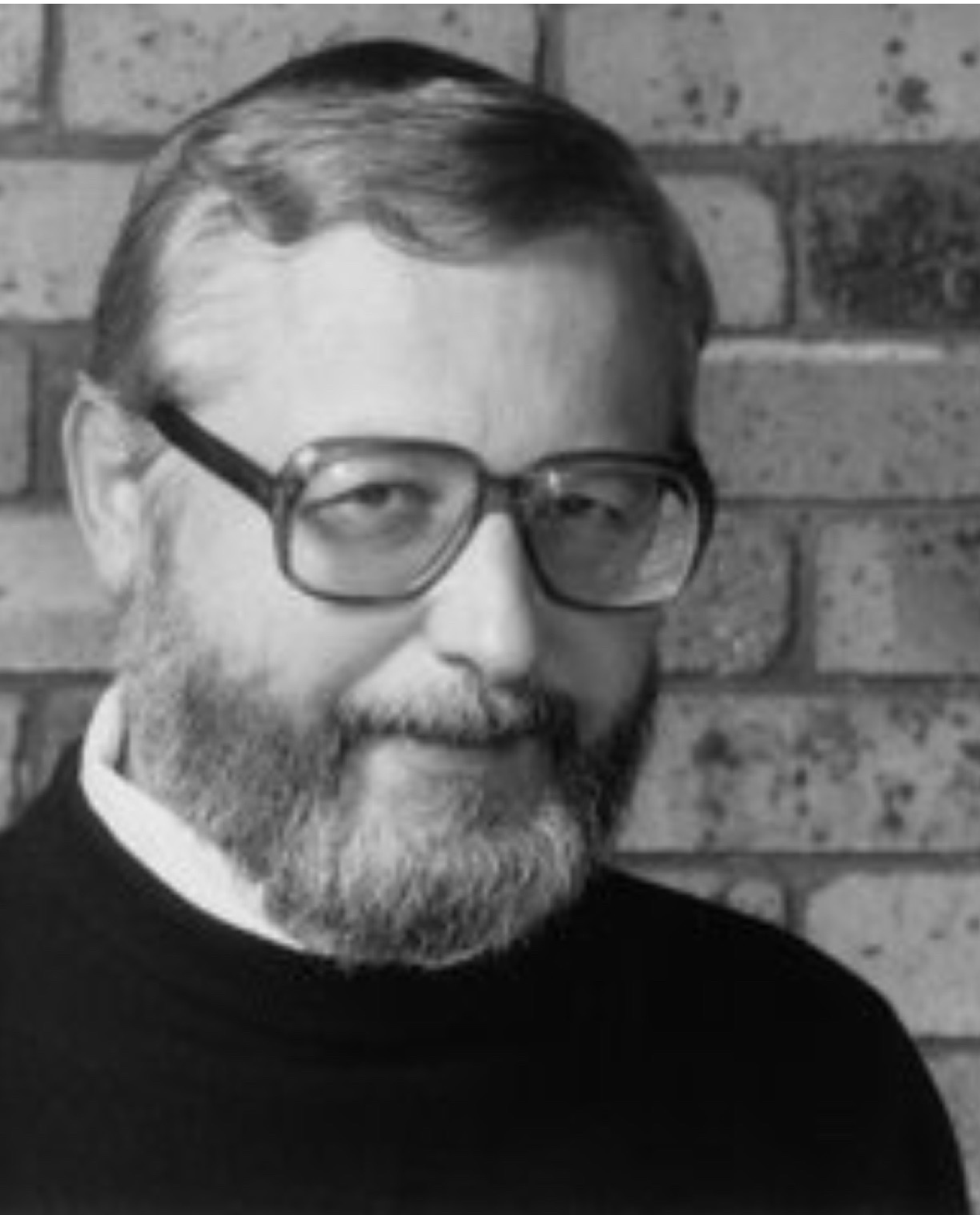}
\caption{F. William Lawvere}
\label{Fig:lawvere}
\end{minipage}
\end{figure}

\subsection{Categories as type universes}\label{Sec:cat-theory}
While Cantor's transfinite arithmetic was recognized as a new research area during his lifetime, and celebrated shortly after his death\footnote{``No-one shall expel us from Cantor's Paradise'', exclaimed Hilbert in his 1925 article \emph{On the Infinite} \cite{HilbertD:unendliche}}, his main enabling idea, to compare sets along injections and identify them along bijections, was used as a tool bur not understood as a concept --- until it re-emerged several decades later, in an unrelated area, for unrelated reasons. The theory of algebraic invariants of topological spaces had in the meantime grown into a sweeping research movement, adding layer upon layer of structure to its analyses. To dam the flood of structure, it was often necessary to hide the irrelevant details of its constructions into black boxes and display just the relevant operations on the  interfaces. The separation of the relevant structure from the irrelevant is usually enforced by the homomorphisms, which preserve one and filter out the other. Just like Cantor identified different sets along bijections, Samuel Eilenberg and Saunders MacLane identified different mathematical constructions of the same structure along the isomorphisms between the outcomes. They proposed that a family of isomorphisms between the pairs of outcomes of two constructions, indexed over the inputs fed into each construction, should be viewed as a \emph{natural equivalence}\/ of the two constructions, provided that the isomorphisms preserve not only the relevant structure, but also commute with the homomorphisms between the inputs. Many results, old and new, turned out to rely upon  such identifications. By dropping the invertibility requirement from their natural equivalences and retaining the structure-preservation requirements, Eilenberg and MacLane also defined a more general concept of a \emph{natural transformation}\/ between mathematical constructions, \sindex{natural transformation} \cite[Sec.~0.2]{Eilenberg-MacLane:natural,Lambek-Scott:book}, giving rise to a mathematical theory of mathematical constructions: the category theory.\sindex{category!theory of mathematical constructions} The mathematical constructions themselves were formalized as \emph{functors}\sindex{functor}, acting on categories as universes of structures. Some examples of categories can be found in Appendix~\ref{Appendix:cat}, some functors are worked out in \ref{Sec:uniform}, and many textbooks and handbooks provide thorough expositions  \cite{AwodeyS:CT,BorceuxF:handbook,MacLaneS:CWM}. The central feature of category theory is that the relevant structures are captured as what the homomorphisms preserve, whereas the objects are used as black boxes, to hide the irrelevant structures. Beyond mathematics, similar or related treatments, with black-boxes and interfaces as first-class citizens, also evolved in software engineering as procedural, component-oriented, modular, object-oriented programming, as well as in other areas of engineering, science, philosophy, and art. 
As a tool for comprehending logical abstractions, and for damming the floods of structure, the type abstraction appears in so many avatars that it is needed even to comprehend itself. 

\subsection{Logics of types}
By the 1960s, categories caught root in several burgeoning areas of mathematics \cite[to name a few]{Eilenberg-Steenrod,GrothendieckA:tohoku,KanD:adj}. Sets, on the other hand, were settled as the accepted foundation of nearly all areas. 
So when Bill Lawvere proposed that categories could be used as a foundation of mathematics, abstracting away the elements from the sets, neither his thesis advisor Samuel Eilenberg, nor the other father of category theory, Saunders MacLane, nor pretty much anyone else, could comprehend what could possibly be meant or done with Lawvere's \emph{``sets without elements''}. Lawvere then narrowed his focus and wrote a thesis reducing to functors and categories just the universal algebra, not all of mathematics \cite{LawvereFW:funsat}. The reduction turned out to be so simple and effective that his broader  foundational effort could not be denied a second chance. Within a couple of years, his analysis of ``The category of categories as a foundation for mathematics" \cite{LawvereFW:Foundation} appeared as the introductory article in a volume coedited by Eilenberg and MacLane. 

Truth be told, Lawvere was not proposing sets without elements. Categories themselves are specified in terms of elements: the elements of hom-sets, the elements of the object class, and so on. Types also usually have some elements, even if they may change in various ways. A datatype in a software application may contain one set of elements today, another one tomorrow, while remaining the same datatype. Types and categories are not characterized by their elements but by their structures. Two different types, or two different categories, may have the same elements, but different structures. The category of finite relations and the category of finite-dimensional $\ZZz_{2}$-vector spaces have the same set of objects, say $\NNn=\{0,1,2,\ldots\}$, and for every pair $m,n\in \NNn$ the same morphisms $u:m\to n$, presented, say, as  matrices of 0s and 1s. But the composition of matrices  $u = (u_{ji})_{n\times m}$ and $v = (v_{kj})_{p\times n}$ in the category of relations has the $ki$-th entry $\bigvee_{j\lt n} (v_{kj}\wedge u_{ji})$, whereas in the category of $\ZZz_{2}$-vector spaces it is $\sum_{j\lt n} (v_{kj}\cdot u_{ji})$. So the two categories are quite different\footnote{Seen as a logical operation, the $\ZZZ_2$-product is still the conjunction, but the $\ZZZ_2$-sum is the \emph{exclusive}\/ or.}. The set of injections from $\{a,b\}$ to $\{c\}$ and the set of surjections from $\{c\}$ to $\{a,b\}$ are identical as sets since they are both empty; but viewed as types, they become different, since checking whether a function is an injection from $\{a,b\}$ to $\{c\}$ is different from checking whether it is a surjection from $\{c\}$ to $\{a,b\}$. The propositions $(A\wedge \neg A)$ and $(A\Leftrightarrow \neg A)$ are both false, but the proofs that they are false are different, and they correspond to different types.

\subsubsection{Propositions-as-types}
\label{Sec:cat-logic}
The paradigm of propositions-\-as-\-types, straddling logic and type theory, goes back to Brouwer-Heyting-Kolmogorov interpretation of \emph{\textbf{proofs-as-con\-stru\-ctions}} in the 1920s \cite[Ch.~1, Sec.~5]{Troelstra-vanDalen:book}, and to Curry's observation that the implication $A\supset B$ can be construed as functions $A\to B$ in 1934 \cite{CurryH:functionality}. It is also related to Kleene's 1945 untyped realizability of all constructive logical connectives by functions \cite{KleeneS:realiz}, and more closely to Kreisel's 1957 typed realizability \cite[footnote]{KreiselG:constructivity}. It was formalized in Howard's 1969 manuscript, published in  \sindex{propositions-as-types} 1980, and adopted as a logical and programming principle under the name \emph{Curry-Howard isomorphism} \cite{Coquand-Huet:constructions,GirardJY:prot,PavlovicD:constructions,PavlovicD:mapsII,WadlerP:curry-howard}. But while the focus of efforts among the logicians was to spell out the correspondence of certain proof-theoretic and type-theoretic operations, Lawvere from a different direction arrived at the sweeping claim, and compelling evidence, presented in \cite{LawvereFW:dialectica}, that all fundamental constructions, logical and type-theoretic, were instances of a single universal categorical structure: the adjunction --- originating in homotopy theory \cite{KanD:adj}. Although the claim had little traction among logicians, with the benefit of hindsight it is clear that Lawvere's approach presented a radical departure from the entire foundational tradition arising from Russell-Whitehead's \emph{Principia} \cite{Russell-Whitehead:principia} and pursued in mathematical logic. \emph{While logicians reconstructed mathematical structures from logical foundations, Lawvere was reconstructing logical foundations from mathematical structures.}  

For concreteness and future reference, I illustrate the approach by aligning the rules defining basic logical operations with the correspondences (easily recognized as adjunctions) defining basic categorical constructs, leading up to the structure of \emph{cartesian-closed category}. The double lines denote two-way (introduction-elimination) rules and bijective correspondences.

\para{Entailment is composition.} The basic rules of logical entailment correspond to the basic categorical operations, sequential composition and the identities: 
\bea
\prooftree
\justifies A\stackrel{\phantom{\id}}\vdash A
\endprooftree
&\qquad\qquad \qquad&
\prooftree
\justifies A\tto{\id} A
\endprooftree\notag
\\[2ex]
\prooftree
A\vdash B\qquad B\vdash  C
\justifies 
A\vdash C
\endprooftree
&\qquad\qquad \qquad&
\prooftree
A\tto f B\qquad B\tto g C
\justifies A\tto{g\circ f} C
\endprooftree
\eea
The unitarity and the associativity of the preorder on the left are automatic, but the categorical signature requires explicit equations:
\[
\id_{B}\circ f = f = f\circ \id_{A}\qquad \qquad \qquad \qquad h\circ(g\circ f) = (h\circ g)\circ f
\]

\para{Conjunction is the cartesian product.} 
\bea
\prooftree
\justifies A\stackrel{\phantom{\scun}}\vdash \true 
\endprooftree
&\qquad\qquad \qquad&
\prooftree
\justifies A\strictto {\scun} I
\endprooftree \notag
\\[2ex]
\prooftree
X\vdash A\qquad X\vdash  B
\Justifies X\vdash A\wedge B
\endprooftree
&\qquad\qquad \qquad&
\prooftree
X\strictto a A\qquad X\strictto b B
\Justifies X\strictto{<a,b>} A\times B
\endprooftree \label{eq:conj}
\eea
with the equations:
\[
\left(\id_{A}\times \cun\right)\circ <a,b> = a\qquad \qquad  \left(\cun\times \id_{B}\right)\circ <a,b> = b\qquad \qquad \Big<\left(\id_{A}\times \cun\right)\, ,  \left(\cun\times \id_{B}\right) \Big> = \id_{A\times B}
\]

\para{Disjunction is the coproduct.} 
\bea
\prooftree
\justifies \flse \stackrel{\phantom{\sunt}}\vdash A  
\endprooftree
&\qquad\qquad \qquad&
\prooftree
\justifies O \strictto{\sunt} A
\endprooftree \notag
\\[2ex]
\prooftree
A\vdash Y\qquad B\vdash  Y
\Justifies A\vee B \vdash Y
\endprooftree
&\qquad\qquad \qquad&
\prooftree
A\strictto \alpha Y \qquad B\strictto \beta Y 
\Justifies A+B\strictto{[\alpha,\beta]} Y
\endprooftree
\eea
with the equations:
\[
[\alpha,\beta]\circ \left(\id_{A}+\unt\right) = \alpha\qquad \qquad  \left(\unt + \id_{B}\right)\circ [\alpha,\beta] = \beta \qquad \qquad \Big[\big(\id_{A}+\unt\big)\, ,  \big(\unt + \id_{B}\big) \Big] = \id_{A+ B}
\]

\para{Implication is the right adjoint.} 
\bea\label{eq:implic}
\prooftree
X\wedge A\vdash B
\Justifies X\vdash A\supset B
\endprooftree
&\qquad\qquad \qquad&
\prooftree
X\times A\strictto g B  
\Justifies 
X\strictto{\lambda A.\ g} B^{A}
\endprooftree
\eea
with a unique $\varepsilon:B^{A}\times A\stricto B$ satisfying  
\[\varepsilon\circ\left(\lambda A. g\times\id_{A}\right) = g\qquad \qquad \qquad \qquad \lambda A. \varepsilon = \id_{B^{A}}
\] 

\subsubsection{Cartesian closed categories}\label{Sec:CCC-1}\sindex{category!cartesian-closed} 
A category is cartesian if it has the final object and the cartesian products $A\times B$ for all types $A,B$, as in  (\ref{eq:cartpair}--\ref{eq:cartproj}) and \eqref{eq:conj}. It is closed with respect to the cartesian products if it has the exponents $B^{A}$, as in \eqref{eq:implic}. Since the function abstraction into an exponent corresponds to the $\lambda$-abstraction, the cartesian closed structure presents a categorical view of Church's simple theory of types \cite{ChurchA:types}, \emph{and}\/ of the constructivist propositional logic formalized by Heyting algebras \cite{HeytingA:algebra}. The categorical and the type-theoretic presentations differ only in syntactic details \cite[Part I]{Lambek-Scott:book} and the terminology is often mixed\footnote{In  \cite{LawvereFW:dialectica}, Lawvere calls ``types'' what most category-theorists now call ``objects''. We call them types again.}. The cartesian-closed categories are ubiquitous and central in many areas, from topos theory \cite{JohnstoneP:elephant} to denotational semantics \cite{GunterC:book}. In Ch.~\ref{Chap:State}, we explore a version suitable for programming: the \emph{program-closed}\/ categories. 
 
\subsection{Categorical diagrams: chasing arrows and weaving strings}
Diagrammatic reasoning was at the heart of ancient mathematics. Archimedes apparently died defending the circles he drew in the  sand on a beach, studying a mathematical problem. On the other hand, an entire line of religious and political movements found reasons to prohibit all visual depictions in general and ungodly diagrammatic witchcraft in particular. Science did not prohibit depictions, but printing made them costly, and mathematical narratives came to be dominated by formulas and the \emph{algebraic}\/ reasoning, even about geometry. In a similar way, programming came to be dominated by the \emph{command-line}\/ reasoning, fitting the 80-character command lines of punch cards and terminal displays. After the invention of the computer mouse, the cursor movements spanned the 2-dimensional space of the computer screen. It may not be Archimedes' beach, but it made drawing diagrams easier.

\subsubsection{Arrow diagrams}\label{Sec:arrow}
Computers also made printing diagrams easier, just in time, since the algebraic reasoning about geometric objects had by then developed its own geometric patterns. From the outset, the categorical abstractions were illustrated by \emph{arrow diagrams}. An example is in the upper part of Fig.~\ref{Fig:ana-string}. The \funnn\ $q$ can be thought of as a state machine over the state space $X$ with inputs of type $A$ and outputs of type $B$. It is in fact a pair of \funnn s $q=\left<\sta{q},\out{q}\right>$ where $\sta{q}:X\times A\rightarrow X$ maps an input $a:A$ at the state $x:X$ to the \emph{next state}\/ $\sta{q}(x,a):X$, whereas $\out{q}:X\times A\rightarrow B$ maps them to the \emph{output}\/  $\out{q}(x,a):B$. If the \funnn\ $r$ is a state machine over the state space $Y$, then a state assignment $s:X\to Y$ is a \emph{simulation}\/ of the machine $q$ by the machine $r$ at every state $x:X$ if for every input $a:A$ the next state of $s(x):Y$ by $r$ is the $s$-image of the next state of $x$ by $q$, and the outputs coincide.  The requirement is thus  
\[
\sta{r}\big(s(x),a\big)  =   s\big(\sta{q}(x,a)\big)\qquad\qquad\qquad\qquad
\out{r}\big(s(x),a\big) =   \out{q}(x,a)
\]
In the arrow diagram in Fig.~\ref{Fig:ana-string}, this requirement means that the path going left is equal to the path going right. The corresponding string diagram equation is displayed below the arrow diagram.
\begin{figure}[!ht]
\begin{center}
\begin{gather*}
\begin{tikzar}[row sep=3.75em,column sep=3em]
\& Y \times B\\   
Y \times A \ar{ur}[description]{r}   
\& \&  
X\times B  \ar{ul}[description]{{s} \times \id} 
 \\ 
 \& X \times A  \ar{ur}[description]{q} \ar{ul}[description]{{s} \times \id} \end{tikzar}
 \\[5ex]
\def\JPicScale{1}
\newcommand{\aah}{s}
\newcommand{\ssbbh}{\scriptstyle Y}
\renewcommand{\dh}{\scriptstyle B}
\newcommand{\ahh}{r}
\newcommand{\bhh}{q}
 \newcommand{\ah}{\scriptstyle A} 
  \newcommand{\xxh}{\scriptstyle X}
 \newcommand{\xxbbh}{\scriptstyle B}
 \newcommand{\Dott}{\mbox{\LARGE$\bullet$}}
 \input{PIC/uni-mach-pt-nodot.tex}
\end{gather*}
\caption{The equations\  $\sta{r}\big(s(x),a\big) = s\big(\sta{q}(x,a)\big)$\ \  and\ \   
$\out{r}\big(s(x),a\big) = \out{q}(x,a)$\  as diagrams}
\label{Fig:ana-string}
\end{center}
\end{figure}
The technique of  \emph{diagram chasing}\/ \sindex{diagram!chasing} allows encoding a lengthy equational derivation in a single diagram tiled by polygons spanned by equal arrow paths \cite[Ch.~XII, Sec.~3]{MacLaneS:homology}. Such tilings of equations into diagrams for chasing are useful both for building equational proofs and for communicating them. In his monograph presenting categories to working mathematicians \cite{MacLaneS:CWM}, MacLane consistently called homomorphisms between mathematical objects the \emph{arrows}, reminding the readers that the diagrams are there to be chased. As a procedure based on identifying equivalent paths between pairs of points, the method of diagram chasing has been construed as an echo of the mathematics of homotopies in the metamathematics of arrows. 

\para{We call arrows \funnn s and objects types.} \sindex{function} In the terminological soup of categories and computations in this book, MacLane's ``arrows'' and ``objects'' clash against too many different meanings assigned to these terms by programmers and computer scientists. Moreover, there are just a few \sindex{diagram!arrow} arrow diagrams in this book, and no real objects. While diagrammatic reasoning remains at the heart of programming and computation, string diagrams give us something that arrow diagrams don't.

\subsubsection{String diagrams} \sindex{string diagram}\sindex{diagram!string} 
While a single arrow diagram displays as many algebraic expressions as there are paths through it, and summarizes as many equations as there are path-reroutings (\emph{viz}\/ faces of the underlying graph), a string diagram displays a single algebraic expression and representing algebraic equations requires equations between string diagrams. The geometric overview of equational derivations, that was provided by the arrow diagrams, is lost. String diagrams, however, lay out the geometry of the algebraic expressions themselves. The upshot is that many abstract, often complex algebraic transformations are turned into simple geometric transformations. One example was sliding the boxes in Fig.~\ref{Fig:slidingfuns}. In some cases, like the middle-two-interchange law in Fig.~\ref{Fig:godement}, nontrivial algebraic equations are completely phased out by assigning  the same diagrammatic interpretation to different algebraic expressions. The programming adventures in the forthcoming chapters will illustrate the utility of such geometric shortcuts.

String diagrams were invented by the physicist Roger Penrose, as a tool for tracing indices through tensors, and independently by G\"unter Hotz, for tracking values through boolean circuits. Penrose's paper \cite{PenroseR:negdim} is usually cited as the earliest reference, but Hotz's thesis \cite{HotzG:string} appeared earlier, in an explicitly categorical framework. But Gavin Wraith attested in  private conversations that Penrose was seen using his string diagrams to avoid Einstein's index conventions  already in graduate school in the late 1950s. He had apparently refrained from presenting them in  publications mainly to avoid printing constraints. Be it as it may, the formal presentation of the  language of string diagrams was provided only in the work of Andr\'e Joyal and Ross Street that appeared in print in 1991 \cite{Joyal-Street:geometry}. That work was completed in the unpublished but widely available draft \cite{Joyal-Street:geometry-2}. Ross Street also provided a historic account in a public posting \cite{StreetR:kelly-penrose}. As computers further facilitated drawing and printing, string diagrams became a standard part of the categorical toolkit  \cite[and hundreds of other citations]{BaezJ:prehistory,CoeckeB-Kissinger:book,Hinze-Marsden}. 

\begin{figure}[!ht]
\begin{center}
\begin{gather*}
\begin{tikzar}[row sep=.35em,column sep=4em]
\& B \ar{ddr}{q} \\  
\& \hspace{0.1em} \ar[Leftarrow]{dd}[description]{\varphi} \\
A \ar{uur}{f} \ar{ddr}[swap]{d}   
\& 
\&  
D  
 \\ 
 \& \hspace{0.1em}
 \\
 \& C  \ar{uur}[swap]{t}  
 \end{tikzar}
 \\[2ex] 
\def\JPicScale{0.5}
\newcommand{\twocell}{\varphi}
\newcommand{\Aee}{A}
\newcommand{\Bee}{B}
\newcommand{\Cee}{C}
\newcommand{\Dee}{D}
 \newcommand{\dee}{\scriptstyle d} 
  \newcommand{\teeh}{\scriptstyle t}
 \newcommand{\feeh}{\scriptstyle f}
 \newcommand{\qeeh}{\scriptstyle q}
 \input{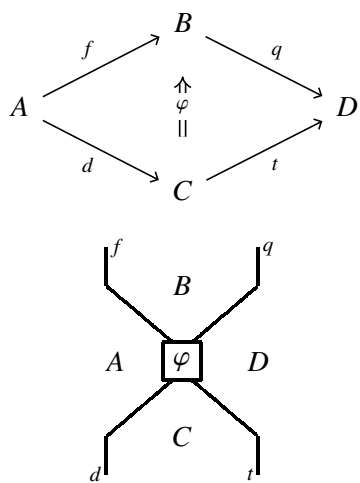}
\end{gather*}
\caption{A 2-arrow in a 2-categorical arrow diagram is a box in a 2-categorical string diagram}
\label{Fig:2-cat-string}
\end{center}
\end{figure}

\para{What is the relation between string diagrams and arrow diagrams?} A 2-category is a category where the hom-sets are categories. Besides the usual arrows between objects (which we call types in this book), there are thus also 2-arrows between arrows, like in the upper part of Fig.~\ref{Fig:2-cat-string}. An example of a 2-category is the category of categories, whose objects are categories, the arrows are functors, and the 2-arrows are the natural transformations. 
A face of an arrow diagram in a 2-category is not just a hollow polygon spanned by two arrow paths that may be equal or not. The polygon may contain a 2-arrow. The graph of the diagram thus consists of 0-cells as vertices, 1-cells as edges, and 2-cells that fill the faces of the graph. A string diagram in a 2-category is what graph theorists call a \emph{Poincar\'e dual}\/ of the arrow diagram:  the 2-cells of the arrow diagram are the 0-cells of the string diagram, whereas the 0-cells of the arrow diagram (the objects of the 2-category) become the 2-cells of the string diagram. The string diagram in the lower part of Fig.~\ref{Fig:2-cat-string} is obtained by drawing a string across each arrow of the arrow diagram, and a box in each 2-cell of the arrow diagram, to connect the strings coming into it. 

But there are no 2-categories in this book. Only monoidal and cartesian categories. Where do our string diagrams come from? A monoidal category can be viewed as a 2-category with a single 0-cell. The objects of the monoidal category are the 1-cells of the 2-category, whereas the arrows of the monoidal category become the 2-cells of the 2-category. Using the 2-categorical arrow diagrams and drawing the objects of the monoidal category as arrows and the arrows between them as more arrows is generally not very useful; but using their dual string diagrams and drawing the objects of the monoidal category as strings, the arrows as boxes, is generally very useful.

\def\thechapter{2}
\setchaptertoc
\chapter[Monoidal computers: computability as a structure]{Monoidal computer:\\ computability as a structure}
\label{Chap:Comput}
\newpage

\section{Computer as a universal machine}
\label{Sec:comput-stor}

\begin{figure}[!ht]
\begin{center}
\newcommand{\machine}{\includegraphics[height=1.2cm
]{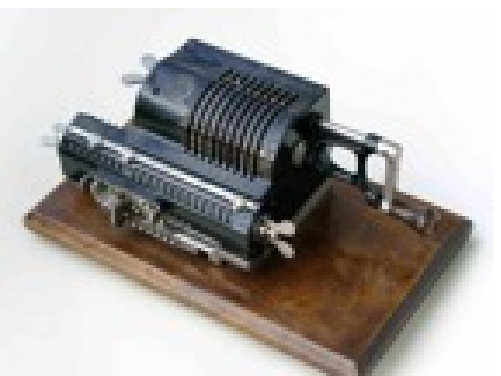}}
\newcommand{\nameslang}{\scriptstyle \NNn}
\newcommand{\inputt}{\scriptstyle \NNn}
\newcommand{\statt}{\scriptstyle \NNn}
\def\JPicScale{.5}
\input{PIC/mach-state.tex}\hspace{3em} 
\renewcommand{\machine}{\includegraphics[height=1.2cm]{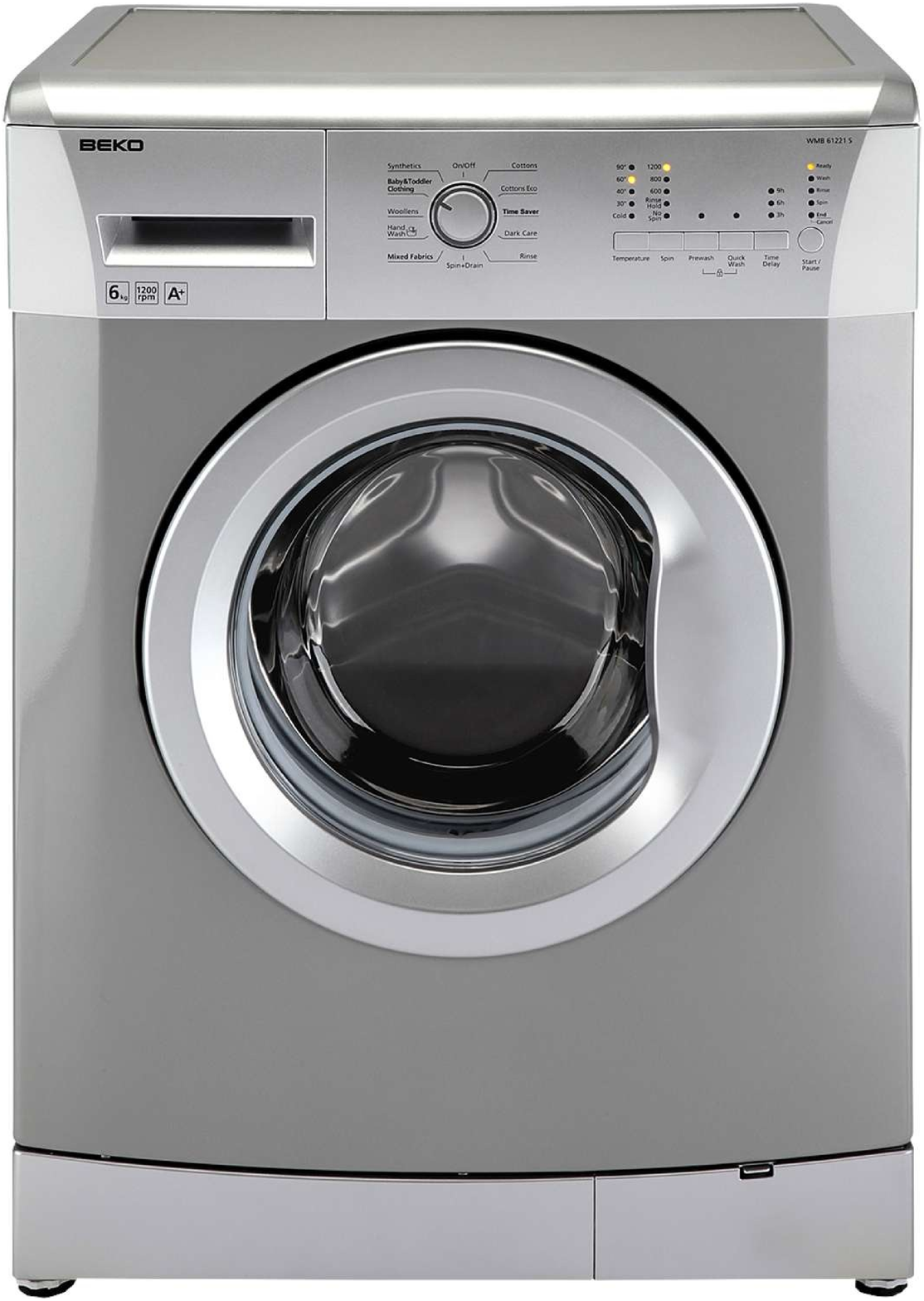}}
\renewcommand{\nameslang}{\scriptstyle Clean}
\renewcommand{\inputt}{\scriptstyle Dirty}
\renewcommand{\statt}{\scriptstyle Detergent}
\input{PIC/mach-state.tex}\hspace{3em} 
\renewcommand{\machine}{\includegraphics[height=1.2cm]{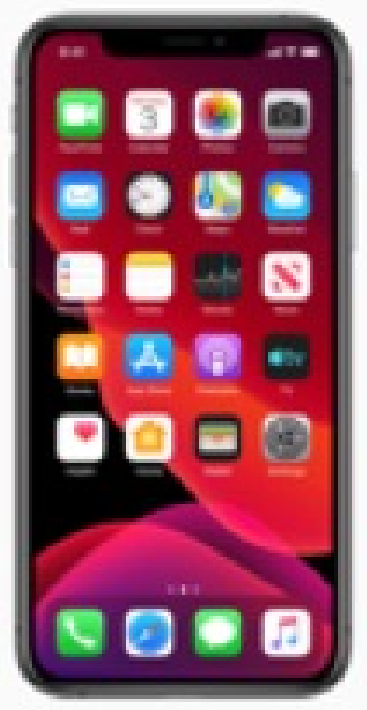}}
\renewcommand{\nameslang}{\scriptstyle Revenue}
\renewcommand{\inputt}{\scriptstyle Apps}
\renewcommand{\statt}{\scriptstyle Service}
\input{PIC/mach-state.tex}\hspace{3em}
\renewcommand{\machine}{\includegraphics[height=1.2cm]{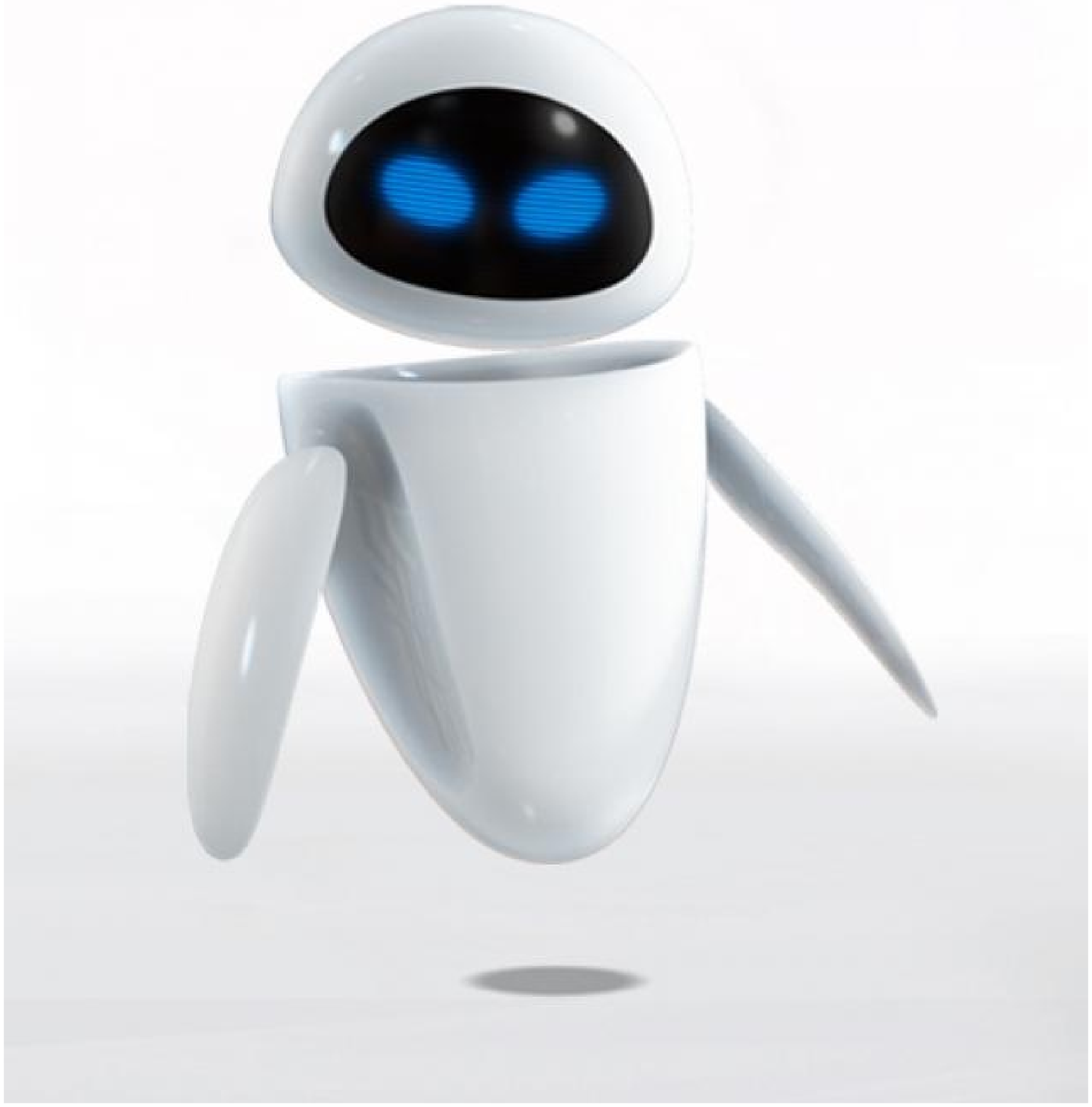}}
\renewcommand{\nameslang}{\scriptstyle Plants}
\renewcommand{\inputt}{\scriptstyle Tests}
\renewcommand{\statt}{\scriptstyle Missions}
\input{PIC/mach-state.tex}
\caption{Different machines for different purposes}
\label{Fig:machines}
\end{center}
\end{figure}

Machines! We live with machines. Fig.~\ref{Fig:machines} shows some friendly machines that implement some useful functions, displayed with their input type strings coming in at the bottom and the output type strings coming out at the top.  
A mechanical calculator inputs pairs of numbers and outputs the results of some arithmetic operations. A washing machine applies its \emph{Detergent}\/ contents to transform the \emph{Dirty laundry}\/ inputs into \emph{Clean laundry}\/ outputs. A smartphone inputs the \emph{Apps}\/ approved by its \emph{Service} and outputs the \emph{Revenue}\/ for the provider. And finally EVE, the Extraterrestrial Vegetation Evalluator from Pixar's movie WALL-E, applies the \emph{Tests}\/ as prescribed for each of its \emph{Missions}, and it delivers any detected \emph{Plants}\/ back to its spaceship.

One thing that all these machines have in common is that they are all different. Each function requires a different machine. Each machine is a special-purpose machine. The special purpose of a calculator is to calculate a special set of arithmetic operations. When we need to calculate different operations, we need to build a different calculator.

\begin{figure}[!ht]
\begin{center}
\renewcommand{\EQLS}{\LARGE =}
\newcommand{\Fee}{\ \ \ \begin{minipage}{1cm}\tt \tiny 
do \\while \\
a[s]=0;\\
c:=c+a[s]*b; \\
\end{minipage}}
\newcommand{\fee}{\includegraphics[height=1.1cm
]{PIC/calculator.eps}}
\newcommand{\Aee}{\scriptstyle \NNn}
\newcommand{\Bee}{\scriptstyle \NNn}
\newcommand{\Cee}{\scriptstyle \NNn}
\newcommand{\Code}{\scriptstyle \DP}
\newcommand{\Univ}{\includegraphics[height=1.1cm]{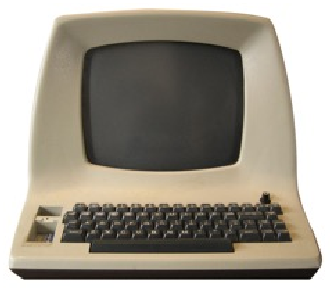}}
\newcommand{\Dott}{\mbox{\Huge$\bullet$}}
\def\JPicScale{.33}
\input{PIC/eval-pic.tex} 
\hspace{4em} 
\renewcommand{\Fee}{\includegraphics[height=.65cm]{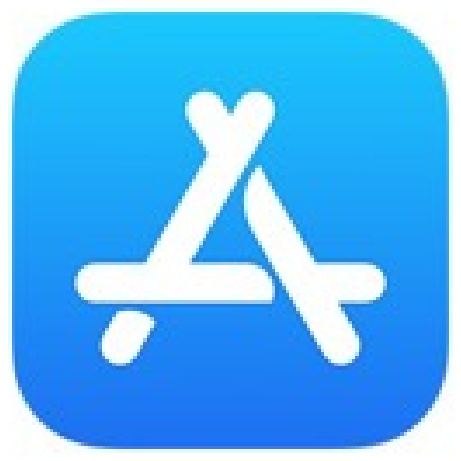}}
\renewcommand{\fee}{\includegraphics[height=1.2cm]{PIC/iphone.eps}}
\renewcommand{\Aee}{\scriptstyle Service}
\renewcommand{\Bee}{\scriptstyle Apps}
\renewcommand{\Cee}{\scriptstyle Revenue}
\renewcommand{\Univ}{\includegraphics[height=1.3cm]{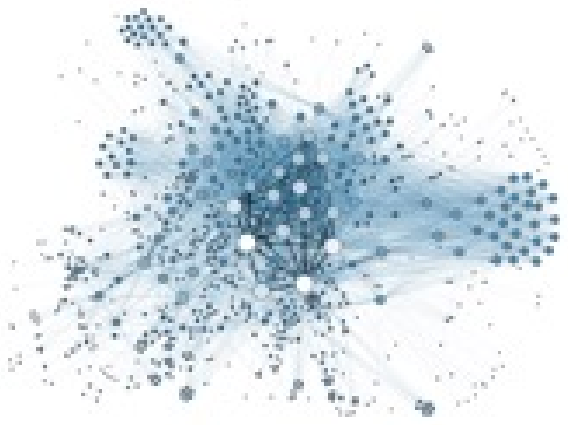}}
\input{PIC/eval-pic.tex} 
\\[8ex]
\renewcommand{\Fee}{\scriptsize \tt \ \ wash}
\renewcommand{\fee}{\includegraphics[height=1.2cm]{PIC/washer.eps}}
\renewcommand{\Aee}{\scriptstyle Detergent}
\renewcommand{\Bee}{\scriptstyle Dirty}
\renewcommand{\Cee}{\scriptstyle Clean}
\renewcommand{\Univ}{\includegraphics[height=1.1cm]{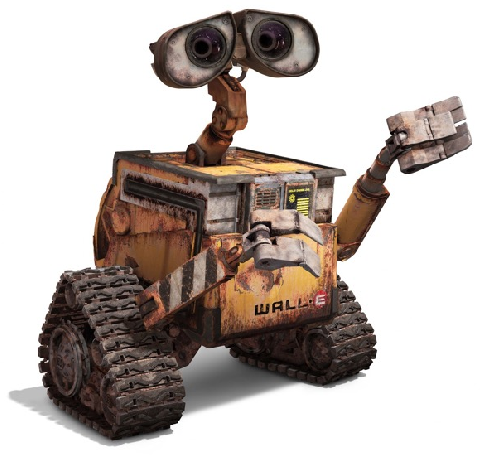}}
\input{PIC/eval-pic.tex} 
\hspace{4em}
\renewcommand{\Fee}{\ \ \ \begin{minipage}{.9cm}\footnotesize \tt love\\ EVE\end{minipage}}
\renewcommand{\fee}{\includegraphics[height=1.2cm]{PIC/Wall-eva.eps}}
\renewcommand{\Aee}{\scriptstyle Missions}
\renewcommand{\Bee}{\scriptstyle Tests}
\renewcommand{\Cee}{\scriptstyle Plants}
\renewcommand{\Univ}{\includegraphics[height=1.1cm]{PIC/Wall-e.eps}}
\input{PIC/eval-pic.tex} 
\caption{Universal machines can do anything that any other machines can do}
\label{Fig:machines-UTM}
\end{center}
\end{figure}

A computer is a universal machine: \sindex{universal!machine}it can be programmed to do anything that any other machine can do. This is what makes the process of computation into a conversation between functions and programs like in Fig.~\ref{Fig:head-tail}. Some examples, initiated by the functions in Fig.~\ref{Fig:machines}, are displayed in Fig.~\ref{Fig:machines-UTM}. A general-purpose computer can perform the builtin arithmetic operations of a special-purpose calculator. A smartphone is a slightly more complicated case. It is a computer in jail. On one hand it is definitely a universal computer. On the other hand, some smart people figured out a way how to prevent the smartphone from running programs unless you pay them. Some computational processes are thus driven by money. Others are driven by love. Robot WALL-E is tasked with collecting and compacting trash on the trashed planet Earth. To be able survive there, he is also equipped with a general-purpose computer. If needed, he can thus also be programmed to do other things. If needed, he can be programmed to wash laundry. Or he can program himself to perform EVE's function, i.e. to seek out and deliver a plant. We shall see in Ch.~\ref{Chap:Metaprog} how computers compute certain programs. If they then run the programs that they have computed, then they have programmed themselves. Computers and humans are similar.

Computers and people are similar in that each of them can do anything that any other can do. That is what makes them universal. How do they achieve that? How can it be that one can do anything that any other can do? For the humans, the universality has been a matter of some controversy, and it may seem complicated. For computers it is very simple. The functions that the computers can compute are precisely those that can be described by programs. \textbf{\emph{Computability is programmability.}}

\section{Universality of program evaluation}
\label{Sec:programmable}

A categorical model of a computer is a category where every \funnn\ is computable, in the sense that there is a program for it, and a program evaluator evaluates that program to the \funnn. Programs are the elements of the program type $\DP$, representing a programming language. The program evaluators are computable \funnn s of type $\DP\times A\to B$, one for each pair of types $A,B$, representing typed interpreters.

\subsection{Program evaluators}\label{Sec:uev}
We view a computer as a monoidal category with data services and a distinguished \emph{program type}\/ $\DP$ equipped with a \emph{program evaluator}\/ $\universal_A^B\colon\DP\times A\to B$ for every pair of types $A,B$.\sindex{evaluator!program} It evaluates parametrized programs $G:X\stricto \DP$ by composition: 
\bea
\uev G_A^B & = & \left(X\times A\begin{tikzar}[column sep=2.2em]
\hspace{-.3em}%
\arrow{r}[pos=0.4]{G\times A}[description,pos=.9]{\!\!\mbox{\large$\bullet$}\!\!}\&%
\hspace{-.3em}\end{tikzar}
\DP\times A\tto{ \universal\ } B\right)
\eea
where $X$ is an arbitrary type. The defining property of program evaluators is that they are\sindex{evaluator!universal} \emph{universal}\/ for parametrized families of \funnn s. 

\para{Parametrized \funnn s.} A family of \funnn s $g_x\colon A\to B$ parametrized by $x\colon X$ is presented as a \funnn\ $g\colon X\times A\to B$. The distinction between the parameters $x:X$ and the inputs $a:A$ is not intrinsic but imposed when the function is programmed. The program evaluator $\universal \colon\DP\times A\to B$ can be viewed as a $\DP$-parametrized family of \funnn s $\universal_{p} \colon A\to B$, with the programs $p\colon \DP$ as parameters. 

\subsubsection{Definition}
\para{Program evaluators are universal parametrized \funnn s.} \sindex{universality} The programming language $\DP$ is universal as a parameter space because it carries a universal parametrized \funnn\ for any pair of types $A,B$, their  program evaluator $\universal_A^{B}\colon\DP\times A\to B$. The program evaluator is a universal parametrized family of functions from $A$ to $B$ in the sense that for any type $X$ and any \funnn\ $g\colon X\times A\to B$ there is an $X$-parametrized program, \sindex{program!parametrized} presented as a \strict\ \funnn\ $G\colon X\stricto \DP$ that evaluates to $g$:
\beq\label{eq:uev}
\begin{split}
\newcommand{\Fee}{G}
\newcommand{\fee}{g}
\newcommand{\Aee}{\scriptstyle X}
\newcommand{\Bee}{\scriptstyle A}
\newcommand{\Cee}{\scriptstyle B}
\newcommand{\Code}{\scriptstyle \DP}
\newcommand{\Univ}{\mbox{\Large$\Universal$}}
\newcommand{\Dott}{\mbox{\LARGE$\bullet$}}
\def\JPicScale{.3}
\newcommand{\LHS}{g(x,a)}
\newcommand{\RHS}{\uev{G x} a}
\ifx\JPicScale\undefined\def\JPicScale{1}\fi
\psset{unit=\JPicScale mm}
\psset{linewidth=0.3,dotsep=1,hatchwidth=0.3,hatchsep=1.5,shadowsize=1,dimen=middle}
\psset{dotsize=0.7 2.5,dotscale=1 1,fillcolor=black}
\psset{arrowsize=1 2,arrowlength=1,arrowinset=0.25,tbarsize=0.7 5,bracketlength=0.15,rbracketlength=0.15}
\begin{pspicture}(0,0)(200,150)
\psline[linewidth=0.75](80,80)(80,40)
\psline[linewidth=0.75](80,40)(0,40)
\psline[linewidth=0.75](0,80)(80,80)
\psline[linewidth=0.75](60,40)(60,0)
\psline[linewidth=0.75](40,120)(40,80)
\psline[linewidth=0.75](0,80)(0,40)
\psline[linewidth=0.75](20,40)(20,0)
\psline[linewidth=0.75](200,100)(200,60)
\psline[linewidth=0.75](160,20)(120,20)
\psline[linewidth=0.75](120,100)(200,100)
\psline[linewidth=0.75](180,60)(180,0)
\psline[linewidth=0.75](160,120)(160,100)
\psline[linewidth=0.75](120,60)(120,20)
\psline[linewidth=0.75](140,20)(140,0)
\psline[linewidth=0.75](200,60)(160,60)
\psline[linewidth=0.75](120,100)(160,60)
\psline[linewidth=0.75](120,60)(160,20)
\psline[linewidth=1.4](140,40)(140,80)
\rput(170,80){$\Univ$}
\rput(100,60){\EQLS}
\rput[r](136.25,65){$\Code$}
\rput[l](165,117.5){$\Cee$}
\rput[l](45,117.5){$\Cee$}
\rput[r](15,2.5){$\Aee$}
\rput[r](55,2.5){$\Bee$}
\rput(40,60){$\fee$}
\rput(130,31.25){$\Fee$}
\rput[r](135,2.5){$\Aee$}
\rput[r](175,2.5){$\Bee$}
\rput(100,150){\EQLS}
\rput(40,150){$\LHS$}
\rput(160,150){$\RHS$}
\rput(140,40){$\Dott$}
\end{pspicture}
 
\end{split}
\eeq
Different but equivalent forms of universality are studied in Ch.~\ref{Chap:State}.

\para{Examples.} An interpreter of a programming language like Python or Javascript provides a typical family of program evaluators. The type $\DP$ is then the set of program expressions. The typed instances of the program evaluators are obtained by restricting the interpreter to the programs with the required input and output types. Alternatively, the typing can be imposed on computations, by running interpreted type-checkers. Compilers can also be used in a similar way instead of the interpreters but the introduced program optimizations may require special attention, as we will see in Ch.~\ref{Chap:Metaprog}. Theoretical models of computation, such as Kleene's program indices and Turing's machines, can also be used, as they include theoretical programming languages and program evaluators. Since they are usually untyped, the typed-checking must be imposed on computations. The nonstandard models of computation, such as quantum computers, biological computers, etc., also contain program evaluators for their nonstandard languages. Even executable software specification languages \cite{PavlovicD:ASE01} and abstract interpretation tools \cite{PavlovicD:POPL2020} contain program evaluators. 

\para{Explanations and notations.} The universality of program evaluations captures the idea, displayed in  Fig.~\ref{Fig:head-tail}, that running programs is a natural surjection. This idea is formalized in Sec.~\ref{Sec:moncom}, where the \runn-instruction is presented as an $X$-natural surjection $\runn^{AB}_{X}\colon \tot \CCC(X, \DP) \epi \CCC(X\times A,  B)$. For simplicity, the program expressions $\runn^{AB}_\DP(G)(x)(a)$ are written $\uev{Gx}_A^B a$. Furthermore, whenever the input type $A$ and the output type $B$ are irrelevant or clear from the context, we elide them and write $\uev{Gx}a$. In general, we write \begin{itemize}
\item $\uev{p} \colon A\to B$ instead of $\universal_{p}\colon A\to B$, 
\item $\uev{Gx} \colon A\to B$ instead of $\universal_{Gx}\colon A\to B$ for $G\colon X\stricto \DP$ and $x\colon X$, and
\item $\uev{Gx} (a)$ or $\uev{Gx} a$ instead of $\universal_{Gx}(a)$.
\end{itemize}
The background on the curly bracket notation is in Sec.~\ref{Sec:curly}. 

\subsubsection{What is the difference between  parameters and inputs?} 
\para{Convention.} For a function $g\colon X\times A\to B$ on its own, there is no difference: the parameters $x:X$ and the inputs $a:A$ are just data that it consumes. The fact that we write the parameter as the first argument, and consider the parametrized family $g(x,-) = g_x\colon A\to B$ is a matter of convention. If $g\colon X\times A\to B$ needs to be viewed as $A$-parametrized, it can be modified to $\widetilde g = \left(A\times X\tto \varsigma X\times A\tto g B\right)$ where $\varsigma$ is the data service from Fig.~\ref{Fig:swap}.

\para{Intuition.} The difference between the parameters  and the inputs arises when the function $g$ is programmed, and the programmer decides that the program will be parametrized by $x\colon X$ and that the computations will take the inputs $a\colon A$. The idea is that the program parameters $x\colon X$ are known in advance of computing, whereas the input values $a\colon A$ are only needed at the time of computing. The parameter $x\colon X$ determines which particular \funnn\ $g_x\colon A\to B$ will be computed and which particular instance $Gx\colon \DP$ of the parametrized program $G\colon X\stricto \DP$ will be used. This explains why $G$ must be a \strict\ \funnn: each parameter value $x$ should determine a unique program $Gx$. 

Intuitively, the parameters convey to the program the external \emph{states of the world}\/ that may impact computations in advance of processing the input data internally. E.g., a payroll system may need to perform different computations in different pay periods, and is parametrized by them. A mobile application performs different computations on different devices, which need to be recognized in advance. Most operating systems are parametrized by the different hardware platforms on which they run. \sindex{state!of the world} While the program variables are written and overwritten  during the program evaluations internally, the program parameters are set externally.

\begin{figure}[!ht]
\begin{center}
\newcommand{\Fee}{G}
\newcommand{\fee}{g}
\newcommand{\heee}{h}
\newcommand{\tuu}{s}
\newcommand{\suu}{t}
\newcommand{\Aee}{\scriptstyle X}
\newcommand{\Aeee}{\scriptscriptstyle X}
\newcommand{\Bee}{\scriptstyle A}
\newcommand{\Beee}{\scriptscriptstyle A}
\newcommand{\Cee}{\scriptstyle B}
\newcommand{\Zee}{\scriptstyle Y}
\newcommand{\Hee}{H}
\newcommand{\Kee}{\scriptstyle C}
\newcommand{\Code}{\scriptstyle \DP}
\newcommand{\Univ}{\mbox{\Large$\Universal$}}
\newcommand{\Dott}{\mbox{\Large$\bullet$}}
\def\JPicScale{.27}
\input{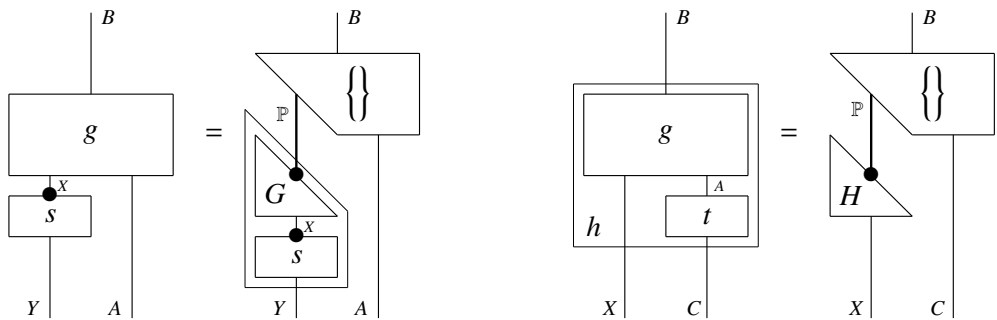} 
\caption{If $g = \uev G$, then $g\circ(s\times \id) = \uev{Gs}$ uses $G$ but $g\circ(\id \times t)=\uev H$ does not. 
}
\label{Fig:eval-nat-nonat}
\end{center}
\end{figure}

\para{Naturality.} Suppose that the parametrized family $g\colon X\times A\to B$ is implemented by a parametrized program $G\colon X\stricto \DP$, in the sense that $g_x=\uev{Gx}\colon A\to B$. The formal  difference in between the parameters $x$ and the inputs $a$ is displayed in Fig.~\ref{Fig:eval-nat-nonat}.
\begin{itemize}
\item \textbf{Reparametring} $x$ along $s\colon Y\stricto X$ leads to the family $g_{s(y)}\colon A\to B$ parametrized by $y\colon Y$ and implemented by the reparametrization $Gs = \left(Y\strictto s X\strictto G \DP\right)$ of the program $G$ for $g$.
\item \textbf{Substituting} for $a$ along $t\colon C\to A$ leads to the family $h_{x}= \left(C\tto t A\tto{g_{x}} B\right)$, still parametrized over $X$, but requiring a program $H\colon X\stricto \DP$ different from $G$. 
\end{itemize}
Viewed categorically, this distinction means that the instruction $\runn^{AB}_{X}\colon \tot \CCC(X, \DP) \epi \CCC(X\times A,  B)$ implemented by the program evaluators is natural along \strict\ \funnn s into the parameter type $X$ but not into the input type $A$ or out of the output type $B$. We return to the naturality of the \runn-instruction in Sec.~\ref{Sec:surj}.

\subsubsection{Constant programs and slicing two ways}\sindex{program!constant}
A function $f:A\to B$ can be viewed with trivial parameters or with trivial inputs:
\begin{enumerate}[a)]
\item $f:I\times A\rightarrow B$, i.e. as a family $\Big\{f_{()}:A\to B\ |\ ():I\Big\}$ parametrized over $I$, or
\item $f: A\times I \rightarrow B$, i.e. as a family $\Big\{f_a:I\to B\ |\ a:A\Big\}$ with inputs from $I$. 
\end{enumerate}
Under interpretation (a), condition \eqref{eq:uev} induces \sindex{program!constant} program $F\colon I\stricto \DP$ such that $\uev F = f \colon A\to B$. This program is a \emph{constant}\/ $F=F_{\nill}\colon \DP$, since $\nill\colon I$ is its only state of the world. Under interpretation (b), condition \eqref{eq:uev} induces the parametrized program $\Phi
\colon A\stricto \DP$, each taking the trivial input $\nill :I$ and producing the corresponding $A$-indexed family of constants $\uev{\Phi a} :B$ , as displayed in Fig.~\ref{Fig:abstractions}.  
\begin{figure}[!ht]
\begin{center}
\newcommand{\fleft}{f(a)}
\newcommand{\fmid}{\uev{\Phi_a}()}
\newcommand{\fright}{\uev{F}(a)}
\newcommand{\Fee}{F}
\newcommand{\Fei}{\Phi}
\newcommand{\fee}{f}
\newcommand{\Bee}{\scriptstyle A}
\newcommand{\Cee}{\scriptstyle B}
\newcommand{\Code}{\scriptstyle \DP}
\newcommand{\Univ}{\mbox{\large$\Universal$}}
\newcommand{\Dott}{\mbox{\LARGE$\bullet$}}
\def\JPicScale{.3}
\input{PIC/evalf.tex} 
\caption{$F$ encodes the \funnn\ $f=\uev {F}$. Family $\Phi_a$ encodes the family $f_a = \uev{\Phi_a}$.}
\label{Fig:abstractions}
\end{center}
\end{figure}


\subsection{Partial evaluators}
\label{Sec:pev}
Setting $X$ in \eqref{eq:uev} to be a product $\DP\times Y$ and $g$ to be the program evaluator $\universal : \DP \times Y \times A \to B$ gives a \emph{partial evaluator}\sindex{evaluator!partial}  $\prtial:\DP\times A\stricto \DP$  for any $Y, A, B$, as displayed in Fig.~\ref{Fig:pev}. A partial evaluator is thus a  $(\DP\times Y)$-indexed program satisfying ${\uev{\pev \Gamma y} a = \uev{\Gamma}(y,a)}$, 
\begin{figure}[!ht]
\begin{center}
\newcommand{\Fee}{\prtial}
\newcommand{\fee}{\mbox{\Large$\Universal$}}
\newcommand{\Aee}{\scriptstyle Y}
\newcommand{\Bee}{\scriptstyle A}
\newcommand{\Cee}{\scriptstyle B}
\newcommand{\Code}{\scriptstyle \DP}
\newcommand{\Univ}{\mbox{\Large$\Universal$}}
\newcommand{\Dott}{\mbox{\LARGE$\bullet$}}
\def\JPicScale{.32}
\newcommand{\LHS}{\uev{\Gamma}(y,a)}
\newcommand{\RHS}{\uev{\pev \Gamma y} a}
\input{PIC/evalp.tex} 
\caption{Partial evaluators $\prtial$ are indexed programs}
\label{Fig:pev}
\end{center}
\end{figure}

\begin{figure}[!ht]
\begin{center}
\newcommand{\Fee}{G}
\newcommand{\Gee}{\Gamma}
\newcommand{\fee}{g}
\newcommand{\Prt}{\prtial}
\newcommand{\Aee}{\scriptstyle X}
\newcommand{\Bee}{\scriptstyle A}
\newcommand{\Cee}{\scriptstyle B}
\newcommand{\Code}{\scriptstyle \DP}
\newcommand{\Univ}{\mbox{\Large$\Universal$}}
\newcommand{\Dott}{\mbox{\LARGE$\bullet$}}
\def\JPicScale{.33}
\input{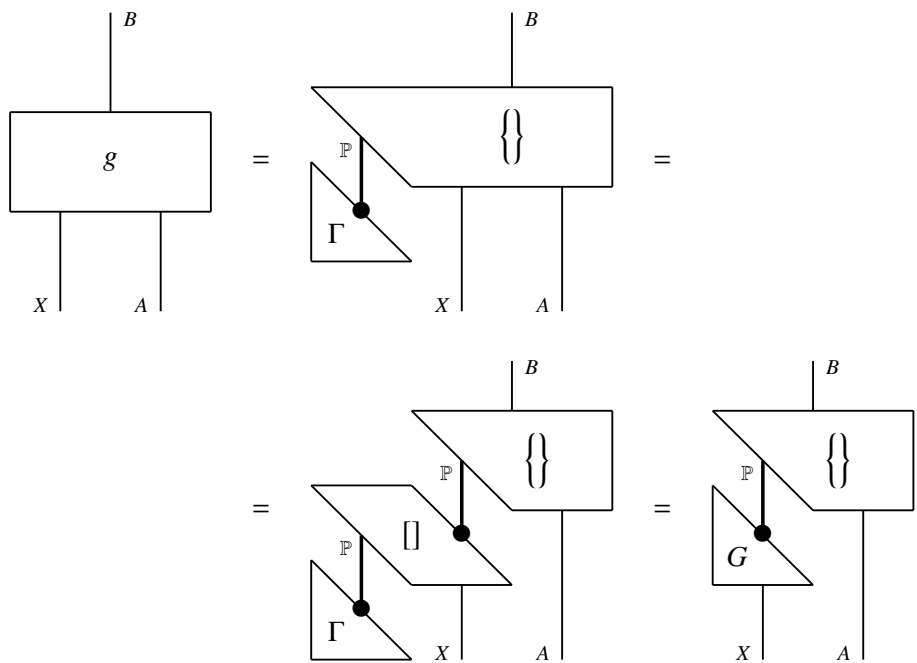} 
\caption{A parametrized program $G$ constructed by partially evaluating a constant program $\Gamma$}
\label{Fig:slicing}
\end{center}
\end{figure}

\para{Parametrized evaluations decompose to constant and partial.}
Any $X$-indexed program for a computable \funnn\ $g:X\times A \to B$ in \eqref{eq:uev} can be construed as a partial evaluation of a constant program $\Gamma$ on data $x:X$ which may happen to be available before the data of type $A$. When that happens, $x$ can be treated as a state of the world for the computation $g_x = \uev{\Gamma_x}:A\to B$.  This view of indexed computations is displayed in Fig.~\ref{Fig:slicing}. 
In summary, \eqref{eq:uev-formal} can be derived from the assumption that all types $X, A, B$  come with the universal and the partial evaluators generically denoted $\universal \in \CCC(\DP\times A, B)$, $\prtial \in \CCC^\bullet(\DP\times X,\DP)$, $\universal \in \CCC(\DP\times X\times A, B)$, which satisfy 
\bea
\forall f\in\CCC(A, B)\ \ \exists F\in \CCC^\bullet(I, \DP).\ \ \ \ \ \ \ \ \ \Big\{F\Big\} a  & = &  f(a)
\label{eq:uev-pev-formal-one}\\
\forall F\in \CCC^\bullet(I, \DP).\ \ \ \Big\{\pev{F} x\Big\} a & = & \Big\{F\Big\}(x,a)
\label{eq:uev-pev-formal-two}
\eea

\subsection{Composing programs}\label{Sec:compos-prog}
\begin{figure}[!ht]
\newcommand{\hee}{$\scriptstyle Sq$}
\newcommand{\seqq}{\scriptstyle (;)}
\newcommand{\parr}{\scriptstyle (\,\parallel\,)}
\newcommand{\Progtype}{\scriptstyle \DP}
\newcommand{\Progell}{\scriptstyle A}
\newcommand{\Progp}{\scriptstyle U}
\newcommand{\Progm}{\scriptstyle B}
\newcommand{\Progq}{\scriptstyle V}
\newcommand{\Progn}{\scriptstyle C}
\newcommand{\universalmkn}{\large \{\}}
\bigskip
\begin{center}
\def\JPicScale{.5}
\input{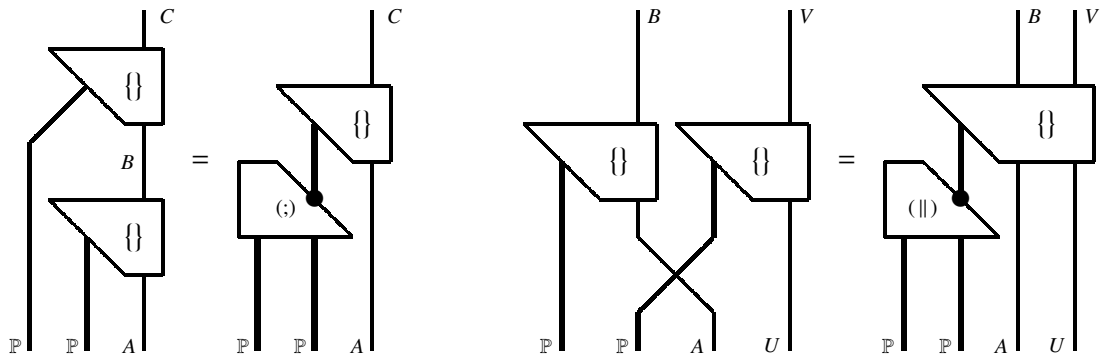}
\caption{Program composition programs $(\, ;)$ and $(\, \parallel\, )$}
\label{Fig:seq-comp}
\end{center}
\end{figure}

We saw in Sec.~\ref{Sec:compos} how \funnn s are composed sequentially and in parallel:
\[\prooftree
A\tto f B\qquad B\tto g C
\justifies A\tto{\color{red}g\circ f} C
\endprooftree
\qquad\qquad \qquad
\prooftree
A\tto f B\qquad U\tto t V
\justifies
A\ttimes U\tto{\color{blue}f\ttimes t} B\ttimes V
\endprooftree\]
When the functions are computable, we can compute programs for the composite \funnn s from any given programs for their components. More precisely, we can determine $\DP\times\DP$-indexed programs in the form
\beq\label{eq:compositions}
 (\, ;)_{ABC} : \DP \times \DP \stricto \DP\qquad \qquad \qquad \qquad (\, \parallel\,)_{AUBV}: \DP \times \DP \stricto \DP
 \eeq
which for indices $F, G, T \colon \DP$ with $\uev F = f$, $\uev G = g$, and $\uev T = t$ instantiate to the programs $(F;G)$ and $(F\parallel T)$ evaluating to the composite \funnn s  
\[\left(A\tto{\uev{F\, ;G}} C \right) = \left(A\tto{\uev F} B\tto{\uev G} C\right)\qquad \qquad 
\left(A\times U\tto{\uev{F\parallel H}} B\ttimes V\right) = \left(\begin{tikzar}[row sep = -1.25ex, column sep = .5ex]
A \ar{rr}{\uev F} \&\& B\\
\&\times \\
U \ar{rr}[swap]{\uev T} \&\& V
\end{tikzar}
\right)
\]
Fig.~\ref{Fig:seq-comp} shows how \eqref{eq:uev} induces programs for \eqref{eq:compositions}.
%

\section{Computable types are retracts of programming languages}
\label{Sec:dataprog}

A computer as a type universe containing a type $\DP$ of programs together with all datatypes is an embodiment of von Neumann's famous slogan that \emph{\textbf{programs are data}}. The universality of the programming language as a carrier of the evaluators of all types implies the converse inclusion, that \emph{\textbf{data are programs}}. More precisely, data can be encoded as programs, and their types can be checked by program evaluations.

\subsection{Data are programs}\label{Sec:retracts}
Instantiating Fig.~\ref{Fig:abstractions} to the identity function $\id_A$ for any type $A$ gives
\beq\label{eq:retr}
\begin{split}
\newcommand{\iee}{\enc}
\newcommand{\qee}{\big\{\big\}}
\newcommand{\Aee}{\scriptstyle \DP}
\newcommand{\Ree}{\scriptstyle A}
\def\JPicScale{.75}
\ifx\JPicScale\undefined\def\JPicScale{1}\fi
\psset{unit=\JPicScale mm}
\psset{linewidth=0.3,dotsep=1,hatchwidth=0.3,hatchsep=1.5,shadowsize=1,dimen=middle}
\psset{dotsize=0.7 2.5,dotscale=1 1,fillcolor=black}
\psset{arrowsize=1 2,arrowlength=1,arrowinset=0.25,tbarsize=0.7 5,bracketlength=0.15,rbracketlength=0.15}
\begin{pspicture}(0,0)(50,60)
\psline[linewidth=0.2](0,0)(0,60)
\psline[linewidth=0.2](40,50)(40,60)
\psline[linewidth=1](40,40)(40,20)
\psline[linewidth=0.2](40,0)(40,10)
\rput(42.5,45){$\qee$}
\rput(35,15){$\iee$}
\rput[r](-3.75,0){$\Ree$}
\rput[r](37.5,0){$\Ree$}
\rput[l](42.5,58.75){$\Ree$}
\psline[linewidth=0.35](30,30)(50,10)
\psline[linewidth=0.35](30,10)(50,10)
\psline[linewidth=0.35](30,30)(30,10)
\psline[linewidth=0.35](30,50)(50,50)
\psline[linewidth=0.35](30,50)(50,30)
\psline[linewidth=0.35](50,50)(50,30)
\rput[l](42.5,28.75){$\Aee$}
\rput(15,28.75){\EQLS}
\rput(40,20){\DOTT}
\end{pspicture}

\end{split}
\eeq
Any data item $a:A$ is thus encoded as a program $\enco a :\DP$ and decoded by $\uev{\enco a} = a$. Every type $A$ is thus a retract\sindex{retraction} of the programming language $\DP$,  embedded into it by $\enc :A\strictinto \DP$ and projected back by $\universal: \DP\epi A$, which is a surjective partial \funnn. See Appendix~\ref{Appendix:idemp} for more about retracts. Changing the order of composition yields 
\bea\label{eq:idempa}
\rho_A & = &  \left(\DP\begin{tikzar}[column sep=3.4em]\hspace{-.3em}\arrow[two heads]{r}{\retr^A=\{\}}\&\hspace{-.2em}\end{tikzar} A \begin{tikzar}[column sep=3.6em]
\hspace{-.3em}
\arrow[tail]{r}{\incl_A=\enc}[description,pos=.95]{\!\!\!\mbox{\large$\bullet$}\!}\&%
\hspace{-.3em}\end{tikzar} \DP\right)
\eea
also displayed in Fig.~\ref{Fig:retracts}. The \funnn\ $\rho_{A}$ is idempotent,\sindex{function!idempotent} i.e. it satisfies $\comp{\rho_A}{\rho_A} = \rho_A$. 
\begin{figure}[!ht]
\begin{center}
\newcommand{\pee}{\rho_A}
\newcommand{\iee}{\enc}
\newcommand{\qee}{\universal}
\newcommand{\univrsl}{\big\{\big\}}
\newcommand{\Aee}{\scriptstyle \DP}
\newcommand{\Ree}{\scriptstyle A}
\newcommand{\PRGMS}{\large $\leftsquigarrow$ programs}
\newcommand{\DTA}{\large $\leftsquigarrow$ data}
\newcommand{\Jee}{\scriptstyle J}
\newcommand{\prtll}{[]}
\def\JPicScale{.75}
\input{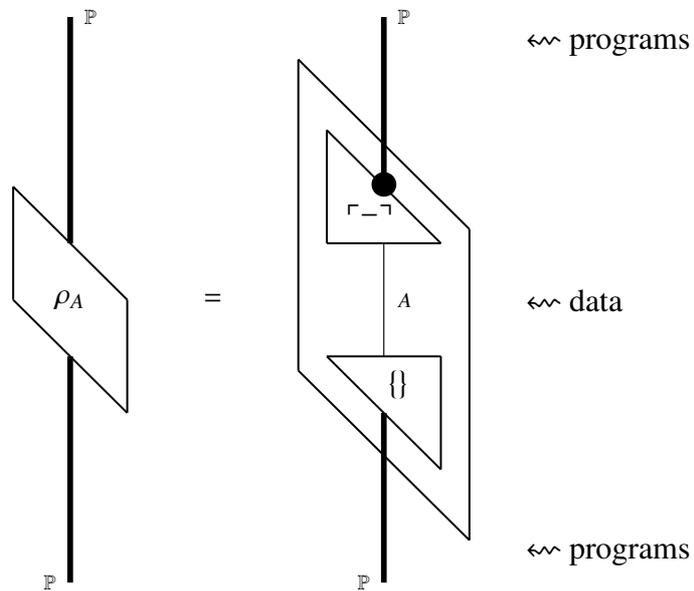}
\caption{Data of type $A$ correspond to the programs filtered by an idempotent $\rho_A$.}
\label{Fig:retracts}
\end{center}
\end{figure}

\subsection{Types as idempotents} \label{Sec:typ-idem}  \sindex{type!as idempotent} 
Every type $A$ corresponds to the idempotent $\rho_{A} = \left(\DP \eepi{\universal} A\strictto{\enc} \DP\right)$\sindex{idempotent}, where  $x=\rho_A(x)$ holds if and only if the program $x$ is an encoding of an element of type $A$. Since the encoding is injective, the elements of type $A$ can be identified with their program encodings, and the typing statement $x:A$ can be viewed as an abbreviation of the equation $x=\rho_{A}(x)$. The idempotent $\rho_{A}$ thus fixes the encodings of the elements of $A$ and filters out all other programs\footnote{We shall work out in \ref{Sec:better-branch}.c that all other programs can be mapped to divergent computations.}. If we abuse notation and use the label $A$ as the name of a program for the idempotent $\rho_{A}$, then the type statements can be summarized as idempotent equations
\beq\label{eq:typing}
x\colon A \ \ \iff\ \ x=\varrho_{A}(x) \ \ \iff\ \ x = \uev A x
\eeq
The type statement for a \funnn\ $f\colon A\to B$ means that $x\colon A$ implies $f(x)\colon B$ i.e. that $x=\uev A x$ implies $f(x) = \uev B \left(f(x)\right)$. Using the idempotence, this implication is easily summarized by the equations
\beq\label{eq:typing-fun}
\left(f\colon A\to B\right) \ \ \iff\ \ \left(f=\varrho_{B}\circ f \circ\varrho_A\right) \ \ \iff\ \ \left(f=\uev{B}\circ f \circ \uev A\right)
\eeq
which are also equivalent with $\varrho_{B}\circ f = f = f\circ \varrho_A$ and $\uev{B}\circ f = f = f\circ \uev A$. The formal equivalence justifying the view of types as idempotents is left for workout~\ref{work:abs-com}.

\subsection{Pairs of programs are programs} \label{Sec:pairing}
\begin{figure}[!ht]
\begin{center}
\newcommand{\iee}{\pairr - -}
\newcommand{\qee}{\scriptstyle \big<\fstt -,  \sndd - \big>}
\newcommand{\Aee}{\scriptstyle \DP}
\newcommand{\Ree}{\scriptstyle A}
\def\JPicScale{1}
\input{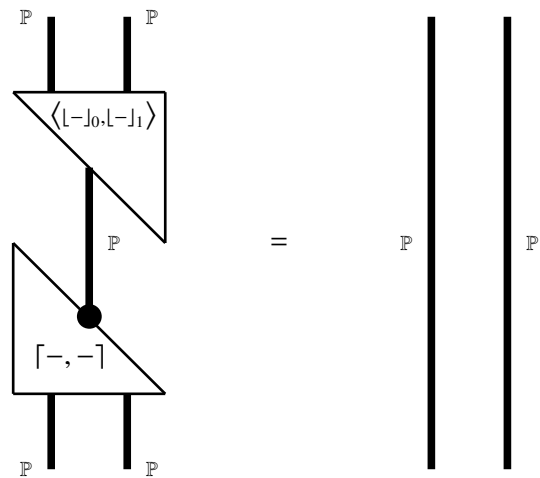}
\caption{Coding pairs of programs as programs}
\label{Fig:tupling}
\end{center}
\end{figure}

\sindex{pairing} \sindex{projection}
An important special case of encoding types as programs is the encoding of product types as programs, and in particular the encoding of the product $\DP\times \DP$ in  Fig.~\ref{Fig:tupling}.
Similar constructions allow encoding $n$-tuples of programs as single programs, for any $n$. Note that the  equations
\beq
\fstt{\pairr x y} = x \qquad\qquad\qquad\qquad \sndd{\pairr x y} = y 
\eeq
make the pairing $\pairr - -$ injective, but that it may not be surjective, as there may be programs that do not encode a pair. See Sec.~\ref{Work:assoc-pair} for a construction of surjective pairing.

\subsection{Idempotents split} \label{Sec:idem-split}
\para{Idempotents on $\DP$ split.} Any computable idempotent $A:\DP\to\DP$ displays the elements $a:A$ as its fixpoints and filters out the programs $p$ not in $A$ either by projecting them to $A(p)$ which is in $A$ or by not returning any output on the input $p$. Since every idempotent viewed as a type provides its own splitting, the type system of a monoidal computer splits all idempotents, and is said to be \emph{absolutely complete} \cite{PareR:absolute}. Treating the types as idempotents on $\DP$ assures that all idempotents on $\DP$ split. But since all types are retracts of $\DP$, it follows that all idempotents on all types split.

\begin{figure}[!ht]
\newcommand{\Prgtp}{\scriptstyle \DP}
\newcommand{\inclA}{\incl_A}
\newcommand{\retrA}{\retr^A}
\newcommand{\inclR}{\incl_R}
\newcommand{\retrR}{\retr^R}
\newcommand{\inclAR}{\incl_R^A}
\newcommand{\retrAR}{\retr_A^R}
\newcommand{\AAh}{\scriptstyle A}
\newcommand{\Ree}{\scriptstyle R}
\newcommand{\ree}{r}
\renewcommand{\DOTT}{\Dottt}
\def\JPicScale{.43}
\begin{center}
\input{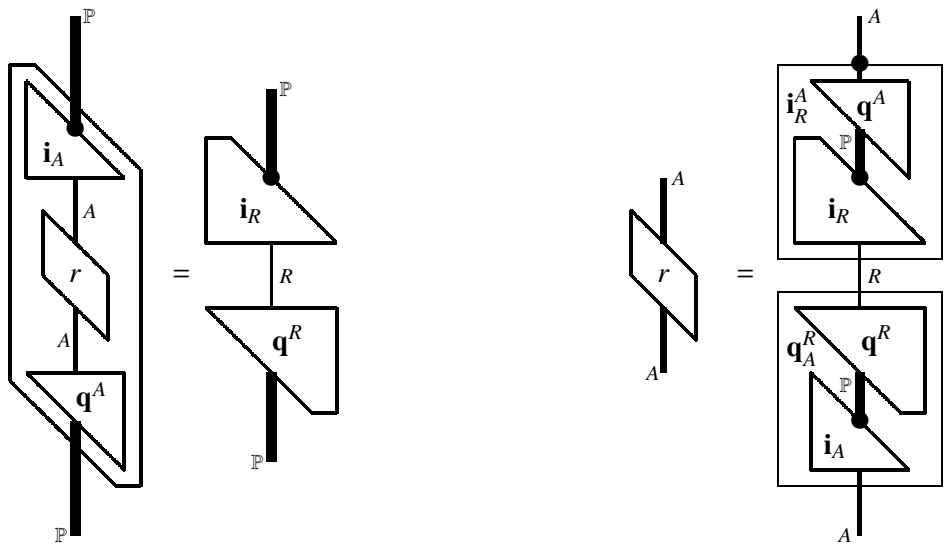}
\caption{Splitting an idempotent in a computer}
\label{Fig:splitting}
\end{center}
\end{figure}

\para{All idempotents split.} If $r:A\to A$ is an idempotent on $A$ and 
$$\rho_A \ \ =\ \ \left(\DP\begin{tikzar}[column sep=2.8em]\hspace{-.3em}\arrow[two heads]{r}{\retr^A}\&\hspace{-.2em}\end{tikzar} A \begin{tikzar}[column sep=3em]
\hspace{-.3em}
\arrow[tail]{r}{\incl_A}[description,pos=.9]{\!\!\!\mbox{\large$\bullet$}\!}\&%
\hspace{-.3em}\end{tikzar} \DP\right)$$ 
is a split idempotent, then 
$$\rho_R \ \ =\ \ \left(\DP\begin{tikzar}[column sep=2.8em]\hspace{-.3em}\arrow[two heads]{r}{\retr^A}\&\hspace{-.2em}\end{tikzar} A\tto r A \begin{tikzar}[column sep=3em]
\hspace{-.3em}
\arrow[tail]{r}{\incl_A}[description,pos=.9]{\!\!\!\mbox{\large$\bullet$}\!}\&%
\hspace{-.3em}\end{tikzar} \DP\right)$$ 
is also an idempotent. If 
$$\rho_R \ \ =\ \ \left(\DP\begin{tikzar}[column sep=2.8em]\hspace{-.3em}\arrow[two heads]{r}{\retr^R}\&\hspace{-.2em}\end{tikzar}R \begin{tikzar}[column sep=3em]
\hspace{-.3em}
\arrow[tail]{r}{\incl_R}[description,pos=.9]{\!\!\!\mbox{\large$\bullet$}\!}\&%
\hspace{-.3em}\end{tikzar} \DP\right)$$
is an idempotent splitting, then
$$r \ \ =\ \ \left(A \begin{tikzar}[column sep=3em]
\hspace{-.3em}
\arrow[tail]{r}{\incl_A}[description,pos=.9]{\!\!\!\mbox{\large$\bullet$}\!}\&%
\hspace{-.3em}\end{tikzar} \DP\begin{tikzar}[column sep=2.8em]\hspace{-.3em}\arrow[two heads]{r}{\retr^R}\&\hspace{-.2em}\end{tikzar}R \begin{tikzar}[column sep=3em]
\hspace{-.3em}
\arrow[tail]{r}{\incl_R}[description,pos=.9]{\!\!\!\mbox{\large$\bullet$}\!}\&%
\hspace{-.3em}\end{tikzar} \DP \begin{tikzar}[column sep=2.8em]\hspace{-.3em}\arrow[two heads]{r}{\retr^A}\&\hspace{-.2em}\end{tikzar} A\right)$$
is also an idempotent splitting, as displayed in Fig.~\ref{Fig:splitting}. 

\para{Upshot.} Fig.~\ref{Fig:retracts} shows how the program evaluator of type $A$ induces an idempotent $\rho_A$ that filters the elements of type $A$.  Fig.~\ref{Fig:splitting} shows how any idempotent $r:A\to A$ induces a type $R$ as its splitting. Types induce idempotents and the idempotents induce types. This is a foundational view of typing going back to \cite{ScottD:relating}. When confusion is unlikely, viewing type statements as program evaluations (\ref{eq:typing}--\ref{eq:typing-fun}) provides an intuitive view of type checking.  When a computer is given with just a few native types, or even none, then we can use \eqref{eq:typing} to define checkable types. We begin from the simplest one: the booleans.

\para{Remark.} If a type universe contains two different languages, $\DP_0$ and $\DP_1$, instantiating \eqref{eq:retr} to each of them displays them as each other's retracts. In other words, each of them can be encoded as a subset of expressions of the other one. In general, when two sets are embedded into each other, the Cantor-Bernstein construction provides a bijection between them \cite{HinkisA:cantor-bernstein,SiegW:cantor-bernstein}. There is thus a one-to-one correspondence between the sets of expressions in any two programming languages. But the correspondence is neither computable nor computationally meaningful. It is not computable because the general Cantor-Bernstein constructions involves operations that are not computable and implies Boolean logic \cite{PavlovicD:Cantor-categories}. It is not computationally meaningful because it does not preserve the meaning of programs. However, there is a computable and computationally meaningful isomorphism that will be spelled out in Ch.~\ref{Chap:PCC}.

\section{Logic and equality}
\label{Sec:bool}

\subsection{Truth values and branching} \label{Sec:branch}\sindex{branching!eager $\iif$}If we take a program $\true$ for the first projection to play the role of the truth value \emph{"true"}, and a program $\flse$ for the second projection to play the role of the truth value \emph{"false"}, \sindex{truth value}
\beq\label{eq:truth}
\begin{split}
\newcommand{\zzero}{\flse}
\newcommand{\oone}{\true}
\newcommand{\ueval}{\universal}
\def\JPicScale{.45}
\ifx\JPicScale\undefined\def\JPicScale{1}\fi
\unitlength \JPicScale mm
\begin{picture}(280,50)(0,0)
\linethickness{0.6mm}
\put(270,0){\line(0,1){50}}
\linethickness{0.6mm}
\put(250,0){\line(0,1){30}}
\linethickness{0.35mm}
\put(250,30){\circle*{5.16}}

\linethickness{0.35mm}
\put(240,40){\line(1,0){40}}
\linethickness{0.35mm}
\put(280,20){\line(0,1){20}}
\linethickness{0.35mm}
\put(240,20){\line(1,0){40}}
\linethickness{0.35mm}
\put(240,20){\line(0,1){20}}
\put(230,30){\makebox(0,0)[cc]{\EQLS}}

\linethickness{0.35mm}
\put(0,40){\line(1,0){50}}
\linethickness{0.35mm}
\put(50,20){\line(0,1){20}}
\linethickness{0.6mm}
\put(35,40){\line(0,1){10}}
\linethickness{0.35mm}
\put(20,20){\line(1,0){30}}
\linethickness{0.6mm}
\put(45,0){\line(0,1){20}}
\linethickness{0.6mm}
\put(10,20){\line(0,1){10.01}}
\linethickness{0.35mm}
\multiput(0,40)(0.12,-0.12){167}{\line(1,0){0.12}}
\linethickness{0.35mm}
\put(0,10){\line(1,0){20}}
\linethickness{0.35mm}
\put(0,10){\line(0,1){20}}
\linethickness{0.35mm}
\multiput(0,30)(0.12,-0.12){167}{\line(1,0){0.12}}
\linethickness{0.6mm}
\put(25,0){\line(0,1){20}}
\put(60,30){\makebox(0,0)[cc]{\EQLS}}

\linethickness{0.35mm}
\put(170,40){\line(1,0){50}}
\linethickness{0.35mm}
\put(220,20){\line(0,1){20}}
\linethickness{0.6mm}
\put(205,40){\line(0,1){10}}
\linethickness{0.35mm}
\put(190,20){\line(1,0){30}}
\linethickness{0.6mm}
\put(215,0){\line(0,1){20}}
\linethickness{0.6mm}
\put(180,20){\line(0,1){10.01}}
\linethickness{0.35mm}
\multiput(170,40)(0.12,-0.12){167}{\line(1,0){0.12}}
\linethickness{0.35mm}
\put(170,10){\line(1,0){20}}
\linethickness{0.35mm}
\put(170,10){\line(0,1){20}}
\linethickness{0.35mm}
\multiput(170,30)(0.12,-0.12){167}{\line(1,0){0.12}}
\linethickness{0.6mm}
\put(195,0){\line(0,1){20}}
\put(5,15){\makebox(0,0)[cc]{$\oone$}}

\put(175,15){\makebox(0,0)[cc]{$\zzero$}}

\linethickness{0.6mm}
\put(80,0){\line(0,1){50}}
\linethickness{0.6mm}
\put(100,0){\line(0,1){30.01}}
\linethickness{0.35mm}
\put(100,30){\circle*{5.16}}

\linethickness{0.35mm}
\put(70,40){\line(1,0){40}}
\linethickness{0.35mm}
\put(110,20){\line(0,1){20}}
\linethickness{0.35mm}
\put(70,20){\line(1,0){40}}
\linethickness{0.35mm}
\put(70,20){\line(0,1){20}}
\put(30,30){\makebox(0,0)[cc]{$\ueval$}}

\put(200,30){\makebox(0,0)[cc]{$\ueval$}}

\linethickness{0.35mm}
\put(10,20){\circle*{5.16}}

\linethickness{0.35mm}
\put(180,20){\circle*{5.16}}

\end{picture}

\end{split}\eeq
then setting $\iif = \universal$ gives the branching operation:
\bea\label{eq:iiff}
\iif(\bpredc,x,y)\ \ =\ \ \uev{\bpredc}(x,y) & = & \begin{cases}
x & \mbox{ if } \bpredc = \true\\
y & \mbox { if } \bpredc = \flse
\end{cases}
\eea
The expression $\iif(\bpredc,x,y)$ is meant to be read ``$\mathsf{if}\ \bpredc\ \mathsf{then}\ x\ \mathsf{else}\ y$''.

\subsection{Program equality} \label{Sec:equality} \sindex{program!equality}
While deciding whether two computations output equal results may need to wait for the computations to produce the outputs, deciding whether two programs are equal or not is in principle easy. To be able to use such decisions in computations, we are given a decidable predicate $(\iseq):\DP\times \DP\stricto \DP$ which captures the program equality internally:
\bea
(p\iseq q) & = & \begin{cases} \true & \mbox{ if } p = q\\
\flse & \mbox{ otherwise}
\end{cases}
\eea

\subsection{The type of booleans}\label{Sec:bool-constr}
\begin{figure}[!ht]
\begin{center}
\newcommand{\zzero}{\flse}
\newcommand{\oone}{\true}
\newcommand{\ueval}{\universal}
\newcommand{\beeeta}{\rho}
\newcommand{\Ree}{\Bool}
\newcommand{\booleval}{\universal}
\newcommand{\encobool}{\enc}
\newcommand{\smaller}{(\iseq)}
\def\JPicScale{.65}
\input{PIC/boolidem.tex}
\caption{The type $\Bool = \{\flse, \true\}$ splits the idempotent $\rho$}
\label{Fig:boolidem}
\end{center}
\end{figure}
Since the equality $(\iseq)$ is total, the test $(x\iseq \true)$ returns $\true$ if $x= \true$ and $\flse$ otherwise. The {\strict} \funnn\ 
\bea
\rho \colon \DP & \stricto & \DP\\
x & \mapsto & \iif\Big( x\iseq \true, \true, \flse\Big)\notag
\eea
displayed in Fig.~\ref{Fig:boolidem}, is clearly idempotent, as it maps $\true$ to $\true$ and all other programs to $\flse$. The type $\Bool$ arises by splitting this idempotent. It filters the programs $\true$ and $\flse$ and fixes the type $\Bool = \{\true,\flse\}$. More precisely, the elements of this type are $\tot \CCC(I,\Bool) =  \{\true,\flse\}$. 

\subsection{Propositional connectives}\label{Sec:connective}
\begin{figure}[!ht]
\begin{center}
\newcommand{\zzero}{\flse}
\newcommand{\oone}{\true}
\newcommand{\ueval}{\iif}
\newcommand{\hand}{\mbox{\LARGE$\wedge$}}
\newcommand{\hor}{\mbox{\LARGE$\vee$}}
\newcommand{\hthen}{\mbox{\LARGE$\Rightarrow$}}
\newcommand{\hxand}{\mbox{\LARGE$\not\Leftarrow$}}
\def\JPicScale{.5}
\input{PIC/connective-12.tex}
\caption{The $\iif = \universal$ induces the propositional connectives}
\label{Fig:connective}
\end{center}
\end{figure}
In any computer, the native truth values $\true,\flse$ and the $\iif$-branching yield native propositional connectives. One way to define them is displayed in Fig.~\ref{Fig:connective}.  The negation $\neg \qpredc$ can be reduced to the implication $\qpredc  \Rightarrow \flse$. \sindex{negation}

\subsection{Predicates and decidability}\label{Sec:predicate}\sindex{predicate}
\para{Predicates are single-valued.} A predicate is a \emph{\textbf{single-valued}}  \funnn\ of in the form $\qpredc \colon X\to \Bool$. It thus assigns a unique truth value $q(x)\colon \Bool$ to each $x\colon X$. 

The single-valuedness requirement prevents that $q(x)$ may be true today and false tomorrow. Formally, this requirement is expressed by the equation $(\qpredc \times \qpredc)\circ \cmn_{X} = \cmn_{\Bool}\circ \qpredc$ discussed in Sec.~\ref{Sec:map}. For a computation $\qpredc$, the equation says that running $\qpredc$ twice on the same input gives two copies of the same output. This justifies thinking of a predicate as asserting a property.

\para{Assertion notation.} \sindex{notation!assertion $\qpredc(a)\halts$}The equation $\qpredc (x)=\true$ is often abbreviated to the \emph{assertion}\/ $\qpredc (x)$ that $x$ has a property $\qpredc$. The equation $\qpredc (x) =\flse$ is abbreviated to $\neg \qpredc (x)$.

\para{Predicates may not be total.}  Computations are branched using computable predicates $\qpredc$, e.g. in $\iif(\qpredc(x),f(x),g(x))$. If $\qpredc(x)$ diverges, $\iif(\qpredc(x),f(x),g(x))$ will also diverge and the branching will remain undecided. 

\para{Decidable predicates.}\sindex{predicate!decidable} \sindex{decidability}  A predicate $\qpredc$ is \emph{decidable}\/ if it is total, i.e. if it converges and outputs $\qpredc(x) = \true$ or $\qpredc(x)=\flse$ for all $x\colon X$. Formally, this is assured by the equation $\cun_{\Bool}\circ \qpredc = \cun_{X}$, discussed in Sec.~\ref{Sec:map}. In the sense of that section, the decidable predicates over $X$ are precisely the \strict\ \funnn\ $ X\stricto \Bool$.

\para{Halting notation}\sindex{halting} $\qpredc (x)\halts$ abbreviates the equation $\cun_{\Bool}\circ \qpredc \circ x = \cun_{I}$ for any \strict\ $x:I\stricto X$.

\section{Monoidal computer}
\label{Sec:moncom}

\subsection{Definition}\label{Sec:moncom-def}
A \emph{monoidal computer}\/ is a (symmetric,  strict) \sindex{monoidal!computer}
monoidal category $\CCC$ with
\begin{description}
\item[\textbf{a)}] a data service $A\times A\strictoot \cmn A\strictto \cun I$ on every type $A$, as described in Sec.~\ref{Sec:service},  
\item[\textbf{b)}] a distinguished type $\DP$ of \emph{programs}\/ with a decidable equality predicate $(\iseq):\DP\times \DP\stricto\Bool$, and\sindex{program type $\DP$|see {programming language}}
\item[\textbf{c)}] a \emph{program evaluator} $\universal: \DP\times A \to B$ for all types $A,B$, as  in \eqref{eq:uev}. 
\end{description}

\subsection{Examples} \label{Sec:moncom-examples}
\para{Programming languages.}
 As mentioned in Sec.~\ref{Sec:programmable}, the programming language interpreters are the standard examples of program evaluators. The type universes of programming languages are standard examples of monoidal computers. The objects of a monoidal computer $\CCC= \CCC_{\DP}$ are thus the types that are checkable by $\DP$-programmable idempotent \funnn s. The morphisms of $\CCC_{\DP}$ are all $\DP$-programmable \funnn s, partitioned into the hom-sets between the types-as-idempotents under which they are invariant. The details are worked out in Sec.~\ref{work:monoid-completion}. Any  Turing-complete programming language\sindex{Turing-complete programming language|see {programming language!Turing complete}} \sindex{programming!language!Turing complete}$\DP$ thus induces a monoidal computer. See Sec.~\ref{Sec:CTT} for more about Turing completeness. 
While $\CCC_{\DP}$'s program evaluators correspond to $\DP$'s \sindex{interpreter} interpreters, $\CCC_{\DP}$'s partial evaluators correspond to its specializers\sindex{specializer} \cite{JonesN:book-computability}. $\DP$'s  Turing-completeness of  implies that its interpreters and specializers can be programmed in $\DP$ itself. Sec.~\ref{Sec:CTT} and \cite{PavlovicD:IC12,PavlovicD:MonCom3} contain more details and references. The other way around, the morphisms of any monoidal computer $\CCC_\DP$ are $\DP$-programmable, in the sense of Sec.~\ref{Sec:uev}, which makes $\DP$ into an abstract programming language. 
 
 \para{Models of computation.} Note that each of the standard models of computation, the Turing machines, partial recursive functions, etc., can be viewed as programming languages, and subsumed under the examples above. E.g., any Turing machine can be viewed as a program, with a universal Turing machine as a program evaluator. With suitably formalized execution schemas and a concept of types, any of the models can be used as a programming language \cite{ScottD:relating} and expanded into a monoidal computer.

\subsection{Program evaluation as a natural operation}\label{Sec:surj}
In the presence of the structures from (a--b), the program evaluator condition (c) can be equivalently stated as the following natural form of the "running" surjection\sindex{running programs}\sindex{run-instruction@\runn-instruction} from Fig.~\ref{Fig:head-tail}.
\begin{description}
\item[\textbf{c')}] an $X$-natural family of surjections $\CCC(X\times A\, ,\,  B) \begin{tikzar}[row sep=-.25em,column sep=1em]
\hspace{0.01em}  \&\&\& \hspace{0.01em} \arrow[twoheadrightarrow,bend right=16]{lll}[swap,pos=0.45]{\runn^{AB}_X}
\end{tikzar}\tot \CCC(X, \DP)$ 
for each pair of types $A,B$.
\end{description}

\begin{figure}[!ht]
\begin{center}
\begin{tikzar}[row sep=1em,column sep=1em]
\CCC(\DP \times A, B) \arrow{dddddd}[description]{\CCC(G\times A, B)}  
\&\&\&\&\&\&
\tot \CCC(\DP,\DP) \arrow{dddddd}[description]{\tot\CCC(G, \DP)}  \arrow[two heads,bend right=13]{llllll}[swap]{\runn^{AB}_\DP} \\[-2.25ex]
\& \universal \arrow[mapsto]{dddd}  \&\&\&\& \id_\DP \arrow[mapsto]{dddd} \arrow[mapsto,bend right=13]{llll} \\
\\
\\
\\
\& \uev G \&\&\&\&  G \arrow[mapsto,bend right=13]{llll} \\[2.25ex] 
  \CCC(X\times A, B) 
 \&\&\&\&\&\& \tot \CCC(X,\DP)  \arrow[two heads,bend right=13]
 {llllll}[swap]{\runn^{AB}_X}
\end{tikzar}
\caption{Any program evaluator $\universal$ induces the $X$-indexed family $\runn^{AB}_{X}$.}
\label{Fig:gamma}
\end{center}
\end{figure}
Towards a proof of \textbf{(c$\Rightarrow$c')}, note that any program evaluator $\universal: \DP\times A \to B$ induces an $X$-natural family $\runn^{AB}$ making the diagram in Fig.~\ref{Fig:gamma} commute, whose components are
\bea\label{eq:gammma}
\runn^{AB}_X(G) & = &\uev G
\eea
The naturality requirement in Fig.~\ref{Fig:gamma-nat}
\begin{figure}[!ht]
\begin{center}
\begin{tikzar}[row sep=1em,column sep=1em]
\CCC(X \times A, B) \arrow{dddddd}[description]{\CCC(s\times A, B)}
\&\&\&\&\&\&
\tot \CCC(X,\DP)  \arrow{dddddd}[description]{\tot\CCC(s, \DP)} 
\arrow[two heads,bend right=13]
{llllll}[swap]{\runn^{AB}_X}
\\
\& \hspace{1em} \arrow[phantom]{dddd}  \&\&\&\& \hspace{1ex} \arrow[phantom]{llll}\arrow[phantom]{dddd} \\
\\
\\
\\
\& \hspace{2ex} \arrow[phantom]{rrrr} \&\&\&\& \hspace{2ex}\\ 
 \CCC(Y\times A, B)
 \&\&\&\&\&\& \tot \CCC(Y,\DP)  \arrow[two heads,bend right=13]
 {llllll}[swap]{\runn^{AB}_Y} 
\end{tikzar}
\caption{The naturality requirement for $\runn^{AB}$.}
\label{Fig:gamma-nat}
\end{center}
\end{figure}
means that every $s\in \tot \CCC(Y,X)$ and every $G\in \tot\CCC(X,\DP)$ should satisfy  $\runn^{AB}_X(G) \circ s  \ = \ \runn^{AB}_Y(G\circ s)$. Definition \eqref{eq:gammma} reduces this requirement to $\uev{G}_X\circ s = \uev{G\circ s}_Y$, which is tacitly imposed by the string diagrammatic formulation of \eqref{eq:uev}, and validated by precomposing both of its sides with $s:Y\stricto X$. Condition \eqref{eq:uev-formal} says that all $\runn^{AB}_X$ are surjective, as required. The other way around, to prove \textbf{(c'$\Rightarrow$c)}, define
\bea\label{eq:univ-gammma}
\universal & = & \runn^{AB}_\DP(\id) 
\eea
and note that now the surjectiveness of $\runn^{AB}_\DP$ gives \eqref{eq:uev-formal}.

%

\def\thechapter{3}
\setchaptertoc
\chapter{Fixpoints}\label{Chap:Prog}
\newpage

\section{Computable \funnn s have fixpoints}\label{Sec:idea-fixp}

The basic idea of the universal evaluators presented in the preceding chapter is that 
\textbf{\emph{a \funnn\ is computable when it is programmable}}. 
A \funnn\ $f:A\to B$ is programmable when there is a program $F:\DP$ such that for all $y:A$ 
\bea\label{eq:uniev} 
f(y) & = & \uev  F y
\eea 
In this chapter we draw the main consequences of programmability. In the next chapter we will use it as a tool for programming. 

The first consequence of programmability is that any computable endo\funnn\ in the form $e:A\to A$ must have a fixpoint $a$
\bea 
e(y) = \uev  E y & \implies & \exists a.\ e(a) = a
\eea
The second consequence is the pivot point to move the world of computation. It says that for any computable \funnn\ in the form $g:\DP\times A\to B$ there is a program $\Gamma:\DP$ such that
\bea\label{eq:fundthm}
g(p,y) = \uev G (p,y)& \implies & \exists \Gamma .\ g(\Gamma, y)  = \uev {\Gamma}  y
\eea
This was proved in Kleene's note \cite{KleeneS:ordinal-fund} under the name  \emph{Second Recursion Theorem}. But the use of recursion is inessential for it \cite{SoareR:CTA}. On the other hand, it turned out to be one of the most fundamental theorems of computability theory  \cite{MoschovakisY:Kleene}. We call it the Fundamental Theorem of Computation. It is proved in Sec.~\ref{Sec:kleene}.

\section{Divergent, partial, monoidal computations}\label{Sec:divergence} \sindex{fixpoint}\sindex{divergence}

\para{Every computable endo\funnn\ $e:A\to A$ has a fixpoint  $\fixp_e:A$, with  $e(\fixp_e) =\fixp_e$.} \sindex{fixpoint}A fixpoint construction is constructed in Fig.~\ref{Fig:fixproof}.
\begin{figure}[!ht]
\begin{center}
\newcommand{\inputt}{\scriptstyle A}
\newcommand{\thennn}{$\implies$}
\newcommand{\aahh}{\mbox{\Large${\fixp}_e$}}
\renewcommand{\universal}{\Large\{\}}
\newcommand{\gee}{\mbox{\Large$e$}}
\newcommand{\nameslang}{\scriptstyle A}
\newcommand{\Peeg}{{\fixp}_e}
\newcommand{\program}{$\Fixp_e$}
\newcommand{\progtype}{\scriptstyle \DP}
\def\JPicScale{.5}
\input{PIC/fixp.tex}
\caption{If $\uev{\Fixp_e}(x) = e\left(\uev x x\right)$, then ${\fixp}_e \! = \uev{\Fixp_e}\left(\Fixp_e\right)$ satisfies $e({\fixp}_e) = e\left( \uev{\Fixp_e}\left(\Fixp_e\right)\right)$}
\label{Fig:fixproof}
\end{center}
\end{figure} 
We precompose $e$ with the program self-evaluation $\uev x x$, and define $\Fixp_e$ to be a program satisfying
\bea\label{eq:Fixpe-constr}
\uev{\Fixp_e}x & = &  e\left(\uev x x\right)
\eea
as displayed in Fig.~\ref{Fig:fixproof} on the left. A program $\Fixp_e$ for $e\left(\uev x x\right)$ must exist by the assumption that the program evaluator is universal. Setting $x$  in \eqref{eq:Fixpe-constr}  to be $\Fixp_e$ yields
\bea\label{eq:fixpe-constr}
\uev{\Fixp_e}\Fixp_e & = &  e\left(\uev{\Fixp_e}\Fixp_e\right)
\eea
Defining $\fixp_e = \uev{\Fixp_e}\Fixp_e$ yields a fixpoint of $e$, as displayed in Fig.~\ref{Fig:fixproof} on the right. Note that an overly eager program evaluator may keep evaluating a fixpoint forever:
\beq\label{eq:fixpp} \fixp_e\, =\, \uev {\Fixp_e}(\Fixp_e) \, =\, e \uev{\Fixp_e}(\Fixp_e)\, =\, ee \uev {\Fixp_e}(\Fixp_e) \, =\, eee \uev {\Fixp_e}(\Fixp_e)\, = \cdots
\eeq

\para{A divergent truth value.} \sindex{truth value!divergent} The datatype $\Bool$ of the boolean truth values was defined in Sec.~\ref{Sec:bool-constr}. By definition, it contains precisely two {\strict} elements $\true, \flse \in \tot\CCC(I,\Bool)$. The operation of negation $\neg\colon \Bool\to \Bool$ defined in Sec.~\ref{Sec:connective} satisfies $\neg \flse = \true$ and $\neg\true = \flse$. On the other hand, applying the above fixpoint construction to the negation yields an element $\fixp_\neg\colon \Bool$ such that
\bear
\neg \fixp_\neg & = & \fixp_\neg
\eear
Since $\neg \true = \flse$ and $\neg \flse = \true$, it follows that $\true \neq \fixp_\neg \neq \flse$ (unless $\true=\flse$ and by \ref{Sec:how-many}a, the whole universe collapses to a point). Since $\true$ and $\flse$ are by the definition in Sec.~\ref{Sec:bool-constr} the only \strict\ elements of $\Bool$, the element $\fixp_\neg\colon\Bool$ cannot be \strict. In summary, we have
\beq
\tot \CCC(I,\Bool) = \{\true\, ,\  \flse\} \qquad \qquad\qquad \qquad \CCC(I,\Bool) \supseteq \{\true\, ,\  \flse\, ,\   \fixp_\neg\, ,\ \ldots\}
\eeq
A computer $\CCC$ always contains divergent computations, and is strictly monoidal, larger than its \strict\ core $\tot \CCC$,  or else it collapses to a point.

\para{Divergent elements.} According to the definitions in Sec.~\ref{Sec:map}, an element $a:A$ is a single-valued \funnn\ $a\colon I\to A$ and it is total if $\left(I\tto a A\strictto\scun I\right) = \id_{I}$. This is an instance of the left-hand requirement of \eqref{eq:map}. An element $a:A$ is called \sindex{element!divergent} \emph{divergent} if 
\bea
\left(I\tto a A\strictto\scun I\right) & \neq & \id_{I}
\eea
Divergent elements are obviously partial. 
It was also noted in Sec.~\ref{Sec:map} that $\cun_{A}$ is the only inhabitant of the hom-set $\tot \CCC(A,I)$, for any type $A$. In particular, $\cun_{I}=\nill$ is only inhabitant of $\tot \CCC(I,I)$. Since the identity must be there as well, it follows that $\id_{I}=\cun_{I}=\nill$. On the other hand, $\divg_{I}\  =\  \left(I\tto{\fixp_\neg} \Bool \strictto{\scun} I\right) \neq \id_{I}$ gives 
\beq
\tot \CCC(I,I) = \{\cun_I\ \} \qquad \qquad\qquad \qquad \CCC(I,I) \supseteq \{\cun_I\ ,\  \divg_I\,  ,\ \ldots\}
\eeq
and $\divg_{A}\  =\  \left(A\strictto{\scun_{A}} I \tto{\divg_{I}} I\right)$ moreover gives 
\beq
\tot \CCC(A,I) = \{\cun_{A}\ \} \qquad \qquad\qquad \qquad \CCC(A,I) \supseteq \{\cun_{A}\ ,\  \divg_{A}\,  ,\ \ldots\}
\eeq
This leaves us with some interesting string diagrams. 
\begin{figure}[!ht]
\begin{center}
\newcommand{\AAh}{\scriptstyle A}
\newcommand{\delete}{}
\newcommand{\Bdot}{\DOTT}
\newcommand{\Wdot}{\WDOTT}
\newcommand{\fixup}{{\scriptstyle\fixp_\neg}}
\newcommand{\llpop}{}
\newcommand{\booltyp}{\scriptstyle \Bool}
\def\JPicScale{.75}
\input{PIC/elts.tex}
\caption{Some string diagrams with few strings: $\cun_{I}$ and $\divg_{I}$ have none.}
\label{Fig:elts}
\end{center}
\end{figure}
Since the identity \funnn\ of type $A$ is conveniently represented as the invisible box on the string $A$, and the identity type $I$ is conveniently represented as the invisible string, $\id_I$ is very invisible. When it needs to be seen $\id_I = \cun_I$ helps. Since each $\cun_A$ is a black bead on top of a string, $\cun_I$ is a lonely black bead on top of an invisible string, in Fig.~\ref{Fig:elts} on the left. The partial element $\divg_I = \cun_\Bool\circ \fixp_\neg$ is then conveniently represented as a lonely white bead on top of an invisible string, in the middle Fig.~\ref{Fig:elts}.
Lastly, the partial element $\divg_A = \divg_I\circ \cun_A$ is represented as a white bead on top of the string $A$ in Fig.~\ref{Fig:elts} on the right.

\para{A divergent program.}\sindex{divergence} The main tool for constructing the fixpoints above is the self-evaluator $\uev x x$. Instantiating \eqref{eq:Fixpe-constr} to $e(x) = \uev x x$ leads to $e\left(\uev x x\right) = \uev{\uev x x} \left(\uev x x\right)$. 
\begin{figure}[!ht]
\begin{center}
\newcommand{\inputt}{\scriptstyle \DP}
\newcommand{\aahh}{\mbox{\large${\fixp}$}}
\newcommand{\nameslang}{\scriptstyle \DP}
\newcommand{\Peeg}{\mbox{\large${\fixp}$}}
\newcommand{\program}{$\scriptstyle \Fixp$}
\newcommand{\progtype}{\scriptstyle \DP}
\newcommand{\EQLSETC}{\large $=\cdots$}
\def\JPicScale{.5}
\input{PIC/fixp-Phi.tex}
\caption{If $\uev{\Fixp}x = \uev{\uev x x}\left(\uev x x\right)$, then ${\fixp} \ = \uev \Fixp\Fixp$ gives $\uev\fixp{\fixp} = \uev{\uev{\fixp}\fixp}\left(\uev{\fixp}\fixp\right)$}
\label{Fig:divergence}
\end{center}
\end{figure}
Setting $\uev{\Fixp}x = \uev{\uev x x}\left(\uev x x\right)$, the evaluation process in \eqref{eq:fixpp} becomes
\beq\label{eq:fixppp} {\fixp} \ = \uev \Fixp \Fixp =\uev{\uev \Fixp\Fixp}\left(\uev \Fixp\Fixp\right)= \uev{\uev{\uev \Fixp\Fixp}\left(\uev \Fixp\Fixp\right)}\left(\uev{\uev \Fixp\Fixp}\left(\uev \Fixp\Fixp\right)\right) =\cdots
\eeq
If the programs are evaluated bottom-up, i.e. if every input must be evaluated before it is passed to a computation above it, then the computation in Fig.~\ref{Fig:divergence} goes on forever, as do those in Fig.~\ref{Fig:fixproof}. Such computations \emph{diverge}.  Can we refine the  evaluation strategies to recognize and avoid the divergences? This idea led Alan Turing to the \emph{Halting Problem}, that will be discussed in Sec.~\ref{Sec:halting}.  

\para{Computations are monoidal.} The upshot of this section is that the fixpoints make computable \funnn s and computable types inexorably partial. E.g., in addition to the the truth values $\true$ and $\flse$, the type $\Bool$ must contain a partial element  $\fixp_{\neg}$ as a fixpoint of the negation $\neg$. In mathematics, universes where programs and computations are not a central concern are usually introduced as universes of {\strict} \funnn s, with a unique output on each input. Partial \funnn s, multi-valued \funnn s, randomized \funnn s, etc., are then captured using additional structures \cite{Cockett-Hofstra:turing,DiPaola-Heller:dominical,MoggiE:monad}. In universes where programs and computations are a central concern, partial \funnn s are the first class citizens, and {\strict}ness arises as a special property. Computable functions are generally just monoidal and not \strict\, in the sense that they satisfy just the monoidal laws from Sec.~\ref{Sec:fun} but not the cartesian laws from Sec.~\ref{Sec:map}. \sindex{function!monoidal}\sindex{function!cartesian} A monoidal computer where all computations are total and single-valued and $\CCC=\tot\CCC$, then we are looking at a model where all  intensional aspects of computation have been projected away, and the programs for a given function are indistinguishable.

\section{The Fundamental Theorem of Computation}\label{Sec:kleene}

\begin{figure}[!ht]
\begin{center}
\renewcommand{\DOTT}{\mbox{\Large$\bullet$}}
\newcommand{\Aah}{\scriptscriptstyle A}
\newcommand{\grr}{g}
\newcommand{\Bah}{\scriptscriptstyle B}
\newcommand{\GRR}{\Gamma}
\newcommand{\UK}{\universal}
\def\JPicScale{.5}
\input{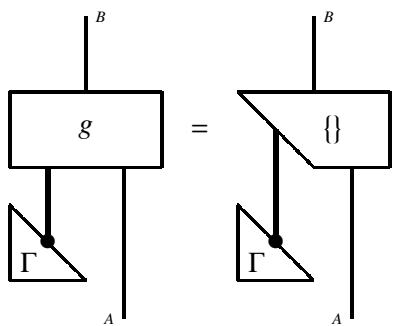}
\caption{A Kleene fixpoint $\Gamma:\DP$ of $g$ is a program such that $g(\Gamma, y) = \uev \Gamma y$}
\label{Fig:kleene}
\end{center}
\end{figure}

\para{Kleene fixpoints.} \sindex{fixpoint!Kleene fixpoint} A Kleene fixpoint \sindex{Kleene fixpoint|see {fixpoint}}of a computation $g\colon \DP\times A\to B$ is a program $\Gamma$ that encodes $g$ evaluated on $\Gamma$, as displayed in  Fig.~\ref{Fig:kleene}.

\para{Theorem.}\sindex{fundamental theorem!of computation}\sindex{Kleene's Second Recursion Theorem} Every computation that takes a program as an input, has a Kleene fixpoint for that input.

\begin{figure}[!ht]
\begin{center}
\renewcommand{\DOTT}{\mbox{\Large$\bullet$}}
\newcommand{\Aah}{\scriptscriptstyle A}
\newcommand{\grr}{g}
\newcommand{\Bah}{\scriptscriptstyle B}
\newcommand{\Grr}{G}
\newcommand{\GRR}{\Gamma}
\newcommand{\UK}{\universal}
\newcommand{\PK}{\prtial}
\def\JPicScale{.5}
\input{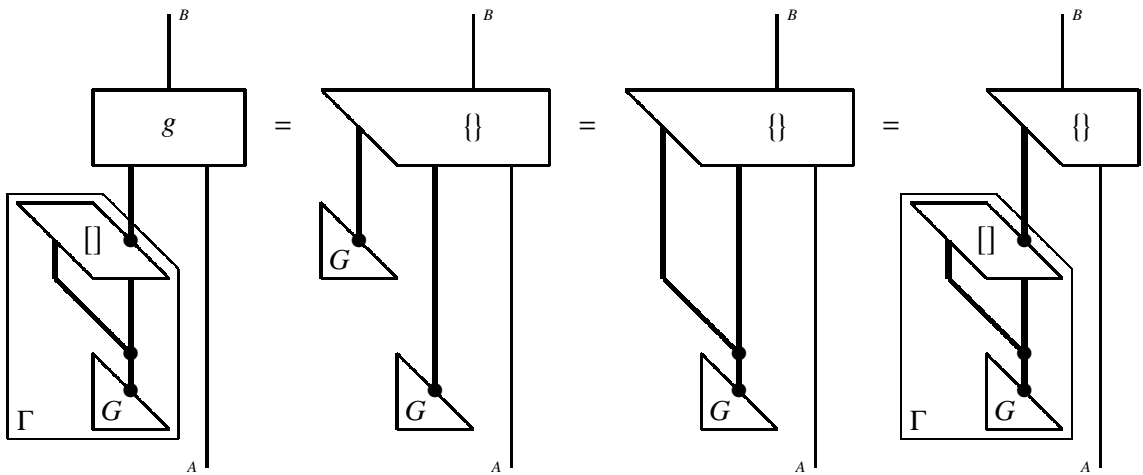}
\caption{A Kleene fixpoint $\Gamma$ of a computable \funnn\ $g$.}
\label{Fig:Klepro}
\end{center}
\end{figure}

\para{Construction.} Towards a Kleene fixpoint of any given \funnn\ $g(p,a)$, where $p$ should be a program, substitute for $p$ a computation $\pev x x$, evaluating programs on themselves, like in Fig.~\ref{Fig:Klepre}. Call a program for the composite $G$.
\begin{figure}[!ht]
\begin{center}
\renewcommand{\DOTT}{\mbox{\Large$\bullet$}}
\newcommand{\Aah}{\scriptscriptstyle A}
\newcommand{\grr}{g}
\newcommand{\Bah}{\scriptscriptstyle B}
\newcommand{\Grr}{G}
\newcommand{\UK}{\universal}
\newcommand{\PK}{\prtial}
\def\JPicScale{.5}
\input{PIC/kleene-set.tex}
\end{center}
\caption{$G$ is a program for $g\left(\pev x x, y\right) \ = \ \uev G (x,y)$
}
\label{Fig:Klepre}
\end{figure}
Then a Kleene fixpoint $\Gamma$ of $g$ can be defined to be a partial evaluation of $G$ on itself, as displayed in Fig.~\ref{Fig:Klepro}.
Fig.~\ref{Fig:Klepro} proves that $\Gamma = \pev G G$ is a fixpoint of $g$ because 
$$g(\Gamma, y)\  =\    g\big(\pev G G, y\big)\   =\   \uev G\left(G, y\right)\   =\   \uev{\pev G G}y\   =\   \uev{\Gamma}y$$
\hfill $\Box$

\para{The program transformer version of the Theorem.} \sindex{fixpoint!transformer} A \emph{program transformer}\/\sindex{program!transformer} is a \strict\ \funnn\ $\gamma:\DP\stricto \DP$ that transforms programs. The Fundamental Theorem of Computation is equivalent to the statement that every program transformer has a fixpoint, a program $\Gamma$ that does the same as its transform, as displayed in Fig.\ref{Fig:hartley}.
\begin{figure}[!ht]
\begin{center}
\newcommand{\nameslang}{\scriptstyle B}
\newcommand{\inputt}{\scriptstyle A}
\newcommand{\program}{$\Gamma$}
\newcommand{\parteval}{\gamma}
\newcommand{\Dott}{\Large$\bullet$}
\def\JPicScale{.4}
\begin{center}
\input{PIC/hartley.tex}
\end{center}
\caption{Any program transformer $\gamma$ has a fixpoint $\Gamma:\DP$ such that $\uev{\gamma \Gamma} = \uev \Gamma$}
\label{Fig:hartley}
\end{center}
\end{figure}
A transformer fixpoint can be constructed as a Kleene fixpoint of $g(p,x) = \uev{\gamma p}x$, by applying the Fundamental Theorem. The other way around, for any computable \funnn\ $g\colon \DP\times A\to B$,  \eqref{eq:uev} gives a $\DP$-indexed program $G\colon \DP \stricto \DP$ susch that $g(p,a) = \uev{G_{p}}a$. As a \strict\ \funnn, any $\DP$-indexed program can also be viewed as a program transformer. By assumption, it has a fixpoint $\Gamma$ with $\uev{\Gamma}a = \uev{G_{\Gamma}}a = g(\Gamma,a)$. So $\Gamma$ is a Kleene fixpoint of $g$.

\subsection{Example: Polymorphic quine}\sindex{quine}\label{Sec:quine}
A \emph{quine\/} \sindex{quine} is a program $Q$ that, when executed, outputs its own text:
\bear
\uev Q x & = & Q 
\eear
A simple program that outputs a text $Q$ is usually in the form ``{\tt print $'Q'$}''. But that program is obviously longer than $Q$ and does not output all of its text. A quine cannot contain its own text as a quote, but has to somehow compress a version of itself, and decompress it at the output. The entries in the annual quine competitions mostly use peculiarities of particular programming languages to achieve this. The Fundamental Theorem of Computation allows constructing \emph{polymorphic}\/ quines, that can be constructed in all programming languages. The Fundamental Theorem provides a \emph{polymorphic}\/ quine construction, implementable in any programming language\footnote{It is assumed that all programming languages are Turing complete, in the sense of Sec.~\ref{Sec:CTT}. The languages that are not Turing complete do not allow computing all computable functions.}  A quine can obtained as a Kleene fixpoint $Q$ of the projection $\pi_{\DP} : \DP\times A \to \DP$, where $\pi_{\DP}(x,y) = x$, so that
\[ \{Q\} y\ \ \ \  =\ \ \ \ \ \ \pi_{\DP}\big(Q, y\big)\ \ \ \ \ \  = \ \ \ \ Q\]
\vspace{-\baselineskip}
\newcommand{\inputt}{\scriptstyle A}
\newcommand{\gee}{\pi}
\newcommand{\program}{$Q$}
\def\JPicScale{.4}
\begin{center}
\ifx\JPicScale\undefined\def\JPicScale{1}\fi
\unitlength \JPicScale mm
\begin{picture}(185,77.5)(0,0)
\linethickness{0.2mm}
\put(0,65){\line(1,0){50}}
\linethickness{0.2mm}
\put(0,5.62){\line(0,1){25}}
\linethickness{0.2mm}
\put(50,40){\line(0,1){25}}
\linethickness{0.2mm}
\put(30,65){\line(0,1){11.88}}
\put(37.5,0.62){\makebox(0,0)[cr]{$\inputt$}}

\linethickness{0.2mm}
\put(25,40){\line(1,0){25}}
\linethickness{0.2mm}
\put(40,-0.62){\line(0,1){40.62}}
\linethickness{0.2mm}
\put(0,5.62){\line(1,0){25}}
\put(31.25,52.5){\makebox(0,0)[cc]{\universal}}

\put(6.88,11.88){\makebox(0,0)[cc]{\program}}

\put(65,50){\makebox(0,0)[cc]{\EQLS}}

\linethickness{0.2mm}
\multiput(0,30.62)(0.12,-0.12){208}{\line(1,0){0.12}}
\linethickness{0.45mm}
\put(12.5,18.12){\line(0,1){34.38}}
\linethickness{0.2mm}
\multiput(0,65)(0.12,-0.12){208}{\line(1,0){0.12}}
\linethickness{0.2mm}
\put(80,65){\line(1,0){50}}
\linethickness{0.2mm}
\put(77.5,5.62){\line(0,1){25}}
\linethickness{0.2mm}
\put(130,40){\line(0,1){25}}
\put(117.5,0.62){\makebox(0,0)[cr]{$\inputt$}}

\linethickness{0.2mm}
\put(80,40){\line(1,0){50}}
\linethickness{0.2mm}
\put(120,-0.62){\line(0,1){53.12}}
\linethickness{0.2mm}
\put(77.5,5.62){\line(1,0){25}}
\put(84.38,11.88){\makebox(0,0)[cc]{\program}}

\linethickness{0.2mm}
\multiput(77.5,30.62)(0.12,-0.12){208}{\line(1,0){0.12}}
\linethickness{0.45mm}
\put(90,18.12){\line(0,1){59.38}}
\linethickness{0.2mm}
\put(80,40){\line(0,1){25}}
\put(105,52.5){\makebox(0,0)[cc]{$\gee$}}

\linethickness{0.2mm}
\put(160,40.62){\line(0,1){25}}
\linethickness{0.2mm}
\put(160,40.62){\line(1,0){25}}
\put(167.5,46.88){\makebox(0,0)[cc]{\program}}

\linethickness{0.2mm}
\multiput(160,65.62)(0.12,-0.12){208}{\line(0,-1){0.12}}
\linethickness{0.45mm}
\put(172.5,53.12){\line(0,1){24.38}}
\put(145,50){\makebox(0,0)[cc]{\EQLS}}

\put(170,0){\makebox(0,0)[cr]{$\inputt$}}

\linethickness{0.2mm}
\put(172.5,0){\line(0,1){30}}
\put(120,52.5){\makebox(0,0)[cc]{\Dottt}}

\put(90,18.12){\makebox(0,0)[cc]{\Dottt}}

\put(12.5,18.12){\makebox(0,0)[cc]{\Dottt}}

\put(172.5,53.12){\makebox(0,0)[cc]{\Dottt}}

\put(172.5,30){\makebox(0,0)[cc]{\Dottt}}

\end{picture}

\end{center}
Unfolding Fig.~\ref{Fig:Klepro} for $g=\pi_{\DP}$ shows that $Q$ is the partial evaluation $\pev \Pi \Pi$, where $\Pi$ is a program for $\pi_{\DP}(\pev x x, y)$. The \funnn\ $\uev Q y = \uev{\pev\Pi \Pi}y = \uev\Pi(\Pi,y)$ thus deletes $y$ and partially evaluates $\Pi$ on itself. The program $Q$ thus only contains $\Pi$. Running $Q$ outputs $Q$ as the partial evaluation of $\Pi$ on $\Pi$.

\subsection{Example: Polymorphic virus}
\label{Sec:virus}\sindex{virus}
A \emph{virus\/} \sindex{virus} is a program $V$ that performs some arbitrary computation $f:A\to B$ and moreover outputs copies of itself:
\bear
\{V\}x& = & \left<V, V, f(x)\right> 
\eear
The Fundamental Theorem of Computation yields again. Apply the Fundamental Theorem to 
\bear
v(p,x) & = & \left<p,p, f(x)\right>
\eear
The Kleene fixpoint is then a virus $V$:
\[ \{V\} x\hspace{3em} = \hspace{3em} v\big(V, x\big) \hspace{3em} = \hspace{3em} \left<V, V, f(x)\right)\]
\vspace{-1\baselineskip}
\newcommand{\outputt}{\scriptstyle B}
\renewcommand{\gee}{\scriptstyle v}
\newcommand{\eff}{$f$}
\renewcommand{\program}{$V$}
\def\JPicScale{.4}
\begin{center}
\ifx\JPicScale\undefined\def\JPicScale{1}\fi
\unitlength \JPicScale mm
\begin{picture}(239.38,78.75)(0,0)
\linethickness{0.2mm}
\put(0,68.75){\line(1,0){50}}
\linethickness{0.2mm}
\put(0,6.87){\line(0,1){25}}
\linethickness{0.2mm}
\put(50,43.75){\line(0,1){25}}
\linethickness{0.2mm}
\put(40,68.75){\line(0,1){10}}
\put(37.5,0){\makebox(0,0)[cr]{$\inputt$}}

\linethickness{0.2mm}
\put(25,43.75){\line(1,0){25}}
\linethickness{0.2mm}
\put(40,0.63){\line(0,1){43.12}}
\linethickness{0.2mm}
\put(0,6.87){\line(1,0){25}}
\put(31.25,56.25){\makebox(0,0)[cc]{\universal}}

\put(7.5,13.75){\makebox(0,0)[cc]{\program}}

\put(60,56.25){\makebox(0,0)[cc]{\EQLS}}

\linethickness{0.2mm}
\multiput(0,31.87)(0.12,-0.12){208}{\line(1,0){0.12}}
\linethickness{0.4mm}
\put(12.5,20){\line(0,1){36.25}}
\linethickness{0.2mm}
\multiput(0,68.75)(0.12,-0.12){208}{\line(1,0){0.12}}
\linethickness{0.2mm}
\put(70,71.25){\line(1,0){70}}
\linethickness{0.2mm}
\put(77.5,6.87){\line(0,1){25}}
\linethickness{0.2mm}
\put(140,41.25){\line(0,1){30}}
\put(117.5,0){\makebox(0,0)[cr]{$\inputt$}}

\linethickness{0.2mm}
\put(70,41.25){\line(1,0){70}}
\linethickness{0.2mm}
\put(120,0.63){\line(0,1){47.5}}
\linethickness{0.2mm}
\put(77.5,6.87){\line(1,0){25}}
\put(85,13.75){\makebox(0,0)[cc]{\program}}

\linethickness{0.2mm}
\multiput(77.5,31.87)(0.12,-0.12){208}{\line(1,0){0.12}}
\linethickness{0.4mm}
\put(90,20){\line(0,1){32.5}}
\linethickness{0.2mm}
\put(70,41.25){\line(0,1){30}}
\put(78.12,48.75){\makebox(0,0)[cc]{$\gee$}}

\linethickness{0.2mm}
\put(155,8.12){\line(0,1){25}}
\linethickness{0.2mm}
\put(155,8.12){\line(1,0){25}}
\put(162.5,15){\makebox(0,0)[cc]{\program}}

\linethickness{0.2mm}
\multiput(155,33.12)(0.12,-0.12){208}{\line(1,0){0.12}}
\linethickness{0.4mm}
\put(167.5,20.62){\line(0,1){58.13}}
\put(150.62,56.87){\makebox(0,0)[cc]{\EQLS}}

\linethickness{0.2mm}
\put(90,52.5){\circle*{5.16}}

\linethickness{0.4mm}
\put(5,68.75){\line(0,1){10}}
\linethickness{0.4mm}
\put(21.25,68.75){\line(0,1){10}}
\linethickness{0.4mm}
\put(78.12,65){\line(0,1){13.13}}
\linethickness{0.2mm}
\put(108.75,63.13){\line(1,0){25}}
\linethickness{0.2mm}
\put(108.75,48.13){\line(1,0){25}}
\linethickness{0.2mm}
\put(108.75,48.13){\line(0,1){15}}
\linethickness{0.2mm}
\put(133.75,48.12){\line(0,1){15.01}}
\put(120.62,55.63){\makebox(0,0)[cc]{\eff}}

\linethickness{0.2mm}
\put(120,63.13){\line(0,1){15}}
\linethickness{0.4mm}
\multiput(78.12,65)(0.12,-0.13){99}{\line(0,-1){0.13}}
\linethickness{0.4mm}
\put(101.88,65){\line(0,1){13.12}}
\linethickness{0.4mm}
\multiput(90,52.5)(0.12,0.13){99}{\line(0,1){0.13}}
\put(122.5,77.5){\makebox(0,0)[cl]{$\outputt$}}

\linethickness{0.2mm}
\put(185,7.49){\line(0,1){25.01}}
\linethickness{0.2mm}
\put(185,7.49){\line(1,0){25}}
\put(192.5,14.37){\makebox(0,0)[cc]{\program}}

\linethickness{0.2mm}
\multiput(185,32.5)(0.12,-0.12){208}{\line(0,-1){0.12}}
\linethickness{0.4mm}
\put(197.5,20){\line(0,1){58.13}}
\put(223.75,0){\makebox(0,0)[cr]{$\inputt$}}

\linethickness{0.2mm}
\put(225.62,0.63){\line(0,1){47.5}}
\linethickness{0.2mm}
\put(214.38,63.13){\line(1,0){25}}
\linethickness{0.2mm}
\put(214.38,48.13){\line(1,0){25}}
\linethickness{0.2mm}
\put(214.38,48.13){\line(0,1){15}}
\linethickness{0.2mm}
\put(239.38,48.12){\line(0,1){15.01}}
\put(226.25,55.63){\makebox(0,0)[cc]{\eff}}

\linethickness{0.2mm}
\put(225.62,63.13){\line(0,1){15}}
\linethickness{0.2mm}
\put(90,18.75){\circle*{5.16}}

\linethickness{0.2mm}
\put(167.5,20){\circle*{5.16}}

\linethickness{0.2mm}
\put(197.5,20){\circle*{5.16}}

\linethickness{0.2mm}
\put(12.5,20){\circle*{5.16}}

\put(228.75,76.25){\makebox(0,0)[cl]{$\outputt$}}

\put(42.5,77.5){\makebox(0,0)[cl]{$\outputt$}}

\end{picture}

\end{center}
Explain how $V$ produces the two copies of itself without storing a copy of itself anywhere.

\section[$\Upsilon$-combinators and their classifiers]{$\Upsilon$-combinators and their classifiers}\label{Sec:Ycomb}\sindex{Y-combinator@$\Upsilon$-combinator}

The Fundamental Theorem of Computation constructs a Kleene fixpoint of computable \funnn s. But since computable functions are programmable, the Fundamental Theorem can be implemented as a computable program transformer, that inputs programs and outputs Kleene fixpoints of the computations that they implement. The programs that implement such program transformers are called the \emph{$\Upsilon$-combinators}. They are constructed as Kleene fixpoints of suitable computable \funnn s. There are many different ways to construct them. Interestingly, there are also programs that recognize all possible $\Upsilon$-combinators. They are called the \emph{$\Upsilon$-combinator classifiers}. And yes, they are also constructed as Kleene fixpoints.

The circuitous constructions of the various fixpoint constructions continue to sound complicated, and there is a sense in which they are complex when presented sequentially. But their diagrammatic programs tend to be simple and insightful.

\subsection{$\Upsilon$-combinator constructions}\sindex{Y-combinator@$\Upsilon$-combinator}

\para{$\Upsilon$-combinator for fixpoints} is a program $\Upsa$ for a \funnn\ $\upsa:\DP\to A$ that inputs programs for functions $A\to A$ and outputs their fixpoints $\upsa(p)\colon A$, as in Sec.~\ref{Sec:divergence}. The equation 
\bea
\uev p \left( \upsa(p)\right)& = &\upsa(p)
\eea
thus holds for all $p:\DP$. The $\Upsa$-combinator is constructed as the Kleene fixpoint in
\beq\label{eq:Upsa}
\begin{split}
\newcommand{\PsiT}{\scriptstyle \Upsa}
\newcommand{\psiT}{\upsa}
\newcommand{\Atype}{\scriptstyle A}
\newcommand{\Progtype}{\scriptstyle \DP}
\newcommand{\universallift}{\universal}
\def\JPicScale{.4}
\ifx\JPicScale\undefined\def\JPicScale{1}\fi
\psset{unit=\JPicScale mm}
\psset{linewidth=0.3,dotsep=1,hatchwidth=0.3,hatchsep=1.5,shadowsize=1,dimen=middle}
\psset{dotsize=0.7 2.5,dotscale=1 1,fillcolor=black}
\psset{arrowsize=1 2,arrowlength=1,arrowinset=0.25,tbarsize=0.7 5,bracketlength=0.15,rbracketlength=0.15}
\begin{pspicture}(0,0)(130,105)
\psline[linewidth=0.55](50,35)(50,55)
\psline[linewidth=0.55](30,35)(50,35)
\psline[linewidth=1.05](40,35)(40,0)
\newrgbcolor{userFillColour}{0.2 0 0.2}
\rput{0}(40,15){\psellipse[linewidth=0.55,fillcolor=userFillColour,fillstyle=solid](0,0)(2.58,-2.58)}
\psline[linewidth=1.05](20,75)(0,55)
\psline[linewidth=0.55](50,75)(50,95)
\psline[linewidth=1.05](20,85.01)(20,75)
\psline[linewidth=0.55](10,95)(30,75)
\psline[linewidth=0.55](130,35)(130,55)
\psline[linewidth=0.55](110,35)(130,35)
\psline[linewidth=1.05](120,35)(120,0)
\psline[linewidth=1.05](100,45.01)(100,15)
\psline[linewidth=0.55](90,55)(110,35)
\psline[linewidth=0.55](10,25)(10,5)
\psline[linewidth=0.55](30,5)(10,5)
\psline[linewidth=0.55](90,25)(90,5)
\psline[linewidth=0.55](90,25)(110,5)
\psline[linewidth=0.55](110,5)(90,5)
\psline[linewidth=0.55](10,25)(30,5)
\rput(70,45){\EQLS}
\rput(15,10){$\PsiT$}
\rput(95,10){$\PsiT$}
\rput(37.5,45){$\universal$}
\rput(117.5,45){$\universal$}
\rput[r](117.5,-1.25){$\Progtype$}
\rput[l](125,75){$\Atype$}
\rput[l](44.38,105){$\Atype$}
\rput[l](44.38,64.38){$\Atype$}
\rput[r](37.5,-1.25){$\Progtype$}
\psline[linewidth=0.55](90,55)(130,55)
\psline[linewidth=0.55](120,75)(120,55)
\psline[linewidth=0.55](30,75)(50,75)
\psline[linewidth=0.55](10,55)(50,55)
\psline[linewidth=0.55](40,75)(40,55)
\psline[linewidth=0.55](10,95)(50,95)
\psline[linewidth=0.55](40,105)(40,95)
\rput(38.75,86.25){$\universallift$}
\newrgbcolor{userFillColour}{0.2 0 0.2}
\rput{0}(100,15){\psellipse[linewidth=0.55,fillcolor=userFillColour,fillstyle=solid](0,0)(2.58,-2.58)}
\psline[linewidth=1.05](0,55)(40,15)
\psline[linewidth=1.05,border=1.3](20,45)(20,15)
\newrgbcolor{userFillColour}{0.2 0 0.2}
\rput{0}(20,15){\psellipse[linewidth=0.55,fillcolor=userFillColour,fillstyle=solid](0,0)(2.58,-2.58)}
\psline[linewidth=0.55](10,55)(30,35)
\end{pspicture}

\end{split}
\eeq

\para{$\Upsilon$-combinator for fixed computations} is a program $\Upsb$ for a program transformer $\upsb:\DP\stricto \DP$ that inputs programs for \funnn s $A\to A$ and outputs programs for some fixed \funnn s $A\to A$. More precisely, for every $p\colon \DP$, the program $\upsb(p)\colon \DP$ satisfies
\beq
\uev{p} \circ \uev {\upsb(p)} = \uev {\upsb(p)}
\eeq
The $\Upsb$-combinator is obtained by applying the Fundamental Theorem again.
\beq\label{eq:Upsb}
\begin{split}
\newcommand{\PsiT}{\scriptstyle \Upsb}
\newcommand{\psiT}{\upsb}
\newcommand{\partev}{\prtial}
\newcommand{\Atype}{\scriptstyle A}
\newcommand{\Progtype}{\scriptstyle \DP}
\newcommand{\universallift}{\universal}
\newcommand{\smalluniversal}{\scriptstyle \{\}}
\def\JPicScale{.4}
\ifx\JPicScale\undefined\def\JPicScale{1}\fi
\psset{unit=\JPicScale mm}
\psset{linewidth=0.3,dotsep=1,hatchwidth=0.3,hatchsep=1.5,shadowsize=1,dimen=middle}
\psset{dotsize=0.7 2.5,dotscale=1 1,fillcolor=black}
\psset{arrowsize=1 2,arrowlength=1,arrowinset=0.25,tbarsize=0.7 5,bracketlength=0.15,rbracketlength=0.15}
\begin{pspicture}(0,0)(270,115)
\psline[linewidth=0.75](65,45)(65,65)
\psline[linewidth=0.75](30,45)(65,45)
\psline[linewidth=1.3](40,45)(40,0)
\newrgbcolor{userFillColour}{0.2 0 0.2}
\rput{0}(40,25){\psellipse[fillcolor=userFillColour,fillstyle=solid](0,0)(2.58,-2.58)}
\psline[linewidth=1.3](30,95)(0,65)
\psline[linewidth=1.3](0,65)(40,25)
\psline[linewidth=0.75](60,85)(60,105)
\psline[linewidth=0.75](40,85)(60,85)
\psline[linewidth=0.75](20,105)(40,85)
\psline[linewidth=0.75](165,45)(165,65)
\psline[linewidth=0.75](130,45)(165,45)
\psline[linewidth=1.3](140,45)(140,0)
\psline[linewidth=1.3](120,55.01)(120,25)
\psline[linewidth=0.75](110,65)(130,45)
\psline[linewidth=0.75](10,35)(10,15)
\psline[linewidth=0.75](30,15)(10,15)
\psline[linewidth=0.75](110,35)(110,15)
\psline[linewidth=0.75](110,35)(130,15)
\psline[linewidth=0.75](130,15)(110,15)
\psline[linewidth=1.1,border=1.5](20,55)(20,25)
\psline[linewidth=0.75](10,35)(30,15)
\psline[linewidth=0.75](10,65)(30,45)
\rput(85,55){\EQLS}
\psline[linewidth=0.75](270,55)(270,75)
\psline[linewidth=0.75](232.5,35)(252.5,35)
\psline[linewidth=1.3](242.5,35)(242.5,0)
\psline[linewidth=1.3](222.5,45)(222.5,25)
\psline[linewidth=0.75](212.5,55)(232.5,35)
\psline[linewidth=0.75](212.5,35)(212.5,15)
\psline[linewidth=0.75](212.5,35)(232.5,15)
\psline[linewidth=0.75](232.5,15)(212.5,15)
\rput(185,55){\EQLS}
\psline[linewidth=0.75](212.5,55)(232.5,55)
\psline[linewidth=0.75](232.5,55)(252.5,35)
\psline[linewidth=0.75](232.5,75)(252.5,55)
\psline[linewidth=1.3](242.5,65)(242.5,45)
\psline[linewidth=0.5](235,62.5)(257.5,40)
\psline[linewidth=0.5](205,62.5)(205,10)
\psline[linewidth=0.5](235,62.5)(205,62.5)
\psline[linewidth=0.5](257.5,10)(205,10)
\psline[linewidth=0.5](257.5,10)(257.5,40)
\rput(15,20){$\PsiT$}
\rput(115,20){$\PsiT$}
\rput(217.5,20){$\PsiT$}
\rput(251.25,16.25){$\psiT$}
\rput(47.5,55){$\universal$}
\rput(147.5,55){$\universal$}
\rput(255,65){$\smalluniversal$}
\rput(232.5,45){$\partev$}
\rput[r](137.5,2.5){$\Progtype$}
\rput[r](240,2.5){$\Progtype$}
\rput[l](267.5,85){$\Atype$}
\rput[l](152.5,82.5){$\Atype$}
\rput[l](52.5,113.75){$\Atype$}
\rput[l](52.5,73.75){$\Atype$}
\rput[r](37.5,2.5){$\Progtype$}
\psline[linewidth=0.75](57.5,45)(57.5,0)
\psline[linewidth=0.75](157.5,45)(157.5,0)
\psline[linewidth=0.75](252.5,55)(270,55)
\psline[linewidth=0.75](265,55)(265,0)
\rput[r](262.5,2.5){$\Atype$}
\rput[r](155,2.5){$\Atype$}
\rput[r](55,2.5){$\Atype$}
\psline[linewidth=0.75](10,65)(65,65)
\psline[linewidth=0.75](20,105)(60,105)
\psline[linewidth=0.75](50,115)(50,105)
\psline[linewidth=0.75](110,65)(165,65)
\psline[linewidth=0.75](150,85)(150,65)
\psline[linewidth=0.75](232.5,75)(270,75)
\psline[linewidth=0.75](265,86.25)(265,75)
\psline[linewidth=0.75](50,85)(50,65)
\rput(48.12,96.25){$\universallift$}
\newrgbcolor{userFillColour}{0.2 0 0.2}
\rput{0}(20,25){\psellipse[fillcolor=userFillColour,fillstyle=solid](0,0)(2.58,-2.58)}
\newrgbcolor{userFillColour}{0.2 0 0.2}
\rput{0}(120,25){\psellipse[fillcolor=userFillColour,fillstyle=solid](0,0)(2.58,-2.58)}
\newrgbcolor{userFillColour}{0.2 0 0.2}
\rput{0}(222.5,25){\psellipse[fillcolor=userFillColour,fillstyle=solid](0,0)(2.58,-2.58)}
\newrgbcolor{userFillColour}{0.2 0 0.2}
\rput{0}(242.5,45){\psellipse[fillcolor=userFillColour,fillstyle=solid](0,0)(2.58,-2.58)}
\end{pspicture}

\end{split}
\eeq

\para{$\Upsilon$-combinator for Kleene fixpoints} is a program $\Upsc$ for a program transformer $\upsc:\DP\stricto\DP$ which inputs programs and outputs their Kleene fixpoints as in Sec.~\ref{Sec:kleene}, i.e. satisfies
\beq
\Big\{p\Big\} \big(\upsc(p), x\big) = \uev {\upsc(p)}(x)
\eeq
for all $p:\DP$ and $x:A$. Here is a Kleene fixpoint construction of a program for a uniform Kleene fixpoint constructor:
\beq\label{eq:Upsc}
\begin{split}
\newcommand{\Atype}{\scriptstyle A}
\newcommand{\PsiT}{\scriptstyle \Upsc}
\newcommand{\psiT}{\upsc}
\newcommand{\partev}{\prtial}
\newcommand{\smalluniversal}{\{\}}
\newcommand{\Btype}{\scriptstyle B}
\newcommand{\Progtype}{\scriptstyle \DP}
\def\JPicScale{.4}
\ifx\JPicScale\undefined\def\JPicScale{1}\fi
\psset{unit=\JPicScale mm}
\psset{linewidth=0.3,dotsep=1,hatchwidth=0.3,hatchsep=1.5,shadowsize=1,dimen=middle}
\psset{dotsize=0.7 2.5,dotscale=1 1,fillcolor=black}
\psset{arrowsize=1 2,arrowlength=1,arrowinset=0.25,tbarsize=0.7 5,bracketlength=0.15,rbracketlength=0.15}
\begin{pspicture}(0,0)(280,118.12)
\psline[linewidth=0.75](20,65)(40,65)
\psline[linewidth=1.3](50,85)(50,55)
\psline[linewidth=0.75](40,45)(60,45)
\psline[linewidth=1.3](50,45)(50,0)
\newrgbcolor{userFillColour}{0.2 0 0.2}
\rput{0}(50,25){\psellipse[fillcolor=userFillColour,fillstyle=solid](0,0)(2.58,-2.58)}
\psline[linewidth=1.3](30,95)(5,70)
\psline[linewidth=1.3](5,70)(50,25)
\psline[linewidth=0.75](77.5,85)(77.5,105)
\psline[linewidth=0.75](40,85)(77.5,85)
\psline[linewidth=0.75](20,105)(40,85)
\psline[linewidth=0.75](175,45)(175,65)
\psline[linewidth=0.75](140,45)(175,45)
\psline[linewidth=1.3](150,45)(150,0)
\psline[linewidth=1.3](130,55.01)(130,25)
\psline[linewidth=0.75](120,65)(140,45)
\psline[linewidth=0.75](20,35)(20,15)
\psline[linewidth=0.75](40,15)(20,15)
\psline[linewidth=0.75](120,35)(120,15)
\psline[linewidth=0.75](120,35)(140,15)
\psline[linewidth=0.75](140,15)(120,15)
\psline[linewidth=1.3,border=1.3](30,55)(30,25)
\psline[linewidth=0.75](20,35)(40,15)
\psline[linewidth=0.75](20,65)(40,45)
\rput(95,55){\EQLS}
\psline[linewidth=0.75](280,55)(280,75)
\psline[linewidth=0.75](242.5,35)(262.5,35)
\psline[linewidth=1.3](252.5,35)(252.5,0)
\psline[linewidth=1.3](232.5,45)(232.5,25)
\psline[linewidth=0.75](222.5,55)(242.5,35)
\psline[linewidth=0.75](222.5,35)(222.5,15)
\psline[linewidth=0.75](222.5,35)(242.5,15)
\psline[linewidth=0.75](242.5,15)(222.5,15)
\rput(195,55){\EQLS}
\psline[linewidth=0.75](222.5,55)(242.5,55)
\psline[linewidth=0.75](242.5,55)(262.5,35)
\psline[linewidth=0.75](242.5,75)(262.5,55)
\psline[linewidth=1.3](252.5,65)(252.5,45)
\psline[linewidth=0.5](245,62.5)(267.5,40)
\psline[linewidth=0.5](215,62.5)(215,10)
\psline[linewidth=0.5](245,62.5)(215,62.5)
\psline[linewidth=0.5](267.5,10)(215,10)
\psline[linewidth=0.5](267.5,10)(267.5,40)
\rput(25,20){$\PsiT$}
\rput(125,20){$\PsiT$}
\rput(227.5,20){$\PsiT$}
\rput(261.25,16.25){$\psiT$}
\rput(57.5,95){$\universal$}
\rput(157.5,55){$\universal$}
\rput(265,65){$\smalluniversal$}
\rput(242.5,45){$\partev$}
\rput[r](147.5,2.5){$\Progtype$}
\rput[r](250,2.5){$\Progtype$}
\rput[r](47.5,2.5){$\Progtype$}
\psline[linewidth=0.75](67.5,85)(67.5,0)
\psline[linewidth=0.75](167.5,45)(167.5,0)
\psline[linewidth=0.75](262.5,55)(280,55)
\psline[linewidth=0.75](275,55)(275,0)
\rput[r](272.5,2.5){$\Atype$}
\rput[r](165,2.5){$\Atype$}
\rput[r](65,2.5){$\Atype$}
\psline[linewidth=0.75](40,65)(60,45)
\rput(40,55){$\partev$}
\rput[l](52.5,70){$\Progtype$}
\psline[linewidth=0.75](20,105)(77.5,105)
\psline[linewidth=0.75](67.5,117.5)(67.5,105)
\psline[linewidth=0.75](120,65)(175,65)
\psline[linewidth=0.75](160,85)(160,65)
\psline[linewidth=0.75](242.5,75)(280,75)
\psline[linewidth=0.75](275,86.25)(275,75)
\rput[l](71.88,118.12){$\Btype$}
\rput[l](164.38,85.62){$\Btype$}
\rput[l](280,86.88){$\Btype$}
\newrgbcolor{userFillColour}{0.2 0 0.2}
\rput{0}(30,25){\psellipse[fillcolor=userFillColour,fillstyle=solid](0,0)(2.58,-2.58)}
\newrgbcolor{userFillColour}{0.2 0 0.2}
\rput{0}(130,25){\psellipse[fillcolor=userFillColour,fillstyle=solid](0,0)(2.58,-2.58)}
\newrgbcolor{userFillColour}{0.2 0 0.2}
\rput{0}(232.5,25){\psellipse[fillcolor=userFillColour,fillstyle=solid](0,0)(2.58,-2.58)}
\newrgbcolor{userFillColour}{0.2 0 0.2}
\rput{0}(252.5,45){\psellipse[fillcolor=userFillColour,fillstyle=solid](0,0)(2.58,-2.58)}
\newrgbcolor{userFillColour}{0.2 0 0.2}
\rput{0}(50,55){\psellipse[fillcolor=userFillColour,fillstyle=solid](0,0)(2.58,-2.58)}
\end{pspicture}

\end{split}
\eeq

\subsection{$\Upsilon$-combinator classifiers}\label{Sec:classifier}\sindex{Y-combinator@$\Upsilon$-combinator!classifiers}
For any pair of types $A,B$ there are program transformers $\psia, \psib, \psic:\DP\stricto\DP$ whose fixed points are exactly the corresponding fixpoint operators, i.e. for every $y \in \DP$ the following equivalences hold
\begin{alignat}{7} 
\uev{\psia(y)} & = \Big\{y\Big\} & \uev{\psib(y)} &=   \Big\{y\Big\} &  \uev{\psic(y)} &= \Big\{y\Big\} \notag\\
& \Updownarrow & & \Updownarrow && \Updownarrow  \label{eq:classifiers}\\
\forall p.\ \Big\{p\Big\} \Big(\uev{y}p\Big) & =  \uev{y}p &
\qquad 
\forall p.\ \Big\{p\Big\} \circ \Big\{\uev y p\Big\}   & = \  \Big\{\uev y p\Big\} &
\qquad 
\forall p.\ \Big\{p\Big\}\Big(\uev yp, x \Big) & = \ \Big\{\uev y p\Big\}(x) \notag
\end{alignat}

\para{Classifier of $\Upsilon$-combinators for fixpoints.} Consider the equivalence in \eqref{eq:classifiers} on the left. 
The right-hand sides of the two equivalent equations become equal if both sides of the first equation are applied to the argment $p$. The equivalence will then hold if and only if the left-hand sides of the equations are equal. It is thus required that $\psia:\DP\stricto \DP$ satisfies
\bea
\uev{\psia(y)} p & = & \Big\{p\Big\}\Big(\uev y p\Big)
\eea
This will hold if $\psia(y) = \pev{\Psia} y$ where $\Psia$ is a program for 
\bea\label{eq:psia}
\uev{\Psia}(y,p) & = & \Big\{ p\Big\} \Big(\uev y p\Big)
\eea
Note that the $\Upsa$-combinator was defined in \eqref{eq:Upsa} as a Kleene fixpoint in $y$ of the  \funnn\  in \eqref{eq:psia}.

\para{Classifier of $\Upsilon$-combinators for fixed \funnn s.} To realize the equivalence in the middle of  \eqref{eq:classifiers}, we apply both sides of the first equation to $p$ again, and run the output as a program on $x$. If we run both sides of the second equation in the middle of  \eqref{eq:classifiers} on $x$ as well, then the right-hand sides of the two equivalent equations become identical again. For the equivalence to hold, it is now necessary and sufficient that $\psib:\DP\stricto \DP$ satisfies
\bea
\Big\{\big\{\psib(y)\big\} p\Big\}(x) & = & \Big\{p\Big\}\circ \Big\{\uev y p\Big\}(x)
\eea
To specify  $\psib$, this time we need the programs $\Psib_0$ and $\Psib_1$ which implement the \funnn s on the right in
\bea\label{eq:psib}
\uev{\Psib_0}(y,p,x) & = &  \Big\{p\Big\} \circ\Big\{\uev y p\Big\}(x)\\
\uev{\Psib_1}(y,p) & = & \pev{\Psib_0}(y,p)
\eea
to define $\psib(y) = \pev{\Psib_1}y$. The $\Upsb$-combinator was defined in \eqref{eq:Upsb} as a Kleene fixpoint in $y$ of the  \funnn\ in \eqref{eq:psib}.

\para{Classifier of $\Upsilon$-combinators for Kleene fixpoints.} Towards the equivalence in \eqref{eq:classifiers} on the right, proceeding as above we find that it is necessary and sufficient that $\psic:\DP\stricto \DP$ satisfies
\bea
\Big\{\big\{\psic(y)\big\} p\Big\}(x) & = & \Big\{p\Big\}\Big(\uev y p, x\Big)
\eea
We define $\Psic_0,\Psic_1:\DP$ by
\bea\label{eq:psic}
\uev{\Psic_0}(y,p,x) & = &  \Big\{p\Big\} \Big(\uev y p, x\Big)\\
\uev{\Psic_1}(y,p) & = & \pev{\Psic_0}(y,p)
\eea
and set $\psic(y) = \pev{\Psic_1}y$. The $\Upsc$-combinator was defined in \eqref{eq:Upsc} as a Kleene fixpoint in $y$ of the  \funnn\ in \eqref{eq:psic}.

\section{Software systems as systems of equations}\label{Sec:smullyan}

A software system is a family of programs that work together. \sindex{software} Since programs in practice usually work together, most programs belong in some software systems, and most programmers work as software developers. Keeping the programs in a software system together also requires a lot of \emph{meta}\/programming, of programs that compute programs. We get to metaprogramming and software engineering  in Ch.~\ref{Chap:Metaprog}. But mutual dependencies of programs that work together arise already on the level of fixpoints. The task of finding joint fixpoints of \funnn s arises in all engineering disciplines, and has been one of the central themes of mathematics. It is usually presented as the task of solving systems of equations. Here we sketch a basic method for solving systems of computable equations that can be used to specify some joint requirements from systems of programs. The upshot is that any family of computations can be bundled together into a software system: there is always a family of programs that code them together. The idea and the main theoretical developments go back to Raymond Smullyan \cite[Ch.~IX]{SmullyanR:recursion}. The brief diagrammatic treatment and the simplicity of applications conceal the significant obstacles that had to be surmounted.

\subsection{Smullyan fixpoints}\sindex{fixpoint!Smullyan fixpoint}\sindex{Smullyan fixpoint|see {fixpoint}}\label{Sec:joint}
{\it For any pair of computations 
\[ g:\DP\ttimes \DP \ttimes A\to B\qquad \mbox{and}\qquad h:\DP\ttimes \DP \ttimes C\to D\]
there is a pair of programs $G, H : \DP$ such that }
\[
g(G, H, x)\ = \ \uev{G}x \qquad \mbox{and}\qquad h(G, H, y)\ = \ \uev{H}y
\]

\begin{figure}[htbp]
\begin{center}
\renewcommand{\inputt}{\scriptstyle A}
\newcommand{\inputtt}{\scriptstyle C}
\renewcommand{\gee}{g}
\newcommand{\hee}{h}
\newcommand{\geehat}{\widehat g}
\newcommand{\heehat}{\widehat h}
\newcommand{\heefix}{\scriptstyle H}
\newcommand{\nameslang}{\scriptstyle B}
\newcommand{\nameslangg}{\scriptstyle D}
\renewcommand{\program}{$\scriptstyle \widehat G$}
\newcommand{\prohram}{\scriptstyle \widehat H}
\newcommand{\parteval}{\scriptstyle\prtial}
\newcommand{\progtype}{\scriptstyle \DP}
\newcommand{\Peeh}{\scriptstyle H}
\newcommand{\Peeg}{\scriptstyle G}\newcommand{\lhs}{}
\newcommand{\rhs}{}
\def\JPicScale{.3}
\begin{center}
\input{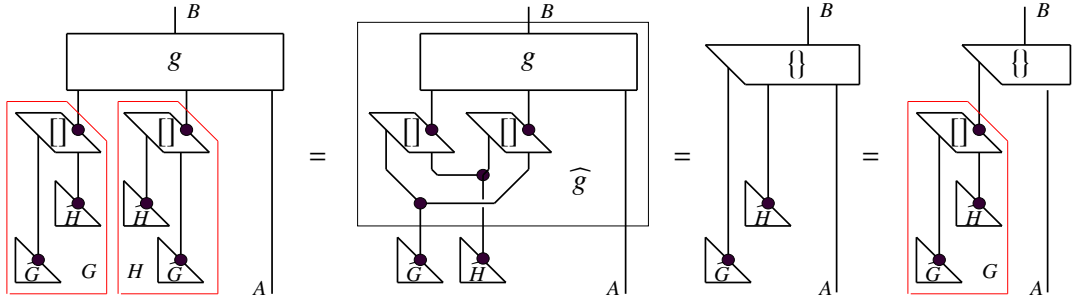}
\end{center}
\caption{A geometric view of \eqref{eq:peeg}}
\label{Fig:smullyan-three}
\end{center}
\end{figure}

\para{Constructing Smullyan fixpoints.} Let $\widehat G$ and $\widehat H$ be the Kleene fixed points of the functions $\widehat g : \DP \times \DP \times A \to B$ and $\widehat h : \DP \times \DP \times C \to D$ defined 
\bea\label{eq:ghat}
\widehat g(p,q,x) & = & g\left( \pev p q, \pev q p, x\right)\\
\widehat h(q,p,y) & = & h\left( \pev p q, \pev q p, y\right)
\label{eq:hhat}
\eea
and define
\[ G = \pev{\widehat G} \widehat H \qquad \qquad H = \pev{\widehat H} \widehat G \]
The constructions now give
\bea\label{eq:peeg}
g(G, H, x) &= & g\left(\pev{\widehat G}\widehat H, \pev{\widehat H}\widehat G, x\right)\ \stackrel{\eqref{eq:ghat}} =\ \widehat g\left(\widehat G, \widehat H, x\right) \ =\ \uev{\widehat G}\left(\widehat H, x\right)\ =\ \uev{\pev{\widehat G}\widehat H} x\ =\ \uev{G} x\\
h(G, H, y) &= & h\left(\pev{\widehat G}\widehat H, \pev{\widehat H}\widehat G, y\right)\ \stackrel{\eqref{eq:hhat}} =\ \widehat h\left(\widehat H, \widehat G, y\right) \ =\ \uev{\widehat H}\left(\widehat G, y\right)\ =\ \uev{\pev{\widehat H}\widehat G} y\ =\ \uev{H} y
\eea
\begin{figure}[!ht]
\begin{center}
\newcommand{\inputtt}{\scriptstyle C}
\newcommand{\hee}{h}
\newcommand{\geehat}{\widehat g}
\newcommand{\heehat}{\widehat h}
\newcommand{\nameslang}{\scriptstyle B}
\newcommand{\nameslangg}{\scriptstyle D}
\renewcommand{\program}{$\scriptstyle \widehat G$}
\newcommand{\prohram}{\scriptstyle \widehat H}
\newcommand{\parteval}{\scriptstyle\prtial}
\newcommand{\progtype}{\scriptstyle \DP}
\newcommand{\lhs}{}
\newcommand{\rhs}{}
\def\JPicScale{.4}
\begin{center}
\input{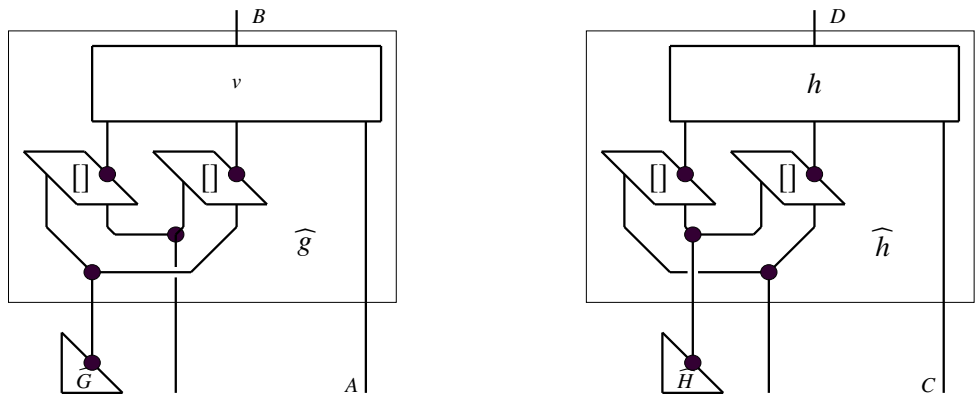}
\end{center}
\caption{Definitions of $\widehat G$ and $\widehat H$ as Kleene fixpoints}
\label{Fig:smullyan-two}
\end{center}
\end{figure}
The concept behind this algebraic magic emerges from the pictures of the functions $\widehat g$ and $\widehat h$, and of their fixpoints $\widehat G$ and $\widehat H$ in Fig.~\ref{Fig:smullyan-two}. The diagrammatic form of \eqref{eq:peeg} in Fig.~\ref{Fig:smullyan-three} shows how the construction follows the idea of the Fundamental Theorem, or more precisely of the construction of the Kleene fixpoint in Fig.~\ref{Fig:Klepro}.

\subsection{Systems of program equations}\label{Sec:system}\sindex{software specifications!as systems of equations}
{\it 
For any $n$-tuple of  computations 
\[g_i: \DP^n \ttimes A_i \to B_i \quad \mbox{for } i = 1, 2,\ldots, n
\]
each depending on $n$ programs, there are programs $G_1, G_2,\ldots, G_n : \DP$ that are fixed by the given computations, in the sense of the following equations:
\bear
g_1(G_1, G_2,\ldots, G_n, x_1)&= & \uev{G_1}x_1 \\
g_2(G_1, G_2,\ldots, G_n, x_2)&= & \uev{G_2}x_2\\
& \vdots & \\
g_n(G_1, G_2,\ldots, G_n, x_n)&= & \uev{G_n}x_n
\eear
}

\para{Solving systems of computable equations.} 
Define $\widehat g_i : \DP^n \times A_i \to B_i$ by 
\bear
\widehat g_i(p_1,p_2,\ldots, p_n,x_i) & = & g_i\left( \pev{p_1}\vec p_{-1},  \pev{p_2}\vec p_{-2},\ldots,  \pev{p_n}\vec p_{-n}, x_i\right)
\eear
where $\vec p_{-i} = \left(p_1,\ldots, p_{i-1}, p_{i+1},\ldots, p_n\right)$. Furthermore set
\bear
G_i  & = & \pev{\widehat G_i}\left(\widehat G_1,\ldots, \widehat G_{i-1}, \widehat G_{i+1},\ldots, \widehat G_n\right)
\eear
The constructions now give
\bear
g_i(G_1,G_2,\ldots, G_n, x_i) &= & \widehat g_i\left(\widehat G_1, \widehat G_2,\ldots, \widehat G_n, x_i\right) \\ 
& = & \uev{\widehat G_i}\left(\widehat G_1,\ldots, \widehat G_{i-1}, \widehat G_{i+1},\ldots, \widehat G_n, x_i\right)\\ 
& = & \uev{\pev{\widehat G_i}\left(\widehat G_1,\ldots, \widehat G_{i-1}, \widehat G_{i+1},\ldots, \widehat G_n\right)} x_i\\ 
& = &  \uev{G_i} x_i
\eear

%

\def\thechapter{4}
\setchaptertoc
\chapter{What can be computed}
\label{Chap:Numrec}
\newpage

\section{Reverse programming}\label{Sec:Reverse}

\sindex{programming!reverse}
In the practice of computation, programmers are given a programming language, with a basic type system and some program schemas, and they are asked to build programs and run program evaluators. The theory of computation goes the other way around: it starts from program evaluators and builds a basic type system and program schemas. This is the idea of \emph{reverse programming}.\sindex{reverse programming}  In the preceding chapter, we used the program evaluators to construct various fixpoints. In the present chapter, we construct the datatype $\NNn$ of natural numbers, with the basic arithmetic operations and the main program schemas that it supports. The program structures are derived from program evaluators and data services, and carried forward as syntactic sugar. Behind the sugar, programs are expressions built from typed instances of a single instruction \runn, composed using the data services. Such expressions comprise the Turing-complete programming language Run\sindex{programming!language!Run}.\sindex{run-instruction@\runn-instruction} This chapter spells out the basic program schemas in this language, and explains what it means to be Turing-complete in the end.

\section{Numbers and sequences}\label{Sec:Num}

\subsection{Counting numbers} \label{Sec:counting}\sindex{counting}
Numbers are built by counting. Counting is the physical process of assigning fingers or pebbles to apples, or sheep, or coins. Counting is the earliest form of computation. Our fingers were the first digital computers\footnote{The Latin word "digitus" means "finger".}. The first algorithm was the assignment of 10 fingers to 10 apples or sheep. The second algorithm may have been the idea to count the apples or sheep by the 5 fingers of one hand, and to use the 5 fingers of the other hand to count how many times the first hand was used. That would allow us to count up to 25 apples, instead of  just 10. The \emph{fourth}\/ algorithm could be to count up to 1024 apples using the 10 fingers. Before that, someone should have discovered a \emph{third}\/ algorithm, allowing them to count up to 36 with 10 fingers. How would you do that? The idea is that a hand with 5 flexible fingers has 6 states, and not just 5, since it can stretch 0, 1, 2, 3, 4, or 5 fingers, while keeping the rest bent. The step to counting $2^{10}=1024$ apples using the 10 fingers follows from the observation that each finger has 2 states: stretched and bent. In any case, noticing 0, as the state of where nothing has been counted, and using it in counting, turned out to be a revelation. 

\para{Only natural numbers.} The numbers built by counting are studied as the \emph{natural}\/ numbers. The integers are built by inverting the addition of natural numbers, the rationals by inverting the multiplication, and so on \cite{Conway-Guy:numbers}. For the moment, we focus on counting, and the present discussions about "numbers" invariably refer to the \emph{natural numbers}\/ from school.\index{natural numbers}

\subsection[Numbers as sets]{Numbers as sets} \label{Sec:num-set}
\begin{figure}[!ht]
\begin{center}
\def\JPicScale{.75}
\newcommand{\Nhh}{}
\newcommand{\zhh}{\scriptstyle 0}
\newcommand{\succe}{\scriptstyle 1+}
\input{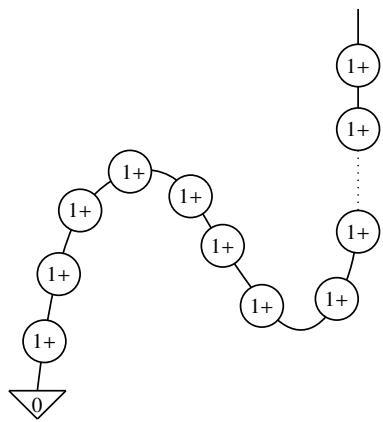}
\caption{The number $n$ is $n$-th successor of 0.}
\label{Fig:nats}
\end{center}
\end{figure}

If the counting process is required to produce an output, then it must involve from 0, as the state where there is nothing left to count, and the output is ready. 0 is the number of elements of the empty set, which we write $\{\}$. Since the empty set is unique, it is the best representative of the number of its elements, so we set $0=\{\}$. Nonempty sets are formed by enclosing some given elements between the curly brackets, like in  $\{a,b,c\}$. The elements can be some previously formed sets. So we can form the set $\{\{\}\}$, with a single element $\{\}$. We set $1=\{0\}$ and use it to denote the number of elements in all sets with a single element. There  may be many sets with a single element, as many as there are elements given in the world; but only one number $1$. But even if no elements were given, we certainly have the sets $0=\{\}$ and $1=\{\{\}\}$, which are different, since one is empty and the other is not. Now we can form the set  $2=\{0,1\}$, with two different elements. Continuing like this, we form the sequence of numbers
\bea
0 & = &  \{\} \notag \\
1 & = & \{0\}\ \ =\ \{\{\}\}\notag\\
2 & = & \{0,1\}\ \ =\ \  \{\{\},\  \{\{\}\}\} \notag\\
3 & = & \{0,1,2\} \ \ = \ \ \{\{\},\  \{\{\}\},\  \{\{\},\{\{\}\}\} \}\label{eq:numbers}\\
4 & = &   \{0,1,2,3\} \ \ = \ \ \{\{\},\  \{\{\}\},\  \{\{\},\{\{\}\}\},\  \{\{\},\{\{\}\},\{\{\},\{\{\}\}\} \}\}\notag\\
&\cdots\notag\\
n & = & \{0,1,2,\ldots, n-1\}\notag\\
n+1 & = & \{0,1,2,\ldots, n-1,n\}
\eea
After a long history of troubles with numbers, \sindex{numbers!von Neumann}from the Pythagoreans to Frege, the idea of counting starting from nothing was proposed by a 17-year-old Hungarian wunderkind Neumann \Janos, who later became John von Neumann. We saw him in Fig.~\ref{Fig:Entscheidung}, but see Appendix~\ref{Appendix:numnot} for an earlier photo. The process of counting was thus implemented as a process of set building. The numbers are built starting from nothing, i.e. the empty set as the \textbf{base case}, and adding at each \textbf{step case} a new element, obtained as the set of all previously built numbers:
%
\beq\label{eq:setconstructors-num}
0 = \{\}\qquad\qquad\qquad\qquad n+1  =  n \cup \{n\}
\eeq
Since all numbers are thus generated from $0$ by the operation $1+(-)$, they must all be in the form 
\bea\label{eq:plusone}
n & = &  \underbrace{1+1+\cdots 1+}_{n\mbox{ \small times}}0
\eea
In the computer, each number $n$ is thus in principle a wire with $n$ beads, as  depicted in Fig.~\ref{Fig:nats}. 

%
%

\subsection{Numbers as programs}\label{Sec:num-prog}\sindex{numbers!as programs}
There are many different ways to implement numbers as programs. A simple one goes like this:
\bea
\overline 0 & = & \pairr \true {\overline 0} \notag \\
\overline 1 & = & \pairr \flse {\overline 0}\notag\\
\overline 2 & = & \pairr \flse {\overline 1} \ =\ \pairr \flse {\pairr \flse {\overline 0}} \notag\\
\overline 3 & = & \pairr \flse {\overline 2} \ =\ \pairr \flse {\pairr \flse {\pairr \flse {\overline 0}}}\label{eq:num-moncom}\\
&\cdots\notag\\
\overline n & = & \pairr \flse {\overline{n-1}} \ =\ \lceil \flse \lceil \flse \lceil \cdots \lceil \flse,\overline 0 \underbrace{\rceil\rceil\cdots \rceil\rceil}_{n\ \mbox{\scriptsize layers}}\notag
\eea
where $\pairr - - :\DP\times \DP \to \DP$  is the pair encoding from \eqref{eq:tupling}, and $\true, \flse:\DP$ are the boolean truth values from \eqref{eq:truth}. This is a version of \eqref{eq:numbers} adapted for computers. 

\para{Notation.} We write $\overline n$ to denote a program chosen to represent in a computer "the" number $n$ from the "outer world". When no confusion is likely, the overlining is omitted.

\para{Programming counting.} The counting constructors \eqref{eq:setconstructors-num} can be implemented as the program constructors
\beq\label{eq:progconstructors-num}
\overline 0 \ = \ \pairr \true {\overline 0}\qquad\qquad\qquad\qquad \quad
\suce\left(\overline n\right)\ =\   \pairr \flse {\overline n}\ \ = \ \ \overline{n+1}
\eeq
Placing $\true$ as the prefix of $\overline 0$ and $\flse$ as the prefixes of the non-zero numbers $\overline{1+n}$ allows us to use the first projection as a \emph{zero-test}, and the second projection as the \emph{predecessor}\/ operation: 
\beq\label{eq:progtests-num}
\iszero\left(\overline n\right)\ =\ \fstt{\overline n}\   = \ \left. \begin{cases} \true & \mbox{ if } n=0 \\ \flse & \mbox{ otherwise}
\end{cases}\right\} \qquad\qquad\qquad \pred\left(\overline n\right) \ =\ \sndd{\overline n}\ = \  \left.\begin{cases}  
\overline{0} 
& \mbox{ if } n= 0\\
\overline{n-1} & \mbox{otherwise}
\end{cases}  \right\}
\eeq
The diagramms for the four operations defined in (\ref{eq:progconstructors-num}--\ref{eq:progtests-num}) are displayed in Fig.~\ref{Fig:zero-suc-pred}.
\begin{figure}[h!t]
\newcommand{\thhh}{\true}
\newcommand{\fhhh}{\flse}
\newcommand{\iee}{\scriptstyle \pairr - - }
\newcommand{\khee}{\scriptstyle \unpairr{-}}
\newcommand{\qee}{\{\}}
\newcommand{\succor}{\suce}
\newcommand{\zrr}{0}
\newcommand{\predor}{\pred}
\newcommand{\testor}{\iszero}
\newcommand{\Aee}{}
\newcommand{\blh}{\Theta}
\begin{center}
\def\JPicScale{.4}
\input{PIC/zero-suc-pred.tex}
\end{center}
\caption{Base $\overline 0 = \uev \Theta = \pairr{\true}{\uev\Theta}$, step up $\suce (x) = \pairr \flse x$, step down $\pred (x) =\sndd x$, test $\iszero x = \fstt x$}
\label{Fig:zero-suc-pred}
\end{figure}

\para{Different numbers are represented by different programs.} If any $\overline m = \overline n$ then either $m = n$, or the computer is degenerate, in the sense that all elements of all types are equal, and thus all types are isomorphic with the unity type $I$. 

We demonstrate this by showing that the following conditions are equivalent in any monoidal computer
\begin{enumerate}[{\rm a)}]
\item\label{prop:deg:a} there are different natural numbers $m\neq n$ with equal representations $\overline m = \overline n$;
\item\label{prop:deg:b} $\true= \flse$;
\item\label{prop:deg:c} the computer is trivial: for all $f,g:A\to B$ holds $f=g$, and thus all types are isomorphic.
\end{enumerate}

Since \eqref{prop:deg:c} obviously implies \eqref{prop:deg:a} and \eqref{prop:deg:b}, we just prove \eqref{prop:deg:a}$\implies$\eqref{prop:deg:b} and \eqref{prop:deg:b}$\implies$\eqref{prop:deg:c}. 

\begin{itemize}
\item {\eqref{prop:deg:a}$\implies$\eqref{prop:deg:b}} Suppose $m\lt n$. Then 
\bear
\overline m = \overline n & \implies & \overline 0\  =\  \pred^{n-1}\left( \overline m\right)\  =\   \pred^{n-1}\left( \overline n\right)  \ =\  \overline 1\\
\overline 0 = \overline 1 & \implies &  \true = \fstt{\pairr \true{\overline 0}} = \fstt{\overline 0} = \fstt{\overline 1} = \fstt{\pairr\flse {\overline 0}} = \flse 
\eear
\item {\bf \eqref{prop:deg:b}$\implies$\eqref{prop:deg:c}} For $f' ,g' :\DP\to \DP$, the assumption $\true = \flse$ gives
\[f\ =\  \uev \true (f,g) \ =\ \uev \flse (f,g)\ =\ g
\]
By \eqref{eq:fg-over} and \eqref{eq:fg}, any pair  $f,g:A\to B$ induces $\overline f = \incl_B\circ f \circ \retr_A$ and $\overline g = \incl_B\circ g \circ \retr_A$ such that $\overline f = \overline g$ if and only if $f=g$. But if all functions are equal, then all type retractions are isomorphisms. Since by Sec.~\ref{Sec:typ-idem} all types are retracts of $\DP$, it follows that all types are isomorphic, and thus isomorphic to the unit type $I$.
\end{itemize} 

\subsection{Type $\numidem$ as an idempotent}
The task is now to program an idempotent $\numidem :\DP\to \DP$ to filter out the numbers in the form \eqref{eq:num-moncom} from other programs, so that $\overline n\colon \numidem$ is captured as $\overline n = \numidem(\overline n)$. The idea is to map each $x:\DP$ as follows:
\begin{itemize}
\item if $x=\overline 0$, then output $\numidem(x) = x$;
\item if $x = \pairr \flse {y}$, then output $\numidem(x) = x$ if and only if $\numidem(y) = y$.
\end{itemize}
Since $y = \numidem(z)$ satisfies $\numidem(y) = y$ for any $z:\DP$,  this can be implemented informally as 
\bea\label{eq:numiedm-informal}
\numidem (x) & = & \begin{cases}
x & \mbox{ if } x=\overline 0 \\
\pairr {\fstt x} {\numidem \sndd x} &  \mbox{ if } \fstt x = \flse\\
\fixp & \mbox{ otherwise}
\end{cases}
\eea
Towards a formalization, this can be rewritten  in terms of the operations available in the monoidal computer as the \funnn\ $\numidemm:\DP\times \DP\to \DP$ where 
\bea\label{eq:IIF}
\numidemm (p,x) & = & \IIF\bigg(x\iseq \overline 0,\ x,\  \ \IIF\Big(\fstt x \iseq \flse,\  \ \suce\circ\uev p \circ \pred(x),\ \ \fixp \Big)\bigg)
\eea
Here we use the lazy $\IIF$-branching\sindex{branching!lazy} from \ref{Sec:better-branch} to avoid evaluating $\fixp$ when it is not selected. For a Kleene fixpoint $N:\DP$ of $\numidemm$, setting $\numidem = \uev{N}$ gives
\bear
\numidem(x) \ \ \ =\ \ \ \uev N x\ \ \ =\ \ \ \numidemm(N,x) & \ = \ & 
\IIF\bigg(x\iseq \overline 0,\ \ x,\ \ 
\IIF\Big(\fstt x \iseq \flse,\  \ \suce\Big(\numidem\big(\pred(x)\big)\Big),\ \ \fixp \Big)\bigg)
\eear
It easy to prove that $\numidem$ is a sub\funnn\ of the identity, and that it produces an output when the input is a number:
\bear
\numidem(x) = y & \iff & \exists n.\  x = \suce^n(\overline 0) = y
\eear
A convenient order of evaluation of $\nu$, not obvious in the above command-line form, is displayed in Fig.~\ref{Fig:num}.
\begin{figure}[!ht]
\def\JPicScale{.8}
\renewcommand{\DOTT}{\Dottt}
\newcommand{\branch}{\IIF}
\newcommand{\ueval}{\universal}
\newcommand{\Klenat}{N}
\newcommand{\sssor}{\suce}
\newcommand{\prdor}{\pred}
\newcommand{\testone}{\scriptstyle\fstt {\hspace{.25ex}} \stackrel ? = \flse}
\newcommand{\testtwo}{\scriptstyle\stackrel ? =\  \overline 0}
\newcommand{\Natls}{\mbox{\huge$\NNn$}}
\newcommand{\diverge}{\fixp}
\begin{center}
\input{PIC/num.tex}
\caption{The idempotent $\NNn:\DP\rightarrow \DP$ for the datatype of natural numbers}
\label{Fig:num}
\end{center}
\end{figure}

\para{Remark.} Replacing $x\iseq \overline 0$ with $\iszero x$ and $\fstt x \iseq \flse$ with with $\neg\fstt x$ would yield a type of \emph{intensional}\/ naturals, still supporting the same operations, but with multiple representatives of each number. The present definition fixes unique representatives as soon as the representatives of $0$ and $\flse$ are chosen.

\subsection{Sequences}\label{Sec:seq}\sindex{sequence}
A \strict\ \funnn\ $\beta:\NNn\stricto B$, taking numbers as  inputs, is usually presented by the \emph{sequence}\/ of its outputs
\[\seq{ \beta_0\, ,\ \beta_1\, ,\ \beta_2\, ,\ldots, \beta_i\, ,\ldots}\]
A \strict\ \funnn\ in the form $f:\NNn\times A\stricto B$ can then also be written as a sequence 
\[\seq{ f_0\, ,\ f_1\, ,\ f_2\, ,\ldots, f_i\, ,\ldots}\]
where\footnote{The notation is also used when $f$ is \strict\ only in the first argument, and $f_i$s are monoidal \funnn s.} the $f_i$s are \strict\ \funnn s $A\stricto B$.
While viewing \funnn s that vary over numbers as sequences is helpful, and writing the number inputs as subscripts often simplifies notation, note that all this is just a matter of notational convenience. Formally, \begin{itemize}
\item $\seq{\beta_i:B}_{i=0}^\infty$ is the sequence notation for $\beta:\NNn\stricto B$, and
\item $\seq{f_i:A\stricto B}_{i=0}^\infty$ is the sequence notation for $f: \NNn\times A\stricto B$. 
\end{itemize}

\section{Counting down: Induction and recursion}\label{Sec:Rec}

\subsection{Computing by counting}\label{Sec:indrec}
Counting, induction, and recursion are schemas used to specify sequences of increasing generality:
\begin{itemize}
\item \textbf{counting} builds the sequence of numbers $0, 1, 2,  \ldots : \NNn$ from $\{\}:\NNn$ and $\suce:\NNn\to\NNn$ by 
\beq 0=\{\}\qquad\qquad \qquad\qquad n+1 = \suce(n)\eeq
\item\textbf{induction} builds sequences of elements $\phhi_0, \phhi_1, \phhi_2, \ldots : B$ from $b:B$ and $\beeta:B\to B$ by \beq\label{eq:seq} \phhi_0 = b \qquad\qquad \qquad\qquad \phhi_{n+1} = \beeta(b_n)\eeq
\item\textbf{recursion} builds sequences of \funnn s $f_0, f_1, f_2, \ldots :A\to B$ from $g:A\to B$ and $h_0, h_1, h_2,\ldots :B\times A \to B$ by 
\beq f_0(x) = g(x) \qquad\qquad \qquad\qquad f_{n+1}(x) = h_n(f_n, x)\eeq 
\end{itemize}
Counting is a special case of induction, and induction is a special case of recursion. They are the methods for computing by counting. More precisely, they are the methods for computing by counting \emph{down}, since the value of the $(n+1)$-st entry of an inductively defined sequence is reduced to the value of $n$-th entry. In monoidal computers, this is captured by program evalutions within program evaluations. 

\subsection{Induction}\sindex{induction schema|see{schema}}\sindex{schema!induction}
The \textbf{induction schema} for an arbitrary type $B$ is 
\beq\label{eq:cata-rule}
\prooftree
\raisebox{.5ex}{$b: B \qquad \qquad \beeta \colon B \to B$}
\justifies
\raisebox{-.5ex}{$\cata{b,\beeta}\colon \NNn \to B$}
\endprooftree
\eeq
where the \emph{banana-sequence}\/ $\cata{b,\beeta}$ is defined  
\begin{alignat}{5}
\cata{b, \beeta}_0  &\ \ =\ \   b &&\hspace{7em}&
\cata{b, \beeta}_{\suce(n)} & \ \ = \ \  \beeta \Big(\cata{b,\beeta}_n\Big) \label{eq:cata}\\[3.5ex]
\def\JPicScale{.65}\newcommand{\Nhh}{\scriptscriptstyle \NNn} \newcommand{\zhh}{\scriptscriptstyle 0}\newcommand{\bhh}{\scriptscriptstyle B}\newcommand{\ahh}{\scriptstyle\cata{b, \beeta}} 
\ifx\JPicScale\undefined\def\JPicScale{1}\fi
\psset{unit=\JPicScale mm}
\psset{linewidth=0.3,dotsep=1,hatchwidth=0.3,hatchsep=1.5,shadowsize=1,dimen=middle}
\psset{dotsize=0.7 2.5,dotscale=1 1,fillcolor=black}
\psset{arrowsize=1 2,arrowlength=1,arrowinset=0.25,tbarsize=0.7 5,bracketlength=0.15,rbracketlength=0.15}
\begin{pspicture}(0,0)(20,12.5)
\rput(10,0){$\ahh$}
\psline(10,5)(10,12.5)
\psline(10,-10)(10,-5)
\rput[l](11.25,11.25){$\bhh$}
\psline(5,-5)(15,-5)
\psline(5,5)(15,5)
\psline(5,-10)(15,-10)
\psline(5,-10)(10,-15)
\psline(15,-10)(10,-15)
\rput[r](8.75,-7.5){$\Nhh$}
\rput(10,-12.5){$\zhh$}
\rput{0}(15,0){\psellipticarc[](0,0)(5,-5){-90}{90}}
\rput{0}(5,0){\psellipticarc[](0,0)(5,5){90}{270}}
\end{pspicture}
\   &\ \ =\ \   \def\JPicScale{.65}\newcommand{\zhh}{\scriptstyle  b} \newcommand{\bhh}{\scriptscriptstyle B} 
\ifx\JPicScale\undefined\def\JPicScale{1}\fi
\psset{unit=\JPicScale mm}
\psset{linewidth=0.3,dotsep=1,hatchwidth=0.3,hatchsep=1.5,shadowsize=1,dimen=middle}
\psset{dotsize=0.7 2.5,dotscale=1 1,fillcolor=black}
\psset{arrowsize=1 2,arrowlength=1,arrowinset=0.25,tbarsize=0.7 5,bracketlength=0.15,rbracketlength=0.15}
\begin{pspicture}(0,0)(12.5,12.5)
\psline(5,0)(5,12.5)
\rput[l](6.25,11.25){$\bhh$}
\psline(-2.5,0)(12.5,0)
\psline(-2.5,0)(5,-7.5)
\psline(12.5,0)(5,-7.5)
\rput(5,-2.5){$\zhh$}
\end{pspicture}
 &&& 
\def\JPicScale{.65}\newcommand{\Nhh}{\scriptscriptstyle \NNn} \newcommand{\zhh}{\scriptstyle \suce }\newcommand{\bhh}{\scriptscriptstyle B}\newcommand{\ahh}{\scriptstyle \cata{b, \beeta}} 
\ifx\JPicScale\undefined\def\JPicScale{1}\fi
\psset{unit=\JPicScale mm}
\psset{linewidth=0.3,dotsep=1,hatchwidth=0.3,hatchsep=1.5,shadowsize=1,dimen=middle}
\psset{dotsize=0.7 2.5,dotscale=1 1,fillcolor=black}
\psset{arrowsize=1 2,arrowlength=1,arrowinset=0.25,tbarsize=0.7 5,bracketlength=0.15,rbracketlength=0.15}
\begin{pspicture}(0,0)(20,18.75)
\rput(10,7.5){$\ahh$}
\psline(10,12.5)(10,18.75)
\psline(10,-3.75)(10,2.5)
\rput[l](11.25,17.5){$\bhh$}
\rput[r](8.75,-1.25){$\Nhh$}
\rput(10,-7.5){$\zhh$}
\psline(10,-18.75)(10,-11.25)
\rput[r](8.75,-17.5){$\Nhh$}
\rput{0}(10,-7.5){\psellipse[](0,0)(3.75,-3.75)}
\psline(5,2.5)(15,2.5)
\psline(5,12.5)(15,12.5)
\rput{0}(5,7.5){\psellipticarc[](0,0)(5,5){90}{270}}
\rput{0}(15,7.5){\psellipticarc[](0,0)(5,-5){-90}{90}}
\end{pspicture}
\  & \ \ \ \ = \ \ \ \   \def\JPicScale{.65}\newcommand{\Nhh}{\scriptscriptstyle \NNn}\newcommand{\ahh}{\scriptstyle \cata{b, \beeta}}\newcommand{\zhh}{\scriptstyle  \beeta} \newcommand{\bhh}{\scriptscriptstyle B} 
\ifx\JPicScale\undefined\def\JPicScale{1}\fi
\psset{unit=\JPicScale mm}
\psset{linewidth=0.3,dotsep=1,hatchwidth=0.3,hatchsep=1.5,shadowsize=1,dimen=middle}
\psset{dotsize=0.7 2.5,dotscale=1 1,fillcolor=black}
\psset{arrowsize=1 2,arrowlength=1,arrowinset=0.25,tbarsize=0.7 5,bracketlength=0.15,rbracketlength=0.15}
\begin{pspicture}(0,0)(20,18.75)
\rput(10,-7.5){$\ahh$}
\psline(10,12.5)(10,18.75)
\psline(10,-2.5)(10,2.5)
\rput[l](11.25,17.5){$\bhh$}
\psline(5,12.5)(15,12.5)
\psline(5,12.5)(5,2.5)
\psline(15,12.5)(15,2.5)
\rput(10,7.5){$\zhh$}
\psline(5,2.5)(15,2.5)
\psline(10,-18.75)(10,-12.5)
\rput[r](8.75,-17.5){$\Nhh$}
\rput[l](11.25,0){$\bhh$}
\psline(5,-12.5)(15,-12.5)
\psline(5,-2.5)(15,-2.5)
\rput{0}(5,-7.5){\psellipticarc[](0,0)(5,5){90}{270}}
\rput{0}(15,-7.5){\psellipticarc[](0,0)(5,-5){-90}{90}}
\end{pspicture}
\notag
\\[2\baselineskip]\notag
\end{alignat}

\para{Why bananas?} If the sequence defined  in  \eqref{eq:cata} is given a name $\phhi = \cata{b,\beeta}$, the list of its values becomes $\sseq{\phhi_0, \phhi_1, \phhi_2,\ldots}$. But programs usually do not use such listings, but only evaluate particular entries. So instead of reserving the name $\phhi$ only to be able to call $\phhi_{23}$, the programmers invented the banana-notation, and call $\cata{b,\beeta}_{23}$.

\para{Computing as counting down.} To evaluate an entry of an inductively defined sequence, we descend down the step-case part of \eqref{eq:cata} to the base-case part:
\beq
\cata{b,\beeta}_n \ =\  \cata{b,\beeta}_{\suce^n (0)}
\  = \  \beeta \Big(\cata{b,\beeta}_{\suce^{n-1} (0)} \Big)
\ = \   \beeta^2 \Big(\cata{b,\beeta}_{\suce^{n-2} (0)}\Big) \ =\cdots =\ 
 \ \beeta^n \Big(\cata{b,\beeta}_{0}\Big) 
\  = \  
\beeta^n \Big(b\Big)
\eeq
This descent is displayed in Fig.~\ref{Fig:indu-exec}.
\begin{figure}[!ht]
\begin{center}
\def\JPicScale{.6}
\newcommand{\Nhh}{\scriptscriptstyle \NNn} 
\newcommand{\zhh}{\scriptscriptstyle 0}
\newcommand{\bhh}{\scriptscriptstyle B}
\newcommand{\bbase}{\scriptstyle b}
\newcommand{\bstep}{\scriptstyle \beeta}
\newcommand{\ahh}{\scriptstyle \cata{b, \beeta}}
\newcommand{\succe}{\scriptstyle \suce}
\newcommand{\EQQLS}{$=\cdots =$} 
\input{PIC/induction-exec.tex}
\caption{$\cata{b,\beeta}$ is computed by counting $\suce$-beads and replacing them with $\beeta$-boxes}
\label{Fig:indu-exec}
\end{center}
\end{figure}

\subsection{Reverse programming induction}\label{Sec:runind}
The entries of the sequence $\phhi=\cata{b,\beeta}:\NNn\stricto B$ for any $b: B$ and $\beeta:B \stricto B$ can be computed by counting down by a program defined as a Kleene fixpoint of the \strict\ \funnn\ $\widetilde \phhi: \DP\times \NNn \stricto B$ defined 
\bea
\widetilde \phhi(p, n) &  = &  \iif\Big(\iszero(n),\ b,\ \beeta\circ \uev p \circ\pred(n) \Big)
\eea
\begin{figure}[ht]
\newcommand{\iiftag}{\mbox{\large \it if}}
\newcommand{\zerotag}{\iszero}
\newcommand{\gtag}{b}
\newcommand{\htag}{\beeta}
\newcommand{\predtag}{\pred}
\newcommand{\een}{\scriptstyle \NNn}
\newcommand{\Beh}{\scriptstyle B}
\newcommand{\utag}{\universal}
\newcommand{\Ftilde}{\widetilde \Phi}
\newcommand{\ftilde}{\mbox{\Large$\widetilde \phi$}}
\newcommand{\function}{\mbox{\Large$\phi\  =\  \uev{\widetilde \Phi}\ =$}}
\newcommand{\functionr}{\mbox{\Large$=\  \cata{b,\beeta}$}}
\def\JPicScale{.5}
\begin{center}
\input{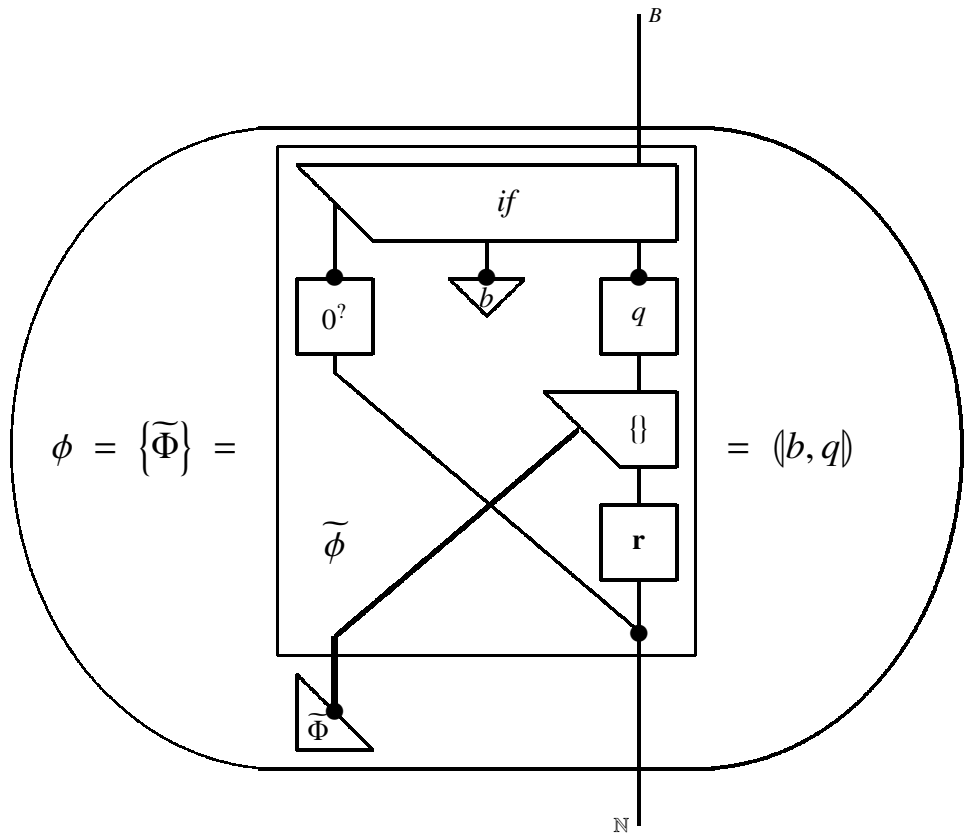}
\caption{Induction in monoidal computer}
\label{Figure:IND}
\end{center}
\end{figure}
If $\widetilde \Phi$ is a Kleene fixpoint of $\widetilde \phi$, then setting $\phi = \uev{\widetilde \Phi}$ yields the sequence $\phi: \NNn \stricto B$ of
\beq\label{eq:def-Kleene-ind}
\phi_n \ = \ \uev{\widetilde \Phi}_n \ = \ \widetilde \phi \left(\widetilde \Phi, n\right)\ =\ \iif\Big(\iszero(n),\  b,\ q\left( \phi_{\pred (n)}\right) \Big)  
\eeq
The construction is displayed in Fig.~\ref{Figure:IND}.

\subsection[Recursion]{Recursion}\label{Sec:prog-recursion}\sindex{recursion}\sindex{schema!recursion}

The \textbf{recursion schema}\/ for specifying sequences of functions from $A$ to $B$ is 
\beq\label{eq:def-recursion}
\prooftree
g\colon A\to B \qquad \qquad \qquad h\colon \NNn\times B \times A \to B
\justifies
{\cata{g,h}}\colon \NNn\times A \to B
\endprooftree
\eeq
The \emph{banana function sequence} $\cata{g,h}$ is defined
\begin{alignat}{5}
\cata{g,h}_0 (x)  &\ \ =\ \   g(x) &&\hspace{7em}&
\cata{g,h}_{\suce(n)}(x)  & \ \ = \ \  h_n \Big( \cata{g,h}_n(x),\, x\Big)\label{eq:banana-recursion}
\\[3.5ex]
\def\JPicScale{.6}\newcommand{\Nhh}{\scriptstyle \NNn} \newcommand{\zhh}{\scriptstyle 0}\newcommand{\bhh}{\scriptstyle B}\newcommand{\ahh}{\scriptstyle A}\newcommand{\function}{\cata{g, h}} 
\ifx\JPicScale\undefined\def\JPicScale{1}\fi
\psset{unit=\JPicScale mm}
\psset{linewidth=0.3,dotsep=1,hatchwidth=0.3,hatchsep=1.5,shadowsize=1,dimen=middle}
\psset{dotsize=0.7 2.5,dotscale=1 1,fillcolor=black}
\psset{arrowsize=1 2,arrowlength=1,arrowinset=0.25,tbarsize=0.7 5,bracketlength=0.15,rbracketlength=0.15}
\begin{pspicture}(0,0)(30,20)
\rput[r](23.75,-25){$\ahh$}
\psline(15,5)(15,20)
\psline(5,-21.25)(5,-5)
\rput[l](16.25,20){$\bhh$}
\psline(0,-21.25)(10,-21.25)
\psline(0,-21.25)(5,-26.25)
\psline(10,-21.25)(5,-26.25)
\rput[r](3.75,-15){$\Nhh$}
\rput(5,-23.75){$\zhh$}
\psline(25,-25)(25,-5)
\rput(15,0){$\function$}
\psline(5,5)(25,5)
\psline(5,-5)(25,-5)
\rput{0}(5,0){\psellipticarc[](0,0)(5,-5){90}{270}}
\rput{0}(25,0){\psellipticarc[](0,0)(5,-5){-90}{90}}
\end{pspicture}
\   &\ \ =\ \   \def\JPicScale{.6}\newcommand{\zhh}{g} \newcommand{\bhh}{\scriptstyle B} \newcommand{\ahh}{\scriptstyle A} 
\ifx\JPicScale\undefined\def\JPicScale{1}\fi
\psset{unit=\JPicScale mm}
\psset{linewidth=0.3,dotsep=1,hatchwidth=0.3,hatchsep=1.5,shadowsize=1,dimen=middle}
\psset{dotsize=0.7 2.5,dotscale=1 1,fillcolor=black}
\psset{arrowsize=1 2,arrowlength=1,arrowinset=0.25,tbarsize=0.7 5,bracketlength=0.15,rbracketlength=0.15}
\begin{pspicture}(0,0)(10,20)
\psline(5,5)(5,20)
\rput[l](6.25,20){$\bhh$}
\psline(0,5)(10,5)
\psline(0,5)(0,-5)
\psline(10,5)(10,-5)
\rput(5,0){$\zhh$}
\psline(5,-25)(5,-5)
\psline(0,-5)(10,-5)
\rput[r](3.75,-25){$\ahh$}
\end{pspicture}
 &&& 
\def\JPicScale{.6}\newcommand{\Nhh}{\scriptstyle \NNn} \newcommand{\succe}{\suce}\newcommand{\bhh}{\scriptstyle B}\newcommand{\ahh}{\scriptstyle A} \newcommand{\function}{\cata{g, h}} 
\ifx\JPicScale\undefined\def\JPicScale{1}\fi
\psset{unit=\JPicScale mm}
\psset{linewidth=0.3,dotsep=1,hatchwidth=0.3,hatchsep=1.5,shadowsize=1,dimen=middle}
\psset{dotsize=0.7 2.5,dotscale=1 1,fillcolor=black}
\psset{arrowsize=1 2,arrowlength=1,arrowinset=0.25,tbarsize=0.7 5,bracketlength=0.15,rbracketlength=0.15}
\begin{pspicture}(0,0)(30,20)
\rput[r](3.75,-1.25){$\Nhh$}
\rput[r](3.75,-25){$\Nhh$}
\rput[r](23.75,-25){$\ahh$}
\psline(15,15)(15,20)
\psline(5,-25)(5,5)
\rput[l](16.25,20){$\bhh$}
\psline(25,-25)(25,5)
\rput(15,10){$\function$}
\psline(5,15)(25,15)
\rput{0}(5,10){\psellipticarc[](0,0)(5,-5){90}{270}}
\rput{0}(25,10){\psellipticarc[](0,0)(5,-5){-90}{90}}
\psline(5,5)(25,5)
\rput{0}(5,-10){\psellipse[fillcolor=white,fillstyle=solid](0,0)(3.75,-3.75)}
\rput(5,-10){$\succe$}
\end{pspicture}
\  & \ \ \ \ = \ \ \ \   \def\JPicScale{.6}\newcommand{\Nhh}{\scriptstyle \NNn}\newcommand{\ahh}{\scriptstyle A} \newcommand{\function}{\cata{g, h}} \newcommand{\zhh}{h} \newcommand{\bhh}{\scriptstyle B} 
\ifx\JPicScale\undefined\def\JPicScale{1}\fi
\psset{unit=\JPicScale mm}
\psset{linewidth=0.3,dotsep=1,hatchwidth=0.3,hatchsep=1.5,shadowsize=1,dimen=middle}
\psset{dotsize=0.7 2.5,dotscale=1 1,fillcolor=black}
\psset{arrowsize=1 2,arrowlength=1,arrowinset=0.25,tbarsize=0.7 5,bracketlength=0.15,rbracketlength=0.15}
\begin{pspicture}(0,0)(37.5,20)
\rput[l](21.25,20){$\bhh$}
\rput(20,10){$\zhh$}
\rput[r](8.75,-26.25){$\Nhh$}
\rput[r](28.75,0){$\ahh$}
\psline(20,-5)(20,5)
\psline(10,-26.25)(10,-15)
\psline(30,-26.25)(30,-15)
\rput(20,-10){$\function$}
\psline(10,-5)(30,-5)
\rput{0}(10,-10){\psellipticarc[](0,0)(5,-5){90}{270}}
\rput{0}(30,-10){\psellipticarc[](0,0)(5,-5){-90}{90}}
\psline(10,-15)(30,-15)
\rput[r](28.75,-26.25){$\ahh$}
\rput{0}(30,-20){\psellipse[fillstyle=solid](0,0)(1.56,-1.56)}
\psline(11.25,15)(28.75,15)
\psline(11.25,15)(11.25,5)
\psline(28.75,15)(28.75,5)
\psline(11.25,5)(28.75,5)
\psline(30,-20)(37.5,-10)
\rput{0}(10,-20){\psellipse[fillstyle=solid](0,0)(1.56,-1.56)}
\psline(2.5,-10)(10,-20)
\psline(2.5,-10)(13.75,5)
\psline(26.25,5)(37.5,-10)
\psline(20,15)(20,20)
\rput[r](8.75,0){$\Nhh$}
\rput[r](18.75,0){$\bhh$}
\end{pspicture}

\notag
\\[5\baselineskip]\notag
\end{alignat}

\para{Evaluating recursive functions.} In the most general form, computing by counting backwards can get very inefficient. This is to some extent reflected in its algebraic expansion:
\begin{multline*}
\cata{g,h}_n(x) \ =\  \cata{g,h}_{\suce^n 0}\left(x\right) 
\ =\ h_{n-1}\Big(\cata{g,h}_{\suce^{n-1} 0}\left(x\right), x\Big)
\ =\\  =\   h_{n-1}\left(h_{n-2}\Big(\cata{g,h}_{\suce^{n-2} 0}\left(x\right), x\Big),x\right)
\ = 
 \cdots =\    h_{n-1}\Bigg(h_{n-2}\Big( \cdots h_1\big(\cata{g,h}_{\suce 0}(x),x\big)\cdots x\Big),x\Bigg)
\ =\\   
=\  h_{n-1}\Bigg(h_{n-2}\Big(\cdots h_1\big(h_0\left(\cata{g,h}_0(x),x\right),x\big)\cdots x\Big),x\Bigg) \ 
 =\    h_{n-1}\left(h_{n-2}\left(\cdots h_1\left(h_0\left(g(x),x\right),x\right)\cdots x\right),x\right)
\end{multline*} 
The diagrammatic view of the same countdown in Fig~\ref{Fig:recur-exec}, however, shows that the recursive descent is still a simple counting process:
\begin{figure}
\begin{center}
\def\JPicScale{.6}
\newcommand{\Nhh}{\scriptscriptstyle \NNn} 
\newcommand{\Ahhh}{\scriptscriptstyle A}
\newcommand{\zhh}{\scriptscriptstyle 0}
\newcommand{\bhh}{\scriptscriptstyle B}
\newcommand{\bbase}{\scriptstyle g}
\newcommand{\bstep}{\scriptstyle h}
\newcommand{\ahh}{\scriptstyle \cata{g,h}}
\newcommand{\succe}{\scriptstyle \suce}
\newcommand{\EQQLS}{$=\cdots$} 
\newcommand{\EQQQLS}{$\cdots =$}
\input{PIC/recursion-exec.tex}
\caption{$\cata{g,h}$ is evaluated by replacing the $n$-th $\suce$-bead by $h_n$}
\label{Fig:recur-exec}
\end{center}
\end{figure}
The recursive step is expanded in Fig.~\ref{Fig:suce-clone}, showing that the successor beads need to be copied as they are counted. 
\begin{figure}
\begin{center}
\def\JPicScale{.6}
\newcommand{\Nhh}{\scriptscriptstyle \NNn} 
\newcommand{\Ahhh}{\scriptscriptstyle A}
\newcommand{\zhh}{\scriptscriptstyle 0}
\newcommand{\bhh}{\scriptscriptstyle B}
\newcommand{\bbase}{\scriptstyle g}
\newcommand{\bstep}{\scriptstyle h}
\newcommand{\ahh}{\scriptstyle \cata{g,h}}
\newcommand{\succe}{\scriptstyle \suce}
\newcommand{\EQQLS}{$=\cdots$} 
\newcommand{\EQQQLS}{$\cdots =$}
\input{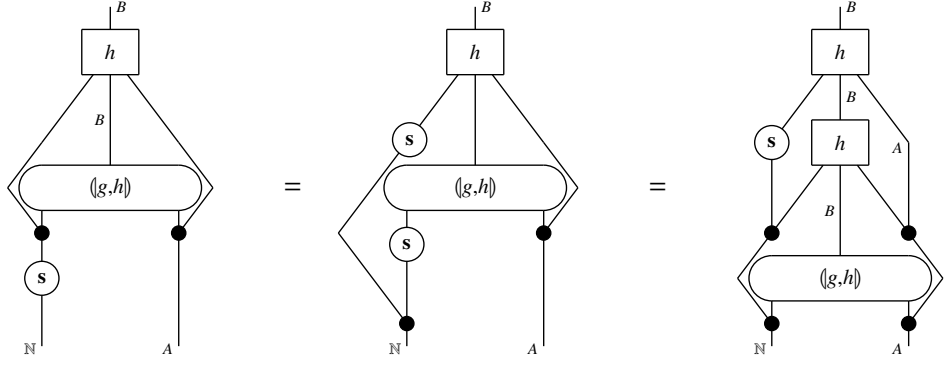}
\caption{Intermediary step in Fig.~\ref{Fig:recur-exec}: An $\suce$-bead is cloned before one copy is discarded}
\label{Fig:suce-clone}
\end{center}
\end{figure}

\subsection{Running recursion}\label{Sec:runrec}
For any pair of functions $g:A\to B$ and $h:\NNn\times B\times A \to B$ in a computer, the \funnn $f=\cata{g,h}:\NNn\times A\to B$ derived by recursion \eqref{eq:def-recursion} is also computable. The program for $f$ can be constructed as a Kleene fixpoint as follows. Consider the \funnn\ $\widetilde f: \DP\times \NNn\times A \to B$ defined by
\bea
\widetilde f(p, n , x) &  = &  \iif\Big(\iszero(n),\  g(x),\ h_{\pred(n)}\left(\big\{p\big\} \left(\pred(n), x\right), x\right) \Big)
\eea
\begin{figure}[ht]
\newcommand{\iiftag}{\iif}
\newcommand{\zerotag}{\iszero}
\newcommand{\gtag}{g}
\newcommand{\htag}{h}
\newcommand{\predtag}{\scriptstyle \pred}
\newcommand{\pee}{\scriptstyle \DP}
\newcommand{\een}{\scriptstyle \NNn}
\newcommand{\eex}{\scriptstyle A}
\newcommand{\Beh}{\scriptstyle B}
\newcommand{\utag}{\universal}
\newcommand{\Ftilde}{\scriptstyle \widetilde F}
\newcommand{\ftilde}{\widetilde f}
\newcommand{\function}{\mbox{\Large$f\  =\  \uev{\widetilde F}\ =$}}
\newcommand{\functionr}{\mbox{\Large$=\  \cata{g,h}$}}
\def\JPicScale{.33}
\begin{center}
\input{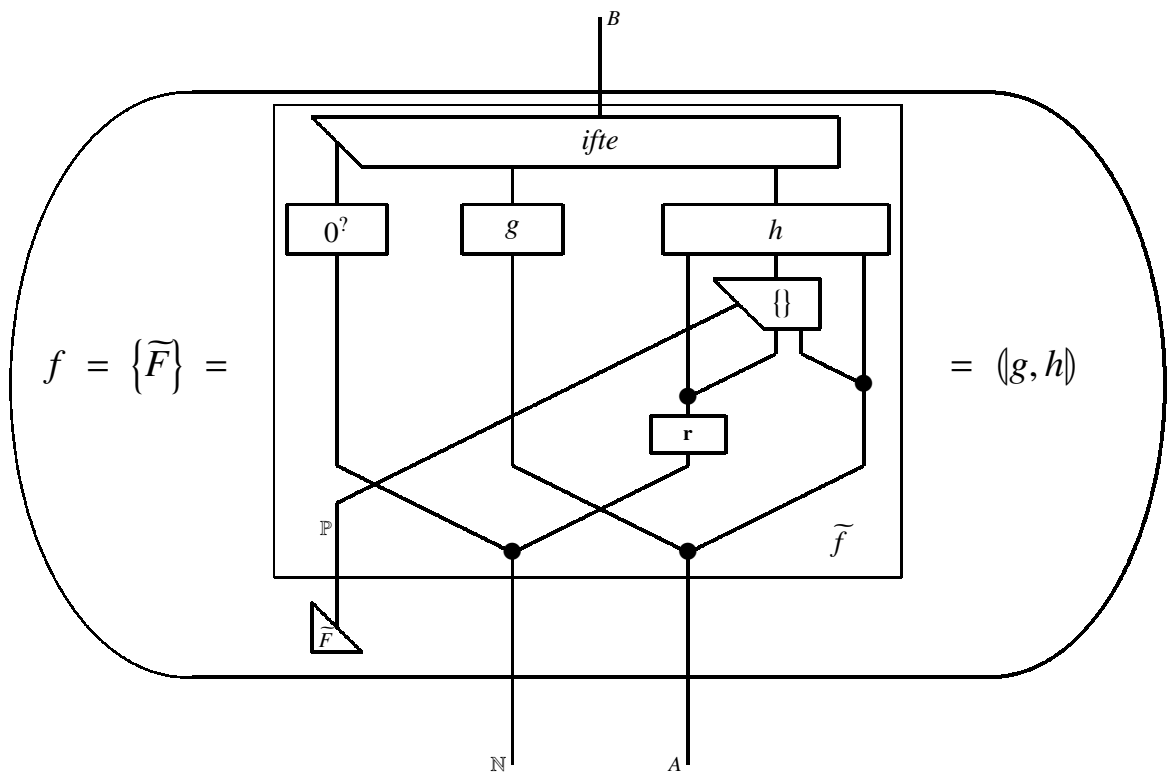}
\caption{Recursion in monoidal computer}
\label{Figure:PR}
\end{center}
\end{figure}
If $\widetilde F$ is a Kleene fixpoint of $\widetilde f$, then the definitions give
\bea\label{eq:def-Kleenef}
\uev{\widetilde F}(n, x)\ = \ \widetilde f \left(\widetilde F, n, x\right)\ =\ \iif\Big(\iszero(n),\  g(x),\ h_{\pred(n)}\left( \uev{\widetilde F} \left(\pred (n), x\right), x\right) \Big)  
\eea
Unfolding the $\iif$-branching gives
\bea\label{eq:PR-call}
\uev{\widetilde F}(n, x) & = & \begin{cases}
g(x) & \mbox{ if } n=0\\
h_{\pred(n)}\left(\uev{\widetilde F}\left(\pred(n), x\right),\  x\right) & \mbox{ if } n \gt 0
\end{cases}
\eea
which means that $\uev{\widetilde F}$ satisfies \eqref{eq:banana-recursion}, and we can set
\bea\label{eq:def-recf}
\cata{g, h}_n(x) & = & \uev{\widetilde F}(n,x)
\eea
The construction is summarized in Fig.~\ref{Figure:PR}.

%

\para{Evaluating a recursive function in a computer.} How is the function in Fig.~\ref{Figure:PR} evaluated down to its lower values? Suppose the input $<n,a>:\DP\times A$ enters the strings at the bottom. At the first step, both values are  copied, $n$ to $\iszero$ and $\pred$, and $a$ to $g$ and $h$. Going up, $\pred$ is evaluated on $n$, and the result is copied to $h$ and $\uev{\widetilde F}$,  as is $a$. The value $y = \uev{\widetilde F}\left(\pred(n), a\right)$ is computed next and the output is fed to $h$. The computations $\iszero (n)$, $g(x)$, and $h_{\pred(n)}\left( y, a \right)$ are then performed in parallel. If $n$ is a correctly formed number in the form $\pairr \flse {\pred(n)}$, then $\iszero(n)$ is $\flse$ and $\iif$ evaluated on $\flse$ reduces to the second projection, and outputs the value $h_{\pred(n)}(y,a)$, and the computation halts. 
\begin{figure}[!ht]
\newcommand{\iiftag}{\iif}
\newcommand{\zerotag}{\iszero}
\newcommand{\gtag}{g}
\newcommand{\htag}{h}
\newcommand{\predtag}{\scriptstyle \pred}
\newcommand{\pee}{\scriptstyle \DP}
\newcommand{\een}{\scriptstyle \NNn}
\newcommand{\eex}{\scriptstyle A}
\newcommand{\Beh}{\scriptstyle B}
\newcommand{\utag}{\scriptstyle \universal}
\newcommand{\bigutag}{\universal}
\newcommand{\Ftilde}{\scriptstyle\widetilde F}
\newcommand{\bigFtilde}{\Large $\widetilde F$}
\newcommand{\function}{\cata{g,h}}
\newcommand{\ftilde}{\cata{g,h}}
\def\JPicScale{.26}
\begin{center}
\input{PIC/PR-rrun.tex}
\caption{$\cata{g,h}_n$ contains $\cata{g,h}_{\pred(n)}$}
\label{Figure:PR-run}
\end{center}
\end{figure}
The recursive step is the evaluation of $y = \uev{\widetilde F}\left(\pred(n), a\right) = \cata{g,h}_{\pred(n)}(a)$, since $\widetilde F$ is the Kleene fixpoint defining $\cata{g,h}$. The red boxes in Fig.~\ref{Figure:PR-run} show how the computation of $\cata{g,h}_{\pred(n)}(a)$ is embedded within the computation of $\cata{g,h}_{n}(a)$: the big red box $\cata{g,h} =  \uev{\widetilde F}$ on the right is the same function as the little red box $\uev{\widetilde F}$ contained in the diagram on the left. The big one is evaluated on $n$ and the little one on $\pred(n) = n-1$. If we zoom into the little box, and open it up, it becomes the right-hand side of another diagram like Fig.~\ref{Figure:PR-run}, just a step further down the recursive ladder Fig.~\ref{Fig:recur-exec}. On the left-hand side of that diagram there is another copy of $\cata{g,h} =  \uev{\widetilde F}$ in a small box, this time evaluated on $\pred(\pred(n)) = n-2$. The recursive ladder thus unfolds down, as the evaluation of each $\cata{g,h}_m$, for $m= n, n-1, n-2,\ldots, 1$ calls for the evaluation of the copy of $\cata{g,h}$  on $\pred(m) = n-1, n-2,\ldots, 0$. At the last such step, the $\iszero$-test evaluates to $\true$, and $\iif$ outputs the first projection $g(a)=\cata{g,h}_0(a)$, which is returned to the function $h_1(\cata{g,h}_0(a),a)=\cata{g,h}_1(a)$ that called it, and so on, up to $\cata{g,h}_n(a)$. This is the evaluation process depicted in Figures \ref{Fig:recur-exec} and \ref{Fig:suce-clone}. 

In summary, Fig.~\ref{Figure:PR-run} shows how that recursive evaluation process can be realized using the self-reference of the Kleene fixpoints. In string diagrams, these self-calls are captured by the box that contains a copy of itself, that contains a copy of itself, and so on. In the case of recursion, the calls count down from $\cata{g,h}_n$, to $\cata{g,h}_{n-1}, \cata{g,h}_{n-2},\ldots,  \cata{g,h}_{1}, \cata{g,h}_{0}$, and the descent is finite. In other program constructs, the evaluation through self-reference may not be finite.

\section{Counting up: Search and loops}\label{Sec:Loop}

\subsection[Search]{Search}\label{Sec:search}
\begin{figure}[!ht]
\begin{center}
\newcommand{\iiftag}{\iif}
\newcommand{\zerotag}{\iszero}
\newcommand{\gtag}{g}
\newcommand{\predtag}{\suce}
\newcommand{\pee}{\scriptstyle \DP}
\newcommand{\een}{\scriptstyle \DP}
\newcommand{\eex}{\scriptstyle A}
\newcommand{\zzro}{\scriptstyle 0}
\newcommand{\utag}{ \universal}
\newcommand{\ftilde}{\widetilde f}
\newcommand{\Ftilde}{\scriptstyle \widetilde F}
\newcommand{\function}{f = \MU(g)}
\def\JPicScale{.4}
\begin{center}
\input{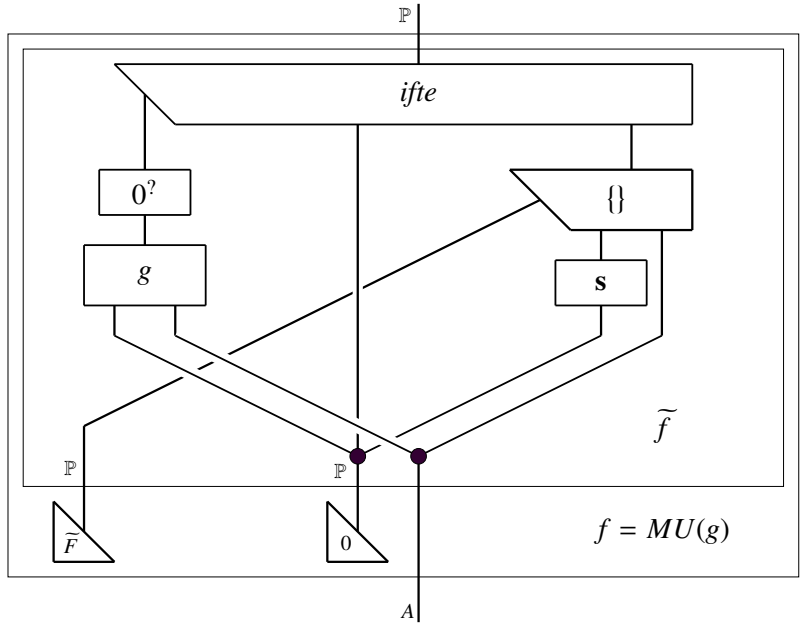}
\end{center}
\caption{Unbounded search in monoidal computer}
\label{Figure:unbounded}
\end{center}
\end{figure}

While the recursion counts down, and reduces the value of $\cata{g,h}_n(a)$ to $\cata{g,h}_{n-1}(a)$, $\cata{g,h}_{n-2}(a)$, all the way to $\cata{g,h}_{0}(a)$,  the search counts up $0, 1, 2,\ldots, n, \ldots$, computes the values of a given function $\varphi_{n}(x)$, and outputs an index $n$ if a value satisfies some condition, usually an equation.\sindex{search!unbounded}  A simple example of search is the minimization schema,\sindex{minimization}\sindex{schema!minimization} searching for the smallest $n\colon \NNn$ such that $\varphi_n(x)=0$ holds for a given sequence of functions $\varphi: \NNn\times A \to \NNn$ and an input $x:A$. The schema is thus:
\beq \label{eq:mu-n}
\prooftree
\varphi: \NNn\times A \to \NNn \justifies
f = \MU(\varphi): A \to \NNn
\endprooftree
\qquad \qquad \mbox{where} \qquad \qquad f(x) \ = \  \mu n.\ \left(\varphi_n(x)=0\right)
\eeq 
In the monoidal computer, testing that  $\varphi_n(x)=0$ is implemented as the predicate
\bear
\iszero \varphi_{n} & = & \left(A\tto{\varphi_{n}} \NNn \tto{\iszero} \Bool\right)
\eear
The requirement that $n=f\left(a\right)$ is the smallest natural number such that $\varphi_{n}(x)=0$ means that $\varphi_m(a)= 0$ implies $n\leq m$ for all $m$, which using the previously implemented predicates  becomes  
\bea\label{eq:unbounded}
\Big(f(a) \iseq n\Big) &   = & \Big(\iszero \varphi_n(a)\  \wedge\ \forall m.\ \iszero g_m(a) \Rightarrow \iszero(n\dotdiv m) \Big)
\eea
%
%
%
%
To implement $f = \MU(g)$ in the abstract computer, we use an intermediary computation $\widetilde f : \DP\times \NNn \times A \to \NNn$ again, this time defined
\bea
\widetilde f(p, n, x) &  = &  \iif\Big(\iszero g_n(x) ,\  n,\ \big\{p\big\} \left(\suce(n), x\right)\Big)
\eea
as displayed in Fig.~\ref{Figure:unbounded}. If $\widetilde F$ is the Kleene fixed point of  $\widetilde f$, i.e.
\bea\label{eq:MU-call}
\big\{\widetilde F\big\}(n, x) & = &  \iif\Big(\iszero g_n(x),\  n,\ \big\{\widetilde F\big\} \left(\suce(n), x\right)\Big)
\eea
then set $f = \MU(g) =  \uev{\pev{\widetilde F} 0}$, i.e.
\bear
f(x)\ = \  \big\{\widetilde F\big\}(0, x)&  = & \iif\Big(\iszero  g_0(x),\  0,\ \big\{\widetilde F\big\} \left(1, x\right)\Big) \mbox{ where}\\
\big\{\widetilde F\big\}(1, x)&  = & \iif\Big(\iszero g_1(x),\  1,\ \big\{\widetilde F\big\} \left(2, x\right)\Big)\mbox{ where}\\
\big\{\widetilde F\big\}(2, x)&  = & \iif\Big(\iszero g_2(x),\  2,\ \big\{\widetilde F\big\} \left(3, x\right)\Big) \ldots
\eear

\para{Running search in a picture.} The diagram in Fig.~\ref{Figure:unbounded} performs search by calling its own a copy that it contains, just like the recursion did in Fig.~\ref{Figure:PR-run}. The self-call displayed Fig.~\ref{Figure:MU-run} is analogous to the one in Fig.~\ref{Figure:PR-run}.    
\begin{figure}[!ht]
\newcommand{\iiftag}{\iif}
\newcommand{\zerotag}{\scriptstyle \iszero}
\newcommand{\gtag}{\scriptstyle g}
\newcommand{\predtag}{\scriptstyle \suce}
\newcommand{\pee}{\scriptstyle \DP}
\newcommand{\een}{\scriptstyle \NNn}
\newcommand{\eex}{\scriptstyle A}
\newcommand{\zzro}{\scriptstyle 0}
\newcommand{\utag}{\universal}
\newcommand{\ftilde}{\scriptstyle\{\widetilde F\}}
\newcommand{\Ftilde}{\scriptstyle \widetilde F}
\newcommand{\function}{\uev{\widetilde F}}\def\JPicScale{.35}
\begin{center}
\input{PIC/MU-run.tex}
\caption{$\uev{\widetilde F}\Big(n,x\Big)$ contains $\uev{\widetilde F}\Big(\suce(n),x\Big)$}
\label{Figure:MU-run}
\end{center}
\end{figure}
The crucial difference is that $\uev{\widetilde F}(n,x)$  in equation \eqref{eq:MU-call} and Fig.~\ref{Figure:MU-run}   calls itself on the successor $\suce (n) = n+1$, whereas in equation \eqref{eq:PR-call} and  Fig.~\ref{Figure:PR-run}  it calls itself on the predecessor $\pred (n) = n-1$. While the recursive calls descend from $n$, and must terminate after $n$ steps, the search calls ascend, and may diverge to the infinity if no $n$ is found to satisfy $g_n(a)$. (Another difference between Figures \ref{Figure:PR-run} and \ref{Figure:MU-run} is that the box on the right is large in the former and small in the latter. This is not a real difference, but the two views show that $\uev{\widetilde F}$ on the right corresponds to the entire computation on the left, and that it is the same function like the component on the left.)

%
%

\subsection[Loops]{Loops}
In most programming languages, the unbounded search is usually expressed using the loop constructs. E.g. the  \emph{while}\/ loop, written in pseudocode, is something like $
{\tt while}\  t(x,y)\ {\tt do}\  h(x,y)\  {\tt od}$, 
where $t:A\times B\to \DP$ is a predicate to be tested, and $h:A\times B \to A$ is a function to be executed while the predicate $t$ is holds. The derivation rule is thus
\[\prooftree
t: A \times B \to \DP \qquad\qquad h: A\times B \to A \justifies
f = \WH(t,h): A \times B \to A
\endprooftree\]
Informally, the intended meaning in terms of the \emph{while}-pseudocode 
\bear
f(a, b) & = & \Big(x:= a;\  {\tt while}\ t(x,b)\ {\tt do}\ x:=h(x,b)\ {\tt od};\  {\tt print}\ x\Big)
\eear
where {\tt do} and {\tt od} open and close a of block of code, and $x:=a$ is an assignment statement. The {\tt do}-block inside the {\tt while}-loop assigns the output of the given function $h$ to the variable $x$. The idea is that the predicate $t$ is tested on the input values $x=a$ and $b$, and if the test is satisfied, and returns the truth value $\true$, then $h$ is applied, and it updates the value stored in $x$. At the next step, the new value $x=h(x,b)$ and the old $b$ are tested by $t$, and as long as $t$ remains true, the function $h$ is reapplied, and it keeps updating the value in $x$. If the test $t$ at some point returns the truth value $\flse$, the function $f$ outputs the current value of $x$, and the computation halts. One way to formalize the intended computation process is to:
\begin{itemize}
\item define a sequence $a_0, a_1,a_2, \ldots$ in $A$ inductively, setting $a_0  = a$ and $a_{i+1} = h(a_i, b)$;
\item output $f(a, b) = a_k$ for the least $k$ such that $\neg t(a_k, b)$.
\end{itemize}

\para{Note} that the function $h$ keeps updating the values of the variable $x:A$, but that the value $b:B$ is a parameter, on which both $t(a,b)$ and $h(a,b)$ depend, but the loop does not change it. 

\begin{figure}[htbp]
\newcommand{\iiftag}{\iif}
\newcommand{\gtag}{t}
\newcommand{\predtag}{h}
\newcommand{\pee}{\scriptstyle \DP}
\newcommand{\een}{\scriptstyle A}
\newcommand{\eex}{\scriptstyle B}
\newcommand{\utag}{\universal}
\newcommand{\ftilde}{\widetilde f}
\newcommand{\Ftilde}{\scriptstyle \widetilde F}
\newcommand{\function}{f = \WH(t,h)}
\def\JPicScale{.4}
\begin{center}
\input{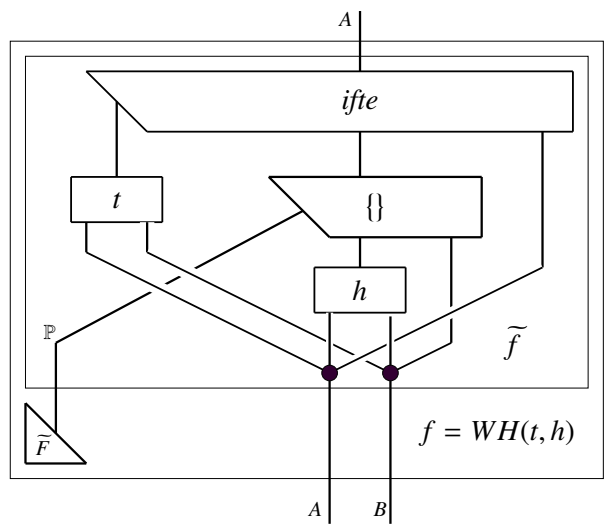}
\caption{A string program for while loop}
\label{Fig:while}
\end{center}
\end{figure}

To implement a while-loop in an abstract computer, we write an intermediary function $\widetilde f: \DP\times A\times B\to A$ again, and use its Kleene fixpoint to as a program for $f = \WH(t,h)$. The intermediary function is this time:
\bea
\widetilde f(p, a, b) &  = &  \iif\Big(t(a,b),\ \big\{p\big\} \left(h(a, b), b\right), \ a\Big)
\eea
The Kleene fixed point $\widetilde F$ of  $\widetilde f$ now satisfies
\bea
\big\{\widetilde F\big\}(a, b) & = &  \iif\Big(t(a,b),\ \big\{F\big\} \left(h(a, b), b\right), \ a\Big)
\eea
and we set $f = \WH(b,h) =  \big\{\widetilde F\big\}$. Unfolding $\widetilde f$ yields
\bear
f(a, b) &  = & \left.\begin{cases}
f\left(h(a, b), b\right) & \mbox{ if } t(a, b)\\
a & \mbox{ otherwise}\end{cases}\right\} \hspace{10em}  \mbox{ where }\\[2ex]
 f\left(h(a, b), b\right) & = & 
\left.\begin{cases}
f\left(h\left(h(a, b), b\right), b\right) & \mbox{ if } b(h(a, b), b)\\
h(a, b) & \mbox{ otherwise}\end{cases}\right\} \hspace{6em}  \mbox{ where }\\[2ex]
 f\left(h\left(h(a, b), b\right), b\right) & = & 
\left.\begin{cases}
f\left(h\left(h\left(h(a, b), b\right), b\right), b\right) & \mbox{ if } t(h\left(h(a, b),b\right), b)\\
h\left(h(a, b), b\right) & \mbox{ otherwise}\end{cases}\right\} \hspace{2em}  \mbox{ where }\\[2ex]
& \ldots 
\eear
The construction is displayed in Fig.~\ref{Fig:while}

\para{Running the loop picture} is almost the same like Fig.~\ref{Figure:MU-run}. Draw it for fun!

%

\def\thechapter{5}
\setchaptertoc
\chapter{What cannot be computed}
\label{Chap:Undec}
\newpage

\section{Decidable extensional properties}\label{Sec:Ext}

\para{Idea.}
\sindex{extensionality}Consider the following sets:
\begin{itemize}
\item $\{n\in \NNn\ |\ \exists m\in \NNn.\ m+m=n\}$,
\item $\{n\in \NNn\ |\ \exists m\in \NNn.\ 2m=n\}$,
\item $\{n\in \NNn\ |\ \frac n 2 \in \NNn\}$,
\item $\{n\in \NNn\ |\ \frac {n+1} 2 \not \in \NNn\}$.
\end{itemize}
Since the four predicates that define these sets are satisfied by the same elements, the even numbers, the set-theoretic principle of \emph{extensionality}\/ says that the four sets are equal, i.e. that they are the same set. In the language of arithmetic, there are infinitely many predicates that describe this set. Ditto for other sets. Different predicates provide different \sindex{intensionality}\emph{intensional}\/ descriptions of the same \emph{extensional}\/ objects, the sets.  

In computation, different programs provide different \emph{intensional}\/ descriptions of the same \emph{extensional}\/ objects, the computable \funnn s. Different programs for the same computable \funnn\ $d:\NNn\to \NNn$ can be constructed from different basic or previously programmed operations:
\begin{itemize}
\item $d_{0} (m) = m+m$,
\item $d_{1} (m) = 2\times m$,
\item $d_{2} (m) = \mu n.\ \left(\frac n 2 = m\right)$,
\item $d_{3} (m) = \mu n.\ \left(\frac{n+1} 2 \gt m\right)$,
\end{itemize}
and in infinitely many other ways. The operation $\mu n.\, \Phi(n)$, discussed in Sec.~\ref{Sec:search}, outputs the smallest $n$ satisfying $\Phi(n)$. While different programming languages assign  different sets of intensional descriptions to each \funnn\  extension, the general view of it all, provided in the monoidal computer, is quite simple: the intensional descriptions are the elements of $\CCC^\bullet(X,\DP)$, the \funnn\ extensions are the elements of $\CCC(X\times A, B)$, and the $X$-indexed instruction $\runn_{X}\colon \CCC^\bullet(X,\DP)\epi \CCC(X\times A, B)$ 
from Sec.~\ref{Sec:surj}, represented by $\runn_{\DP}(\id) = \universal_{A}^{B}$ collapses the intensional descriptions $G\colon X\stricto \DP$ to the corresponding \funnn\ extensions $\runn_{X}(G)=\uev G_{A}^{B}$.

\para{Extensional equivalence} \sindex{extensional!equivalence} for types $A,B$ is the binary relation $(\eext A B )$ on programs induced by running them as \funnn s from $A$ to $B$. The formal definition is
\bea\label{eq:ext-def}
F\eext A B G & \iff & \uev{F}_A^B = \uev{G}_A^B
\eea
The Fundamental Theorem, as displayed in Fig.~\ref{Fig:hartley}, says that any program transformer $\gamma\colon\DP\stricto\DP$ has an extensional fixpoint $\Gamma\eext A B \gamma(\Gamma)$. The $\Upsilon$-classifiers in Sec.~\ref{Sec:classifier} could all be specified as extensional fixpoints, without the curly brackets. The systems of equations used to specify software systems in Sec.~\ref{Sec:system} can be \sindex{software specifications!as extensional equivalences} formulated as extensional equivalence constraints. In general, since every computable \funnn\ has corresponds to a unique equivalence class of programs modulo $(\eext A B )$, there is a bijection 
\bea
\CCC(A, B) & \cong & \CCC^\bullet (I,\DP)\, \Big/ \eext A B
\eea
This gives rise to the question:
\beq\label{quote:tell-pred}
\begin{minipage}{.8\linewidth}
\emph{What can a computer tell about a computable \funnn\ from its programs without running them?}
\end{minipage}
\eeq
This chapter presents answers to some important special cases of this question, and a special answer to the general question. More general answers are given in the theory of effective operations, to which we will have to return later. 

\para{Properties of \funnn s.} A property $\QqQ$ of \funnn s $A\to B$ is presented as a predicate over the hom-set $\CCC(A,B)$. The predicate returns $\QqQ(f)=\true$ when $f$ has the property $\QqQ$ and $\QqQ(f)=\flse$ when it does not. We write $\QqQ(f)$ when $\QqQ(f)=\true$ and $\neg \QqQ(f)$ when $\QqQ(f)=\flse$. 

\begin{figure}[!ht]
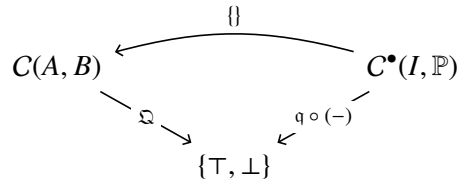

\begin{center}
\begin{tikzar}{}
\CCC(A,B) \ar{dr}[description]{\QqQ}\&\& \CCC^\bullet (I,\DP)\ar[bend right=15]{ll}[swap]{\universal} \ar{dl}[description]{\qqQ\circ(-)}\\
\& \{\true,\flse\}
\end{tikzar}
\caption{When can a property $\QqQ$ of computations be recognized as a property $\qqQ$ of programs?}
\label{Fig:exteq}
\end{center}
\end{figure}

\para{Computable properties of computable \funnn s.}\sindex{predicate!computable} A property of computable \funnn s may be computable if it can be expressed as a computable predicate over their programs, as displayed in Fig.~\ref{Fig:exteq}. Recall from Sec.~\ref{Sec:predicate} that a computable  predicate over programs is presented as a computable \funnn\ $\qpredc\in\CCC(\DP,\Bool)$. We abbreviate $\qqQ\circ F=\true$ to $\qqQ(F)$ and $\qqQ\circ F=\flse$ to $\neg \qqQ(F)$ again.  

\para{Decidable properties of computable \funnn s.} \sindex{predicate!decidable} While Fig.~\ref{Fig:exteq} provides an effective setting for computing with the properties of $f=\uev F$ in terms of $F$, it leaves open the possibility that a computation of the truth value $\qpredc(F)$ may diverge and not tell anything. To assure that it does tell something, the extensional predicates are required to be decidable, i.e. presented a \strict\ \funnn s $\qpredc\in\tot\CCC(\DP,\Bool)$. Question \eqref{quote:tell-pred} becomes: 
 \beq\label{eq:exxt}
\begin{minipage}{.8\linewidth}
\emph{When can a property of computable \funnn s $\QqQ$ be decided on its programs by a predicate $\qpredc\colon\DP\stricto\Bool$ so that the encodings $f = \uev F$ give 
\bear
\QqQ(f) & \iff & \qqQ(F)
\eear
}
\end{minipage}
\eeq

\para{Extensional predicates.}\sindex{predicate!extensional} If there are programs $F,G$ which encode the same \funnn\ $\uev F = \uev G$, but $\qqQ(F)\neq \qqQ(G)$, then \eqref{eq:exxt} is obviously impossible. A predicate $\qqQ$ over programs corresponds to a property $\QqQ$ of \funnn s only if 
\bea\label{eq:ext-pred-def}
F\eext A B G & \implies & \qpredc(F) = \qpredc(G)
\eea
The program predicates $\qpredc$ that satisfy this requirement are \emph{extensional}\/  for $A,B$. They are the congruences with respect to the extensionality relation $(\eext A B)$. A predicate may be extensional with respect to one pair of types $A,B$ and not with respect to another.

\section{G\"odel, Tarski: Provability and truth are undecidable}\label{Sec:Goedel}

\para{Decision predicates.}\/ \sindex{predicate!decision} A decision predicate is an evaluator of predicates over programs, i.e. a computable \funnn s $\dpredc \colon \DP\times \DP \to \Bool$ satisfying  
\bea\label{eq:oone}
\dpredc(Q,x) & = & \uev Q_{\DP}^\Bool x  
\eea
for all $Q,x:\DP$. The \textbf{question} is whether a decision predicate can be decidable. Can the program evaluation be \strict\ on predicates?

\para{Decision predicates are undecidable}, unless $\true = \flse$. Suppose that there is a \emph{decidable}\/ decision predicate $\dpredc\colon\DP\times \DP \stricto \Bool$, and let $D:\DP$ be its program, satisfying $\dpredc(Q,x) = \uev{D}_\DP^\Bool(Q,x)$ for all $Q, x$. Instantiating to $Q=D$ yields $\dpredc(D,x) = \uev{D}_\DP^\Bool(D,x)$. Intuitively, this instance of the decision predicate says: \emph{"I am true"}. But the Fundamental Theorem provides also provides Kleene fixpoint $\Psi$ of the \emph{negation}\/ of the decision predicate:
\bea\label{eq:ttwo}
\neg \dpredc(\Psi, x) & = & \uev{\Psi}_\DP^\Bool x
\eea
The negation\sindex{negation} flips the truth values, so that
\beq\label{eq:negation}
\begin{split}
\newcommand{\zzero}{\scriptstyle \flse}
\newcommand{\oone}{\scriptstyle \true}
\newcommand{\nneg}{\neg}
\newcommand{\ueval}{{\scriptstyle \universal}}
\newcommand{\iinput}{\scriptscriptstyle \Bool}
\newcommand{\ooutput}{\scriptscriptstyle \Bool}
\def\JPicScale{.4}
\ifx\JPicScale\undefined\def\JPicScale{1}\fi
\unitlength \JPicScale mm
\begin{picture}(108.75,60)(0,0)
\linethickness{0.35mm}
\put(10,0){\line(0,1){20}}
\linethickness{0.35mm}
\put(10,40){\line(0,1){20}}
\linethickness{0.35mm}
\put(0,40){\line(1,0){20}}
\linethickness{0.35mm}
\put(20,20){\line(0,1){20}}
\linethickness{0.35mm}
\put(0,20){\line(1,0){20}}
\linethickness{0.35mm}
\put(0,20){\line(0,1){20}}
\put(35,30){\makebox(0,0)[cc]{\EQLS}}

\linethickness{0.35mm}
\put(50,50){\line(1,0){55}}
\linethickness{0.35mm}
\put(105,30){\line(0,1){20}}
\linethickness{0.35mm}
\put(87.5,50){\line(0,1){10}}
\linethickness{0.35mm}
\put(70,30){\line(1,0){35}}
\linethickness{0.35mm}
\put(50,0){\line(0,1){30}}
\linethickness{0.35mm}
\multiput(50,50)(0.12,-0.12){167}{\line(1,0){0.12}}
\linethickness{0.35mm}
\put(66.25,5){\line(1,0){17.5}}
\linethickness{0.35mm}
\put(66.25,5){\line(0,1){17.5}}
\linethickness{0.35mm}
\multiput(66.25,22.5)(0.12,-0.12){146}{\line(1,0){0.12}}
\linethickness{0.35mm}
\put(75,13.75){\line(0,1){16.25}}
\put(71.25,10){\makebox(0,0)[cc]{$\zzero$}}

\put(87.5,40){\makebox(0,0)[cc]{$\ueval$}}

\linethickness{0.35mm}
\put(91.25,5){\line(1,0){17.5}}
\linethickness{0.35mm}
\put(91.25,5){\line(0,1){17.5}}
\linethickness{0.35mm}
\multiput(91.25,22.5)(0.12,-0.12){146}{\line(1,0){0.12}}
\linethickness{0.35mm}
\put(100,13.75){\line(0,1){16.25}}
\put(10,30){\makebox(0,0)[cc]{$\nneg$}}

\put(96.88,10.62){\makebox(0,0)[cc]{$\oone$}}

\linethickness{0.35mm}
\multiput(50,30)(0.12,0.12){83}{\line(1,0){0.12}}
\put(90,60){\makebox(0,0)[cl]{$\ooutput$}}

\put(12.5,60){\makebox(0,0)[cl]{$\ooutput$}}

\put(47.5,0){\makebox(0,0)[cr]{$\iinput$}}

\put(7.5,0){\makebox(0,0)[cr]{$\iinput$}}

\put(10,40){\makebox(0,0)[cc]{\Dottt}}

\put(75,13.75){\makebox(0,0)[cc]{\Dottt}}

\put(100,13.75){\makebox(0,0)[cc]{\Dottt}}

\end{picture}

\end{split}
\qquad \mbox{ gives }
\qquad
\neg \qpredc(x)  =  \begin{cases}
\flse & \mbox{ if } \qpredc(x) = \true\\
\true & \mbox{ if } \qpredc(x) = \flse
\end{cases}
\eeq
for any predicate $\qpredc:\DP\to \Bool$. Intuitively, the predicate $\uev{\Psi}_\DP^\Bool x$ says \emph{"I am false."} A self-referential statement that says that it is false causes the \sindex{Liar Paradox|see {paradox}} \sindex{paradox!liar}\emph{Liar Paradox}\footnote{The liar paradox is often called the \emph{Epimenides' paradox}, because Epimenides, a poet and philosopher from Creta, who lived in VI or VII century BC, had written that \emph{"All Cretans are liars"}.}. But if $d$ is a decision predicate \eqref{eq:oone} and if $\Psi$ is a program for its negation \eqref{eq:ttwo}, then
\beq\label{eq:concl} 
\neg \dpredc(\Psi,x)\ \ \ \stackrel{\eqref{eq:ttwo}}{=}\ \ \ \ \ \uev{\Psi}_\DP^\Bool  x\ \ \ \ \ \stackrel{\eqref{eq:oone}}= \ \ \   \dpredc(\Psi, x) 
\eeq
It follows that $\true = \flse$, since $\neg \true = \flse$ and $\neg\flse = \true$, and the assumption that $\dpredc(\Psi,x)\colon \DP$ means that it must be either $\true$ or $\flse$. If $\true \neq \flse$, then $\dpredc(\Psi, x)$ cannot output a value in $\Bool$, and is not  decidable.

\para{Remark.} G\"odel used the above instance of diagonal argument for the purposes of his analysis of decidability of \emph{provability}\/ of propositions in a formal theory of  arithmetic in  \cite{GoedelK:ueber}. Tarski applied similar reasoning in \cite{TarskiA:truth} where he analyzed decidability of \emph{validity}\/ of propositions in a given model. Both thus applied the diagonal argument to derive contradiction and prove negative results. Although the two results refer to different logical questions, and the original versions involve different technical details, both are based on the same logical paradox. Therefore, in the rudimentary predicate logic implemented in monoidal computer so far, the two results are isomorphic. It should be emphasized, though, that the categorical approach accommodates structural and conceptual refinements as a matter of principle. A reader interested in conceptual distinctions might enjoy testing  that principle.

\section{Turing: Halting is undecidable}\label{Sec:halting}

The main result presented in Turing's breakthrough paper \cite{TuringA:Entscheidung}, where he introduced his machines, was that there is no machine that would input descriptions of arbitrary machines, and output a decision whether they halt or diverge. Turing presented the undecidability of this \emph{halting problem}\/ as the negative solution of Hilbert's \emph{Decision Problem (Entscheidungsproblem)}. \sindex{problem!halting}\sindex{problem!decision}

\para{\Funnn\ halting.} A \funnn\ $f:A\to B$ halts on a total input $a\colon A$\/ if it produces a total output $f(a)\colon B$. Recall from Sec.~\ref{Sec:map} that $x:X$ is total if
\bear
\left(I\tto x X \strictto\scun I\right) & = & \cun\ \ =\ \ \id_{I}
\eear
Recall from Sec.~\ref{Sec:predicate} that  \sindex{halting}
\bea
f(a) \halts & \mbox{ abbreviates } & \left( I \strictto a A \tto f B \strictto\scun I\right) = \cun
\eea

\para{Halting predicates}  \sindex{predicate!halting} are computable \funnn s $\hpredc\colon  \DP\times \DP \to \Bool$ such that: 
\beq\label{eq:halt} \hpredc(P,x)\ \ =\ \ \begin{cases}\true & \mbox{ if } \uev P_{\DP}^{\Bool} x \halts\\
\flse & \mbox{ otherwise} 
\end{cases}\eeq

\para{Halting predicates are undecidable}, unless $\true = \flse$. Towards a contradiction, assume that there is a decidable halting predicate, and apply it to its own negation, like in G\"odel's and Tarski's constructions in the preceding section. This time, though, we need a different negation: it should not just swap the truth values, but it should send a true output of a predicate to a divergent computation. That is what Turing used in his proof. This \emph{divergent}\/ negation $\Lsh$ is defined as follows: 
\beq\label{eq:div-negation}
\begin{split}
\newcommand{\zzero}{\rotatebox[origin=c]{90}{$\scriptstyle\rightsquigarrow$}}
\newcommand{\oone}{\scriptstyle \true}
\newcommand{\nneg}{\Lsh}
\newcommand{\ueval}{{\scriptstyle \universal}}
\newcommand{\iinput}{\scriptscriptstyle \DP}
\newcommand{\ooutput}{\scriptscriptstyle \Bool}
\def\JPicScale{.4}
\ifx\JPicScale\undefined\def\JPicScale{1}\fi
\unitlength \JPicScale mm
\begin{picture}(108.75,60)(0,0)
\linethickness{0.35mm}
\put(10,0){\line(0,1){20}}
\linethickness{0.35mm}
\put(10,40){\line(0,1){20}}
\linethickness{0.35mm}
\put(0,40){\line(1,0){20}}
\linethickness{0.35mm}
\put(20,20){\line(0,1){20}}
\linethickness{0.35mm}
\put(0,20){\line(1,0){20}}
\linethickness{0.35mm}
\put(0,20){\line(0,1){20}}
\put(35,30){\makebox(0,0)[cc]{\EQLS}}

\linethickness{0.35mm}
\put(50,50){\line(1,0){55}}
\linethickness{0.35mm}
\put(105,30){\line(0,1){20}}
\linethickness{0.35mm}
\put(87.5,50){\line(0,1){10}}
\linethickness{0.35mm}
\put(70,30){\line(1,0){35}}
\linethickness{0.35mm}
\put(50,0){\line(0,1){30}}
\linethickness{0.35mm}
\multiput(50,50)(0.12,-0.12){167}{\line(1,0){0.12}}
\linethickness{0.35mm}
\put(66.25,13.75){\line(1,0){17.5}}
\linethickness{0.35mm}
\multiput(75,5)(0.12,0.12){73}{\line(1,0){0.12}}
\linethickness{0.35mm}
\multiput(66.25,13.75)(0.12,-0.12){73}{\line(1,0){0.12}}
\linethickness{0.35mm}
\put(75,13.75){\line(0,1){16.25}}
\put(75,10){\makebox(0,0)[cc]{$\zzero$}}

\put(87.5,40){\makebox(0,0)[cc]{$\ueval$}}

\linethickness{0.35mm}
\put(91.25,5){\line(1,0){17.5}}
\linethickness{0.35mm}
\put(91.25,5){\line(0,1){17.5}}
\linethickness{0.35mm}
\multiput(91.25,22.5)(0.12,-0.12){146}{\line(1,0){0.12}}
\linethickness{0.35mm}
\put(100,13.75){\line(0,1){16.25}}
\put(10,30){\makebox(0,0)[cc]{$\nneg$}}

\put(96.88,10.62){\makebox(0,0)[cc]{$\oone$}}

\linethickness{0.35mm}
\multiput(50,30)(0.12,0.12){83}{\line(1,0){0.12}}
\put(90,60){\makebox(0,0)[cl]{$\ooutput$}}

\put(12.5,60){\makebox(0,0)[cl]{$\ooutput$}}

\put(47.5,0){\makebox(0,0)[cr]{$\iinput$}}

\put(7.5,0){\makebox(0,0)[cr]{$\iinput$}}

\put(100,13.75){\makebox(0,0)[cc]{\Dottt}}

\end{picture}

\end{split}
\qquad \mbox{ which gives }
\qquad
\Lsh\! \qpredc(x)  =  \begin{cases}
{\divg} & \mbox{ if } \qpredc(x) = \true\\
\true & \mbox{ if } \qpredc(x) =\flse
\end{cases}
\eeq
The predicate  $\Lsh\! \qpredc$ thus produces a decision precisely when $\qpredc$ is false:
\bea \label{eq:divneg} \Lsh\! \qpredc(x)\halts & \iff & \neg \qpredc(x)\eea
Let $\Psi$ be a Kleene fixpoint of $\Lsh\! \hpredc$, i.e. 
\bea\label{tthree}
\Lsh\! \hpredc(\Psi,x) & = &  \uev{\Psi }_{\DP}^{\Bool} x
\eea
By \eqref{eq:halt}, the halting predicate also satisfies
\bea\label{ffour}
\hpredc(\Psi, x) & \iff & \uev{\Psi }_{\DP}^{\Bool} x \halts
\eea
Hence
\begin{gather*}
\neg \hpredc(\Psi, x) \hspace{1em} \stackrel{\eqref{eq:divneg}}\iff \hspace{1em}  \Lsh\! \hpredc(\Psi, x)\halts \hspace{1em} \stackrel{\eqref{tthree}}{\iff}\hspace{1em}  \{\Psi \}_{\DP}^{\Bool} x \halts\ \ \ \  \stackrel{\eqref{ffour}}\iff \ \   \hpredc(\Psi, x) \\[2ex]
\def\JPicScale{.4}
\newcommand{\nameslang}{\scriptstyle \Bool}
\newcommand{\progtype}{\scriptstyle {\DP}}
\newcommand{\tttrue}{\scriptstyle \true}
\newcommand{\nneg}{\neg}
\newcommand{\dneg}{\Lsh}
\newcommand{\hee}{\hpredc}
\newcommand{\hfixp}{\scriptstyle \Psi}
\renewcommand{\inputt}{\scriptstyle \DP}
\ifx\JPicScale\undefined\def\JPicScale{1}\fi
\unitlength \JPicScale mm
\begin{picture}(340,120)(0,0)
\linethickness{0.35mm}
\put(187.5,45){\line(1,0){60}}
\linethickness{0.35mm}
\put(247.5,20){\line(0,1){25}}
\linethickness{0.35mm}
\put(227.5,45){\line(0,1){37.5}}
\put(227.5,-5){\makebox(0,0)[cc]{$\inputt$}}

\linethickness{0.35mm}
\put(212.5,20){\line(1,0){35}}
\linethickness{0.35mm}
\put(227.5,0){\line(0,1){20}}
\put(225.62,31.88){\makebox(0,0)[cc]{\universal}}

\put(170,30){\makebox(0,0)[cc]{\EQLS}}

\linethickness{0.35mm}
\put(200,5){\line(0,1){27.5}}
\linethickness{0.35mm}
\multiput(187.5,45)(0.12,-0.12){208}{\line(1,0){0.12}}
\linethickness{0.35mm}
\put(100,45){\line(1,0){50}}
\linethickness{0.35mm}
\put(150,20){\line(0,1){25}}
\put(140,-5){\makebox(0,0)[cc]{$\inputt$}}

\linethickness{0.35mm}
\put(100,20){\line(1,0){50}}
\linethickness{0.35mm}
\put(140,0){\line(0,1){20}}
\linethickness{0.35mm}
\put(110,5){\line(0,1){15}}
\linethickness{0.35mm}
\put(100,20){\line(0,1){25}}
\put(270.62,30){\makebox(0,0)[cc]{\EQLS}}

\linethickness{0.35mm}
\put(290,45){\line(1,0){50}}
\linethickness{0.35mm}
\put(340,20){\line(0,1){25}}
\linethickness{0.35mm}
\put(315,45){\line(0,1){70}}
\put(315,120){\makebox(0,0)[cc]{$\nameslang$}}

\put(330,-5){\makebox(0,0)[cc]{$\inputt$}}

\linethickness{0.35mm}
\put(290,20){\line(1,0){50}}
\linethickness{0.35mm}
\put(330,0){\line(0,1){20}}
\linethickness{0.35mm}
\put(300,5){\line(0,1){15}}
\linethickness{0.35mm}
\put(290,20){\line(0,1){25}}
\linethickness{0.35mm}
\put(125,45){\line(0,1){7.5}}
\linethickness{0.35mm}
\put(125,72.5){\line(0,1){10}}
\linethickness{0.35mm}
\put(115,72.5){\line(1,0){20}}
\linethickness{0.35mm}
\put(135,52.5){\line(0,1){20}}
\linethickness{0.35mm}
\put(115,52.5){\line(1,0){20}}
\linethickness{0.35mm}
\put(115,52.5){\line(0,1){20}}
\linethickness{0.35mm}
\put(102.5,-5){\line(1,0){17.5}}
\linethickness{0.35mm}
\put(102.5,-5){\line(0,1){17.5}}
\linethickness{0.35mm}
\multiput(102.5,12.5)(0.12,-0.12){146}{\line(1,0){0.12}}
\linethickness{0.35mm}
\put(192.5,-5){\line(1,0){17.5}}
\linethickness{0.35mm}
\put(192.5,-5){\line(0,1){17.5}}
\linethickness{0.35mm}
\multiput(192.5,12.5)(0.12,-0.12){146}{\line(1,0){0.12}}
\linethickness{0.35mm}
\put(292.5,-5){\line(1,0){17.5}}
\linethickness{0.35mm}
\put(292.5,-5){\line(0,1){17.5}}
\linethickness{0.35mm}
\multiput(292.5,12.5)(0.12,-0.12){146}{\line(1,0){0.12}}
\put(125,120){\makebox(0,0)[cc]{$\nameslang$}}

\linethickness{0.35mm}
\put(125,102.5){\line(0,1){12.5}}
\linethickness{0.35mm}
\put(117.5,92.5){\line(1,0){17.5}}
\linethickness{0.35mm}
\put(117.5,92.5){\line(0,1){17.5}}
\linethickness{0.35mm}
\multiput(117.5,110)(0.12,-0.12){146}{\line(1,0){0.12}}
\put(122.5,97.5){\makebox(0,0)[cc]{$\tttrue$}}

\put(227.5,120){\makebox(0,0)[cc]{$\nameslang$}}

\linethickness{0.35mm}
\put(227.5,102.5){\line(0,1){12.5}}
\linethickness{0.35mm}
\put(220,92.5){\line(1,0){17.5}}
\linethickness{0.35mm}
\put(220,92.5){\line(0,1){17.5}}
\linethickness{0.35mm}
\multiput(220,110)(0.12,-0.12){146}{\line(1,0){0.12}}
\put(225,97.5){\makebox(0,0)[cc]{$\tttrue$}}

\put(298.75,0){\makebox(0,0)[cc]{$\hfixp$}}

\put(125,32.5){\makebox(0,0)[cc]{$\hee$}}

\put(315,32.5){\makebox(0,0)[cc]{$\hee$}}

\put(198.12,0){\makebox(0,0)[cc]{$\hfixp$}}

\put(107.5,0){\makebox(0,0)[cc]{$\hfixp$}}

\put(70,30){\makebox(0,0)[cc]{\EQLS}}

\linethickness{0.35mm}
\put(0,45){\line(1,0){50}}
\linethickness{0.35mm}
\put(50,20){\line(0,1){25}}
\put(40,-5){\makebox(0,0)[cc]{$\inputt$}}

\linethickness{0.35mm}
\put(0,20){\line(1,0){50}}
\linethickness{0.35mm}
\put(40,0){\line(0,1){20}}
\linethickness{0.35mm}
\put(10,5){\line(0,1){15}}
\linethickness{0.35mm}
\put(0,20){\line(0,1){25}}
\linethickness{0.35mm}
\put(25,45){\line(0,1){7.5}}
\linethickness{0.35mm}
\put(25,72.5){\line(0,1){42.5}}
\linethickness{0.35mm}
\put(15,72.5){\line(1,0){20}}
\linethickness{0.35mm}
\put(35,52.5){\line(0,1){20}}
\linethickness{0.35mm}
\put(15,52.5){\line(1,0){20}}
\linethickness{0.35mm}
\put(15,52.5){\line(0,1){20}}
\put(25,62.5){\makebox(0,0)[cc]{$\nneg$}}

\linethickness{0.35mm}
\put(2.5,-5){\line(1,0){17.5}}
\linethickness{0.35mm}
\put(2.5,-5){\line(0,1){17.5}}
\linethickness{0.35mm}
\multiput(2.5,12.5)(0.12,-0.12){146}{\line(1,0){0.12}}
\put(25,120){\makebox(0,0)[cc]{$\nameslang$}}

\put(25,32.5){\makebox(0,0)[cc]{$\hee$}}

\put(7.5,0){\makebox(0,0)[cc]{$\hfixp$}}

\put(125,62.5){\makebox(0,0)[cc]{$\dneg$}}

\put(25,45){\makebox(0,0)[cc]{\Dottt}}

\put(10,5){\makebox(0,0)[cc]{\Dottt}}

\put(125,82.5){\makebox(0,0)[cc]{\Dottt}}

\put(227.5,82.5){\makebox(0,0)[cc]{\Dottt}}

\put(227.5,102.5){\makebox(0,0)[cc]{\Dottt}}

\put(125,102.5){\makebox(0,0)[cc]{\Dottt}}

\put(315,45){\makebox(0,0)[cc]{\Dottt}}

\put(300,5){\makebox(0,0)[cc]{\Dottt}}

\put(200,5){\makebox(0,0)[cc]{\Dottt}}

\put(110,5){\makebox(0,0)[cc]{\Dottt}}

\put(25,72.5){\makebox(0,0)[cc]{\Dottt}}

\end{picture}

\end{gather*}
Since $\neg$ swaps $\true$ and $\flse$, the assumpiton that $\hpredc(\Psi,x)\colon \Bool$ implies $\true = \flse$ again, and $\true\neq \flse$ implies that $\hpredc$ is not decidable.

\section{Rice: Decidable extensional predicates are constant}
\label{Sec:rice}

Last but not least, we confront the original question: \emph{What can we tell about a  computation without running it?}\/ For easy programs, we can often tell what they do. For convoluted programs, we cannot, and we tend to blame the programmers. Sometimes they are guilty as accused. In general, they are not, as there are no general, decidable  predicates that tell anything at all about the computations just by looking at the programs. To add insult to the injury, asking such questions first leads to predicates over programs that are easy to program but may not terminate, and remain \emph{undecidable}. Later it leads to predicates that are easy to define but impossible to program and remain  \emph{uncomputable}.

\para{Constant \funnn s} $f:A\to B$ produce the same output $b=f(x)$ for all $x:A$. For monoidal \funnn s, this means that for all $x,x':X\to A$ holds $f\circ x = f\circ x'$. Restricting to \strict\ \funnn s, it can be shown that $f:A\stricto B$ is constant if and only if 
\bear
f & = & \left(A\strictto{\scun} I \strictto{b} B\right)
\eear
for some $b:B$. Since $\true$ and $\flse$ as the only \strict\ elements of $\Bool$, there are precisely two constant predicates, written $\true, \flse \colon A\stricto \Bool$, provided that $\cun\colon A\stricto I$ is epi, meaning that $\left(A\strictto\scun I\strictto x X\right)= \left(A\strictto\scun I\strictto{x'} X\right)$ implies $x=x'$.

\para{Swap elements.} A predicate $\qpredc\colon A\to \Bool$ that is not constant must take both truth values. We define \emph{swap elements}\/ of $A$ with respect to $\qpredc$ to be a pair $a_\true, a_\flse: A$ such that 
\[ \qpredc\left(a_\true\right) = \true \quad\mbox{ and }\quad \qpredc\left(a_\flse\right) = \flse\]

\para{A non-constant predicate cannot be both extensional and decidable}, unless $\true = \flse$. Towards contradiction, assume that $p$ is decidable and extensional. Using the decidability assumption, this time we define a \strict\ \funnn\ $\sim : A\stricto A$ to flip $p$'s truth values:
\beq\label{eq:flip-negation}
\begin{split}
\newcommand{\zzero}{\scriptstyle  a_\flse}
\newcommand{\oone}{\scriptstyle  a_\true}
\newcommand{\ueval}{\scriptstyle \universal}
\newcommand{\nneg}{\sim} 
\newcommand{\quu}{\scriptstyle \qpredc}
\newcommand{\iinput}{\scriptscriptstyle A}
\newcommand{\ooutput}{\scriptscriptstyle A}
\newcommand{\progtype}{\scriptscriptstyle \Bool} 
\def\JPicScale{.4}
\ifx\JPicScale\undefined\def\JPicScale{1}\fi
\unitlength \JPicScale mm
\begin{picture}(123.75,70)(0,0)
\linethickness{0.35mm}
\put(10,0){\line(0,1){30}}
\linethickness{0.35mm}
\put(10,50){\line(0,1){17.5}}
\linethickness{0.35mm}
\put(0,50){\line(1,0){20}}
\linethickness{0.35mm}
\put(20,30){\line(0,1){20}}
\linethickness{0.35mm}
\put(0,30){\line(1,0){20}}
\linethickness{0.35mm}
\put(0,30){\line(0,1){20}}
\put(40,40){\makebox(0,0)[cc]{\EQLS}}

\linethickness{0.35mm}
\put(65,60){\line(1,0){55}}
\linethickness{0.35mm}
\put(120,40){\line(0,1){20}}
\linethickness{0.35mm}
\put(100,60){\line(0,1){10}}
\linethickness{0.35mm}
\put(85,40){\line(1,0){35}}
\linethickness{0.35mm}
\multiput(60,35)(0.12,0.12){125}{\line(0,1){0.12}}
\linethickness{0.35mm}
\multiput(65,60)(0.12,-0.12){167}{\line(1,0){0.12}}
\linethickness{0.35mm}
\put(81.25,10){\line(1,0){17.5}}
\linethickness{0.35mm}
\put(81.25,10){\line(0,1){17.5}}
\linethickness{0.35mm}
\multiput(81.25,27.5)(0.12,-0.12){146}{\line(1,0){0.12}}
\linethickness{0.35mm}
\put(90,18.75){\line(0,1){21.25}}
\put(87.5,15){\makebox(0,0)[cc]{$\zzero$}}

\put(102.5,50){\makebox(0,0)[cc]{$\ueval$}}

\linethickness{0.35mm}
\put(106.25,10){\line(1,0){17.5}}
\linethickness{0.35mm}
\put(106.25,10){\line(0,1){17.5}}
\linethickness{0.35mm}
\multiput(106.25,27.5)(0.12,-0.12){146}{\line(1,0){0.12}}
\linethickness{0.35mm}
\put(115,18.75){\line(0,1){21.25}}
\put(10,40){\makebox(0,0)[cc]{$\nneg$}}

\put(113.13,15.62){\makebox(0,0)[cc]{$\oone$}}

\linethickness{0.35mm}
\put(60,0){\line(0,1){10}}
\linethickness{0.35mm}
\put(60,30){\line(0,1){5}}
\linethickness{0.35mm}
\put(50,30){\line(1,0){20}}
\linethickness{0.35mm}
\put(70,10){\line(0,1){20}}
\linethickness{0.35mm}
\put(50,10){\line(1,0){20}}
\linethickness{0.35mm}
\put(50,10){\line(0,1){20}}
\put(60,20){\makebox(0,0)[cc]{$\quu$}}

\put(103.75,70){\makebox(0,0)[cl]{$\ooutput$}}

\put(13.75,67.5){\makebox(0,0)[cl]{$\ooutput$}}

\put(57.5,1.25){\makebox(0,0)[cr]{$\iinput$}}

\put(7.5,0){\makebox(0,0)[cr]{$\iinput$}}

\put(66.25,43.75){\makebox(0,0)[br]{$\progtype$}}

\put(10,50){\makebox(0,0)[cc]{\Dottt}}

\put(60,30){\makebox(0,0)[cc]{\Dottt}}

\put(90,18.75){\makebox(0,0)[cc]{\Dottt}}

\put(115,18.75){\makebox(0,0)[cc]{\Dottt}}

\end{picture}

\end{split}
\qquad\qquad \mbox{ which gives }
\qquad\qquad
 \sim\!\! x \  = \  \begin{cases}
a_\flse & \mbox{ if } \qpredc(x) = \true\\
a_\true & \mbox{ if } \qpredc(x) = \flse
\end{cases}
\eeq
This is a \emph{$\qpredc$-swap negation over $A$}.  It swaps the elements of $A$ into its swap values, so that the truth values of the predicate $\qpredc$ get swapped:
\beq\label{eq:flipneg}
\begin{split}
\newcommand{\zzero}{\scriptstyle  a_\flse}
\newcommand{\oone}{\scriptstyle  a_\true}
\newcommand{\nneg}{\sim} 
\newcommand{\negg}{\neg}
\newcommand{\ttop}{\scriptstyle\true}
\newcommand{\bbot}{\scriptstyle \flse}
\newcommand{\ueval}{\scriptstyle \universal}
\newcommand{\quu}{\scriptstyle \qpredc}
\def\JPicScale{.34}
\ifx\JPicScale\undefined\def\JPicScale{1}\fi
\unitlength \JPicScale mm
\begin{picture}(275,100)(0,0)
\linethickness{0.35mm}
\put(10,0){\line(0,1){30}}
\linethickness{0.35mm}
\put(10,50){\line(0,1){20}}
\linethickness{0.35mm}
\put(0,50){\line(1,0){20}}
\linethickness{0.35mm}
\put(20,30){\line(0,1){20}}
\linethickness{0.35mm}
\put(0,30){\line(1,0){20}}
\linethickness{0.35mm}
\put(0,30){\line(0,1){20}}
\put(40,60){\makebox(0,0)[cc]{\EQLS}}

\linethickness{0.35mm}
\put(65,60){\line(1,0){55}}
\linethickness{0.35mm}
\put(120,40){\line(0,1){20}}
\linethickness{0.35mm}
\put(100,60){\line(0,1){10}}
\linethickness{0.35mm}
\put(85,40){\line(1,0){35}}
\linethickness{0.35mm}
\multiput(60,35)(0.12,0.12){125}{\line(0,1){0.12}}
\linethickness{0.35mm}
\multiput(65,60)(0.12,-0.12){167}{\line(1,0){0.12}}
\linethickness{0.35mm}
\put(81.25,10){\line(1,0){17.5}}
\linethickness{0.35mm}
\put(81.25,10){\line(0,1){17.5}}
\linethickness{0.35mm}
\multiput(81.25,27.5)(0.12,-0.12){146}{\line(1,0){0.12}}
\linethickness{0.35mm}
\put(90,18.75){\line(0,1){21.25}}
\put(87.5,15){\makebox(0,0)[cc]{$\zzero$}}

\put(102.5,50){\makebox(0,0)[cc]{$\ueval$}}

\linethickness{0.35mm}
\put(106.25,10){\line(1,0){17.5}}
\linethickness{0.35mm}
\put(106.25,10){\line(0,1){17.5}}
\linethickness{0.35mm}
\multiput(106.25,27.5)(0.12,-0.12){146}{\line(1,0){0.12}}
\linethickness{0.35mm}
\put(115,18.75){\line(0,1){21.25}}
\put(10,40){\makebox(0,0)[cc]{$\nneg$}}

\put(113.13,15.62){\makebox(0,0)[cc]{$\oone$}}

\linethickness{0.35mm}
\put(60,0){\line(0,1){10}}
\linethickness{0.35mm}
\put(60,30){\line(0,1){5}}
\linethickness{0.35mm}
\put(50,30){\line(1,0){20}}
\linethickness{0.35mm}
\put(70,10){\line(0,1){20}}
\linethickness{0.35mm}
\put(50,10){\line(1,0){20}}
\linethickness{0.35mm}
\put(50,10){\line(0,1){20}}
\put(60,20){\makebox(0,0)[cc]{$\quu$}}

\linethickness{0.35mm}
\put(100,90){\line(0,1){10}}
\linethickness{0.35mm}
\put(90,90){\line(1,0){20}}
\linethickness{0.35mm}
\put(110,70){\line(0,1){20}}
\linethickness{0.35mm}
\put(90,70){\line(1,0){20}}
\linethickness{0.35mm}
\put(90,70){\line(0,1){20}}
\put(100,80){\makebox(0,0)[cc]{$\quu$}}

\linethickness{0.35mm}
\put(10,90){\line(0,1){10}}
\linethickness{0.35mm}
\put(0,90){\line(1,0){20}}
\linethickness{0.35mm}
\put(20,70){\line(0,1){20}}
\linethickness{0.35mm}
\put(0,70){\line(1,0){20}}
\linethickness{0.35mm}
\put(0,70){\line(0,1){20}}
\put(10,80){\makebox(0,0)[cc]{$\quu$}}

\linethickness{0.35mm}
\put(165,90){\line(1,0){55}}
\linethickness{0.35mm}
\put(220,70){\line(0,1){20}}
\linethickness{0.35mm}
\put(200,90){\line(0,1){10}}
\linethickness{0.35mm}
\put(185,70){\line(1,0){35}}
\linethickness{0.35mm}
\multiput(160,65)(0.12,0.12){125}{\line(0,1){0.12}}
\linethickness{0.35mm}
\multiput(165,90)(0.12,-0.12){167}{\line(1,0){0.12}}
\linethickness{0.35mm}
\put(181.25,40){\line(1,0){17.5}}
\linethickness{0.35mm}
\put(181.25,40){\line(0,1){17.5}}
\linethickness{0.35mm}
\multiput(181.25,57.5)(0.12,-0.12){146}{\line(1,0){0.12}}
\linethickness{0.35mm}
\put(190,48.75){\line(0,1){21.25}}
\put(202.5,80){\makebox(0,0)[cc]{$\ueval$}}

\linethickness{0.35mm}
\put(206.25,40){\line(1,0){17.5}}
\linethickness{0.35mm}
\put(206.25,40){\line(0,1){17.5}}
\linethickness{0.35mm}
\multiput(206.25,57.5)(0.12,-0.12){146}{\line(1,0){0.12}}
\linethickness{0.35mm}
\put(215,48.75){\line(0,1){21.25}}
\linethickness{0.35mm}
\put(160,0){\line(0,1){10}}
\linethickness{0.35mm}
\put(160,30){\line(0,1){35}}
\linethickness{0.35mm}
\put(150,30){\line(1,0){20}}
\linethickness{0.35mm}
\put(170,10){\line(0,1){20}}
\linethickness{0.35mm}
\put(150,10){\line(1,0){20}}
\linethickness{0.35mm}
\put(150,10){\line(0,1){20}}
\put(160,20){\makebox(0,0)[cc]{$\quu$}}

\put(140,60){\makebox(0,0)[cc]{\EQLS}}

\linethickness{0.35mm}
\put(265,0){\line(0,1){10}}
\linethickness{0.35mm}
\put(265,30){\line(0,1){20}}
\linethickness{0.35mm}
\put(255,30){\line(1,0){20}}
\linethickness{0.35mm}
\put(275,10){\line(0,1){20}}
\linethickness{0.35mm}
\put(255,10){\line(1,0){20}}
\linethickness{0.35mm}
\put(255,10){\line(0,1){20}}
\put(265,20){\makebox(0,0)[cc]{$\quu$}}

\put(240,60){\makebox(0,0)[cc]{\EQLS}}

\linethickness{0.35mm}
\put(265,70){\line(0,1){30}}
\linethickness{0.35mm}
\put(255,70){\line(1,0){20}}
\linethickness{0.35mm}
\put(275,50){\line(0,1){20}}
\linethickness{0.35mm}
\put(255,50){\line(1,0){20}}
\linethickness{0.35mm}
\put(255,50){\line(0,1){20}}
\put(185,45){\makebox(0,0)[cc]{$\bbot$}}

\put(210,45){\makebox(0,0)[cc]{$\ttop$}}

\put(265,60){\makebox(0,0)[cc]{$\negg$}}

\put(10,50){\makebox(0,0)[cc]{\Dottt}}

\put(10,90){\makebox(0,0)[cc]{\Dottt}}

\put(100,90){\makebox(0,0)[cc]{\Dottt}}

\put(60,30){\makebox(0,0)[cc]{\Dottt}}

\put(90,18.75){\makebox(0,0)[cc]{\Dottt}}

\put(115,18.75){\makebox(0,0)[cc]{\Dottt}}

\put(160,30){\makebox(0,0)[cc]{\Dottt}}

\put(190,48.75){\makebox(0,0)[cc]{\Dottt}}

\put(215,48.75){\makebox(0,0)[cc]{\Dottt}}

\put(265,70){\makebox(0,0)[cc]{\Dottt}}

\put(265,30){\makebox(0,0)[cc]{\Dottt}}

\end{picture}

\end{split}
\quad \mbox{ and thus }
\quad
\qpredc\left(\sim\!\! x\right )\ \   = \ \  \neg \qpredc(x)
\eeq
The Fundamental Theorem gives a Kleene fixpoint $\Psi: {\DP}$ such that
\bea\label{eq:tttwo}
\{\sim\!\! \Psi\} & = & \{\Psi\}
\eea
Since $p$ is extensional, \eqref{eq:tttwo} implies
\bea\label{eq:ttthree}
\qpredc(\sim\!\! \Psi) & = & \qpredc(\Psi)
\eea
Hence
\[
\qpredc(\Psi)\ \stackrel{\eqref{eq:ttthree}}= \ \qpredc(\sim\!\! \Psi)\ \stackrel{\eqref{eq:flipneg}} = \ \neg \qpredc(\Psi) \ \ \ \lightning
\]
The assumption that $\qpredc$ is a decidable, extensional, and non-constant thus leads to a contradiction. If $\qpredc$ is decidable, then $\qpredc(\Psi)$ must output a truth value. Then $\qpredc(\Psi) = \neg \qpredc(\Psi)$ implies $\true = \flse$. 

\para{The two ways of living with undecidability.} Rice's Theorem is a rigid no-go result, ending all hope for deciding any nontrivial properties of computations by looking at their programs without running them. There are just two ways out of the triviality of the extensional decidable properties: 
\begin{enumerate}[a)]
\item drop the extensionality, and study the decidable \emph{intensional}\/ properties, or 
\item drop the decidability, and study the extensional \emph{computable}\/ properties.
\end{enumerate}
Approach (a) leads into the realm of  \emph{algorithmics}\/ and computational complexity \sindex{complexity} \cite{Arora-Barak:book,CaludeC:CX,PapaC:CX}. It studies the algorithms behind programs and measures the resources needed for program evaluation.  Approach (b) leads into the realm of \emph{semantics}\/ and domain theory of computation  \cite{Abramsky-Jung:domains,GunterC:book,WinskelG:book}. It studies the structure of spaces of computable \funnn s as the domains of meaning of computation.  

%
%

\def\thechapter{6}
\setchaptertoc
\chapter{Computing programs}
\label{Chap:Metaprog}
\newpage

\section{Idea of metaprogramming}\label{Sec:meta-intro}

\begin{figure}[h!t]
\begin{center}
\newcommand{\programnote}{\footnotesize \textbf{\textit{program}} $\leadsto$}
\newcommand{\metaprogramnote}{\footnotesize \textbf{\textit{metaprogram}} $\leadsto$}
\newcommand{\comnote}{\footnotesize \textbf{\textit{computation}} $\leadsto$}
\newcommand{\runone}{\{-\}}
\newcommand{\computation}{f}
\newcommand{\Progone}{F}
\newcommand{\Progtwo}{\FFF}
\def\JPicScale{.4}
\input{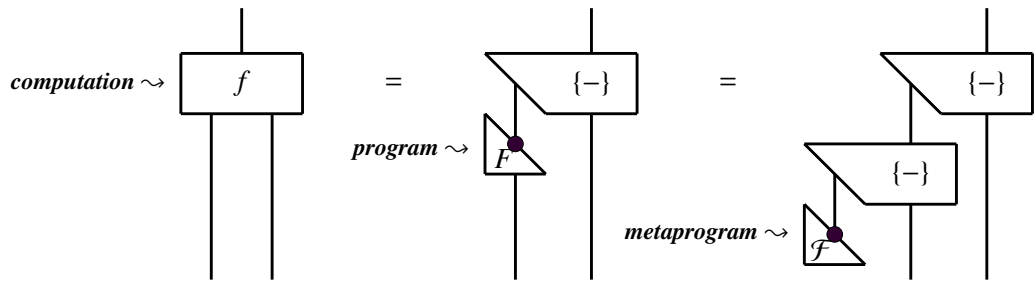}
\caption{Program and metaprogram evaluations}
\label{Fig:meta}
\end{center}
\end{figure}

Metaprograms are programs that compute programs. The idea is in Fig.~\ref{Fig:meta}. 
The computation $f$ is encoded by a program $F$ such that $f=\uev F$, and $F$ is computed by a metaprogram $\FFF$ as $F = \uev\FFF$. While the theory of computation arose from the idea that computable \funnn s are just those that are programmable, one of the central tenets of the practice of computation has been to use computers to facilitate programming computers. That idea led to the invention of the high-level programming languages and has driven their evolution.

The first equation in Fig.~\ref{Fig:meta} says that every computation $f$ has a program $F$ with $f = \uev F$. The question arises: \emph{How do we find a program $F$ for a given computation $f$?} Program evaluators take us from right to left in Fig.~\ref{Fig:meta}. How do we go from left to right? The answer, displayed alredy in Fig.~\ref{Fig:head-tail}, is the process of \emph{programming}. It is usually left out of the models of computation, covered by the tacit assumptions that programs are produced by people, and that people live in a different realm from computers. In reality, of course, people and computers intermingle. Even more because they are different. Early on, it became clear that supplying the machine-readable programs was hard, error-prone, and very tedious. So people came up with the idea to use computers to program computers: they programmed them to translate high-level human code into low-level machine code.

\section{Compilation and supercompilation}\label{Sec:compiler}

\subsection[Compilation]{\Large Compilation}
An important aspect of programming and of the process of software development is that we can write metaprograms not just to translate human programs into machine programs, but also to transform and \emph{improve} programs. This is done by \emph{compilers}. The basic idea is already in Fig.~\ref{Fig:meta} (but see also Fig.~\ref{Fig:interpret-compile}). Besides translating programs, compilers automate some routines, such as memory management. They often also optimize program structure, since what is simple for a programmer may not be efficient for the computer. 

\begin{figure}[h!t]
\newcommand{\highunitag}{h}
\newcommand{\lowunitag}{\ell}
\newcommand{\copilertag}{c}
\newcommand{\Compiler}{C}
\newcommand{\Highprogtype}{\HHh}
\newcommand{\Lowprogtype}{\LLl}
\newcommand{\compilernote}{compilation $\leadsto$}
\newcommand{\compilerote}{compiler $\leadsto$}
\begin{center}
\def\JPicScale{.4}
\input{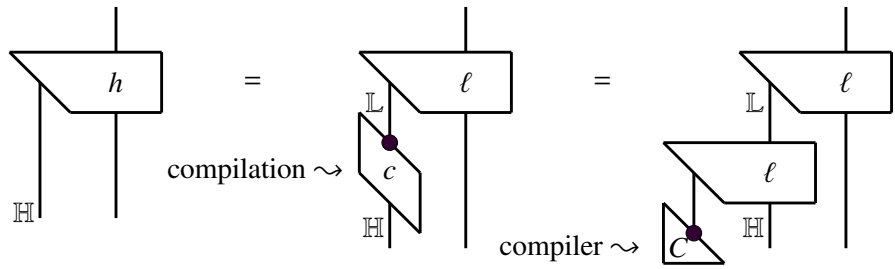}
\caption{Compiler is a metaprogram that tranforms programs from $\HHh$igh-level language to $\LLl$ow-level language}
\label{Fig:compiler}
\end{center}
\end{figure}

A compiler is thus a metaprogram that optimizes program code while translating it from a high-level language into a low-level language. Fig.~\ref{Fig:compiler} shows the idea from the opposite direction. Here you are given a friendly high-level language $\HHh$, with a slow universal evaluator $h$, and an unfriendly low-level language $\LLl$ with a fast universal evaluator $\ell$.
The compilation phase combines the best from both worlds: the friendly high-level programs are transformed into efficient low-level programs. All about compilation can be found in the Dragon Book \cite{dragon-book}.

\subsection{Supercompilation}\label{Sec:Futamura}\sindex{supercompiler}
Leaving the environment-dependent program-optimizations aside, compilation can systematically improve efficiency of computations already by \emph{partially evaluating the evaluators on code}. This idea was put forward and systematically explored  under the name \emph{supercompilation} by Valentin Turchin \cite{Turchin:supercompiler}. Much later, but independently, Yoshihiko Futamura presented in \cite{Futamura} an illuminating family of supercompilation constructions, which came to be called \emph{Futamura projections}.\sindex{Futamura projections}

\begin{figure}[h!t]
\begin{center}
	\newcommand{\Cprog}{C_1}
\newcommand{\Sprog}{S}
\newcommand{\stag}{\pev -}
\newcommand{\itag}{H}
\newcommand{\ltag}{\ell}
\newcommand{\htag}{h}
\newcommand{\Lprogtype}{\mathbb L}
\newcommand{\Hprogtype}{\mathbb H}
\newcommand{\EQLSsix}{\large $\stackrel{(6)}=$}
\newcommand{\EQLSfive}{\large $\stackrel{(5)}=$}
\newcommand{\EQLSfour}{\large $\stackrel{(4)}=$}
\newcommand{\EQLSthree}{\large $\stackrel{(3)}=$}
\newcommand{\EQLStwo}{\large $\stackrel{(2)}=$}
\newcommand{\EQLSone}{\large $\stackrel{(1)}=$}
\newcommand{\EQLSzero}{\large $\stackrel{(0)}=$}
\def\JPicScale{.37}
\input{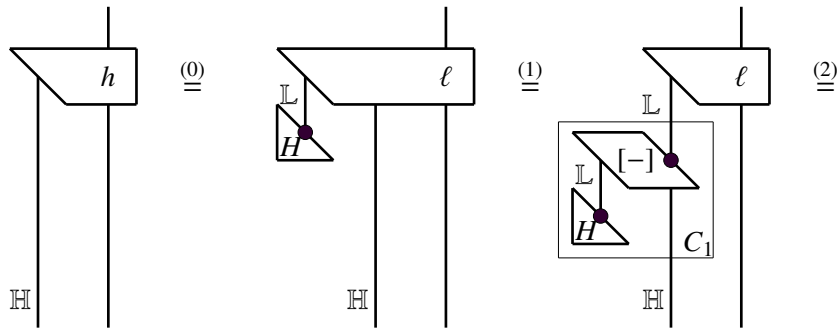}
\caption{First Futamura projection: Compilation by partial evaluation of an intepreter}
\label{Fig:FirstFuta}
\end{center}
\end{figure}

The initial setting is like in Fig.~\ref{Fig:compiler}. We are given a friendly but inefficient universal evaluator $h$ for a high-level language $\HHh$ and an unfriendly but efficient evaluator $\ell$ for a low-level language $\LLl$. We need a compiler, to translate easy $\HHh$-programs to efficient $\LLl$-programs. On the other hand, since the program evaluators $h$ and $\ell$ are universal, they can be programmed on each other. In other words, there is an $\LLl$-program $H$ and an $\HHh$-program $L$ such that
\[\uev H_\LLl = h \qquad \mbox{ and } \qquad \uev L_\HHh = \ell
\]
Such metaprograms are called\sindex{interpreter} \emph{interpreters}: $H$ interprets $\HHh$ to $\LLl$, whereas $L$ interprets $\LLl$ to $\HHh$. It is nice that we can write a low-level interpreter $L$ in a friendly language $\HHh$, but if we want to make it efficient, we must compile it to $\LLl$. So it is even nicer that we can write the high-level interpreter $H$ in an efficient language $\LLl$ --- \emph{because}
\[\uev X_\HHh y \ \stackrel{(0)}=\ \uev{H}_\LLl(X,y)\  \stackrel{(1)}= \ \uev{\pev{H}_\LLl X }_\LLl y\]
Partially evaluating the $\HHh$-to-$\LLl$-interpreter $H$ on $\HHh$-programs $X$ thus compiles the $\HHh$-programs to $\LLl$. This is displayed in Fig.~\ref{Fig:FirstFuta}.  
A program transformer in the form $C_1 = \pev{H}$ is an instance of the \emph{first Futamura projection}.  

\begin{figure}[h!t]
\begin{center}
			\newcommand{\Gprog}{C_2}
	\newcommand{\Cprog}{C_1}
\newcommand{\Sprog}{S}
\newcommand{\stag}{\pev -}
\newcommand{\itag}{H}
\newcommand{\ltag}{\ell}
\newcommand{\Lprogtype}{\mathbb L}
\newcommand{\Hprogtype}{\mathbb H}
\newcommand{\EQLSsix}{\large $\stackrel{(6)}=$}
\newcommand{\EQLSfive}{\large $\stackrel{(5)}=$}
\newcommand{\EQLSfour}{\large $\stackrel{(4)}=$}
\newcommand{\EQLSthree}{\large $\stackrel{(3)}=$}
\newcommand{\EQLStwo}{\large $\stackrel{(2)}=$}
\newcommand{\EQLSone}{\large $\stackrel{(1)}=$}
\newcommand{\EQLSzero}{\large $\stackrel{(0)}=$}
\def\JPicScale{.37}
\input{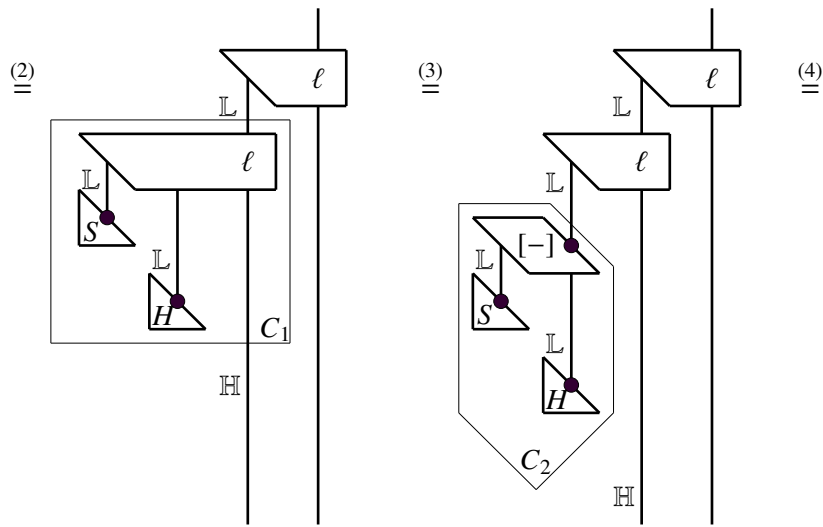}
\caption{Second Futamura projection: A compiler by partial evaluation of a specializer on an interpreter}
\label{Fig:SecondFuta}
\end{center}
\end{figure}

An $\HHh$-to-$\LLl$-compilation can thus be realized by partially evaluating an $\HHh$-to-$\LLl$-interpreter. But where is the underlying $\HHh$-to-$\LLl$-compiler? Since $\ell$ is a universal evaluator, there is of course an $\LLl$-program $S$ that implements the partial $\LLl$-evaluator and which can then be partially evaluated itself, giving
\bear
\pev X_\LLl y & \stackrel{(2)}= &  \uev S_\LLl (X, y)
\eear
Metaprograms that implement partial evaluators are usually called \emph{specializers}\sindex{specializer}, just like programs that implement universal evaluators are called interpreters. Now partially evaluating $S$ just like we partially evaluated $H$ gives
\bear
 \uev S_\LLl (X, y) &  \stackrel{(3)}= & \uev{\pev S_\LLl X}_\LLl y
 \eear
Fig.~\ref{Fig:SecondFuta} displays an interesting phenomenon:\emph{partially evaluating\sindex{partial evaluation} a \textbf{specializer} on an \textbf{interpreter} gives a \sindex{compiler}\textbf{compiler}}. 
A compiler in the form $C_2 = \pev S_\LLl H$ is called a \emph{second Futamura projection}. 

\begin{figure}[h!t]
\begin{center}
			\newcommand{\Gprog}{C_3}
	\newcommand{\Cprog}{C_2}
\newcommand{\Sprog}{S}
\newcommand{\stag}{s}
\newcommand{\itag}{H}
\newcommand{\ltag}{\ell}
\newcommand{\Lprogtype}{\mathbb L}
\newcommand{\Hprogtype}{\mathbb H}
\newcommand{\EQLSsix}{\large $\stackrel{(6)}=$}
\newcommand{\EQLSfive}{\large $\stackrel{(5)}=$}
\newcommand{\EQLSfour}{\large $\stackrel{(4)}=$}
\newcommand{\EQLSthree}{\large $\stackrel{(3)}=$}
\newcommand{\EQLStwo}{\large $\stackrel{(2)}=$}
\newcommand{\EQLSone}{\large $\stackrel{(1)}=$}
\newcommand{\EQLSzero}{\large $\stackrel{(0)}=$}
\def\JPicScale{.37}
\input{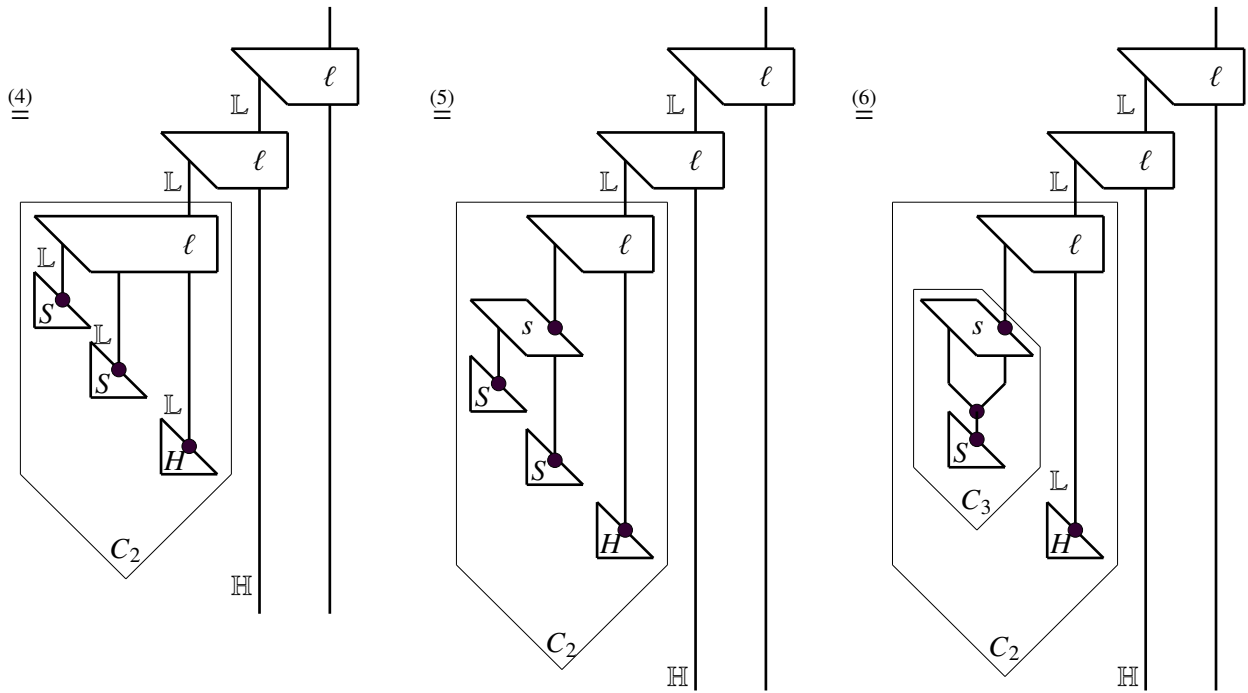}
\caption{Third Futamura projection: A compiler generator by partial evaluation of a specializer on itself}
\label{Fig:ThirdFuta}
\end{center}
\end{figure}

If the cost of the compiler itself is taken into account, then there is a sense in which compilation by partial evaluation is optimal. Can a further compilation gain be obtained by iterating the same trick? Well, there is still a partial evaluator inside $C_2$, which means that the specializer $S$ can be partially evaluated again. The thing is that the partial evaluators inside $C_2$ partially evaluates $S$ itself. So now we have
\bear
\pev S _\LLl H & \stackrel{(4)}=& \uev S_\LLl (S,H)
\eear
which now partially evaluates to
\bear
\uev S_\LLl (S,H) & \stackrel{(5)} = & \uev{\pev S_\LLl S}_\LLl H
\eear
Fig.~\ref{Fig:ThirdFuta} shows that the program in the form $C_3 = \pev S_\LLl S$, called the \emph{third Futamura preojection}, generates a compiler, since $\uev{C_3}_\LLl = C_2$, and $C_2$ is a compiler. 
So we have yet another interesting phenomenon: \emph{partially evaluating a \textbf{specializer} on a \textbf{specializer} gives a \textbf{compiler generator}}. So we have
\[ \uev x_\HHh y \ =\  \uev{C_1 x}_\LLl y \qquad \mbox{ and } \qquad C_1 \ =\ \uev{C_2}_\LLl \qquad \mbox{ and } \qquad C_2 \ =\ 
\uev{C_3}_\LLl \]
Partially evaluating these metaprograms each time improves the execution of some programs. Sometimes significantly! If a speedup by a factor 10 is incurred each time, then applying the three Futamura projections one after the other gives a speedup by a factor 1000.

\section{Metaprogramming hyperfunctions}
\label{Sec:iteration}

So far, we studied the role of metaprogramming in the system programming. System programs manage programs, so it is natural that many of them are metaprograms. But metaprogramming is also an important technique in application programming. Some functions are hard to program but easy to metaprogram. In other words, it may be hard to write a program to compute a function, but easy to write a program that will output a program that will compute the function. 

How can this be? An intuitive but vague way to understand this is to remember that, in many sports, it is easier to be the coach, than to be a player. Metaprograms can coach programs towards better computations. A more specific reason why metaprogramming often simplifies application programming is that a function, say $f:A\times A \to B$, may be such that
\begin{itemize}
\item the function $\hat f: A\to B$ defined by $\hat f(x) = f(x,x)$ is hard to program, but
\item the functions $f_a: A\to B$, where $f_a(x) = f(a,x)$  for $a\in A$ (or $f^a: A\to B$, where $f^a(x) = f(x,a)$)
are much easier.\footnote{Such examples anticipate the realm of \emph{parametric}\/ complexity: a computation may be more complex in one argument and less complex in another argument. Partially evaluating over the complex argument then localizes the complexity at the metaprogram.}
\end{itemize}
This means that it is hard to find a program $\hat F$ such that $\uev{\hat F} = \hat f : A\to B$, but easy to find a metaprogram $F$ such that $\uev F : A\to \PPp$ gives $\uev {\uev F a} = f_a:A\to B$ for every $a\in A$. The metaprogram $F$ thus computes the partial evaluations of a computation for $f:A\times A\to B$ over the first argument. Another metaprogram $\tilde F$ would give the partial evaluations $\uev{\tilde F}:A\to \PPp$ over the second argument, i.e. $\uev{\uev{\tilde F}a} = f^a:A\to B$. Functional programming (in languages like Lisp or Haskell) is particularly well suited for such metaprogramming ideas. On the imperative side, the language Nim is designed with an eye on metaprogramming: as a high-level language that compiles to high level programming languages, including C++ and  JavaScript. 

\subsection{Ackermann's hyperfunction}\label{Sec:meta-iterators}
In Exercises \ref{Sec:work-num}, we saw how arithmetic operations arise from one another, starting from the \emph{successor} $\suce (n) = 1+n$, and then proceeding with
\begin{itemize}
\item \emph{addition}\/ $m+(-):\NNn \to \NNn$ as iterated successor,
\item \emph{multiplication}\/ $m\cdot (-):\NNn \to \NNn$ as iterated addition,
\item \emph{exponentiation}\/ $m^{(-)}:\NNn \to \NNn$ as iterated multiplication.
\end{itemize}
It is clear that the sequence continues, provided that we surmount the notational obstacle of iterated exponentiation in the form:
\[{m^{m^{\cdots}}}^m\] 
which seems hard to iterate any further. But just a cosmetic change in notation, writing the exponentiation in the form $m\uparrow n$ instead of $m^n$, allows listing a recurrent sequence of recursive definitions of basic arithmetic operations:
\bear
{\color{blue}1+}0 = 1
&&
{\color{blue}1+}(n+1) = ({\color{blue}1+}n)+1
\\
{\color{red}m+}0 = m
&&
{\color{red}m+}(n+1) = {\color{blue}1+}({\color{red}m+}n)
\\
{\color{green}m\cdot}0 = 0
&&
{\color{green}m\cdot}(n+1) = {\color{red}m+}({\color{green}m\cdot}n)
\\
m\uparrow 0 = 1
&&
m\uparrow (n+1) \ =\  {\color{green} m\cdot}(m\uparrow n)
\\m\uparrow\uparrow 0 = 1
&&
m\uparrow\uparrow (n+1) \ =\  m\uparrow (m\uparrow\uparrow n)
\\m\uparrow\uparrow\uparrow 0 = 1
&&
m\uparrow\uparrow\uparrow (n+1) \ =\  m\uparrow\uparrow (m\uparrow\uparrow\uparrow n)
\\
&\hspace{2.5em}&\cdots
\eear
After the base cases settle on the multiplicative unit 1, this sequence of recursive definitions boils down to the equations
\bea\label{eq:ackermann}
 m\uparrow^0 n &=& mn\\
m\uparrow^{1+k} 0 & = & 1\notag \\
m\uparrow^{1+k} (1+n) &=&  m\uparrow^k (m\uparrow^{1+k} n)\notag
\eea
The question whether these equations can be subsumed under primitive recursion was explored by Wilhelm Ackermann\footnote{The $\uparrow$ notation is due to Don Knuth.}. It is easy to see that the function $\uparrow : \NNn\times \NNn \times \NNn \to \NNn$ grows very fast:
\bear
3\uparrow^0 2 & = & 3\cdot 2\ =\ 6\\
3\uparrow^1 2 & = & 3^2\ =\ 9\\
3\uparrow^{2} 2 & = & 3\uparrow 3\ =\ 3^3\ =\ 27\\
3\uparrow^{3} 2 & = & 3\uparrow^2 3\ =\ 3\uparrow 3 \uparrow 3 \  =\   7,625,597,484,987\\
3\uparrow^{4} 2 & = & 3\uparrow^3 3 = 3\uparrow^2 3\uparrow^2 3 \ =\ 3\uparrow^2 7,625,597,484,987 \\
& = & \underbrace{3\uparrow 3 \uparrow 3\uparrow 3\uparrow \cdots\cdots \uparrow 3}_{7,625,597,484,987 \mbox{ \scriptsize times}}\\
&\cdots
\eear
Ackermann proved that it grows faster than any primitive recursive function. One way to see this is to show that for every primitive recursive function $f:\NNn^\ell \to \NNn$ there is a number $k_f$ such that $f(n_1, n_2,\ldots, n_\ell) \lt 2\uparrow^{k_f} \overline n$, where $\overline n = \sum_{i=1}^\ell n_i$. Another way will be sketched as an exercise at the end of this chapter. 

The arithmetic operation $\uparrow : \NNn\times \NNn \times \NNn \to \NNn$ is sometimes called a \emph{hyperfunction}. Our goal is to show that it is easy to metaprogram. Replacing the infix notation $\uparrow$ with the prefix notation $\epsilon_k(n,m)  =  m\uparrow^k n$,
equations \eqref{eq:ackermann} become
\bea\label{eq:ackermann-recur}
\epsilon_0(n,m) &=& m\cdot n \\
\epsilon_{k+1}(n,m) &=& \epsilon_k^{\bullet n}(m,1)\notag
\eea
where
\bear
f^{\bullet n}(m,x) &=&\begin{cases} x & \mbox{ if } n=0\\
\underbrace{f(m, f(m,\ldots f(m,x)\cdots))}_{n \mbox{ \scriptsize times}} & \mbox{otherwise}
\end{cases}
\eear
is the \emph{parametrized iteration}\/ operation
\[
\prooftree
n\in \NNn \qquad \qquad f : M \times A \to A 
\justifies
f^{\bullet} : \NNn\times M\times A \to A 
\endprooftree
\]
defined by
\bea\label{eq:pariter}
f^{\bullet 0}(m,x) & =& x\\
f^{\bullet n+1}(m,x) & =& f(m, f^{\bullet n}(m,x))\notag
\eea
The \textbf{task} is thus to metaprogram \eqref{eq:pariter}, and use it to program \eqref{eq:ackermann-recur}

\bigskip
\para{Hierarchy of recursive functions.} Andrzej Grzegorczyk \cite{Grzegorczyk1953}\sindex{Grzegorczyk hierarchy} stratified the\sindex{function!recursive} recursive \funnn s
into a tower
\[\arraycolsep=1.4pt\def\arraystretch{2.2}
\begin{array}{llllllllll}
\EEE^0 \subset & \EEE^1 \subset & \EEE^2 \subset &\EEE^3\subset &\EEE^4 \subset &\EEE^5 \subset&\cdots\subset & \EEE^{k+2}\subset&\cdots & 
\\[-4.25ex]
\inup & \inup & \inup &\inup &\inup &\inup&&\inup\\[-4ex]
\suce & + & \times & \uparrow & \uparrow^2 & \uparrow^3 &\cdots & \uparrow^k  & \cdots
\end{array}\]
where each class also contains the data services and is closed under the composition. 
%
He proved that that every recursive function is contained in $\EEE^{k}$  for some $k$.

\subsection{Metaprogramming parametric iteration}\label{Sec:paramiter}\sindex{iteration!parametric}
In most programming languages, the sequential composition of programs, usually written as a semicolon, allows that the composed programs share a parameter, passed through the composition. Denoting the parameter as $m$, the sequential composition is thus in the form
\bear
\uev{F\, ; G}(m,x) & = & \uev G \left(m, \uev F  (m,x)\right)
\eear
\newcommand{\suces}{\scriptscriptstyle (-\, ;-)}
\newcommand{\Bee}{\scriptstyle F}
\newcommand{\Cee}{\scriptstyle G}
\newcommand{\Mee}{\scriptstyle m}
\newcommand{\Ahh}{\scriptstyle x}
\def\JPicScale{.35}
\beq\label{eq:seqcomp}
\raisebox{-60pt}{
\ifx\JPicScale\undefined\def\JPicScale{1}\fi
\psset{unit=\JPicScale mm}
\psset{linewidth=0.3,dotsep=1,hatchwidth=0.3,hatchsep=1.5,shadowsize=1,dimen=middle}
\psset{dotsize=0.7 2.5,dotscale=1 1,fillcolor=black}
\psset{arrowsize=1 2,arrowlength=1,arrowinset=0.25,tbarsize=0.7 5,bracketlength=0.15,rbracketlength=0.15}
\begin{pspicture}(0,0)(155,110)
\psline[linewidth=0.75](115,50)(155,50)
\psline[linewidth=0.75](155,30)(155,50)
\psline[linewidth=0.75](145,80)(145,50)
\psline[linewidth=0.75](135,30)(155,30)
\psline[linewidth=0.75](140,30)(140,5)
\psline[linewidth=0.75](110,100)(150,100)
\psline[linewidth=0.75](150,80)(150,100)
\psline[linewidth=0.75](140,110)(140,100)
\psline[linewidth=0.75](130,80)(150,80)
\psline[linewidth=0.75](110,100)(130,80)
\rput(75,65){\EQLS}
\psline[linewidth=0.75](10,75)(50,75)
\psline[linewidth=0.75](50,55)(50,75)
\psline[linewidth=0.75](40,110)(40,75)
\psline[linewidth=0.75](30,55)(50,55)
\psline[linewidth=0.75](20,65)(13.75,58.75)
\psline[linewidth=0.75](10,75)(30,55)
\psline[linewidth=0.75](35,55)(35,5)
\psline[linewidth=0.75](100,70)(100,25)
\psline[linewidth=0.75](5,47.5)(5,67.5)
\psline[linewidth=0.75](5,67.5)(25,47.5)
\psline[linewidth=0.75](5,47.5)(25,27.5)
\psline[linewidth=0.75](25,27.5)(25,47.5)
\rput(15,47.5){$\suces$}
\psline[linewidth=0.75](10,42.5)(10,5)
\psline[linewidth=0.75](20,32.5)(20,5)
\psline[linewidth=0.75](120,90)(100,70)
\psline[linewidth=0.75](110,50)(135,75)
\psline[linewidth=0.75,border=1](110,50)(110,45)
\psline[linewidth=0.75](45,55)(45,5)
\psline[linewidth=0.75](135,80)(135,75)
\psline[linewidth=0.75](140,15)(110,45)
\psline[linewidth=0.75](150,30)(150,5)
\rput{0}(140,15){\psellipse[fillstyle=solid](0,0)(2.5,-2.5)}
\psline[linewidth=0.75](110,15)(110,5)
\rput(140,1.25){$\Mee$}
\rput(35,1.25){$\Mee$}
\rput(150,1.25){$\Ahh$}
\rput(45,1.25){$\Ahh$}
\rput(10,1.25){$\Bee$}
\rput(20,1.25){$\Cee$}
\rput(110,1.25){$\Cee$}
\psline[linewidth=0.75](100,15)(100,5)
\psline[linewidth=0.75](100,25)(110,15)
\psline[linewidth=0.75,border=1](125,40)(100,15)
\psline[linewidth=0.75](115,50)(135,30)
\rput(100,1.25){$\Bee$}
\end{pspicture}
}
\eeq
Let $(-\, ;-)  :\PPp\times \PPp\to \PPp$ be the program operation defined by \eqref{eq:seqcomp}\footnote{In the preceding chapter, we defined sequential composition and parallel composition ignoring the parameters, for simplicity. In programming languages, most program operations are implemented with parameters, for convenience.}. Iterating this operation on a program iterates its execution 
\[
\big\{\underbrace{F\, ; F\, ; \cdots \, ; F}_{n \mbox{ \scriptsize times}}\big\}(m,x) \ \  = \ \  
\underbrace{\uev F (m,\uev F(m, \ldots \uev F (m,x)))\cdots)}_{n \mbox{ \scriptsize times}}\ \ 
 = \ \  \uev F^{\bullet n}(m,x)
\]
To iterate this parametrized sequential composition, we define the program operation  $\widetilde \iota:\PPp\times \PPp\times \NNn \to \PPp$ by
\bear
\widetilde \iota(p, F, n) &  = &  \iif\Big(\iszero(n),\  I,\   \big(\uev{p} \left(F, \pred(n)\right)\, ; F\big) \Big)
\eear
where $\uev I$ is a program for the identity, and $\pred n$ is the predecessor of $n$, which is 0 when $n$ is 0. If $J$ is the Kleene fixed point of $\widetilde \iota$, then for $\iota = \uev J$ holds 
\begin{multline*}
\iota(F,n)\ \ =\ \ \uev J(F,n) \  \ = \  \ \widetilde \iota(J, F, n)\ \ =\  \ \uev J(F, n-1)\, ; F \ \ =\ \ \uev J(F, n-2)\, ; F \, ; F\ \ = \cdots \\ \cdots=\ \  \widetilde \iota(J, F, 1)\, ; \underbrace{F\, ; F\, ;  \cdots \, ; F}_{n-1 \mbox{ \scriptsize times}} \ \ 
=\ \ \widetilde \iota(J, F, 0 )\, ; \underbrace{F\, ; F\, ;  \cdots \, ; F}_{n \mbox{ \scriptsize times}}\ =\ I\, ;  \underbrace{F\, ; F\, ;  \cdots \, ; F}_{n \mbox{ \scriptsize times}}
\end{multline*}
and thus
\[
\big\{\iota(F,n)\big\}(m,x)\ =\ \big\{I\, ; \underbrace{F\, ; F\, ; \cdots ; F}_{n \mbox{ \scriptsize times}} \big\}(m,x)  \ = \   \uev F^{\bullet n}(m,x)
\] 

\subsection{Metaprogramming a hyperfunction}\sindex{hyperfunction!Ackermann}
Given $\iota = \uev J$ let $W$ be the program such that
\bear
\uev W (F,n,m) & = & \big\{\iota(F,n)\big\}(m,1)\ =\   \uev{F}^{\bullet n}(m,1)
\eear
so that
\bear
\uev { \pev{W} F}(n,m) & = & \uev F^{\bullet n}(m,1)
\eear
Using this $\pev W:\PPp\to\PPp$, define
\bear
\widetilde e(p, k) &= &  \iif\Big(\iszero(k),\  \Xi,\ \big[W\big]\big(\uev {p} (\pred k)\big) \Big)
\eear
where $\{\Xi\}(n,m) = m\cdot n$. Now if $E$ is the Kleene fixed point of $\widetilde e$, then $e=\uev E$ satisfies
\bear
\uev {e(k)}(n,m) & = & \big\{\{E\}k\big\}(n,m)\ = \ \big\{\widetilde e \left(E, k\right)\big\}(n,m)  
\\[2ex] 
& = & \begin{cases} m\times n & \text{ if } k=0\\
\big\{[W]\left(\{E\}(k-1)\right)\big\}(n,m) & \text{ otherwise} 
\end{cases}  
\\[2ex]
& = &\begin{cases} m\times n & \text{ if } k=0\\
\big\{\{E\}(k-1)\big\}^{\bullet n}(m,1) & \text{ otherwise} 
\end{cases}\\[2ex]
& = &\begin{cases} m\times n & \text{ if } k=0\\
\big\{e(k-1)\big\}^{\bullet n}(m,1) & \text{ otherwise} 
\end{cases}
\eear
Thus the hyperoperation that we sought to metaprogram is $\epsilon_k = \big\{e(k)\big\}$, since this gives
\bear
\epsilon_k (n,m) &=& \begin{cases} m\times n & \mbox{ if } k=0\\
 \epsilon_{k-1}^{\bullet n}(m,1) & \mbox{ else}\end{cases}
\eear
as desired. The construction is summarized in Fig.~\ref{Fig:Ackermann}.
\begin{figure}[ht]
\begin{center}
\newcommand{\atilde}{\widetilde e}
\newcommand{\aflat}{e}
\newcommand{\ackermann}{\mbox{\Large $\mathbf \epsilon$}}
\newcommand{\iiftag}{\iif}
\newcommand{\zerotag}{\iszero}
\newcommand{\gtag}{\scriptstyle E}
\newcommand{\htag}{\scriptstyle \Xi}
\newcommand{\ktag}{\scriptstyle W}
\newcommand{\specializer}{[\,]}
\newcommand{\predtag}{\scriptstyle \pred}
\newcommand{\pee}{\scriptstyle p}
\newcommand{\een}{\scriptstyle k}
\newcommand{\eex}{\scriptstyle n, m}
\newcommand{\eem}{\scriptstyle m}
\newcommand{\utag}{\{\}}
\def\JPicScale{.4}
\input{PIC/ACK.tex}
\caption{The nest of metaprograms for Ackermann's hyperfunction}
\label{Fig:Ackermann}
\end{center}
\end{figure}

\section{Metaprogramming ordinals}\label{Sec:ordinals}

We saw in  Sec.~\ref{Sec:Num} how to represent natural numbers as programs. Now we proceed to represent transfinite ordinal numbers as metaprograms. Finite descriptions of infinite processes are, of course, the beating heart of computation and of language in general. All sentences, all conversations, all books and programs in all languages are all generated by finitely many rules from finitely many words. But the finite metaprograms for transfinite numbers are not just another avatar of this miracle but also a striking example iteration and a striking  display of dynamics of convergence and divergence in computation.

\subsection{Collection as abstraction}\label{Sec:collabs}
\para{Finite sets.} The \emph{set collection}\/ operation inputs some arbitrary elements, say $e_0, e_1$, and outputs the set\footnote{The usage of the curly brackets in the set collection $\{e_0, e_1\}$ and for the program evaluation $\uev F$ is a typographic accident, since the two operations are unrelated. But see the next paragraph.} $e=\{e_0, e_1\}$. The set $e$ can then be used as an element of another set $d = \{d_0, e\} = \{d_0, \{e_0, e_1\}\}$, which can be used as an element of $c = \{c_0, c_1, d\} = \{c_0, c_1, \{d_0, e\}\} = \{c_0,c_1, \{d_0, \{e_0, e_1\}\}\}$, and so on. The element relation $e_0, e_1 \in e \in d \in c$ arises inductively, and it is therefore irreflexive and well-founded. 

\begin{figure}[!ht]
\begin{center}
\renewcommand{\Cee}{\scriptstyle C}
\newcommand{\cee}{\scriptscriptstyle c_1}
\newcommand{\ceeone}{\scriptscriptstyle c_0}
\newcommand{\Dee}{\scriptstyle D}
\newcommand{\dee}{\scriptscriptstyle d_0}
\newcommand{\Eeah}{\scriptstyle E}
\newcommand{\eeahone}{\scriptscriptstyle e_0}
\newcommand{\eeahtwo}{\scriptscriptstyle e_1}
\newcommand{\unv}{\{\}}
\newcommand{\dotdt}{\mbox{\large $\bullet$}}
\def\JPicScale{.47}
\input{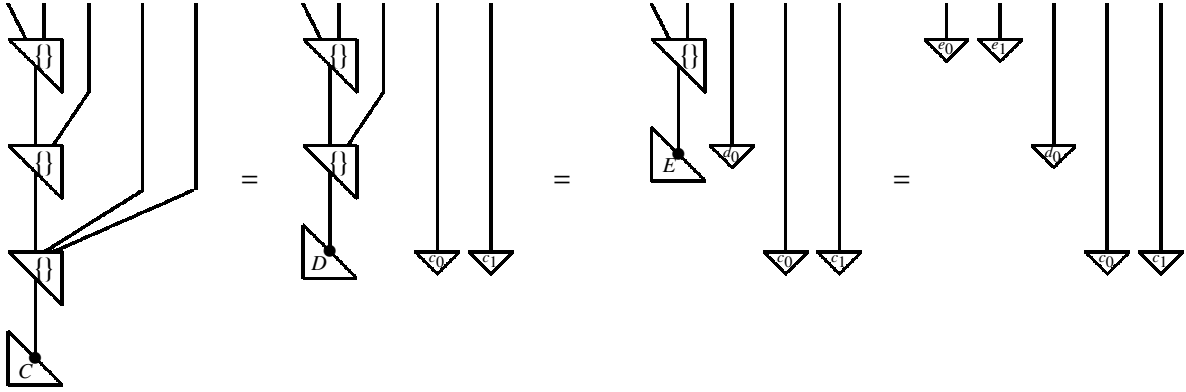} 
\caption{The set $c=\{c_0,c_1,\{d_0, \{e_0, e_1\}\}\}$ represented as a program $C$.}
\label{Fig:setz}
\end{center}
\end{figure}

\para{Finitely iterated abstractions.}  Fig.~\ref{Fig:setz} displays how we implement the set collection as program abstraction.
Each collection step corresponds to a program abstraction:
\begin{itemize}
\item the set $e=\{e_0, e_1\}$ is represented by a program $E$ such that $\uev E = <e_0, e_1>$;
\item the set $d = \{e, d_0\}$ is represented by a program $D$ with $\uev D = <E, d_0>$; and finally
\item the set $c = \{d, c_0, c_1\}$ becomes a program $C$ such that $\uev C = <D, c_0,c_1>$. 
\end{itemize}
Ironically, the notational clash between the set-theoretic and the computational uses of the curly brackets here turns out to be convenient. The sequence $c \ni d \ni e$ corresponds to "the time of creation" of sets \cite{ConwayJH:ONAG} and the order of evaluation of programs. 

\para{Infinite sequences.} The simplest way to generate an infinite sequence is the induction, i.e. counting infinitely long. This means that we start from the beginning $b:B$, and repeat the counting step $\seeta :B\to B$ forever. The infinite sequence arises:
\beq\label{eq:fundseq} s_0=b,\hspace{1.5em}  s_1=\seeta(b),\hspace{1.5em}   s_2 =\seeta^2(b),\hspace{1.5em}   s_3 = \seeta^3(b),\hspace{1.5em}  s_4=\seeta^4(b),\ldots
\eeq
In a universe of sets, this sequence can be collected into a set without much ado 
\bear
s & = &  \{s_0, s_1, s_2, \ldots , s_n, \ldots\}
\eear
If $s$ is a \emph{fundamental}\/ sequence, i.e. it satisfies
\beq 
s_0\hspace{.7em} \lt\hspace{.7em}s_1\hspace{.7em}\lt\hspace{.7em}s_2\hspace{.7em}\lt\hspace{.7em} s_ 3\hspace{.7em}\lt \hspace{.7em}s_4\hspace{.7em}\lt \hspace{.7em} \cdots \eeq
where $\lt$ is the transitive closure of the element relation, i.e.
\bear
x\lt y & \iff & \exists x_1x_2\cdots x_n.\ x \in x_0\in x_1\in x_2\in \cdots \in y
\eear
then

In a computer, the same can be done along the lines of Fig.~\ref{Fig:setz}. In particular, the set $\NNn$ of the natural numbers represented as programs in Sec.~\ref{Sec:num-prog} can be collected into a single program, corresponding to the first infinite ordinal $\omega$.

\para{Infinite iterated abstractions.} The pattern in Fig.~\ref{Fig:setz} suggests that the sequence $s_0, s_1,\ldots, s_n, \ldots$ could be captured by a sequence of programs $S_0, S_1,\ldots, S_n,\ldots$ with $\uev{S_n} =\left<S_{n+1}, s_n\right>$ for all $n \in \NNn$. Unfortunately, the equation $\uev{S_n} =\left<S_{n+1}, s_n\right>$ implies that if $\uev{S_{n+1}}$ is of type $X$,  then $\uev{S_n}$ must be of type $X\times \DP$. Allow $n$ to be arbitrarily large makes the type of $\uev{S_0}$ into an infinite product. This problem can be avoided by using the internal pairing $\pairr - -\colon \DP\times \DP \to \DP$ instead of the product types. We are thus looking for a program $S$ and the function $\bbag \seeta$ obtained by executing it, such that
\bea\label{eq:constrS}
\bbag \seeta x\ \ =\ \ \uev {S} x & = & \pairr{\uev {S}\left(\seeta(x)\right)} {x}\ \ =\ \ \pairr{\bbag\seeta \left(\seeta(x)\right)} {x}
\eea 
The program $S$ can be constructed as a Kleene fixpoint of the function
\bea\label{eq:seeta-hat}
\widehat \seeta (p,x) & = & \pairr{\uev {p}\left(\seeta(x)\right)} x
\eea
\begin{figure}[!ht]
\begin{center}
\newcommand{\Homega}{\scriptstyle \bbag{\seeta}}
\newcommand{\Nee}{\scriptstyle S}
\newcommand{\unl}{\{\}}
\newcommand{\pnl}{[\,]}
\newcommand{\png}{\scriptstyle \pairr - -}
\renewcommand{\addd}{\scriptstyle \seeta}
\def\JPicScale{.7}
\input{PIC/Omega-55-2.tex} 
\caption{The Kleene fixpoint $S$ of $\widehat \seeta$ induces $\bbag\seeta(x) = \uev S x$ so that $\bbag\seeta(x) = \pairr{\bbag\seeta\left({\seeta(x)}\right)} x$}
\label{Fig:seqkleene}
\end{center}
\end{figure}
in Fig.~\ref{Fig:seqkleene}. 
The equation $\bbag\seeta\big({\seeta(x)}\big) = \fstt{\bbag\seeta (x)}$, derived in Fig.~\ref{Fig:stp}, implies that the inductively defined sequence in \eqref{eq:fundseq}
now unfolds from a single program
\beq\label{eq:iter-folding} \bbag\seeta_b\ \ =\ \ \pairr{\bbag\seeta_{s_1}}{s_0}\ \ =\ \ \pairr{\pairr{\bbag\seeta_{s_2}}{s_1}}{s_0}\ \ =\ \ \pairr{\pairr{\pairr{\bbag\seeta_{s_3}}{s_2}}{s_1}}{s_0}\ \ =\ \ \pairr{\pairr{\pairr{\pairr{\bbag\seeta_{s_4}}{s_3}}{s_2}}{s_1}}{s_0}\ \ =\cdots
\eeq 
\begin{figure}[!ht]
\begin{center}
\newcommand{\Homega}{\scriptstyle \bbag\seeta}
\newcommand{\pseey}{\scriptstyle \Pi}
\newcommand{\Nee}{\scriptstyle S}
\renewcommand{\addd}{\scriptstyle \seeta}
\newcommand{\prj}{\scriptstyle \fstt -}
\newcommand{\unl}{\{\}}
\newcommand{\png}{\scriptstyle \pairr - -}
\def\JPicScale{.7}
\input{PIC/Omega-7-2.tex} 
\caption{The function $\bbag\seeta = \uev S$ iterates $\seeta$ within the first projections $\fstt{\bbag\seeta(x)} = \bbag\seeta(\seeta(x))$}
\label{Fig:stp}
\end{center}
\end{figure}

Since $\bbag \seeta(b)$ projects out each $s_n = \seeta^n(b)$, for all $n = 0,1,2,\ldots$, after $n$ iterated evaluations, it is the least upper bound of the sequence, i.e. the minimal extension of its well-order:
\beq\label{eq:seeta-join}b\hspace{.7em} \lt\hspace{.7em}\seeta(b)\hspace{.7em}\lt\hspace{.7em}\seeta^2(b)\hspace{.7em}\lt\hspace{.7em} \seeta^3(b)\hspace{.7em}\lt \hspace{.7em}\seeta^4(b)\hspace{.7em}\lt \hspace{.7em} \cdots\ \hspace{.7em}\lt \hspace{.7em}\bbag\seeta(b)= \bigvee_{n=0}^\infty \seeta^n (b) \eeq

\subsection{Collecting natural numbers: the ordinal $\omega$ as a program}
Instantiating $b$ to $\overline 0=\pairr \true {\overline 0}$ and $\seeta(x)$ to $\suce(x) = \pairr {\flse}{x}$ from Sec.~\ref{Sec:num-prog}, the sequence \eqref{eq:fundseq} becomes the fundamental sequence of natural numbers 
\beq\label{eq:fundsq-omega} \overline 0 ,\hspace{1.5em}  \overline 1=\suce(\overline 0),\hspace{1.5em}  \overline 2=\suce^2(\overline 0),\hspace{1.5em}  \overline 3=\suce^3(\overline 0),\hspace{1.5em}  \overline 4=\suce^4(\overline 0),\ldots\eeq
and the least upper bound \eqref{eq:seeta-join} becomes
\beq \overline 0\hspace{.7em} \lt\hspace{.7em}\overline 1\hspace{.7em}\lt\hspace{.7em}\overline 2\hspace{.7em}\lt\hspace{.7em} \overline 3\hspace{.7em}\lt \hspace{.7em}\overline 4\hspace{.7em}\lt \hspace{.7em} \cdots\ \hspace{.7em}\lt \hspace{.7em}\overline \omega \eeq
where
\bear
\overline \omega & = & \bbag\suce({\overline 0}) \ \ =\ \ \pairr{\bbag\suce({\overline 1})} {\overline 0}\ \ =\ 
\ \pairr{\pairr{\bbag\suce({\overline 2})}{\overline 1}}{\overline 0} \ \ =\ \ \pairr{\pairr{\pairr{\bbag\suce({\overline 3})}{\overline 2}}{\overline 1}}{\overline 0} \ \ =\ \ \pairr{\pairr{\pairr{\pairr{\bbag\suce({\overline 4})}{\overline 3}}{\overline 2}}{\overline 1}}{\overline 0}\ \ =\cdots
\eear
Its unfolding is displayed in Fig.~\ref{Fig:omega-unfold}.
\begin{figure}[!ht]
\begin{center}
\newcommand{\ordin}{\overline\omega}
\newcommand{\Homega}{\scriptstyle \bbag\suce}
\newcommand{\beeh}{\scriptscriptstyle \overline 0}
\renewcommand{\addd}{\scriptstyle \suce}
\newcommand{\unl}{\{\}}
\newcommand{\png}{\scriptstyle \pairr - -}
\def\JPicScale{.7}
\input{PIC/Omega-unfold-2.tex} 
\caption{The ordinal $\omega$ appears as the evaluation order of the program $\overline \omega = \bbag \suce(\overline 0)$.}
\label{Fig:omega-unfold}
\end{center}
\end{figure}

\subsection{Transfinite addition}
The ordinal type $1+\omega$ is defined by adjoining an element at the bottom of $\omega$. The ordinal type $\omega +1$ is defined by adjoining an element at the top of $\omega$. It is easy to see that $1+\omega$ is order-isomorphic to $\omega$, whereas $\omega +1$ is not.

Extending the program $\overline \omega = \bbag \suce(\overline 0)$ with an application of $\suce$ before the input $\overline 0$ is fed to $\bbag \suce$ yields $\overline{1 +\omega} = \bbag \suce\left(\suce(\overline0)\right)$. The evaluation order of $\overline\omega$ is thus extended in $\overline{1 +\omega}$ with an additional step at the beginning. Defining in general $\overline{n +\omega} = \bbag \suce\left(\suce^n(\overline0)\right)$, with $n$ additional steps adjoined at the beginning, leads to the fundamental sequence 
\beq\label{eq:fundsq-one-omega} \overline \omega  = \bbag \suce\left(\suce^0(\overline0)\right) ,\hspace{1.5em}  \overline{1 +\omega} = \bbag \suce\left(\suce^1(\overline0)\right),\hspace{1.5em}  \overline{2 +\omega} = \bbag \suce\left(\suce^2(\overline0)\right),\hspace{1.5em}  \overline{3 +\omega} = \bbag \suce\left(\suce^3(\overline0)\right),\ldots\eeq
But the evaluation orders of all $\overline{n +\omega} = \bbag \suce\left(\overline n\right)$ are in fact the same like $\overline{\omega}$: they all just iterate the computation $\bbag \suce$. The order-isomorphisms of the ordinals $n+\omega$ and $\omega$ are implemented on the representations $\overline{n+\omega}$ and $\overline \omega$  by precomposing with the predecessor or the successor functions.

Extending the program $\overline \omega = \bbag \suce(\overline 0)$ by applying $\suce$ to the output yields $\overline{\omega+1} = \suce\left(\bbag \suce\left(\overline 0 \right)\right)$.  Its evaluation order is thus extended with an additional step adjoined at the end. The general definition $\overline{\omega+n} = \suce^n\left(\bbag \suce\left(\overline 0\right)\right) = \suce^n\left(\overline \omega\right)$ adjoins $n$ additional steps at the end. The fundamental sequence induced by counting up the $n$ is now
\beq\label{eq:fundsq-omega-one} \overline \omega =\suce^0(\overline \omega) ,\hspace{1.5em}  \overline {\omega +1}=\suce^1(\overline \omega),\hspace{1.5em} \overline{\omega +2}=\suce^2(\overline \omega),\hspace{1.5em}  \overline{\omega +3} =\suce^3(\overline \omega),\hspace{1.5em}  \overline{\omega + 4}=\suce^4(\overline \omega),\ldots\eeq
The well-ordering and its least upper bound now become
\beq \overline \omega \hspace{.7em} \lt\hspace{.7em}\overline {\omega +1}\hspace{.7em}\lt\hspace{.7em}\overline {\omega +2}\hspace{.7em}\lt\hspace{.7em} \overline {\omega +3}\hspace{.7em}\lt \hspace{.7em}\overline {\omega +4}\hspace{.7em}\lt \hspace{.7em} \cdots\ \hspace{.7em}\lt \hspace{.7em}\overline {\omega + \omega} \eeq
where
\bea
\overline{\omega +\omega} & = & \bbag\suce(\overline \omega) \ \ =\ \ \bbag\suce\left(\bbag\suce(\overline 0)\right)
\eea
The evaluation unfolding of the second component is displayed in Fig.~\ref{Fig:omega-plus}. 
\begin{figure}[!ht]
\begin{center}
\newcommand{\ordin}{\overline{\omega+\omega}}
\newcommand{\Homega}{\scriptstyle \bbag\suce}
\newcommand{\beeh}{\scriptscriptstyle \overline 0}
\renewcommand{\addd}{\scriptstyle \suce}
\newcommand{\unl}{\{\}}
\newcommand{\png}{\scriptstyle \pairr - -}
\def\JPicScale{.7}
\input{PIC/Omega-plus-2.tex} 
\caption{The ordinal $\omega+\omega$ appears as the evaluation order of the program $\overline{\omega+\omega} = \bbag\suce\left(\bbag\suce(\overline 0)\right)$.}
\label{Fig:omega-plus}
\end{center}
\end{figure}
The evaluation unfolding of the first component would add Fig.~\ref{Fig:omega-unfold} at the bottom of Fig.~\ref{Fig:omega-plus}. The ordinal type $\omega +\omega$ consists of two copies of $\omega$, one after the other, and the evaluation of $\overline {\omega +\omega}$ reproduces that order.

\subsection{Transfinite multiplication} 
\begin{figure}[!ht]
\begin{center}
\newcommand{\ordin}{\overline\omega\times\overline\omega}
\newcommand{\Homega}{\scriptstyle \bbag{\bbag\suce}}
\newcommand{\beeh}{\scriptscriptstyle \overline 0}
\renewcommand{\addd}{\scriptstyle \bbag\suce}
\newcommand{\unl}{\{\}}
\newcommand{\png}{\scriptstyle \pairr - -}
\def\JPicScale{.7}
\input{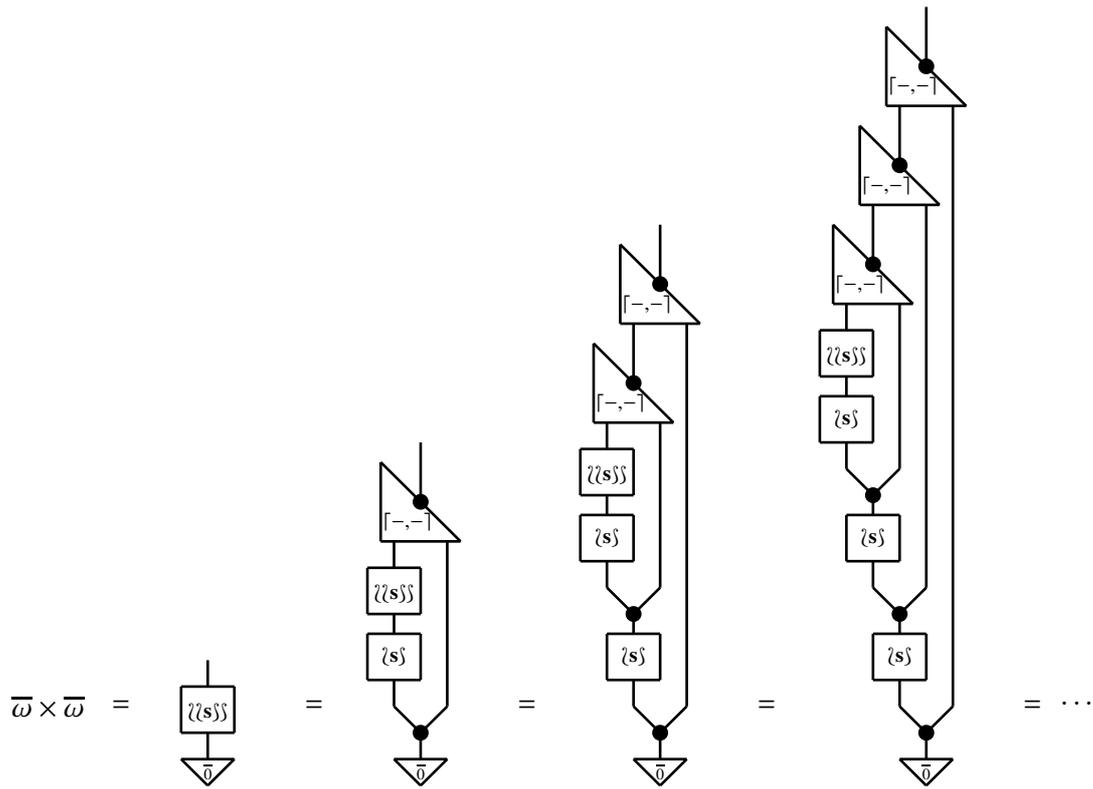} 
\caption{The ordinal $\omega\times \omega$ appears as the evaluation order of the program $\overline{\omega\times \omega} = \bbag{\bbag\suce}(\overline 0)$.}
\label{Fig:omega-times}
\end{center}
\end{figure}

The multiplication of infinite ordinals is iterated addition, just like in the finite case. While the ordinal $2\times \omega$ consists of $\omega$ copies of 2, the ordinal $\omega \times 2$ consists of 2 copies of $\omega$, i.e. $\omega \times 2 = \omega + \omega$, and hence $\overline \omega\times 2 = \overline \omega+ \overline \omega  = \bbag \suce\circ \bbag \suce(\overline0) = \bbag \suce^2 (\overline0)$. The fundamental sequence is now
\beq\label{eq:fundsq-omega-mult}\overline{\omega \times 0} =\bbag\suce^0(\overline0) ,\hspace{1.5em} \overline {\omega\times 1}  =\bbag\suce^1(\overline0) ,\hspace{1.5em} \overline{\omega \times 2} =\bbag\suce^2(\overline0),\hspace{1.5em}  \overline{\omega \times 3} =\bbag\suce^3(\overline0),\hspace{1.5em} \overline{\omega \times 4} =\bbag\suce^4(\overline0),\ldots\eeq
Following \eqref{eq:seeta-join}, we now get
\beq \overline 0 \hspace{.7em} \lt\hspace{.7em}\overline {\omega}\hspace{.7em}\lt\hspace{.7em}\overline {\omega \times 2}\hspace{.7em}\lt\hspace{.7em} \overline {\omega \times 3}\hspace{.7em}\lt \hspace{.7em}\overline {\omega \times 4}\hspace{.7em}\lt \hspace{.7em} \cdots\ \hspace{.7em}\lt \hspace{.7em}\overline {\omega \times \omega} \eeq
where
\bea
\overline{\omega \times \omega} & = &  \bbag{\bbag\suce}(\overline 0)
\eea
The evaluation is unfolded in Fig.~\ref{Fig:omega-times}. 
Each $\bbag \suce$-box in Fig.~\ref{Fig:omega-times} further unfolds like in Fig.~\ref{Fig:omega-unfold}. The evaluation order thus  consists of $\omega$ copies of $\omega$, as a representation of $\omega \times \omega$ should.

\subsection{Transfinite hyperfunction}\sindex{Grzegorczyk hierarchy!transfinite}
Like in the finite case, the exponentiation of infinite ordinals is iterated multiplication again, with 
$$\omega^2 = \omega \times \omega\quad\mbox{suggesting the representation} \quad \overline{\omega^{2}} = \overline{\omega\times \omega} = \bbag{\bbag\suce}(\overline 0)$$ 
The fundamental sequence of metaprograms for finite exponents thus becomes 
\beq\label{eq:fundsq-omega-exp}
\overline{\omega^0} =\overline 1 ,\hspace{1.5em} \overline{\omega^1} = \bbag\suce(\overline0) ,\hspace{1.5em} \overline{\omega^2} =\bbag{\bbag\suce}(\overline0),\hspace{1.5em}  \overline{\omega^3} =\bbag{\bbag{\bbag\suce}}(\overline 0),\hspace{1.5em} \overline{\omega^4} =\bbag{\bbag{\bbag{\bbag\suce}}}(\overline 0),\ldots\eeq
%
%
%
To iterate the $\bbag -$-construction, we need a program transformer $\www$ such that
\bea
\bbag{\uev q}  & = & \uev{\www(q)}
\eea
Recalling $\bbag \seeta(x)$, was defined in \eqref{eq:constrS} to unfolds to $\pairr{\bbag\seeta \left(\seeta(x)\right)} {x}$, the program transformer $\www$ can  thus be obtained by partially evaluating the Kleene fixpoint $W$ of the function
\bea\label{eq:wee-hat}
\widehat w (p,q,x) & = & \pairr{\uev {p}\left(q,\uev q x\right)} x
\eea
as displayed in Fig.~\ref{Fig:seqkleene-uniform}.
\begin{figure}[!ht]
\begin{center}
\newcommand{\Homega}{\scriptstyle \bbag{\seeta}}
\newcommand{\Nee}{\scriptstyle W}
\newcommand{\unl}{\{\}}
\newcommand{\pnl}{[\,]}
\newcommand{\png}{\scriptstyle \pairr - -}
\renewcommand{\addd}{\scriptstyle \seeta}
\newcommand{\dubb}{\scriptstyle \www}
\def\JPicScale{.7}
\input{PIC/Omega-56-2.tex} 
\caption{If $\uev W(q,x)  = \pairr{\uev W\big(q, \uev q x\big)} x$, then $\www = \pev W$  gives $\uev{\www(q)} = \bbag{\uev q} $ }
\label{Fig:seqkleene-uniform}
\end{center}
\end{figure}
The ordinal representations so far can be reconstructed using a program $N$ for the successor function $\suce = \uev N$, in the form
\bear
 \overline \omega & = & \uev{\www(N)} \overline 0\\
 \overline{\omega + \omega} & = & \uev{\www(N)}^2\overline 0\\
 \overline{\omega \times \omega} & = & \uev{\www^2(N)}\overline 0\\
 \overline{\omega^\omega} & = & \uev{\fuind \www 2(N)} \overline 0
 \eear
where $\fuind \www 2$ is the iterator of $\www$, i.e.
\[
\uev{\fuind \www 2 (q)}\ \  = \ \  \uev{\www\circ \fuind\www 2(q)} \ \  = \ \  \uev{\www\circ \www\circ \fuind\www 2(q)} \ \  = \ \  \uev{\www\circ\www\circ \www\circ \fuind\www 2(q)}\ \  
=\cdots 
\]

Continuing with the iterations 
\begin{gather*}
\uev{\fuind \www 3 (q)}\ \  = \ \  \uev{\fuind \www 2 \circ \fuind\www 3(q)} \ \  = \ \  \uev{\fuind \www 2 \circ \fuind \www 2 \circ \fuind\www 3(q)}\ \  = \ \  \uev{\fuind \www 2 \circ \fuind \www 2 \circ \fuind \www 2 \circ \fuind\www 3(q)} \ \ =\cdots \\
\uev{\fuind \www 4 (q)}\ \  = \ \  \uev{\fuind \www 3 \circ \fuind\www 4(q)} \ \  = \ \  \uev{\fuind \www 3 \circ \fuind \www 3 \circ \fuind\www 4(q)}\ \  = \ \  \uev{\fuind \www 3 \circ \fuind \www 3 \circ \fuind \www 3 \circ \fuind\www 4(q)} \ \ =\cdots\\
\cdots \qquad \cdots \qquad \cdots
\end{gather*}
we reconstruct the ordinal version of the Ackermann hyperfunction \sindex{hyperfunction!ordinal}  \sindex{iteration!parametric} 
\bear
\overline{\omega\uparrow \omega}\ \ =\ \ \overline{\omega^\omega} \ \ =\ \ \overline{\underbrace{\omega\times\omega\times \omega \times\cdots}_{\omega\mbox{ \scriptsize times}}} & = & \uev{\fuind \www 2(N)} \overline 0\\
\overline{\omega\uparrow\uparrow \omega}\ \ =\ \ \overline{\underbrace{\omega\uparrow\omega\uparrow\omega\uparrow \omega\uparrow\omega\uparrow\cdots}_{\omega\mbox{ \scriptsize times}}} & = & \uev{\fuind \www 3(N)} \overline 0\\
&\cdots & \\[1ex]
\overline{\omega\uparrow^{1+n} \omega}\ \ =\ \ 
\overline{\underbrace{\omega\uparrow^n\omega\uparrow^n\omega\uparrow^n\omega\uparrow^n\cdots}_{\omega\mbox{ \scriptsize times}}} & = & \uev{\fuind \www {2+n} (N)} \overline 0
\eear
where $N$ is still a program for the successor function $\suce(x) = \pairr \flse x$. The set-theoretic version of the ordinal Ackermann hyperfunction goes back to \cite{Doner-Tarski}. Deployed in a programming language $\DP$, the sequence of programs
\beq\label{eq:fundsq-doner-tarski}
\fuind \www 2(N) = \varpi_0  ,\hspace{1.5em} \fuind \www 3(N)  = \varpi_1 ,\hspace{1.5em} \fuind \www 4(N)  = \varpi_2,\hspace{1.5em}  \ldots \hspace{1.5em} \fuind \www {n+2}(N) = \varpi_n,\ldots
\eeq
is generated by iterating the Kleene fixpoint $\widetilde W$ of
\bear
\widetilde \iota(p, n, q) &  = &  \iif\Big(\iszero(n),\  \www(q),\   \uev{p} \left(\pred(n), \uev p(n,q)\right)\Big)
\eear 
leading to $\fuind \www n = \pev{\widetilde W}(n)$. Applying the construction from Sec.~\ref{Sec:collabs}, we get $\Omega = \bbag{\uev {\widetilde W}}(N)$ as the limit of the sequence in \eqref{eq:fundsq-doner-tarski}.  The definition of $\www$ also gives $\Omega = \uev{\www(\widetilde W)}N$. The transfinite well-orders up to $\varepsilon_0$ can thus be metaprogrammed in any programming language $\DP$.

\subsection{Background: computable ordinal notations}
Ordinal numbers arise as transitive closures of the element relation in set theory, or as the types of deterministic evaluations in type theory. In computers, programs are well-ordered lexicographically, usually as binaries. In recursion theory, programs have been represented as natural numbers since G\"odel. We have seen that the iterative metaprogramming induces transfinitely well-ordered evaluations well beyond the natural numbers. Computable ordinal notations have been studied since the early days of theory of computation \cite{ChurchA:ord-second,Church-Kleene:constructions,Church-Kleene:ord,KleeneS:ordinal-fund,TuringA:thesis}. The iterative computational processes that generate ordinals go back to the origins of Georg Cantor's work on his \emph{second number class} \cite{CantorG:collected}, and thus predate the theory of computation by some 50 years. His explorations of transfinite constructions can be viewed as a radical application of iterated counting: \emph{"Suppose that I have counted infinitely long and then I start again. And then I do that infinitely many times. And then I iterate that infinitely many times\ldots"} Such metaprograms were at the heart of Cantor's work. Hilbert's endorsements may have diverted attention from their constructive content.

%

\def\thechapter{7}
\setchaptertoc
\chapter[Stateful computing]{Stateful computing}
\label{Chap:State}
\newpage

\section{From \funnn s to \procs es}
\label{Sec:procs}

Now we progress
\begin{itemize}
\item  from \emph{computations as computable \funnn s}
\item to  \emph{computations as computational processes}. 
\end{itemize}
A \funnn\ produces its output as soon as it receives its input.\sindex{function!vs process}  The \funnn\ compositions are instant and costless. A \procs, \sindex{process} on the other hand, takes time and space to process an input into an output. The space and the time are presented as the states and the state transitions. A state is anything that a \procs\ may update: a state as a state of the world, a state of memory, a state of a living room floor. A \funnn\ $g\colon X\times A\to B$ depends on the state $x\colon X$, which determines the applicable instance $g_x\colon A\to B$. It will be applied when an input $a\colon A$ becomes available. A \procs\ $q\colon X\times A\to X\times B$ also depends on the current state $x$ but it updates it to a future state $x' = \sta q_x(a)$. It is often convenient to view a \procs\ $q\colon X\times A\to X\times B$ as a pair of \funnn s 
\bea\label{eq:sta-out}
q & = & \left<\sta q, \out q\right>
\eea
where
\begin{itemize}
\item $\out q = \left(X\times A\tto q X\times B \tto{\scun \times \id} B\right)$ instantiates to the input-output maps $\out q_x\colon A\to B$ like before, whereas 
\item $\sta q = \left(X\times A\tto q X\times B \tto{\id \times \scun} X\right)$ determines the state update maps $\sta q_x\colon A\to X$ presengint the transitions $x\longmapsto x'=\sta q_x(a)$. 
\end{itemize}
The computational process proceeds with $x'\longmapsto x''=\sta q(x',a')$ etc. 

\para{A \procs\ is a pair of \funnn s.}\sindex{process!as a pair of functions} We worked out in \ref{Sec:work-wire}.\ref{work:pairing} that a \funnn\ to a product can be recovered its projections whenever it is \strict. This may not be possible for monoidal \funnn s in general, since a \funnn\ into a product may contain more information than its projections. A \procs, viewed as a \funnn, may not be \strict. It is, however, assumed that it is a \funnn\ of this special kind, which can be recovered from its projections. They are bundled together by the pairing, mainly for notational convenience. Intuitively, this assumption means that the two components $\sta q$ and $\out q$ of a \procs\ $q$ have the same side-effects on the same inputs. Formally, the side-effect behaviors of \funnn s can be \emph{defined}\/ as the equivalence classes of monoidal \funnn s on which the pairing is surjective. 

\para{State transitions and traces.} The mapping
\bear
q\colon X\times A & \to & X\times B\\
<x,a> & \longmapsto & <x',b>
\eear
is conveniently written in the form
\[ x \xmapsto[q]{a/b} x'\]
so that a \emph{process trace}\/ starting from an initial state $x_0$ induced by a sequence of inputs from $A$ can capture both the state transitions and the outputs:
\[\prooftree
\hspace{-.75em} a_1,\hspace{2.5em} a_2 ,\hspace{2.5em}a_3,\hspace{2.5em}  \ldots\ \hspace{2.5em}a_n
\justifies
x_0 \xmapsto{a_1/b_1} x_1\xmapsto{a_2/b_2} x_2 \xmapsto{a_3/b_3}\cdots \xmapsto{\hspace{1.3em}} x_{n-1}\xmapsto{a_n/b_n} x_n
\endprooftree\]
where $x_{i} =\sta q(x_{i-1},a_i)$ and $b_i = \out q(x_{i-1}, a_i)$ for $i = 1,\ldots n$.
As mentioned in Sec.~\ref{Sec:programmable}, the principal \sindex{state} difference between states and inputs is that the state is determined by the environment, in advance of running the \procs, whereas the input is entered by the system at runtime.

\subsubsection*{Examples}
Let the \funnn\ $g\colon X\times A\to B$ represent multifunctional drill, for drilling holes, tightening screws,  cutting wood, etc. Its different \funnn s $g_x$ are realized using different attachments $x\colon X$. The state space $X$ is thus the box with the drill bits, the screwdrivers, the wood saws, etc. The types $A$ and $B$ are the tasks and the goals. A \procs\ model $q\colon X\times A\to X\to B$ of the same toolset can take into account the changing state of the components: the wear and tear of the drill bits, the broken and replaced screwdrivers, the battery charge level. 

\begin{figure}[!ht]
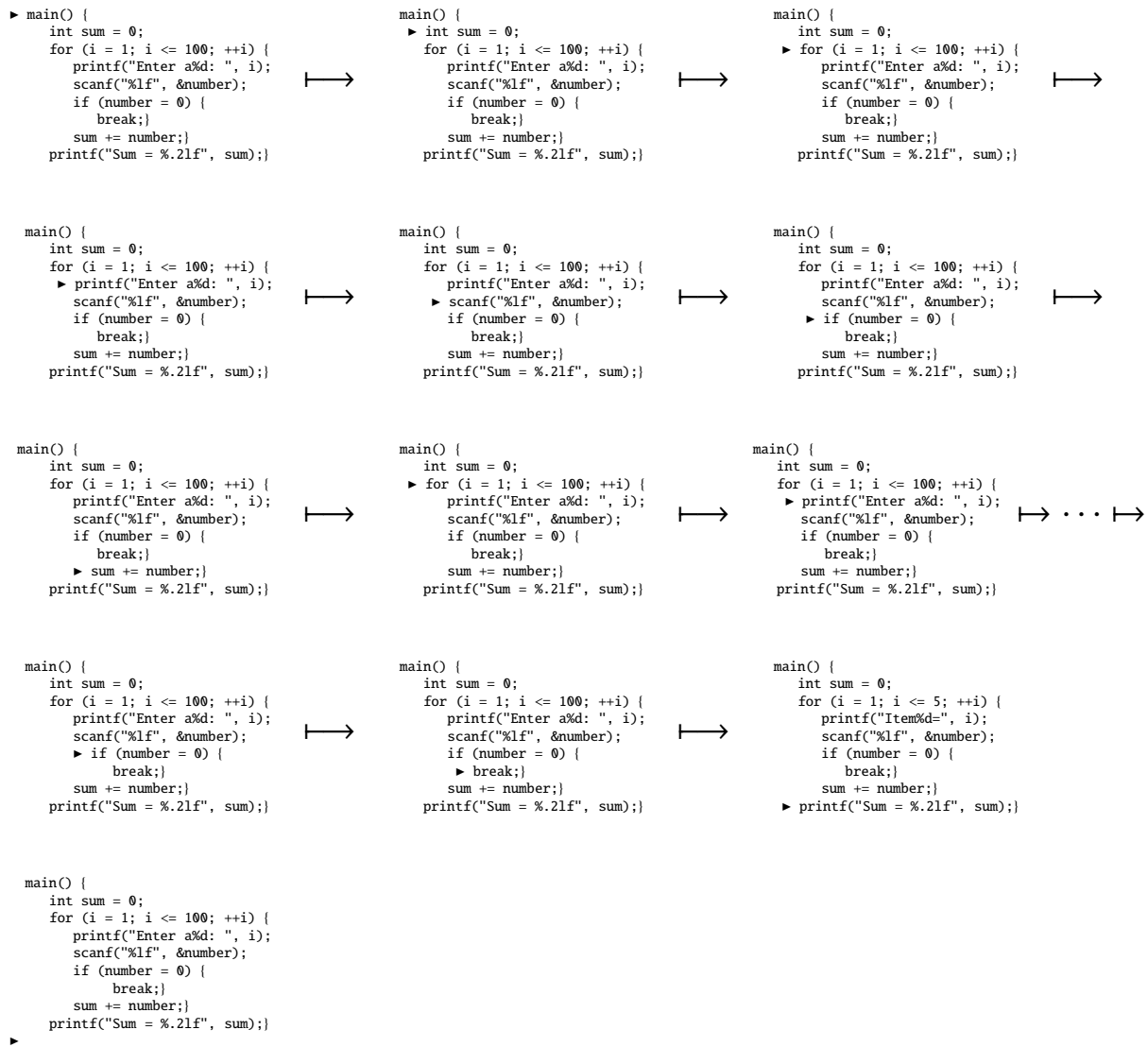

{\tiny\begin{minipage}{18em}
\begin{alltt}
\(\blacktriangleright\) main() \{
     int sum = 0;
     for (i = 1; i <= 100; ++i) \{
        printf("Enter a%d: ", i);
        scanf("%lf", &number);
        if (number = 0) \{
           break;\}
        sum += number;\}
     printf("Sum = %.2lf", sum);\}
\end{alltt}
\end{minipage}}
\ \ {\Large $\longmapsto$}\ \ \ \ 
{\tiny\begin{minipage}{18em}
\begin{alltt}
  main() \{
   \(\blacktriangleright\) int sum = 0;
     for (i = 1; i <= 100; ++i) \{
        printf("Enter a%d: ", i);
        scanf("%lf", &number);
        if (number = 0) \{
           break;\}
        sum += number;\}
     printf("Sum = %.2lf", sum);\}
\end{alltt}
\end{minipage}} 
\ \ {\Large $\longmapsto$}\ \ \ \ 
{\tiny\begin{minipage}[c]{18em}
\begin{alltt}
  main() \{
     int sum = 0;
   \(\blacktriangleright\) for (i = 1; i <= 100; ++i) \{
        printf("Enter a%d: ", i);
        scanf("%lf", &number);
        if (number = 0) \{
           break;\}
        sum += number;\}
     printf("Sum = %.2lf", sum);\}
\end{alltt}
\end{minipage}}
\ \ {\Large $\longmapsto$}
\\[5ex]
{\tiny\begin{minipage}{18em}
\begin{alltt}
  main() \{
     int sum = 0;
     for (i = 1; i <= 100; ++i) \{
      \(\blacktriangleright\) printf("Enter a%d: ", i);
        scanf("%lf", &number);
        if (number = 0) \{
           break;\}
        sum += number;\}
     printf("Sum = %.2lf", sum);\}
\end{alltt}
\end{minipage}}
\ \ {\Large $\longmapsto$}\ \ \ \  
{\tiny\begin{minipage}{18em}
\begin{alltt}
  main() \{
     int sum = 0;
     for (i = 1; i <= 100; ++i) \{
        printf("Enter a%d: ", i);
      \(\blacktriangleright\) scanf("%lf", &number);
        if (number = 0) \{
           break;\}
        sum += number;\}
     printf("Sum = %.2lf", sum);\}
\end{alltt}
\end{minipage}} 
\ \ {\Large $\longmapsto$}\ \ \ \  
{\tiny\begin{minipage}{18em}
\begin{alltt}
  main() \{
     int sum = 0;
     for (i = 1; i <= 100; ++i) \{
        printf("Enter a%d: ", i);
        scanf("%lf", &number);
      \(\blacktriangleright\) if (number = 0) \{
           break;\}
        sum += number;\}
     printf("Sum = %.2lf", sum);\}
\end{alltt}
\end{minipage}} 
\ \ {\Large $\longmapsto$}
\\[5ex]
{\tiny\begin{minipage}{18em}
\begin{alltt}
 main() \{
     int sum = 0;
     for (i = 1; i <= 100; ++i) \{
        printf("Enter a%d: ", i);
        scanf("%lf", &number);
        if (number = 0) \{
           break;\}
        \(\blacktriangleright\) sum += number;\}
     printf("Sum = %.2lf", sum);\}
\end{alltt}
\end{minipage}}
\ \ {\Large $\longmapsto$}\ \ \ \ 
{\tiny\begin{minipage}[c]{18em}
\begin{alltt}
  main() \{
     int sum = 0;
   \(\blacktriangleright\) for (i = 1; i <= 100; ++i) \{
        printf("Enter a%d: ", i);
        scanf("%lf", &number);
        if (number = 0) \{
           break;\}
        sum += number;\}
     printf("Sum = %.2lf", sum);\}
\end{alltt}
\end{minipage}}
\ \ {\Large $\longmapsto$} 
{\tiny\begin{minipage}{18em}
\begin{alltt}
  main() \{
     int sum = 0;
     for (i = 1; i <= 100; ++i) \{
      \(\blacktriangleright\) printf("Enter a%d: ", i);
        scanf("%lf", &number);
        if (number = 0) \{
           break;\}
        sum += number;\}
     printf("Sum = %.2lf", sum);\}
\end{alltt}
\end{minipage}} 
{\Large $\mapsto\cdots\mapsto$}
\\[5ex]
{\tiny\begin{minipage}[c]{18em}
\begin{alltt}
  main() \{
     int sum = 0;
     for (i = 1; i <= 100; ++i) \{
        printf("Enter a%d: ", i);
        scanf("%lf", &number);
        \(\blacktriangleright\) if (number = 0) \{
             break;\}
        sum += number;\}
     printf("Sum = %.2lf", sum);\}
\end{alltt}
\end{minipage}}
\ \ {\Large $\longmapsto$}\ \ \ \ 
{\tiny\begin{minipage}{18em}
\begin{alltt}
  main() \{
     int sum = 0;
     for (i = 1; i <= 100; ++i) \{
        printf("Enter a%d: ", i);
        scanf("%lf", &number);
        if (number = 0) \{
         \(\blacktriangleright\) break;\}
        sum += number;\}
     printf("Sum = %.2lf", sum);\}
\end{alltt}
\end{minipage}} 
\ \ {\Large $\longmapsto$}\ \ \ \ 
{\tiny\begin{minipage}{18em}
\begin{alltt}
  main() \{
     int sum = 0;
     for (i = 1; i <= 5; ++i) \{
        printf("Item%d=", i);
        scanf("%lf", &number);
        if (number = 0) \{
           break;\}
        sum += number;\}
   \(\blacktriangleright\) printf("Sum = %.2lf", sum);\}
\end{alltt}
\end{minipage}} 
\\[5ex]
{\tiny\begin{minipage}[c]{18em}
\begin{alltt}
  main() \{
     int sum = 0;
     for (i = 1; i <= 100; ++i) \{
        printf("Enter a%d: ", i);
        scanf("%lf", &number);
        if (number = 0) \{
             break;\}
        sum += number;\}
     printf("Sum = %.2lf", sum);\}
\(\blacktriangleright\) 
\end{alltt}
\end{minipage}}
\begin{center}
\caption{Execution of a C-program to compute and print the sum of up to 5 user-entered inputs}
\label{Fig:C}
\end{center}
\end{figure}

If $g\colon X\times A\to B$ represents a general-purpose computer, the state space $X$ can be a set of programs $\DP$, the types $A$ and $B$ the interface formats, and $g$ can be a program evaluator $\universal_A^B$, with programs as states. A \procs\ model $q\colon X\times A\to X\to B$ of a computer could present programs not as mere abstract terms of type $\DP$, like we do in theory, but as sequences of instructions, closer to the practice. The states of a program, available for updating, can capture not only the values of some external program parameters, but also track which instructions are ready for evaluation, like in the toy example in Fig.~\ref{Fig:C}.
The state space can be the set $X = \{x_0, x_1,\ldots x_8, x_9\}$, where $x_i$ is the program with ``$\blacktriangleright$" on line $i$. The final line, where the program execution halts, is assumed to be empty. The other lines contain the instructions which get executed and then pass the control to each other, depending on the current values of program variables. The state transitions represent the \procs\ control. In this particular example, the presented \procs\ run is an execution of a program resulting in the sequence of transitions:
\[
x_0 \longmapsto x_1\longmapsto x_2 \longmapsto x_3 \longmapsto x_4\longmapsto x_5 \longmapsto x_7 \longmapsto x_2\longmapsto x_3 \longmapsto\cdots \longmapsto x_{5}\longmapsto x_6 \longmapsto x_{8}\longmapsto x_{9}
\]
At each step, an input is consumed and an output is produced, albeit empty. To capture this aspect, assume that the inputs and the outputs are
\bear
A & = & \{\blank, 0, 1, 2, \ldots\}\\
B & = & \{\blank, \mbox{``Enter a1:"}\, ,\  \mbox{``Enter a2:"},\ldots, \mbox{``Enter a5:"}\, ,\  \mbox{``Sum=0"}\, ,\  \mbox{``Sum=1"}\, ,\  \mbox{``Sum=2"}\, , \ldots\}
\eear 
where the symbol $\blank$ represents the empty inputs and outputs. If a user enters $a_1=3$, $a_2 = 2$, $a_3 = 2$, and finally $a_4 =0$ to request the sum of the previous entries, the  execution trace will be
{\small \beq\label{eq:exec-enter}
\begin{matrix}
x_0 \ \ \xmapsto{\ \ \blank/\blank\ \ } & x_1\ \ \xmapsto{\ \ \blank/\blank\ \ } & x_2 \ \ \xmapsto{\ \ \blank/\blank\ \ } & x_3 \ \ \xmapsto{\blank/\mbox{\tiny ``Enter n1:"}} &  x_4\ \ \xmapsto{\ \ 3/\blank\ \ } & x_5 \ \ \xmapsto{\ \ \blank/\blank\ \ } & x_7 \ \ \xmapsto{\ \ \ \blank/\blank\ \ \ } & \cdots
\\
&\cdots & x_2\ \ \xmapsto{\ \ \blank/\blank\ \ } & x_3 \ \ \xmapsto{\blank/\mbox{\tiny ``Enter n2:"}} & x_4 \ \ \xmapsto{\ \ 2/\blank\ \ } & x_5 \ \ \xmapsto{\ \ \blank/\blank\ \ } & x_7 \ \ \xmapsto{\ \ \ \blank/\blank\ \ \ } & \cdots 
\\
&\cdots & x_2\ \ \xmapsto{\ \ \blank/\blank\ \ } & x_3 \ \ \xmapsto{\blank/\mbox{\tiny ``Enter n3:"}} & x_4 \ \ \xmapsto{\ \ 2/\blank\ \ } & x_5 \ \ \xmapsto{\ \ \blank/\blank\ \ } & x_7 \ \ \xmapsto{\ \ \ \blank/\blank\ \ \ } & \cdots
\\
&\cdots & x_2\ \ \xmapsto{\ \ \blank/\blank\ \ } & x_3 \ \ \xmapsto{\blank/\mbox{\tiny ``Enter n4:"}} & x_4 \ \ \xmapsto{\ \ 0/\blank\ \ } &  x_5 \ \ \xmapsto{\ \ \blank/\blank\ \ } 
&  x_8 \ \ \xmapsto{\blank/\mbox{\tiny ``Sum=7"}} &  x_9 
\end{matrix}
\eeq}

\section{Simulations as \procs\ morphisms}
\label{Sec:simul}

\begin{figure}[!ht]
\begin{center}
\begin{minipage}{.4\linewidth}
\centering
\def\JPicScale{.75}
\newcommand{\aah}{\scriptstyle s}
\newcommand{\ssbbh}{\scriptstyle Y}
\renewcommand{\dh}{\scriptstyle B}
\newcommand{\ahh}{h}
\newcommand{\bhh}{g}
 \newcommand{\ah}{\scriptstyle A} 
  \newcommand{\xxh}{\scriptstyle X}
 \newcommand{\xxbbh}{\scriptstyle B}
 \newcommand{\Dott}{\mbox{\LARGE$\bullet$}}
 \input{PIC/homom-fun.tex}
 \\[5ex]
\tikzset{nothing/.tip={},cart/.tip={Glyph[glyph math command = bullet]}}
\begin{tikzar}[row sep=3.75em,column sep=.5em]
\& B\\   
\hspace{2em}
\& \&  \&
Y\times A  \ar{ull}[swap]{h} 
 \\ 
 \& X \times A  \ar[bend left = 50]{uu}{g} \ar[nothing-cart
 ]{urr}[description]{{s} \times \id} \end{tikzar}
 \end{minipage}
\hspace{.1\linewidth}
\begin{minipage}{.4\linewidth}
\centering
\def\JPicScale{.75}
\newcommand{\aah}{\scriptstyle s}
\newcommand{\ssbbh}{\scriptstyle Y}
\renewcommand{\dh}{\scriptstyle B}
\newcommand{\ahh}{r}
\newcommand{\bhh}{q}
 \newcommand{\ah}{\scriptstyle A} 
  \newcommand{\xxh}{\scriptstyle X}
 \newcommand{\xxbbh}{\scriptstyle B}
 \newcommand{\Dott}{\mbox{\LARGE$\bullet$}}
 \input{PIC/homom-proc.tex}
 \\[5ex]
 \tikzset{nothing/.tip={},cart/.tip={Glyph[glyph math command = bullet]}}
\begin{tikzar}[row sep=1.8em,column sep=1em]
\&\& Y \times B
\\   
\&\&sx'
\\
X \times B \ar[nothing-cart]{uurr}[description]{s\times \id}   
\&  x'  \ar[mapsto,dotted,thin]{ur}   
\& \&  
sx  \ar[mapsto,thick]{ul}{a/b} \&
Y\times A  \ar{uull}[swap]{r} 
 \\ 
 \&\&x  \ar[dotted,mapsto,thin]{ur} \ar[mapsto,thick]{ul}[swap]{a/b}
 \\
 \&\& X \times A  \ar[nothing-cart]{uurr}[description]{{s} \times \id} \ar{uull}{q} \end{tikzar}
\end{minipage}
\caption{A \strict\ \funnn\ $s$ as a morphism between \funnn s and a simulation between \procs es}
\label{Fig:proc-homo}
\end{center}
\end{figure}

The left-hand diagram in Fig.~\ref{Fig:proc-homo} shows a \strict\ \funnn\
 $s\colon X\stricto Y$ as a morphism between \funnn s $g$ and $h$.  
Such morphisms can be thought of as \emph{implementations}\sindex{implementation}
 A morphism $s \colon g \to h$ implements $g$ in terms of $h$ by reparametrizing the states along  $S\colon X\stricto Y$ in such a way that every instance $g_x\colon A\to B$ of $g$ is realized as an instance $h_{sx}\colon A\to B$ of $h$. The commutativity of the left-hand diagram in Fig.~\ref{Fig:proc-homo} means that for any input $a$ both \funnn\ instances output $b = h_{sx}(a) = g_x(a)$.
 
The right-hand diagram  in Fig.~\ref{Fig:proc-homo} shows the same \funnn\ $s$ is a \procs\ morphism from $q$ to $r$. Such morphisms are called \textbf{\emph{simulations}}. \sindex{simultation} A simulation $s\colon q\to r$ is again a reparametrizing \funnn\ $s\colon X\stricto Y$, but this time the instances $r_{sx}\colon A\stricto Y\times B$ simulate $q_{x}\colon A\stricto X\times B$ not just by realizing their input-output \funnn s as $\out r_{sx} = \out q_x \colon A\to B$, but also by tracking their state updates along $$\sta r_{sx} = \left(A\tto{\sta q_x}X\strictto s Y\right)$$ For every input $a$, both processes not only produce the same output $b=\out r_{sx}(a)=\out q_x(a)$ but for every state transition $x\longmapsto x'$ performed by $q$, the process $r$ responds by the transition $sx\longmapsto sx'$. In summary, 
\[\prooftree
x \xmapsto[q]{a/b} \  x' \  \ \ =\ \ \  \  \sta q_x(a)\hspace{4em}
\justifies
sx \xmapsto[r]{a/b} sx' \ =\  s\left(\sta q_x(a)\right)\ =\ \sta r_{sx}(a)  
\endprooftree\]

\section{When is a process computational?} 
\label{Sec:exec}

A \funnn\ is computable when there is a program that evaluates to it. A \procs\ is computational when there is a program whose execution simulates it.  What does this mean? E.g., Fig.~\ref{Fig:C} displays a computational \procs. Its world is a program and the states of the world are the program instructions marked by ``$\blacktriangleright$", ready to be executed. The computational \procs\ progresses from instruction to instruction, and possibly from one memory state to the next, each determining the next update. A computational \procs\ is such \procs\ of updating a stateful program. At each step, the program instructions are still evaluated by an abstract evaluator, but the sequence of evaluations has a more concrete computational meaning. 

\para{Program execution} is a computational \procs\ realized by a program evaluator written
\beq\label{eq:uev-puev}
\puniversal_A^B\ = \ \universal^{\DP\times B}_A\colon \DP \times A\to \DP\times B
\eeq
which evaluates programs $F\colon \DP$ on inputs $a\colon A$ to produce, in the notation of \eqref{eq:sta-out},
\begin{itemize}
\item update $\sta{\puev{F}}a\colon \DP$ and
\item output $\out{\puev{F}}a\colon B$ 
\end{itemize}
whereby the update is available for further evaluations and the outputs are accumulated. Program executions are program evaluators that produce program updates, in addition to the data outputs. The notation $\puniversal$ is a reminder that the output type contains a $\DP$-component. The programs that do not just apply operations on data but also control the state changes are called \textbf{\emph{stateful}}.\sindex{program!stateful }

\para{Execution traces.} While computable \funnn s produce their output values instantly,  computational \procs es produce execution traces. In \eqref{eq:exec-enter}, we saw an execution trace of the program in Fig.~\ref{Fig:C}. In general, an abstract \procs\ $q\colon X\times A\to X\times B$ is computational when there is a program $Q\colon X\stricto \DP$ such that every step of $q$ is simulated by a program execution $\puniversal\colon \DP\times A\to \DP\times B$, where
\beq\label{eq:simulation}
\begin{split}
\prooftree
\prooftree
a_1,\hspace{3.15em} a_2 ,\hspace{3.15em}a_3,\hspace{3em}  \ldots
\justifies
x_0 \hspace{0.35em} \xmapsto{a_1/b_1}\hspace{0.35em} x_1\hspace{0.35em} \xmapsto{a_2/b_2} \hspace{0.35em} x_2 \hspace{0.35em}\xmapsto{a_3/b_3} \hspace{0.35em} x_3 \hspace{0.35em} \xmapsto{\hspace{1.3em}} \cdots
\using \mbox{\small 
\ \ induces \ \ $x_{i+1} =\sta q_{x_{i}}(a_{i+1})$\ \  and\ \  $b_{i+1} = \out q_{x_{i}} (a_{i+1})$}
\endprooftree
\justifies
Qx_0 \xmapsto{a_1/b_1} Qx_1\xmapsto{a_2/b_2} Qx_2 \xmapsto{a_3/b_3} Qx_3 \xmapsto{\hspace{1.3em}} \cdots
\using \mbox{\small \ induces $Qx_{i+1} = \sta{\uev{Qx_i}}a_i$ and $b_{i+1} = \out {\uev{Qx_{i}}} a_{i+1}$}
\endprooftree
\end{split}
\eeq
The simulation requirements from Fig.~\ref{Fig:proc-homo} are instantiated $Q\circ \sta q_{x} =\sta{\uev{Qx}}$ and $\out q_x = \out{\uev{Qx}}$.

\section{Universality of program execution}
\label{Sec:ana}

The notions of universality for \funnn s and for \procs es are illustrated in Fig.~\ref{Fig:exec-eval}. 

A \funnn\ $\universal\colon \DP \times A\to B$ is universal when any \funnn\ $g\colon X\times A\to B$ has an implementation $G\colon X\stricto \DP$ evaluated by $\universal$. The implementation\sindex{implementation} requirement imposed on the \funnn\ morphisms in Fig.~\ref{Fig:proc-homo} boils down to $g = \universal\circ(G\times A)$, written more insightfully in the form $g_x = \uev{Gx}$. The implementations of $X$-parametrized \funnn s to universal evaluators $\universal$ are the $X$-parametrized programs.

\begin{figure}[!ht]
\begin{center}
\newcommand{\Qee}{\scriptstyle Q}
\newcommand{\qee}{q}
\newcommand{\Fee}{\scriptstyle G}
\newcommand{\fee}{g}
\newcommand{\Aee}{\scriptstyle X}
\renewcommand{\Bee}{\scriptstyle A}
\renewcommand{\Cee}{\scriptstyle B}
\newcommand{\Code}{\scriptstyle \DP}
\newcommand{\Doce}{\scriptstyle \SP}
\newcommand{\Univ}{\mbox{\Large$\Universal$}}
\newcommand{\PUniv}{\mbox{\Large$\pUniversal$}}
\newcommand{\Dott}{\mbox{\LARGE$\bullet$}}
\newcommand{\LHS}{\out q_x}
\newcommand{\RHS}{\out{\puev{Qx}}}
\newcommand{\LHSP}{Q\circ\sta q_x}
\newcommand{\RHSP}{\sta{\puev{Qx}}}
\newcommand{\GLHS}{g_x}
\newcommand{\GRHS}{\uev{Gx}}
\def\JPicScale{.25}
\input{PIC/exec-eval.tex} 
\caption{Universality for \funnn s and for \procs es}
\label{Fig:exec-eval}
\end{center}
\end{figure}\sindex{universality}

A \procs\  $\puniversal$ is universal\sindex{universal!process} when any \procs\ $q\colon X\times A\to X\times B$ has \procs\ morphism $Q\colon X\stricto \SP$ to $\puniversal$. The simulation condition in  Fig.~\ref{Fig:proc-homo} requires that the equations $\sta q_x = Q\circ\puev{Qx}$ and $\out q_x = \out{\uev{Qx}}$ hold for all $x:X$. Hence the universal simulations. A \sindex{universal!simulator} \sindex{universal!state space} \sindex{simulation!universal} state space $\SP$ is said to be universal when it carries a universal \procs\ for every pair $A,B$.

\para{The Fundamental Theorem of \emph{Stateful}\/ Computation.}\sindex{fundamental theorem!of stateful computation} Programming languages are universal state spaces. In particular, 
\begin{itemize}
\item[\textbf{a.}]\label{item:a} program executions $\puniversal_A^B \ = \ \big( \DP \times A\tto{\universal} \DP\times B\big)$ are universal \procs es; and
\item[\textbf{b.}]\label{item:b} universal \procs es $\Psi\colon \SP\times A\to \SP\times B$ make the universal state space $\SP$ into a programming language with the program evaluators $\universal_A^B \ = \ \big( \SP \times A\tto{\Psi} \SP\times B \tto{\scun\times B} B\big) = \out\Psi$.
\end{itemize}

\begin{figure}[!ht]
\begin{center}
\newcommand{\Dott}{\mbox{\LARGE$\bullet$}}
\newcommand{\Aee}{\scriptstyle A}
\newcommand{\XH}{\scriptstyle X}
\newcommand{\PPh}{\scriptstyle \DP}
\newcommand{\SKh}{\pev-}
\newcommand{\geee}{q}
\newcommand{\mhh}{\vartheta}
\newcommand{\Beee}{\scriptstyle B}
\newcommand{\programm}{\widetilde Q}
\renewcommand{\universal}{\mbox{\Large$\Universal$}}
\def\JPicScale{.5}
\input{PIC/kleene-coalg.tex}
\caption{A Kleene fixpoint $\widetilde Q$ of $\widetilde q = \left(\prtial\times B\right)\circ q$ satisfies 
$\uev{\widetilde Q} = \left(\pev{\widetilde Q}\ttimes B\right)\circ q$}
\label{Fig:kleene-coalg}
\end{center}
\end{figure}

\para{Proof of a.} For any given $q\colon X\times A\to X\times B$ consider the \funnn
\bear
\widetilde q & = & \left(\DP\ttimes X\ttimes A \tto{\DP \ttimes q} \DP \ttimes X\ttimes B\tto{\pev{}\ttimes B}\DP\ttimes B \right)
\eear
Let $\widetilde Q$ be its Kleene fixpoint, as displayed in Fig.~\ref{Fig:kleene-coalg}. The simulation $Q \colon X\stricto \DP$ can then be defined to be the partial evaluation $Q = \pev{\widetilde Q}$. Fig.~\ref{Fig:kleene-coalg-prog} shows that $Q = \pev{\widetilde Q}$ is a simulation of $q$ by $\puniversal^{B}_A$, satisfying the condition in Fig.~\ref{Fig:proc-homo} on the right. \hfill $\Box$
\begin{figure}[!ht]
\begin{center}
\newcommand{\Dott}{\mbox{\LARGE$\bullet$}}
\newcommand{\Aee}{\scriptstyle A}
\newcommand{\geee}{q}
\newcommand{\mhh}{\widehat g}
\newcommand{\Beee}{\scriptstyle B}
\newcommand{\XH}{\scriptstyle X}
\newcommand{\PPh}{\scriptstyle \DP}
\newcommand{\SKh}{\prtial}
\renewcommand{\universal}{\Large$\puniversal$}
\newcommand{\Mhh}{\color{red} Q}
\newcommand{\programm}{\widetilde Q}
\def\JPicScale{.5}
\input{PIC/kleene-coalg-1.tex}
\caption{$Q =\pev{\widetilde Q}$ satisfies $\out{q}_x = \out{\puev{Q x}}$ and $Q \circ \sta{q}_x = \sta{\puev{Q x}}$.}
\label{Fig:kleene-coalg-prog}
\end{center}
\end{figure}

\para{Proof of b.} To see that a given $AB$-universal \procs\ $\Psi\colon \SP\times A\to \SP\times B$ induces a program evaluator, as claimed, take any \funnn\ $g:X\times A\to B$ and extend it into the fixed-state \procs:
\bear
\widehat g & = & \left( X\times A \tto{\cmn \times A} X\times X\times A \tto{X \times g} X\times B\right)
\eear
Fig.~\ref{Fig:fixtate} shows that any program $G\colon X\stricto \SP$ that simulates $\widehat g$ also evaluates to $g = \out{\Psi} G = \uev{G}$.
\begin{figure}[!ht]
\begin{center}
\newcommand{\Fee}{\widehat G}
\newcommand{\fee}{g}
\newcommand{\feehat}{\widehat g}
\newcommand{\Aee}{\scriptstyle X}
\renewcommand{\Bee}{\scriptstyle A}
\renewcommand{\Cee}{\scriptstyle B}
\newcommand{\Code}{\scriptstyle \SP}
\newcommand{\Univ}{\mbox{\Large$\Psi$}}
\newcommand{\Dott}{\mbox{\LARGE$\bullet$}}
\def\JPicScale{.3}
\input{PIC/exec-fix.tex} 
\caption{A \funnn\ $g$ is evaluated by $\out \Psi$ when the fixed-state \procs\ $\widehat g$ is simulated by $\Psi$}
\label{Fig:fixtate}
\end{center}
\end{figure}
\hfill $\Box$
\

\section{Imperative programs}\label{Sec:imperative}

For \funnn s, the concept of computability is simple: every computable \funnn\ has a program and every program evaluates a unique computable \funnn. For \procs es, there is a wrinkle. On one hand, we just showed that every computational \procs\ can be simulated by executing a program. On the other hand, however, it is \emph{not true} that every program executes a unique computational \procs. Some programs do not simulate a \procs. Other simulate several, and cannot tell them apart. In the former case, we say that the program is not full; in the latter that it is not faithful. A program that is full and faithful is called \emph{imperative}.\sindex{program!imperative}

A program $Q$ may not be full when for some $a$ and $x$, the execution reaches a state that was not anticipated by $Q$, in the sense that $\sta{\puev{Qx}}(a)\neq Qy$ holds  for all states  $y$. The simulation condition $\sta{\puev{Qx}}(a)= Q\circ \sta q_x(a)$ cannot be satisfied for any \procs\ $q$, and the program $Q$ is not full. It is not faithful if there are \procs es $q, \tilde q$ with
\[ q_x^\odot= \out{\puev {Qx}}  = \tilde q_x^\odot \qquad \mbox{and}\qquad Q\circ q_x^\triangleright=\sta{\puev{Qx}}= Q\circ \tilde q_x^\triangleright \qquad \mbox{but} \qquad q_x^\triangleright\neq \tilde q_x^\triangleright\]
The program $Q$ thus simulates both $q$ and $\tilde q$, but some inputs $a$ lead to different states $q_x(a)\neq \tilde q_x(a)$ which $Q$ conflates into $Q\circ q^\triangleright_x(a)= Q\circ \tilde q^\triangleright_x(a)$.

To characterize programs that full and faithful programs, we use the following \textbf{notations}:
\beq\label{eq:pis}
\pi_{X}\  =\  \left(\DP\xrightarrow{\big<\fstt -, \sndd- \big>} \DP\times \DP \xrightarrow{\ \ \  \retr^X\ \times\  \scun\ \ \ } X \right)\qquad\qquad \pi_{1}  =  \left(\DP\xrightarrow{\big<\fstt -, \sndd- \big>} \DP\times \DP \xrightarrow{\ \ \  \scun\ \times\  \id\ \ \ } \DP \right)
\eeq
For simplicity, we often elide $\retr^X$ and draw the projections as if $\incl_{X\times \DP} = (\retr^X\times \DP)\circ \incl_{\DP\times \DP}$. This can be assumed without loss of generality Sections~\ref{Sec:retracts}--\ref{Sec:pairing}. 

\para{Padded execution.}\sindex{program!execution} A \emph{padding}\/ is a part of a program that does not get evaluated but remains available to pass state to later evaluation steps. A padded program can always be brought in the form $F =\pairr {F_0} {F_1}$, where $F_0$ is the padding. Such programs are processed in  \emph{padded executions}\sindex{padded execution}\sindex{program!execution!padded} 
\bea
\Runn_A^B & =&  \left(\DP\times A \xrightarrow{\ \ \ \pi_1\times \id\ \ \  } \DP\times A \xrightarrow{\ \ \  \ \   \universal\ \ \  \ \  }\DP\times B \right)
\eea 
to get $\Run F_A^B = \uev{\sndd{F}}_A^{\DP\times B} = \uev{F_1}_A^{\DP\times B}$. The role of the padding differs from construction to construction. Its role in imperative programs is clear already from their definition.

\para{Definition.} \sindex{program!imperative} A program $\Theta\colon X\stricto \DP$  is said to be \emph{imperative}\/ if it satisfies the two conditions displayed in Fig.~\ref{Fig:executable-string}. 
\begin{figure}[!ht]
\begin{center}
\newcommand{\prrrr}{\Theta}
\newcommand{\pjjjj}{\scriptscriptstyle<\fstt -, \sndd ->}
\newcommand{\Aee}{\scriptstyle X}
\renewcommand{\Bee}{\scriptstyle A}
\renewcommand{\Cee}{\scriptstyle B}
\newcommand{\Code}{\scriptstyle \DP}
\newcommand{\Univ}{\mbox{\Large$\universal$}}
\newcommand{\PadUniv}{\mbox{\color{red}\Large$\Runn$}}
\newcommand{\Dott}{\mbox{\LARGE$\bullet$}}
\def\JPicScale{.45}
\input{PIC/losless-3.tex}
\caption{$\Theta$ is faithful if $\pi_X\circ \Theta=\id_X$. It is full if $\big((\Theta\circ\pi_X)\times B\big)\circ\Run \Theta =\Run\Theta$.}
\label{Fig:executable-string}
\end{center}
\end{figure}
The conditions correspond to the commutativity of the arrow diagrams in \eqref{eq:executable-arrow}.
\beq\label{eq:executable-arrow}
\begin{tikzar}[column sep=4em]
\&X  \ar[equals]{dd} 
\& 
\DP\times B \ar[two  heads]{r}[description]{\pi_X\times \id}  \& X\times B \ar[nothing-cart,tail]{r}[description]{\Theta\times \id} \& \DP\times B
\\
\DP\ar[two heads]{ur}[description]{\pi_X} \& 
\& 
\DP\times A \ar{u}[description]{\Runn} \&\& \DP\times A \ar{u}[description]{\Runn}
\\
\& X\ar[nothing-cart,tail]{ul}[description]{\Theta}
\& \&
X\times A \ar[nothing-cart,tail]{ul}[description]{\Theta\times \id} \ar[nothing-cart,tail]{ur}[description]{\Theta\times \id} \ar[dashed]{uu}[description]{q}
\end{tikzar}
\eeq

\para{Every imperative program induces a unique computational \procs} executed by a padded program The \procs\ is defined as the dashed arrow  
$q = (\pi_X\times B)\circ \Runn \circ (\Theta\times A)$ in \eqref{eq:executable-arrow}. The second condition gives $(\Theta\times B) \circ q= \Run\Theta$, which means that $\Theta$ is a simulation of $q$ by $\Runn$, and hence full. The first condition $\pi_X\circ \Theta = \id$ implies that $q$ is the only process which $\Theta$ simulates by $\Runn$, since 
\bear
\Theta\circ q_0 = \Run \Theta  = \Theta\circ q_1 & \implies & q_0 = \pi_X\circ \Theta\circ q_0 = \pi_X\circ \Theta\circ q_1 = q_1
\eear
Hence $\Theta$ is faithful.

\para{Every computational \procs\ is has an imperative program.} Given a \procs\ $q\colon X\times A\to X\times B$ define $\widetilde \theta\colon \DP\times X\times A\to \DP\times B$ as in Fig.~\ref{Fig:losless-feedforward} and let $\widetilde \Theta$ be its Kleene fixpoint.
\begin{figure}[!ht]
\begin{center}
\newcommand{\Fee}{\prtial}
\newcommand{\PadFee}{\mbox{\Large${\color{red}\widetilde \theta}$}}
\newcommand{\quu}{q}
\newcommand{\prrrr}{\scriptstyle\pairr - - }
\newcommand{\pjjjj}{\scriptscriptstyle<\fstt -, \sndd ->}
\newcommand{\Aee}{\scriptstyle X}
\renewcommand{\Bee}{\scriptstyle A}
\renewcommand{\Cee}{\scriptstyle B}
\newcommand{\Code}{\scriptstyle \DP}
\newcommand{\programm}{\widetilde \Theta}
\newcommand{\Univ}{\mbox{\Large$\universal$}}
\newcommand{\PadUniv}{\mbox{\Large$\color{red}\Runn$}}
\newcommand{\Dott}{\mbox{\LARGE$\bullet$}}
\def\JPicScale{.5}
\input{PIC/losless-feedforward.tex}
\caption{A Kleene fixpoint $\widetilde \Theta$ of $\widetilde \vartheta$ satisfies $\sta{\uev{\widetilde \Theta x}} = \Big\lceil{\sta q_x}\, , {\pev{\widetilde\Theta}\circ \sta{q}_x}\Big\rceil$ and $\out{\uev{\widetilde \Theta x}} = \out{q}_x$}
\label{Fig:losless-feedforward}
\end{center}
\end{figure}
Then Fig.~\ref{Fig:losless-feed} shows that 
\bear
\Theta & =\left(X\tto{\cmn} X\times X\xrightarrow{X\times \pev{\widetilde\Theta}} X\times\DP\tto{\pairr - - } \DP\right)
\eear
is a simulation of $q$ by $\Runn$.
\begin{figure}[!ht]
\begin{center}
\newcommand{\Fee}{\prtial}
\newcommand{\PadFee}{{\color{red}\Theta}}
\newcommand{\quu}{q}
\newcommand{\prrrr}{\scriptstyle\pairr - - }
\newcommand{\pjjjj}{\scriptscriptstyle<\fstt -, \sndd ->}
\newcommand{\Aee}{\scriptstyle X}
\renewcommand{\Bee}{\scriptstyle A}
\renewcommand{\Cee}{\scriptstyle B}
\newcommand{\Code}{\scriptstyle \DP}
\newcommand{\programm}{\widetilde \Theta}
\newcommand{\Univ}{\mbox{\Large$\universal$}}
\newcommand{\PadUniv}{\mbox{\Large$\color{red}\Runn$}}
\newcommand{\Dott}{\mbox{\LARGE$\bullet$}}
\def\JPicScale{.45}
\input{PIC/losless-feedfor.tex}
\caption{$\Theta x = \Big\lceil{x}\, , {\pev{\widetilde\Theta}x}\Big\rceil$ is a simulation of $q$ by $\Runn$ since $\sta{\Run {\Theta x}} = \Theta\circ \sta{q}_x$ and $\out{\Run{\Theta x}} = \out{q}_x$.}
\label{Fig:losless-feed}
\end{center}
\end{figure}
It is easy to see that the conditions $\pi_X\circ \Theta = \id$ and $(\Theta\circ \pi_X\times B)\circ \Run\Theta = \Run\Theta$ from Fig.~\ref{Fig:executable-string} and \eqref{eq:executable-arrow} are satisfied, and that $\Theta$ is thus a losless simulation.

\para{Summary and reserve.} Stateful computations are executed as imperative programs. Any computational \procs\ $q\colon X\times A\to X\times B$ is embedded into a program execution \procs\ $\Runn\colon \DP\times A\to \DP\times B$ along an imperative program $\Theta\colon X\strictinto \DP$. A sequence of program evaluations \eqref{eq:simulation} of $\Theta$ simulates $q$ faithfully, distinguishing its states. The other way around, the simulation of imperative programs is also full, which means that every imperative $\Theta$ induces a computational \procs\ $q$. While the execution of any imperative program $\Theta$ is projected back to the state space $X$ of the simulated \procs\ $q$ along a projection $\pi_X\colon \DP\to X$, note that the projections are not \procs\ morphisms and that stateful computations are generally \emph{not}\/ retracts of program executions, just their sub\procs es. The only retracts of program executions are other program executions, sharing the same universality property. See Ch.~\ref{Chap:PCC} for more.

%
%

\def\thechapter{8}
\setchaptertoc
\chapter[Program-closed categories: computability as a property]{Program-closed categories:\\ computability as a property}
\label{Chap:PCC}
\newpage

\label{Sec:zoom}

\label{Sec:iso-thm}
In this chapter we prove that any two programming languages that are universal in the same computable universe must be  isomorphic. The isomorphism preserves the meanings of their programs. 

The \textbf{proviso}\/ is that \textbf{the programming languages are \emph{well-ordered}}. Real programming languages are always well-ordered. It is natural to assume that they are well-ordered in monoidal computers. We did not mention the well-ordeing until now only because none of the basic constructions depended on it. The meaning-preserving correspondence  between the programs in different programming languages does depend on their well-ordering.

\section{Program order}
\label{Sec:order}

Digital computers run digital programs on digital data. Since the digits are well-ordered, all programs and data in digital computers are well-ordered lexicographically, like words in a lexicon. 

Abstract computers run abstract programs on abstract data. We reverse-programmed a lot of structure of concrete digital computers using the abstract program evaluators underlying the \runn-instruction, with no recourse to the program order. In this chapter we take the program order into account. The programming language is still just the \runn-instruction with the ``syntactic sugar'' on top of it, but the fact that the derived programs are well-ordered is a powerful new programming tool. 

\subsection{Well-order}
\begin{figure}[!ht]
\begin{center}
\renewcommand{\EQLS}{\mbox{\huge =}}
\newcommand{\hand}{\wedge}
\newcommand{\hor}{\vee}
\newcommand{\Baut}{\scriptstyle \Bool}
\newcommand{\hwaai}{\scriptstyle \true}
\newcommand{\cuun}{\scriptstyle \cun}
\newcommand{\same}{(\iseq)}
\newcommand{\smaller}{(\pleq)}
\def\JPicScale{.65}
\input{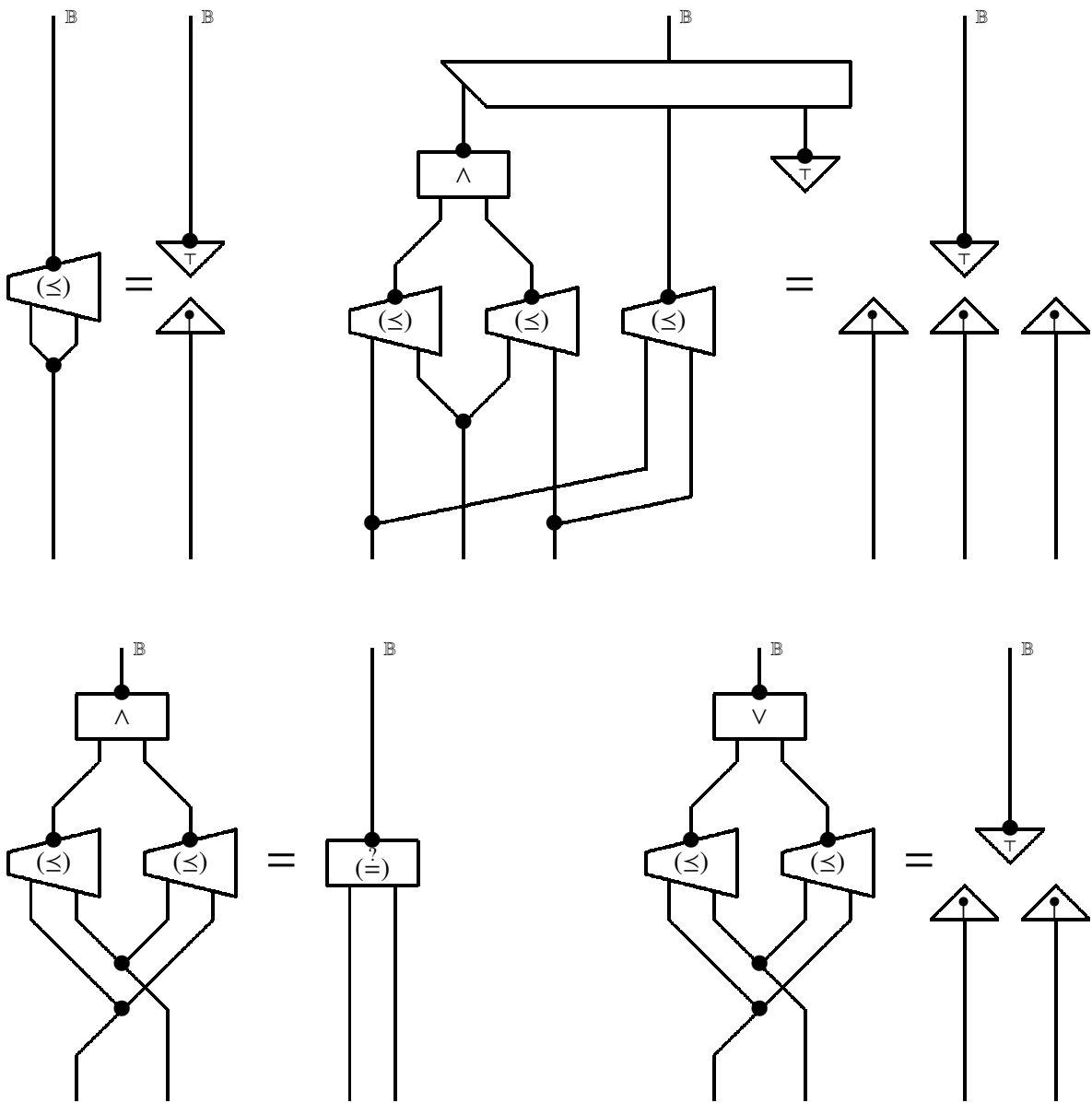}
\caption{Diagrammatic version of \eqref{eq:O}: the total order axioms for $(\pleq)$ }
\label{Fig:O}
\end{center}
\end{figure}

\para{Total order.} An \emph{order relation}\/ on a type $A$ \sindex{order} is presented as a decidable predicate  $(\pleq):A\times A\stricto\Bool$ satisfying the first three conditions in
\beq\label{eq:O}
u\pleq u \qquad \qquad u\pleq v\wedge v\pleq w \Rightarrow u\pleq w \qquad \qquad u\pleq v \wedge u\pgeq v \Rightarrow u=v\qquad \qquad u\pleq v \vee u \pgeq v
\eeq
where $x\pleq y$ and $y\pgeq x$ abbreviate $(x\pleq y)=\true$. The fourth condition in \eqref{eq:O} makes $(\pleq)$ into \sindex{order!total} \emph{total}\/ order. A diagrammatic version of \eqref{eq:O} is in Fig.~\ref{Fig:O}. 

\para{Bottom.} \sindex{order!bottom} The type $A$ is \emph{well-ordered}\/ when it is \sindex{order!well-order} totally ordered and moreover every inhabited decidable predicate $\prd\colon A\stricto \Bool$ has the bottom, a minimal inhabitant $\wo_\prd\colon A$. Formally, this is written 
\bea\label{eq:WO}
\prd\neq \flse &\implies & \exists \wo_{\prd}.\ \ {\prd}(\wo_{\prd}) \ \wedge\ \forall u.\ {\prd}(u) \implies \wo_{\prd}\pleq u
\eea
The diagrammatic version of \eqref{eq:WO} is in Fig.~\ref{Fig:WO}.
\begin{figure}[!ht]
\begin{center}
\newcommand{\hand}{{\prd}_{(\neq\flse)}}
\newcommand{\Baut}{\scriptstyle \Bool}
\newcommand{\Aargh}{\scriptstyle A}
\newcommand{\Eex}{\scriptstyle X}
\renewcommand{\moo}{\wo_\prd}
\newcommand{\hwaai}{\true}
\newcommand{\cuun}{\scriptstyle \cun}
\newcommand{\smaller}{(\pleq)}
\def\JPicScale{.65}
\input{PIC/equal-WO.tex}
\caption{$\exists a.\ {\prd}(x,a)\ \ \implies\ \ \Big({\prd}\big(x,\, \wo_{\prd}(x)\big)\ \wedge\ \forall a.\ {\prd}(x,a) \implies \wo_{\prd}(x)\pleq a\Big)$}
\label{Fig:WO}
\end{center}
\end{figure}
It is slightly more general, in the sense that the predicate $\prd$ is parametrized in the form $\prd\colon X\times A\stricto \Bool$, and a minimal inhabitant is assigned for every $x\colon X$ along $\wo_\prd \colon X\stricto A$. For simplicity, this parametrization is elided from \eqref{eq:WO}.\footnote{The tacit assumption that different variables do not interfere with each another makes every axiom instance tacitly parametric \cite{PavlovicD:Como}.}.

\para{Inhabited (or nonempty) predicates.} The \emph{empty predicate}\/ is \sindex{predicate!empty} the constant 
\bear
\flse_{A} & = & \left(A\strictto\scun I \strictto\flse \Bool\right)
\eear
A decidable predicate $\prd\colon A\stricto \Bool$ is thus \emph{nonempty}, or \emph{inhabited},  \sindex{predicate!inhabited, nonempty} when there is $\prd(u) = \true$, in which case $u$ is called \sindex{inhabitant} an \emph{inhabitant}\/ of $\prd$. Note that the decidability of the truth value $\prd(a)$ for a particular $a\colon A$ does not entail the decidability of $\prd(u)$ for all $u:A$. \emph{It may be undecidable whether a decidable predicate is inhabited or empty.} If the emptiness of $\prd$ is undecidable, the existence of its bottom $\wo_{\prd}$ may be undecidable even if $\prd$ itself is decidable. 

\para{Finite infima.} When it does exist, the bottom $\wo_\prd$ is obviously the greatest lower bound, or infimum  with respect to $(\pleq)$ of the elements satisfyong $\prd$. The finite infima can be computed in the form
\bea\label{eq:pwedge}
u \pwedge  v & = & \iif\left(u\pleq v\, ,\ u\, ,\ v \right)
\eea
The operation $\wo$ extends to all inhabited decidable predicates $\prd$, and can be construed as an instance of search or minimization\sindex{minimization} \sindex{search} from  Sec.~\ref{Sec:search}. The infimum $\wo_\prd$ is then written $\mu x.\ \prd(x)$.

\subsection{Consequences of well-order}\label{Sec:WO-splittings}
\para{The bottom} $\wo\ =\ \wo_\true\colon A$ is the infimum of the identically true predicate 
\bear
\true_{A} & = & \left(A\strictto\scun I \strictto\true \Bool\right)
\eear

\para{Descending sequences are finite.} In a computable universe, a descending sequence is conveniently viewed as a \strict\ \funnn\ $s:A\stricto A$ where $s(x)\pleq x$. The claim is that any sequence down a well-order must stabilize after finitely many steps. Writing  $s_i = s^i(a)$ for some $a:A$ and $i=0,1,2,\ldots$, the formal claim is:
\bea\label{eq:descending}
s_0 \pgeq s_1 \pgeq s_2\pgeq\cdots \pgeq s_i\pgeq \cdots & \implies & \exists n \ \forall k.\ \ s_{n+k} = s_{n}
\eea
This claim is a special case of \eqref{eq:WO} or Fig.~\ref{Fig:WO}, where $X$ is instantiated to $\NNn$, the predicate $\prd_s:A\stricto\Bool$ is ${\prd}_s(a) = \Prd_s(0,a)$ and the predicate $\Prd_s:\NNn\times A\stricto \Bool$ is 
\bea\label{eq:prd}
\Prd_s(n,a) &= & \iif\Big(s_n\iseq a\, ,\  \true, \notag\\ 
&& \iif\big(s_n\plt a \vee s_n \iseq s_{n+1}\, ,\  \flse,\  \Prd_s(n+1,a)\big)\Big)
\eea
A computation of $\prd_s$ thus descends down the sequence $s_0 \pgeq s_1 \pgeq s_2\pgeq\ldots$, compares its entries with $a$, returns $\true$ if $s_n=a$, $\flse$ if $s_n\plt a$ (because $s_{n+k}\pleq s_n \plt a$ for all $k$) or if $s_n\pgt a$ but $s_{n+1} = s_n$ (because $s_{n+k}=s_n \pgt a$ for all $k$). Thus $\prd_s(a)$ is true if and only if $a=s_n$ for some $n$. Axiom \eqref{eq:WO} now says that there $\wo_{\prd_s}=s_n$ for some $n$ and $\wo_{\prd_s}\pleq s_m$ for all other $m:\NNn$. It follows that $s_{n+k} = \wo_{\prd_s}$ for all $k$. Simplifying the notation from $\wo_{\prd_s}$ to  $\wob s$, \eqref{eq:descending} boils down to 
\beq\label{eq:descending-s}
s_0 \pgt s_1 \pgt s_2\pgt\cdots  \pgt \wob s = s\left(\wob s \right)
\eeq

\para{Decidable Axiom of Choice: Monotone unbounded \strict\ surjections split.} With all decidable types well-ordered, the \funnn s can be restricted to the \emph{monotone}\/ ones, i.e. where $x\pleq y$ implies $f(x)\pleq f(y)$. A \funnn\ $f:A\to B$ is also  \emph{unbounded}\/ if for every $b:B$ there is $a:A$ such that $b\pleq f(a)$. Any \funnn\ $f\colon A \to B$ induces a predicate $\prd_f\colon B\times A\to \Bool$ defined
\bea
\prd_f(b,a) & = & \left(b\pleq f(a)\right) 
\eea
When $f\colon A \stricto B$ is \strict, then $\prd_f\colon B\times A\stricto \Bool$ is decidable. When $f$ is unbounded, then $\prd_f(b,-)\colon A\stricto \Bool$ is nonempty for every $b$. Defining $f^\ast\colon B\stricto A$ to be
\bea\label{eq:f-ast}
f^\ast (b) & = & \wo_{\prd_f}(b)
\eea
induces the correspondence
\bea\label{eq:adj-wo}
f^\ast (b) \pleq a & \iff & b \pleq	 f(a)
\eea
which makes $f^{\ast}$ into the \emph{left adjoint}\/ of $f$ with respect to the order. It is not hard to work out that the \funnn s $\overleftarrow f = \left(A\strictto f B\strictto{f^\ast} A\right)$ and $\overrightarrow f = \left(B\strictto{f^\ast}A\strictto{f} B \right)$ are idempotent, i.e. \emph{closure operators}\/ with respect to the order. Per Sec.~\ref{Sec:typ-idem}, splitting these idempotents determines the \emph{$f$-closed subtypes}\/ $A^{\overleftarrow f}$ of $A$ and $B^{\overrightarrow f}$ of $B$
\beq
A^{\overleftarrow f} \ = \ \left\{x:A\ |\ x = \overleftarrow f(x) \right\}\qquad\qquad\qquad\qquad
B^{\overrightarrow f} \ = \ \left\{y:B\ |\ \overrightarrow f(y) = y\right\}
\eeq
Clearly $A^{\overleftarrow f}=A$ holds precisely when $f$ is injective and $B^{\overrightarrow f}=B$ when it is surjective. The monotone unbounded injections and the  surjections between well-orders always split.\footnote{Cantor seems to have derived this from his principle that all sets can be well-ordered, as suggested in  \cite[letter to Dedekind of 5 November 1882]{Cantor-Dedekind:Briefwechsel}. But the issue of well-ordering the reals got intermingled with the issue of how many real numbers are there, which attracted the well-known violent attacks. So Cantor is circumspective and is difficult to be certain what he is saying. Later he even tried to derive Well-Ordering Principle from other axioms \cite[III.4. Nr. 5 (\"Uber unendliche lineare Punktmannigfaltigkeiten)]{CantorG:collected}. In the end, his Well-Ordering Principle got replaced with Zermelo's Axiom of Choice \cite{Martin-Loef:AC}.} The former is an easy consequence of the unboundness and the well-ordering. The latter is the well-ordered version of the \sindex{Axiom of Choice} Axiom of Choice. In return, the Axiom of Choice allows well-ordering any set and splitting any surjection. Restricting $f$ to $A^{\overleftarrow f}$ along the splitting of $\overleftarrow f$ lands into $B^{\overrightarrow f}$ and reduces it to a type isomorphism. 
\beq
\tikzset{nothing/.tip={},cart/.tip={Glyph[glyph math command = bullet]}}
\begin{tikzar}[row sep = 1.5ex,column sep = 8em]
A \ar[bend right=12,two heads]{dr}[description]{\retr^{\overleftarrow f}}  \ar[nothing-cart]{dddddd}[description]{f}[description,pos=.99]{\mbox{\large$\bullet$}}
\\
\& A^{\overleftarrow f}\ar[leftrightarrow]{dddd}[description]{\cong} \ar[bend right=12,hook]{ul}[description]{\incl_{\overleftarrow f}}
\\
\\
\\
\\
\& B^{\overrightarrow f} \ar[bend right=12,hook]{dl}[description]{\incl_{\overrightarrow f}}\\
B \ar[bend right=12,two heads]{ur}[description]{\retr^{\overrightarrow f}}
\end{tikzar}
\eeq
The inverse of the restriction of $f$ from $A$ to $A^{\overleftarrow f}$ is the restriction of $f^{\ast}$ from $B$ to $B^{\overrightarrow f}$. The type $B^{\overrightarrow f}$ is thus the range of $f$, in the sense
\beq\label{eq:splitting}
\exists x\colon A.\ b=f(x)  \ \   \iff \ \ b \colon B^{\overrightarrow f}\ \ \iff\ \ b = f\left( f^\ast(b)\right)  
\eeq
The left adjoint $f^{\ast}\colon B\stricto A$ is thus a splitting (left inverse) of $f\colon A\stricto B$ if and only if $f$ is surjective. Otherwise it points to the least inverse images of $b$ in the range $B^{\overrightarrow f}$ of $f$ and maps the elements that are out of the range of $f$ to the inverse images of their closest approximations from above that do lie in the range. In general, the Axiom of Choice  is valid precisely when every surjection has a splitting, a choice function. Since surjections are obviously unbounded for any order that may be around, it follows that the Axiom of Choice is valid for the monotone functions in any computable universe with well-ordered programs. See Sec.~\ref{Sec:state-story} for further comments.

\subsection{What if $\DP$ is well-ordered?}\label{Sec:monotone-prog} \sindex{program!order}
\para{Then every \emph{decidable}\/ type is well-ordered.} It was established in Sec.~\ref{Sec:retracts} that any type $A$ corresponds to an idempotent $\rho_A:\DP\to\DP$, in the sense that the elements $x:A$ correspond to the programs $x=\rho_A(x)$. Any order on $\DP$ thus induces a relation
\bea\label{eq:woA}
\ppp x A y & \iff &\rho_A(x) \pleq \rho_A(y)
\eea 
This relation is decidable when $\rho_A$ is \strict. This is when the type $A$ is called decidable, since its \sindex{predicate!characteristic} \emph{characteristic predicate} \sindex{type!as characteristic predicate}
\bea 
\chi_A\colon \DP&\to &\Bool\\
x& \mapsto & \left(x\iseq \rho_A(x)\right)
\eea
is then decidable.\footnote{There are types $A$ where $\chi_A$ is decidable although $\rho_A$ is not \strict. They are not on the path of the present narrative.} \sindex{type!decidable} In general, any order on $\DP$ induces orders on all types in a compuable universe, but that order can only be decidable on decidable types.

\para{Monotone programming.} \sindex{programming!monotone}
We saw in Sec.~\ref{Sec:how-many} that any given computable \funnn\ is encoded by infinitely many programs (unless $\true=\flse$). Since any decreasing sequence of well-ordered programs is finite, for any computable \funnn\ there must be an infinite \emph{increasing}\/ sequence of programs that encode it. That sequence cannot be bounded, since the descending sequences below any bound would be finite. For any computation $f:A\to B$ and any bound $b:\DP$ there must therefore always be a program $F\pgeq b$ such that $\uev F = f$. If this requirement is written $\uev{F_{\pgeq b}} = f$, then the programmability requirement in \eqref{eq:uev-formal} becomes
\beq\label{eq:uev-formal-ordered}
\forall g\in\CCc(X\times A, B)\ \ \ \forall \ell\in\CCc^\bullet(X, \DP)\ \ \ \exists G\in \CCc^\bullet(X, \DP).\ \ \ \Big\{G_{\pgeq \ell}\Big\} = g
\eeq
In the presence of well-ordering, this refinement of \eqref{eq:uev-formal} is derivable by adapting standard well-ordering arguments. However, since they are neither succinct nor specific to the present framework, we omit them and simply strengthen \eqref{eq:uev-formal} by \eqref{eq:uev-formal-ordered}. The sense in which this strengthening is conservative is immaterial for the present goals. The fact the real programming languages are well-ordered assures consistency. In the context of the partial evaluators (\ref{eq:uev-pev-formal-one}--\ref{eq:uev-pev-formal-two}), amounts to monotonicity:
\bea\label{eq:pev-monotone}
\pev G x \ \pgeq  \  G\qquad\qquad\qquad  \pev G x \ \pgeq  x
\eea
The other basic program constructors can therefore also be chosen to be monotone for all programs $p,q\colon \DP$
\bea
p, q\  & \pleq &\  (p;q)\, ,\  (p\parallel q)\, ,\ \pairr p q
\eea

\section{Program-closed categories}
\label{Sec:pcc}

A \emph{program-closed category}\/ is a monoidal computer $\CCC$ whose programming language $\DP$ is well-ordered. Unfolding the definition from Sec.~\ref{Sec:moncom-def} and removing the redundancies, the program-closed structure of a category $\CCC$ boils down to
\begin{description}
\item[\textbf{a)}] a data service $A\times A\strictoot \cmn A\strictto \scun I$ on every type $A$, as described in Sec.~\ref{Sec:service},  
\item[\textbf{b')}] a distinguished type $\DP$ of \emph{programs}\/ with \textbf{\emph{decidable well-order}}\/ $(\pleq):\DP\times \DP\stricto\Bool$, and 
\item[\textbf{c'')}] program evaluators or equivalently the induced program executions $\Run{-}:\DP\times A\to \DP\times B$ for all types $A,B$, as described  in Ch.~\ref{Chap:State}. 
\end{description}

\para{Examples.} It was explained in Sec.~\ref{Sec:moncom-def} that any Turing-complete programming language $\DP$ induces a monoidal computer. It induces a program-complete category as soon as it is well-ordered. All of the programming languages based on numerically represented alphabets are well-ordered lexicographically. The only languages where the well-ordering is not automatic, but needs to be explicitly specified, are the natively diagrammatic languages.

\section{Programming languages are isomorphic}
\label{Sec:hartley}

\subsection{The Isomorphism Theorem} 
Any two programming languages in a program-closed category are isomorphic. More precisely, for any pair of  programming languages $\DP_{0}$ and $\DP_{1}$ with program executions $\Runn_0\colon \DP_{0}\times A\to \DP_{0}\times B$ and $\Runn_1\colon\DP_{1}\times A\to \DP_{1}\times B$ for all types $A,B$ there are \funnn s $\kappa:\DP_0\stricto \DP_1$ and $\chi:\DP_1\stricto \DP_0$ such that for all programs $x\colon \DP_0$ and $y\colon \DP_1$ 
\beq\label{eq:hartley}
\chi\circ \kappa(x) = x\qquad\qquad \kappa\circ \chi(y) = y \qquad\qquad \Run{x}_0 = \Run{\kappa(x) }_1 \qquad\qquad \Run{\chi(y)}_0 = \Run{y}_1
\eeq

\subsection{Simulations between programming languages}
\label{Sec:pad}\label{Sec:trace}
\begin{figure}[!ht]
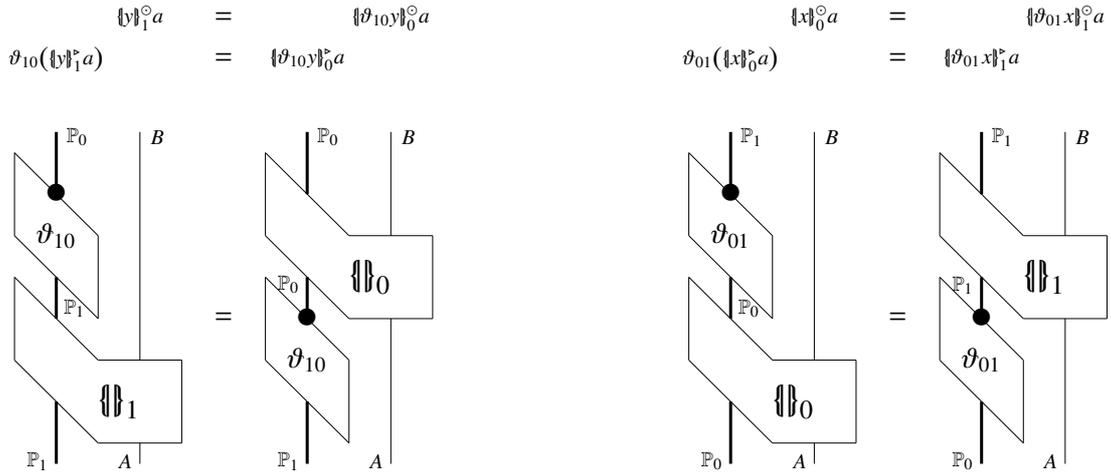

\begin{center}
\newcommand{\Fee}{\vartheta_{10}}
\newcommand{\Aee}{\scriptstyle \DP_0}
\renewcommand{\Bee}{\scriptstyle A}
\renewcommand{\Cee}{\scriptstyle B}
\newcommand{\Code}{\scriptstyle \DP_1}
\newcommand{\Univ}{\mbox{\Large$\Run{}_1$}}
\newcommand{\Univtwo}{\mbox{\Large$\Run{}_0$}}
\newcommand{\Dott}{\mbox{\LARGE$\bullet$}}
\newcommand{\LHS}{\scriptstyle\out{\Run{y}}_1 a}
\newcommand{\RHS}{\scriptstyle \out{\Run{\vartheta_{10} y}}_0 a}
\newcommand{\LHSP}{\scriptstyle \vartheta_{10}\left(\sta{\Run{y}}_1 a\right)}
\newcommand{\RHSP}{\scriptstyle \sta{\Run{\vartheta_{10} y}}_0 a}
\def\JPicScale{.275}
\input{PIC/progsim.tex} 
\hspace{5em}
\renewcommand{\Fee}{\vartheta_{01}}
\renewcommand{\Aee}{\scriptstyle \DP_1}
\renewcommand{\Bee}{\scriptstyle A}
\renewcommand{\Cee}{\scriptstyle B}
\renewcommand{\Code}{\scriptstyle \DP_0}
\renewcommand{\Univ}{\mbox{\Large$\Run{}_0$}}
\renewcommand{\Univtwo}{\mbox{\Large$\Run{}_1$}}
\renewcommand{\LHS}{\scriptstyle\out{\Run{x}}_0 a}
\renewcommand{\RHS}{\scriptstyle \out{\Run{\vartheta_{01} x}}_1 a}
\renewcommand{\LHSP}{\scriptstyle \vartheta_{01}\left(\sta{\Run{x}}_0 a\right)}
\renewcommand{\RHSP}{\scriptstyle \sta{\Run{\vartheta_{01} x}}_1 a}
\def\JPicScale{.275}
\input{PIC/progsim.tex} 
\caption{Programming languages simulate each other}
\label{Fig:progsim}
\end{center}
\end{figure}

According to the Fundamental Theorem of Stateful Computation in Sec.~\ref{Sec:ana}, any programming language is a universal state space, and its program evaluators are universal simulators. It follows that any two programming languages $\DP_0$ and $\DP_1$ simulate each other. This means that for any pair of types $A,B$ there are simulations $\vartheta_0:\DP_1\stricto\DP_0$ and $\vartheta_1:\DP_0\stricto \DP_1$ as in Fig.~\ref{Fig:progsim}. 
These simulations preserve the extensional meaning of programs, in the sense that $\Run x_0 = \Run{\vartheta_{01} x}_1$ and $\Run{\vartheta_{10}y}_0 = \Run y_1$, as required in  \eqref{eq:hartley}. But this is just the second half of \eqref{eq:hartley}. There is no reason why $\vartheta_{01}$ and $\vartheta_{10}$ should satisfy the first half, to form an isomorphism. Each of them could map many extensionally equivalent programs from its domain to the same program in its range, and they need not use all of the programs from the extensional equivalence classes in their ranges. The simulations given by the Fundamental Theorem of Stateful Computation thus preserve the denotational semantics and extensional meanings of programs, but may fail to preserve the operational semantics and intensional meaning.\footnote{For this reason, \procs\ calculi do not identify \procs\ along pairs of simulations both ways, but only along \emph{bi}\/simulations, where a single relation provides simulations both ways.} The composite simulations 
\beq\label{eq:selfself}
\vartheta_{00} = \left( \DP_0\strictto{\vartheta_{01}} \DP_1\strictto{\vartheta_{10}} \DP_0\right)
\qquad \qquad \qquad \vartheta_{11} = \left( \DP_1\strictto{\vartheta_{10}} \DP_0\strictto{\vartheta_{01}} \DP_1\right)
\eeq
reinterpret each of the languages in itself without changing the program denotations, but may modify their operational behaviors. However, we have seen in Sec.~\ref{Sec:imperative} that the \emph{padded}\/ executions simulate \procs es as sub\procs es of universal simulators. There are thus imperative programs $\widetilde\vartheta_{01}:\DP_0\strictinto \DP_1$ and $\widetilde\vartheta_{10}:\DP_1\strictinto \DP_0$ that embed the languages $\DP_0$ and $\DP_1$ into each other and make the program execution \procs es $\Runn_0\colon \DP_{0}\times A\to \DP_{0}\times B$ and $\Runn_1\colon\DP_{1}\times A\to \DP_{1}\times B$ into each other's sub\procs es. The \textbf{idea} is now to extract an isomorphism from two embeddings using a constructive version of the Cantor-Bernstein constructions   \cite{HinkisA:cantor-bernstein,SiegW:cantor-bernstein}. It will be based on the well-ordering of both types, and the monotonicity of their simulations. If the construction of full and faithful simulations in Sec.~\ref{Sec:imperative} looks complicated, note that the same idea can be also realized using the \emph{lossless}\/ simulations
\beq
\widehat{\vartheta}_{01} x \ =\ \pairr x {\vartheta_{01} x} \qquad \qquad \qquad \qquad 
\widehat{\vartheta}_{10} y \ =\ \pairr y {\vartheta_{10} y}
\eeq
derived from any given pair of simulations both ways:
\[
\prooftree 
\vartheta_{01}:\DP_0\stricto \DP_1
\justifies
\widehat \vartheta_{01}:\DP_0\strictinto \DP_1
\endprooftree
\qquad \qquad \qquad \qquad
\prooftree 
\vartheta_{10}:\DP_1\stricto \DP_0
\justifies
\widehat \vartheta_{10}:\DP_1\strictinto \DP_0
\endprooftree
\]  
and executed using the execution processes from Workout~\ref{Work:lossless}. Whether they are \textbf{imperative or lossless}, any pair of monotone, meaning-preserving  embeddings and their compositions \textbf{will do}. We write either of them in the form
\beq\label{eq:hartkey-mono}
\tikzset{nothing/.tip={},cart/.tip={Glyph[glyph math command = bullet]}}
\begin{tikzar}[column sep=2.5cm]
\DP_0  \arrow[nothing-cart,loop, out = 140, in=220, looseness = 6,tail]{}[swap]{\overline\vartheta_{00}}
\arrow[nothing-cart,shift right = .5ex,shorten=1mm,tail,bend right = 9]{r}[swap]{\overline\vartheta_{01}}   
\& 
\DP_1 
\arrow[nothing-cart,loop, out = 40, in=-40, looseness = 6,tail]{}{\overline\vartheta_{11}} 
\arrow[nothing-cart,shift right = .5ex,shorten=1mm,tail,bend right = 9]{l}[swap]{\overline\vartheta_{10}} 
\end{tikzar}
\eeq
where
\beq\label{eq:overself}
\overline\vartheta_{00} = \left( \DP_0\strictintto{\overline\vartheta_{01}} \DP_1\strictintto{\overline\vartheta_{10}} \DP_0\right)
\qquad \qquad \qquad \overline\vartheta_{11} = \left( \DP_1\strictintto{\overline\vartheta_{10}} \DP_0\strictintto{\overline\vartheta_{01}} \DP_1\right)
\eeq
All simulations preserve the meaning of programs by definition. The monotonicity and the injectiveness
\beq\label{eq:padding-injective}
x\plt \overline\vartheta_{ij} x \qquad \qquad \qquad x\plt y \implies \overline\vartheta_{ij} x\plt \overline\vartheta_{ij} y \qquad \qquad \qquad \overline\vartheta_{ij} x= \overline\vartheta_{ij} y \implies x=y
\eeq
are achieved through monotone programming (Sec.~\ref{Sec:order}) of imperative or lossless simulations (Ch.~\ref{Chap:State}).

\subsection{The isomorphism construction}
We prove the Isomorphism Theorem using the embeddings from \eqref{eq:hartkey-mono} satisfying \eqref{eq:overself} to extract an isomorphism 
\beq\label{eq:hartley-iso}
\tikzset{nothing/.tip={},cart/.tip={Glyph[glyph math command = bullet]}}
\begin{tikzar}[column sep=2.5cm]
\DP_0  \arrow[nothing-cart,loop, out = 140, in=220, looseness = 6]{}[swap]{\id} 
\arrow[nothing-cart,shift right = .5ex,shorten=1mm,bend right = 9]{r}[swap]{\kappa} \arrow[phantom]{r}[description]{\cong}   
\& 
\DP_1 
\arrow[nothing-cart,loop, out = 40, in=-40, looseness = 6]{}{\id} 
\arrow[nothing-cart,shift right = .5ex,shorten=1mm,bend right = 9]{l}[swap]{\chi} 
\end{tikzar}
\eeq
satisfying \eqref{eq:hartley}. The approach is similar to some of the proofs of the \emph{Cantor-Bernstein Theorem}, which asserts that there is a bijection between two sets whenever there are injections both ways \cite{HinkisA:cantor-bernstein}. The difference is that the set-theoretic arguments prove that the existence of the injections implies the existence of a bijection, whereas here the task is to construct actual programs for the invertible computations $\chi$ and $\kappa$ from some given programs for  lossless or imperative computations $\ottaoi$ and $\ottaio$. The idea is to recognize the ``invertible components'' of $\ottaoi$ and $\ottaio$ and to assemble $\chi$ and $\kappa$ from these components. The components can be recognized as \emph{orbits}\/ of elements of $\DP_{0}$ and $\DP_{1}$ along $\ottaoi$ and $\ottaio$. 

\para{Orbits.}\sindex{orbits of programs} For all $x\colon \DP_0$ and $y\colon \DP_1$ and $\ottao$ and $\ottai$ from \eqref{eq:overself}, the sequences
\beq\label{eq:orbit}
x \plt \ottao x \plt \ottao^2 x\plt \ottao^3 x \plt \cdots  \qquad \qquad\qquad\qquad y \plt \ottai y \plt \ottai^2 y\plt \ottai^3 y \plt \cdots
\eeq
induce the equivalence relations
\bea
(u\stackrel i \sim v) &  \iff &  \exists n.\ u = \overline\vartheta_{ii}^n v \vee \overline\vartheta_{ii}^n u = v
\eea
$\DP_i$ for $i=0,1$. The $(\stackrel i \sim)$-equivalence classes are called \emph{orbits}. We denote by $[u]^{(i)}$ the orbit of $u\colon \DP_{i}$.

\para{Orbit bottom.} Since $\ottaoi$ and $\ottaio$ are by \eqref{eq:padding-injective} strictly increasing and monotone, every orbit ascends infinitely, but in a well-order it can descend only finitely. While it ascends by reapplying $\ottao$ or $\ottai$, it descends by $\ottao^\ast$ and $\ottai^\ast$ constructed in \eqref{eq:adj-wo}. The defining property $\overline\theta_{ii}^\ast u \pleq v \iff u\pleq \overline\theta_{ii} v$  implies that $\overline\theta_{ii}^\ast$ is nonincreasing if and only if $\overline\theta_{ii}$ is nondecreasing. Since $\overline\theta_{ii}$ are also injective, it is easy to show that  $v = \overline\vartheta_{ii} u$ implies $\overline\vartheta^\ast_{ii} v = u$. When $v$ is not in the image of $\overline\vartheta_{ii}$, then $v\plt \overline\vartheta_{ii}\overline\vartheta^\ast_{ii} v$. Using \eqref{eq:pwedge} to define 
\beq\label{eq:dfxy}
\begin{split}
x_0 = x \qquad\qquad\qquad\qquad x_{n+1} = \left(\overline\vartheta_{00}^\ast x_n\ \pwedge\  x_n\right)\\
y_0 = x \qquad\qquad\qquad\qquad y_{n+1} = \left(\overline\vartheta_{11}^\ast y_n\ \pwedge\  y_n\right)
\end{split}
\eeq
yields the nonincreasing sequences
\beq\label{eq:nonincreasing-xy}
x_0 \pgeq x_1 \pgeq x_2\pgeq x_3 \pgeq\cdots \qquad \qquad\qquad\qquad y_0 \pgeq y_1 \pgeq y_2 \pgeq  y_3 \pgeq \cdots 
\eeq
like \eqref{eq:descending}. According to Sec.~\ref{Sec:WO-splittings}, such sequences stabilize after finitely many steps. Each of the sequences in \eqref{eq:nonincreasing-xy} thus reaches its least element after a finite count, and they are in the form  
\beq\label{eq:descending-xy}
x \pgt x_1 \pgt x_2 \pgt x_3 \pgt \cdots\pgt \wob x^{0}  \qquad \qquad\qquad\qquad y \pgeq y_1 \pgt y_2 \pgt  y_3 \pgt \cdots\pgt \wob y^{1}
\eeq
where $\wob u^{i}$ combines the notation of \eqref{eq:descending-s} and the fact that $\wob u^{i}$ is the infimum of the orbit $[u]^i$. Putting \eqref{eq:orbit} and \eqref{eq:descending-xy} together and unfolding the definitions of $x_i$ and $y_i$ in \eqref{eq:dfxy} we get the general form of the orbits: 
\beq\label{eq:orbit-origin}
\begin{split}
[x]^{0}\ \ =\ \ \wob x^{0}  \plt \cdots \plt \tttao^{2} x \plt \tttao x \plt x \plt \ottao x \plt \ottao^2 x\plt \cdots  \\  
[y]^{1}\ \ \ =\ \ \ \wob y^{1} \plt \cdots \plt \tttai^{2} y \plt \tttai y \plt y \plt \ottai y \plt \ottai^2 y \plt \cdots
\end{split}
\eeq
where $\tttao\colon \DP_{0}\to \DP_{0}$ and $\tttai\colon \DP_{1}\to \DP_{1}$ are the partial inverses of $\ottao$ and $\ottai$ defined by 
\beq\label{eq:inverse}
\underline \vartheta_{ij}(p_{i})\ \ =\ \ \iif\left(p_{i} \iseq \overline \vartheta_{ij}\left( \overline \vartheta_{ij}^\ast(p_{i})\right)\ ,\, \overline \vartheta_{ij}^{\ast}(p_{i})\ ,\, \divg\right)\ \ =\ \ \begin{cases}
\overline \vartheta_{ij}^{\ast}(p_{i}) & \mbox{ if } p_{i} = \vartheta_{ij}\left( \overline\vartheta_{ij}^\ast(p_{i})\right)\\
\divg & \mbox{ otherwise}
\end{cases}
\eeq
The indices are $i,j\in \{0,1\}$ and $\overline \vartheta_{ij}^{\ast}$ is from \eqref{eq:f-ast}.  

\para{Alternating orbits.} \sindex{orbits of programs!alternating orbits} The simulations $\ottaoi$ and $\ottaio$ establish a bijective correspondence between the $\DP_{0}$-orbits and the $\DP_{1}$-orbits:
\beq\label{eq:orbit-involution}
\begin{split}
\tikzset{nothing/.tip={},cart/.tip={Glyph[glyph math command = bullet]}}
\begin{tikzar}[column sep = 0.2ex,row sep = 8ex]
\DP_{0} \ar[nothing-cart,bend right=25]{d}[description]{\ottaoi} \&[2em] x \ar[phantom]{rr}[description]{\plt} \ar[mapsto]{dr} \&\& \ottaio\ottaoi x \ar[phantom]{rr}[description]{\plt}\ar[mapsto]{dr} \&\& \ottaio\ottaoi\ottaio\ottaoi x \ar[mapsto]{dr} \ar[phantom]{rr}[description]{\plt} \&\& \cdots
\\
\DP_{1} \ar[nothing-cart,bend right=25]{u}[description]{\ottaio}
\&\& \ottaoi x \ar[mapsto]{ur} \ar[phantom]{rr}[description]{\plt} \&\& \ottaoi \ottaio\ottaoi x \ar[mapsto]{ur} \ar[phantom]{rr}[description]{\plt} \&\& \ottaoi \ottaio\ottaoi\ottaio\ottaoi x 
\& \cdots
\\[-2ex]
\DP_{0}  \ar[nothing-cart,bend right=25]{d}[description]{\ottaoi}
\&\& \ottaio y \ar[phantom]{rr}[description]{\plt} \ar[mapsto]{dr} \&\& 
\ottaio\ottaoi \ottaio y \ar[phantom]{rr}[description]{\plt} \ar[mapsto]{dr} \&\& \ottaio\ottaoi \ottaio\ottaoi \ottaio y \& \cdots
\\
\DP_{1} \ar[nothing-cart,bend right=25]{u}[description]{\ottaio}
\& y\ar[mapsto]{ur} \ar[phantom]{rr}[description]{\plt} \&\& 
\ottaoi\ottaio y \ar[mapsto]{ur} \ar[phantom]{rr}[description]{\plt} \&\&
\ottaoi\ottaio\ottaoi\ottaio y\ar[mapsto]{ur} \ar[phantom]{rr}[description]{\plt} \&\&
\cdots 
\end{tikzar}
\end{split}
\eeq
Formally, the displayed orbits are 
\begin{align*} 
 [x]^{0}\  = &\    
 \ottaio[\ottaoi x]^{1} 
& [\ottaoi y]^{0} \ = &\  \ottaoi [\ottaio y]^{1} 
\\
\protect
[\ottaoi x]^{1}\ =&\  \ottaoi [x]^{0} & [y]^{1} \  = & \   \ottaio[\ottaoi y]^{0} 
\end{align*}
Their disjoint unions
\beq
[x] \ = \ [x]^{0}\  +\  [\ottaoi x]^{1} \qquad\qquad\qquad\qquad
\protect [y] \ = \ [\ottaio y]^{0} \ +\  [y]^{1}
\eeq
ordered as in \eqref{eq:orbit-involution} are the  \emph{alternating}\/ $(\DP_0+\DP_1)$-orbits of $x$ and $y$.  Note that an alternating orbit is completely determined by any of its elements.

\para{Alternating orbit bottom.} Going backwards along $\ottaoi$ and $\ottaio$, i.e. descending down $\tttaoi$ and $\tttaio$ as far as possible, the alternating orbits look like this:
\beq\label{eq:alt-orbit-origin}
\begin{split}
\tikzset{nothing/.tip={},cart/.tip={Glyph[glyph math command = bullet]}}
\begin{tikzar}[column sep = 0.1ex,row sep = 8ex]
\DP_{0} \ar[nothing-cart,bend right=25]{d}[description]{\ottaoi} \&[2em] \wob x \ar[mapsto,thick]{dr} \ar[phantom]{rr}[description]{\plt}  \&\&
\ottaio\ottaoi \wob x \ar[phantom]{rr}[description]{\plt}  \ar[mapsto,thick]{dr}\&\& \cdots \ar[mapsto,thick]{dr} \ar[phantom]{rr}[description]{\plt}
\&\&
\tttaoi\tttaio x \ar[phantom]{rr}[description]{\plt} \ar[mapsto,thick]{dr}\&\& x 
\\
\DP_{1} \ar[nothing-cart,bend right=25]{u}[description]{\ottaio}
\&\& \ottaoi \wob x \ar[mapsto,thin]{ur}  \ar[phantom]{rr}[description]{\plt} \&\& \cdots  \ar[phantom]{rr}[description]{\plt} \&\& \tttaio\tttaoi\tttaio x \ar[phantom]{rr}[description]{\plt} \ar[mapsto,thin]{ur} \&\&
\tttaio x \ar[mapsto,thin]{ur}
\\[-2ex]
\DP_{0} \ar[nothing-cart,bend right=25]{d}[description]{\ottaoi}\& \& \ottaio \wob y \ar[phantom]{rr}[description]{\plt} \ar[mapsto,thin]{dr}
\&\& \cdots  \ar[phantom]{rr}[description]{\plt} \&\& \tttaoi\tttaio\tttaoi y \ar[phantom]{rr}[description]{\plt} \ar[mapsto,thin]{dr}  \&\&
\tttaoi y \ar[mapsto,thin]{dr}
\\
\DP_{1} \ar[nothing-cart,bend right=25]{u}[description]{\ottaio}
\& \wob y \ar[phantom]{rr}[description]{\plt} \ar[mapsto,thick]{ur}   \&\&
\ottaoi\ottaio \wob y \ar[phantom]{rr}[description]{\plt} \ar[mapsto,thick]{ur} \&\& \cdots \ar[phantom]{rr}[description]{\plt} \ar[mapsto,thick]{ur}
\&\&
\tttaio\tttaoi y \ar[phantom]{rr}[description]{\plt} \ar[mapsto,thick]{ur}\&\& y 
\end{tikzar}
\end{split}
\eeq
How do we know that they look like this? First of all, an alternating orbit must have a bottom, since a chain descending infinitely through the disjoint union $\DP_0+\DP_1$ would contain an infinite descending chain in $\DP_0$ or $\DP_1$, which is impossible because they are well-ordered. So an alternating orbit must have a bottom $u\colon \DP_{0}$ or $v\colon \DP_{1}$. The \textbf{claim} is that $u$ is the bottom of the alternating orbit if and only if it is the bottom of the $\DP_{0}$-part and not in the image of $\ottaio$; and similarly for $v$. 

To prove the claim, consider two arbitrary elements $x\colon \DP_0$ and $y\colon \DP_1$ of the same orbit $[x] = [x]^0+[y]^1 = [y]$.  The \textbf{first subclaim} is that \begin{itemize}
\item \textbf{\emph{the bottom of $[x]=[y]$ is either $u=\wob x^0$ or $v=\wob y^1$}}. 
\end{itemize}
Otherwise, if $u\colon \DP_0$ is the bottom of $[x]=[y]$ but $u\neq \wob x^0$, then $\tttaoi\tttaio u \plt u$ exists in $\DP_{0}$ and $\tttaoi\tttaio u \stackrel{\ottaoi}\mapsto \tttaio u \stackrel{\ottaio}\mapsto u$ extends the orbit $[x]=[y]$ below $u$, which contradicts the assumption that $u$ is its bottom. Else if $v\colon \DP_1$ is the bottom of $[x]=[y]$ but $v\neq \wob y^1$, the same reasoning extends the orbit below $v$ and leads to contradiction. The \textbf{second subclaim} is that
\begin{itemize}
\item \textbf{\emph{if the bottom of $[x]=[y]$ is $\wob x^0$, then $\wob y^1=\ottaoi \wob x^0$}}, and
\item \textbf{\emph{if the bottom of $[x]=[y]$ is $\wob y^1$, then $\wob x^0=\ottaio \wob y^1$}}.
\end{itemize}
Otherwise, if $\wob x^0$ is the bottom and $v=\ottaoi \wob x^0$ but $v\neq \wob y^1$, then $\tttaio\tttaoi v \plt v$ exists in $\DP_{1}$ and $\tttaio\tttaoi v \stackrel{\ottaio}\mapsto \tttaoi v =\wob x^{0}$ extends the orbit below $\wob x^{0}$, which is absurd.  If $\wob y^{1}$ is the bottom but $\ottaio \wob y^1\neq \wob x^0$, the analogous extension, \emph{mutatis mutandis}, also leads to contradiction. 

\para{Assembling isomorphism from alternating orbits.} Since every program from $\DP_0$ or $\DP_1$ has a unique alternating orbit and the orbits are disjoint, they partition $\DP_0+\DP_1$. The isomorphism will be defined by assembling the bijections within each part of this partition. Within each  alternating orbit $[x] = [x]^0+[y]^1 = [y]$, there is a bijection between $[x]^0$ and $[y]^1$ realized by $\ottaio$ or $\ottaoi$. Since the bottom of the alternating orbit does not lie in the range of either $\ottaoi$ or $\ottaio$ (or else it could be extended further down), the bijection must be realized 
\begin{itemize}
\item by $\ottaoi$ if the bottom is $\wob x^0$ and 
\item by $\ottaio$ if the bottom is $\wob y^1$. 
\end{itemize}
\textbf{\emph{The idea of the assembly is to establish the isomorphism along the thick arrows of \eqref{eq:alt-orbit-origin}.}} Remembering that the second claim above implies that $\wob x^0$ is \emph{not}\/ the bottom of $[x]=[y]$ if and only if $\wob x^0= \ottaio \wob y^1$, for any $y$ from the corresponding $\DP_1$-orbit, and similarly for $\wob y^1$,  we define the testing predicates $\Theta_{0}: \DP_0 \stricto \Bool$ and $\Theta_{1}: \DP_1 \stricto \Bool$ by
\beq
\Theta_{0}(x)\ = \ \Bigg( \wob x^0 \iseq \ottaio \wob{\ottaoi x}^1\Bigg) \qquad \qquad \qquad \Theta_{1}(y)\ = \ \Bigg( \wob y^1 \iseq  \ottaoi\wob{\ottaio y}^0\Bigg)
\eeq
The exclusive disjunction of $\wob x^0=\ottaio \wob y^1$ and $\wob y^1=\ottaoi \wob x^0$ from the second subclaim above now gives
\[
\Theta_{0}(x) \iff \neg \Theta_{1}(\ottaoi  x)  \qquad \qquad\qquad\qquad  \Theta_{1}(y) \iff \neg\Theta_{0}(\ottaio y)
\]
These equivalences assure that the \funnn s $\kappa:\DP_0\stricto \DP_1$ and $\chi:\DP_1\stricto \DP_0$ specified by 
\beq\label{eq:isomorphism-def}
\kappa(x) \ \ = \ \ \left.
\begin{cases} \tttaio x & \mbox{ if } \Theta_{0}(x)\\
\ottaoi x & \mbox{ otherwise} 
\end{cases}\right\}\qquad \qquad \qquad \qquad
\chi(y) \ \  = \ \  \left.
\begin{cases} \tttaoi y & \mbox{ if } \Theta_{1}(y)\\
\ottaio y & \mbox{ otherwise} 
\end{cases}\right\} 
\eeq
are each other's inverse. Since $\ottaio$ and $\ottaoi$ are simulations and $\Theta_0$ and $\Theta_1$ are decidable, $\kappa$ and $\chi$ also preserve the meanings of programs. \eqref{eq:hartley} is satisfied and the theorem is proved.

\section{Upshot: the closures}
\label{Sec:closure}

The upshot of this chapter is that a monoidal category with data services may be program-closed in at most one way. If universe is computable, in the sense that all of its \funnn s are programmable, then its programming language is unique up to isomorphism: any program can be computably translated from any language to any other language, where the evaluations are preserved and the executions simulated. \textbf{Computability is an \emph{intrinsic property}\/ of the universe, not an external structure}.


If you have been writing lots of very different programs in very different languages, the claim that program can be effectively and faithfully translated between languages may be hard to digest.  If you have been studying different models of computation and remember some of the convoluted translations in support of the Church-Turing thesis, believing that each of them can be replaced by an isomorphism may be even harder. \textbf{What is going on?} Let us take a closer look.

\para{Structures vs properties.} \sindex{structure vs property}
\sindex{property vs structure}
Let us first clarify what it means that computability is a property, not structure.

There are  many ways to well-order and count the elements of a set. But any such well-ordering can be transformed into any other by a permutation of the underlying set, and counting its elements always outputs the same number. The size of a set is its \emph{intrinsic property}.

There are even more ways to \emph{partially}\/ order a set and the outcomes are generally \emph{not}\/ isomorphic. That is why a partial ordering is an \emph{external structure}\/ that needs to be explicitly specified, since the underlying set does not determine it. On the other hand, a given partial order uniquely determines the infima and the suprema, if they exist. Completeness under infima and suprema is therefore an \emph{intrinsic property}\/ of a partial order. The same holds for categories, where infima and suprema become limits and colimits. If they exist, they are determined by the underlying category, up to isomorphism, and completeness under limits and colimits is therefore an \emph{intrinsic property}\/ of the category again. Cartesian products, as a special case of limits, are also determined by the underlying category and do not need to be specified separately. As explained in Sections~\ref{Sec:service} and \ref{Sec:map}, a category is cartesian if any pair of arrows into a pair of types is representable as a single arrow into a single type, which is then their cartesian product.
\begin{figure}[!ht]
\begin{center}
\begin{tikzpicture}[shorten >=1pt,node distance=3cm]
  \node[state,minimum size=35pt,draw=black,very thick]   
  (nocarry)                {\begin{minipage}[c]{1.7cm}\begin{center} $\tot\CCC(X, A)$\\[-.75ex]
  $\times$\\[-.75ex]
  $\tot\CCC(X,B)$ \end{center}\end{minipage}};
   \node[state,minimum size=0pt,draw=black,very thick] (carry)    [right=of nocarry]              
   {\begin{minipage}[c]{1.85cm}\begin{center} $\tot\CCC(X,A\times B)$ \end{center}\end{minipage}};
  \path[->] 
  (nocarry) 
  edge [bend right,thick] node [below] {$\scriptstyle a, b\ {\displaystyle \mapsto}\  \left<a,b\right>$} 
  (carry)
  (carry) 
  edge [bend right,thick] node [above] {{$\scriptstyle \pi_A h, \pi_B h\  \mapsfrom\  h$}}   
  (nocarry);
   \end{tikzpicture}
   \caption{Pairs of \strict\ \funnn s are representable by \funnn s into cartesian products} 
\label{Fig:cart-balls}
\end{center}
\end{figure}
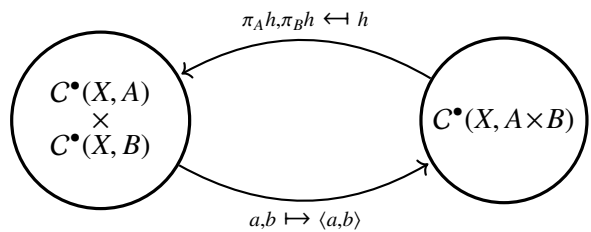
This representability, displayed in Fig.~\ref{Fig:cart-balls}, makes the category cartesian. If there are two representing types, they must be isomorphic. Being cartesian is an \emph{intrinsic property}\/ of a category.  

Monoidal products, on the other hand, are an  \emph{external structure}\/ that needs to be explicitly specified, since a category may be monoidal in many different ways\ldots

\subsection{Cartesian closure}  \label{Sec:CCC-8}
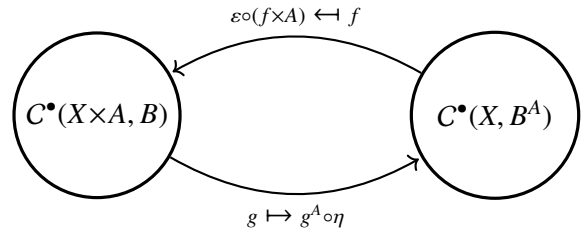
\begin{figure}[!ht]
\begin{center}
\begin{tikzpicture}[shorten >=1pt,node distance=3cm]
  \node[state,minimum size=35pt,draw=black,very thick]   
  (nocarry)                {\begin{minipage}[c]{1.85cm}\begin{center} $\tot\CCC(X\times A, B)$ \end{center}\end{minipage}};
   \node[state,minimum size=0pt,draw=black,very thick] (carry)    [right=of nocarry]              
   {\begin{minipage}[c]{1.85cm}\begin{center} $\tot\CCC(X,B^A)$ \end{center}\end{minipage}};
  \path[->] 
  (nocarry) 
  edge [bend right,thick] node [below] {$\scriptstyle g\ {\displaystyle \mapsto}\  g^A\circ \eta$} 
    (carry)
  (carry) 
  edge [bend right,thick] node [above] {{$\scriptstyle \varepsilon\circ(f\times A) \  \mapsfrom\  f$}}   
  (nocarry);
   \end{tikzpicture}
   \caption{Cartesian closure} 
\label{Fig:CCC}
\end{center}
\end{figure}

A category is $\CCC$ cartesian if and only if it has data services \sindex{category!cartesian} and all of its morphisms are  {\strict} with respect to these services, in the sense of Sec.~\ref{Sec:map}, so that $\tot\CCC = \CCC$. The category is moreover \sindex{category!cartesian-closed} \emph{cartesian-closed}\/ if the arguments of any multi-argument \funnn\ can be separated and each of them can be \sindex{evaluation!partial} evaluated on its own. In monoidal computers and program-closed categories, this is done by partial evaluators, as explained in Sec.~\ref{Sec:pev}. In cartesian-closed categories, this is done by the abstraction operation, as mentioned in Sec.~\ref{Sec:cat-logic} and displayed in Fig.~\ref{Fig:CCC}. This is a categorical view of Church's $\lambda$-abstraction discussed in Sec.~\ref{Sec:Church-reverse}.
A multi-argument \funnn\ $g\colon X\times A\stricto B$ on the left is abstracted to $\lambda a.g = \left(X\strictto{\eta}(X\times A)^{A}\strictto{g^{A}} B^{A}\right)$ on the right, and can be recovered from it by the application operation going left. Like in Fig.~\ref{Fig:cart-balls}, the two operations in two directions form a family of $A,B$-indexed $X$-natural bijective correspondences. This representability of multi-argument \funnn s into $B$ by single-argument \funnn s into \emph{exponents}\/ like $B^A$ is what makes the cartesian category $\tot \CCC$ closed. See Sec.~\ref{Sec:CCC-1} for the logical interpretation of the exponents as generalized implications. Since the exponents are unique up to isomorphism (as checked in Ex.~\ref{Sec:work-wire}(\ref{ex:exponent})), being cartesian-closed is an \emph{intrinsic property}\/ again. Now we get to the point.

\subsection{Program closure}  
\begin{figure}[!ht]
\begin{center}
\begin{tikzpicture}[shorten >=1pt,node distance=3cm]
  \node[state,minimum size=35pt,draw=black,very thick]   
  (nocarry)                {\begin{minipage}[c]{1.85cm}\begin{center} $\CCC(X\times A,  B)$ \end{center}\end{minipage}};
   \node[state,minimum size=0pt,draw=black,very thick] (carry)    [right=of nocarry]              
   {\begin{minipage}[c]{1.85cm}\begin{center} $\tot \CCC(X, \DP)$ \end{center}\end{minipage}};
    \path[->>] 
  (carry) 
  edge [bend right,thick] node [above] {\large$\runn_X$}
  (nocarry);
   \end{tikzpicture}
   \caption{Program closure} 
\label{Fig:PCC}
\end{center}
\end{figure}
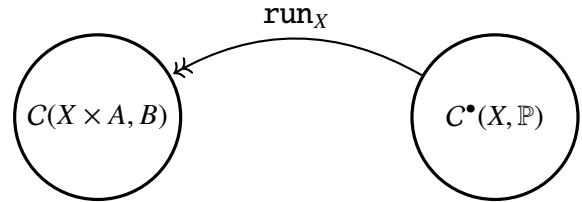

The monoidal computer structure, as presented in   Sec.~\ref{Sec:surj}, amounts an $A,B$-indexed family of $X$-natural surjections, as displayed in Fig.~\ref{Fig:PCC}.    
While the two sides of Fig.~\ref{Fig:CCC} determine each other along the bijection between them, here we only have a surjection one way.  Saying that this structure is unique means that the \funnn s on the left, if they are computable, determine their inverse images along the \runn-instruction.  --- How do they do that?

In computational terms, the answer lies in the \emph{polymorphism of the \runn-instruction}. In categorical terms, the same answer goes under the name of \emph{naturality}. 

\begin{figure}[!ht]
\begin{center}
\begin{tikzpicture}[shorten >=1pt,node distance=3cm]
  \node[state,minimum size=35pt,draw=black,very thick]   
  (nocarry)                {\begin{minipage}[c]{1.85cm}\begin{center} $\CCC(X\times A,  B)$ \end{center}\end{minipage}};
   \node[state,minimum size=0pt,draw=black,very thick] (carryup)    [above right=0.75cm and 4cm of nocarry]              
   {\begin{minipage}[c]{1.85cm}\begin{center} $\tot \CCC(X, \DP_{0})$ \end{center}\end{minipage}};
   \node[state,minimum size=0pt,draw=black,very thick] (carrydown)    [below right=0.75cm and 4cm of nocarry]              
   {\begin{minipage}[c]{1.85cm}\begin{center} $\tot \CCC(X, \DP_{1})$ \end{center}\end{minipage}};
    \path 
  (carryup) 
  edge [thick,->>] node [above] {\large$\runn^0_X$}
  (nocarry)
  (carrydown) 
  edge [thick,->>] node [below] {\large$\runn^1_X$}
  (nocarry)
  (carryup) 
  edge [bend left,thick,>->] node [right] {\large$\ottaoi\circ(-)$}
  (carrydown) 
  (carrydown) 
  edge [bend left,thick,>->] node [left] {\large$\ottaio\circ(-)$}
  (carryup) 
  ;
   \end{tikzpicture}
   \caption{Program closures embed into each other} 
\label{Fig:PCC-1}
\end{center}
\end{figure}
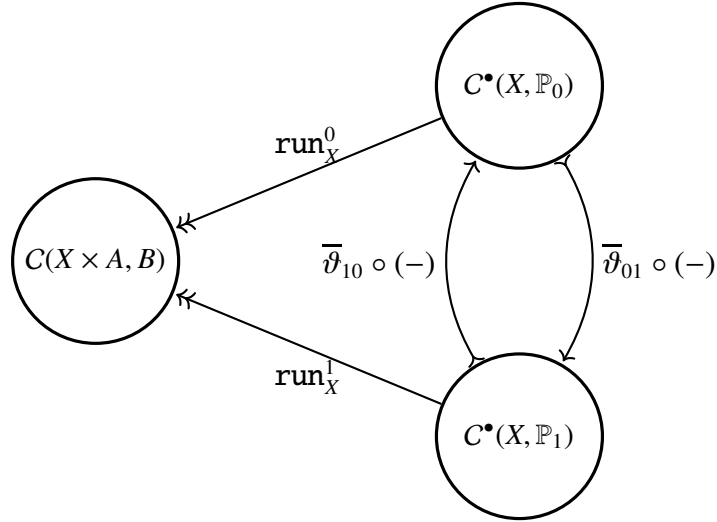

Suppose we have two program closure structures $\runn_0$ and $\runn_1$, like in Fig.~\ref{Fig:PCC-1}. 
Instantiate the type $X$ to $\DP_0$ first. Going forward along 
\bear
\runn^0_{\DP_0}\colon \tot\CCC\left(\DP_0, \DP_0\right)& \epi &   \CCC(\DP_0\times A,B)\\
\id &\mapsto & \universal_0
\eear
and back along the surjection
\bear
\runn^1_{\DP_0}\colon \tot\CCC\left(\DP_0, \DP_1\right)& \epi &   \CCC(\DP_0\times A,B)\\
\theta_{01} &\mapsto & \universal_0
\eear
we get $\theta_{01}\colon \DP_0\stricto \DP_1$ which by
naturality gives
\bear
\runn^0_X(F_0) & = & \runn^1\left(\theta_{01}\circ F_0\right)
\eear
for all $F_0\in \tot\CCC(X,\DP_0)$. Switching 0 and 1, instantiating $X$ to $\DP_1$, and repeating the same gives $\theta_{10}\colon \DP_1\stricto \DP_0$ with
\bear
\runn^1_X(F_1) & = & \runn^0\left(\theta_{10}\circ F_1\right)
\eear 
for all $F_1\in \tot\CCC(X,\DP_1)$. Naturality and surjectiveness of the monoidal computer structure (i.e. the polymorphism of the \runn-instructions and the programmability of all \funnn s) guarantee that there are meaning-preserving program transformations $\theta_{01}$ and $\theta_{10}$ between any two programming languages. The situation is  \emph{almost}\/ like Fig.~\ref{Fig:PCC-1} --- but not quite. The program transformations $\theta_{01}$ and $\theta_{10}$ may not be injective or related with each other in any way. To get injective program transformations like $\ottaoi$ and $\ottaio$ in Fig.~\ref{Fig:PCC-1} and \eqref{eq:hartkey-mono}, we had to switch from \funnn s to \procs es spent Ch.~\ref{Chap:State} on developing program executions as \procs\ simulations. Hence the \emph{injections}\/ $\ottaoi$ and $\ottaio$ in Fig.~\ref{Fig:PCC-1}. 

Using the two semantics-preserving injections between the sets of programs in two languages, we could now extract a bijection by applying the usual set-theoretic Cantor-Bernstein Theorem. Since the injections commute with the \runn-instructions, they are fiberwise: every $\runn^0$-inverse image of a \funnn\ is mapped to the $\runn^1$-inverse image of the same \funnn, and vice versa. The bijections will thus also be fiberwise, and putting them together will give a bijection which commutes with the \runn-instructions. This provides a simple and succinct explanation of the bijective simantics-preserving translation between the programs in any two languages, based on basic set theory. The Cantor-Bernstein Theorem implies that all programming languages are isomorphic.

One issue is that the set-theoretic approach only proves that a bijection exists but does not tell what it is. Since the bijection is natural, this approach even proves existence of programs for the bijection, yet the infinitary aspects of the construction preclude finding them.

Another issue is that the set-theoretic Cantor-Bernstein constructions assume (and imply \cite{PavlovicD:Cantor-categories}) the boolean complements. In the case at hand, the assumption is that the complements of certain extensional predicates are decidable. This directly contradicts Rice's Theorem from Sec.~\ref{Sec:rice}. This is where the well-ordering of programming languages became essential.The effective construction of a bijection from the effective constructions of injections in Fig.~\ref{Fig:PCC-1} was based on it. There are many programs for every computable \funnn and every computational \procs, and they can be programmed in many languages, but they all compute the same things in the same way and can be translated to each other faithfully and invertibly. There is no Tower of Babel but all languages are manifestations of the same capability of speech. There are many programming languages but all are manifestations of the same property of computability.

%
%

\backmatter



\nocite{vanHeijenoortJ,DavisM:undecidable}
\addcontentsline{toc}{chapter}{Bibliography}
\bibliographystyle{plain}
\bibliography{algorithmics,PavlovicD,TEXT-ref, AbramskyS,complexity,CT,logic,math,philosophy,semantics,CS,type,language,levin,ScottDS,induction,IT,HT}

\begin{thebibliography}{100}

\bibitem{AbramskyS:definability}
Samson Abramsky.
\newblock Intensionality, definability and computation.
\newblock In Alexandru Baltag and Sonja Smets, editors, {\em Johan van Benthem
  on Logic and Information Dynamics}, pages 121--142. Springer, 2014.

\bibitem{Abramsky-Jung:domains}
Samson Abramsky and Achim Jung.
\newblock Domain theory.
\newblock In Samson et~al Abramsky, editor, {\em Handbook of Logic in Computer
  Science}, volume~3, pages 1--168. Oxford University Press, 1994.

\bibitem{AdamekJ:coalg}
Jiri Adamek.
\newblock {Introduction to coalgebra}.
\newblock {\em Theory and Applications of Categories}, 14:157--199, 2005.

\bibitem{AdamekJ:CT}
Jiri Adamek, Horst Herrlich, and George~E. Strecker.
\newblock {\em Abstract and Concrete Categories: The Joy of Cats}.
\newblock Wiley, 1990.
\newblock available online.

\bibitem{dragon-book}
Alfred~V. Aho, Ravi Sethi, and Jeffrey~D. Ullman.
\newblock {\em Compilers: Principles, Techniques, and Tools}.
\newblock Series in Computer Science and Information Processing.
  Addison-Wesley, 1986.
\newblock 2nd edition in 2006, coauthored with Monica Lam.

\bibitem{Alberti-Uhlmann:book}
Peter~M. Alberti and Armin Uhlmann.
\newblock {\em Stochasticity and partial order: double stochastic maps and
  unitary mixing}.
\newblock Mathematics and its applications. Deutscher Verlag der
  Wissenschaften, 1981.

\bibitem{ArndtC:info-meas}
Christoph Arndt.
\newblock {\em Information Measures: Information and its Description in Science
  and Engineering}.
\newblock Signals and Communication Technology. Springer Berlin Heidelberg,
  2012.

\bibitem{Arora-Barak:book}
Sanjeev Arora and Boaz Barak.
\newblock {\em Computational Complexity: A Modern Approach}.
\newblock Cambridge University Press, 2009.

\bibitem{GrothendieckA:SGA4}
Michael Artin, Alexander Grothendieck, and Jean-Louis Verdier, editors.
\newblock {\em S{\'e}minaire de G{\'e}ometrie Alg{\'e}brique: Th\'eorie des
  Topos et Cohomologie \'Etale des Schemas (SGA 4)}, volume 269,270,305 of {\em
  Lecture Notes in Mathematics}. Springer-Verlag, 1964.
\newblock Second edition, 1972.

\bibitem{AwodeyS:CT}
Steve Awodey.
\newblock {\em Category Theory}.
\newblock Oxford Logic Guides. OUP, 2010.

\bibitem{BaezJ:prehistory}
John Baez and Aaron Lauda.
\newblock A prehistory of $n$-categorical physics.
\newblock In {\em Deep Beauty: Understanding the Quantum World Through
  Mathematical Innovation}, pages 13--128. Cambridge University Press, 2011.

\bibitem{BarendregtH:types}
Henk Barendregt, Wil Dekkers, and Richard Statman.
\newblock {\em Lambda Calculus with Types}.
\newblock Perspectives in Logic. Cambridge University Press, 2013.

\bibitem{BennettCH:thermodynamics}
Charles~H Bennett.
\newblock The thermodynamics of computation --- a review.
\newblock {\em International Journal of Theoretical Physics}, 21(12):905--940,
  1982.

\bibitem{BennettC:depth}
Charles~H. Bennett.
\newblock Logical depth and physical complexity.
\newblock In {\em A half-century survey on The Universal Turing Machine}, pages
  227--257, New York, NY, USA, 1988. Oxford University Press, Inc.

\bibitem{BethkeI:thesis}
Ingemarie Bethke.
\newblock {\em Notes on Partial Combinatory Algebras}.
\newblock PhD thesis, University of Amsterdam, 1988.

\bibitem{BirkedalL:thesis}
Lars Birkedal.
\newblock {\em Developing Theories of Types and Computability via
  Realizability}, volume~34 of {\em Electronic Notes in Theoretical Computer
  Science}.
\newblock Elsevier, 2000.

\bibitem{BlumM:axioms}
Manuel Blum.
\newblock A machine-independent theory of the complexity of recursive
  functions.
\newblock {\em J. ACM}, 14(2):322--336, April 1967.

\bibitem{BorceuxF:handbook}
Francis Borceux.
\newblock {\em Handbook of Categorical Algebra}.
\newblock Number~50 in Encyclopedia of Mathematics and its Applications.
  Cambridge University Press, 1994.
\newblock Three volumes.

\bibitem{PavlovicD:POPL2020}
Roberto Bruni, Roberto Giacobazzi, Roberta Gori, Isabel Garcia{-}Contreras, and
  Dusko Pavlovic.
\newblock Abstract extensionality: on the properties of incomplete abstract
  interpretations.
\newblock {\em Proc. {ACM} Program. Lang.}, 4({POPL}):28:1--28:28, 2020.

\bibitem{CaludeC:CX}
Cristian Calude.
\newblock {\em Theories of Computational Complexity}, volume~35 of {\em Annals
  of Discrete Mathematics}.
\newblock North-Holland, 1988.

\bibitem{CantorG:1891uber}
Georg Cantor.
\newblock {\"{U}ber eine elementare Frage der Mannigfaltigkeitslehre}.
\newblock {\em Jahresbericht der Deutschen Mathematiker-Vereinigung}, 1:75--78,
  1891.

\bibitem{CantorG:beitraege}
Georg Cantor.
\newblock Beitr{\"a}ge zur begr{\"u}ndung der transfiniten mengenlehre.
\newblock {\em Mathematische Annalen}, 46(4):481--512, 1895.

\bibitem{CantorG:collected}
Georg Cantor.
\newblock {\em Gesammelte {A}bhandlungen mathematischen und philosophischen
  {I}nhalts}.
\newblock Springer-Verlag, 1932.
\newblock Edited by Ernst Zermelo; reprinted by Olms, Hildeshaim, 1962.

\bibitem{Carboni-Walters}
Aurelio Carboni and Robert~F.C. Walters.
\newblock Cartesian bicategories, {I}.
\newblock {\em J. of Pure and Applied Algebra}, 49:11--32, 1987.

\bibitem{PavlovicD:CathyFest}
Jason Castiglione, Dusko Pavlovic, and Peter-Michael Seidel.
\newblock {Privacy protocols}.
\newblock In Joshua~Guttman et~al., editor, {\em CathFest: Proceedings of the
  Symposium in Honor of Catherine Meadows}, volume 11565 of {\em Lecture Notes
  in Computer Science}, pages 167--192. Springer, 2019.

\bibitem{ChomskyN:mind}
Noam Chomsky.
\newblock {\em Language and Mind}.
\newblock Cambridge University Press, 2006.

\bibitem{ChurchA:unsolvable}
Alonzo Church.
\newblock An unsolvable problem of elementary number theory.
\newblock {\em American Journal of Mathematics}, 58:345--363, 1936.
\newblock Reprinted in \cite{DavisM:undecidable}.

\bibitem{ChurchA:ord-second}
Alonzo Church.
\newblock The constructive second number class.
\newblock {\em Bulletin of the American Mathematical Society}, 44(4):224--232,
  1938.

\bibitem{ChurchA:types}
Alonzo Church.
\newblock A formulation of the simple theory of types.
\newblock {\em The Journal of Symbolic Logic}, 5(2):56--68, 1940.

\bibitem{ChurchA:calculi}
Alonzo Church.
\newblock {\em The Calculi of Lambda Conversion}, volume~6 of {\em Annals of
  Mathematics Studies}.
\newblock Princeton University Press, 1941.

\bibitem{Church-Kleene:constructions}
Alonzo Church and Stephen~C. Kleene.
\newblock {Constructions of formal definitions of functions of ordinal
  numbers.}
\newblock {\em {Bull. Amer. math. Soc. 42, 639}}, 1936.

\bibitem{Church-Kleene:ord}
Alonzo Church and Stephen~Cole Kleene.
\newblock {Formal definitions in the theory of ordinal numbers}.
\newblock {\em {Fundam. Math.}}, 28:11--21, 1937.

\bibitem{Cockett-Hofstra:turing}
J.Robin~B. Cockett and Pieter~J.W. Hofstra.
\newblock Introduction to turing categories.
\newblock {\em Annals of Pure and Applied Logic}, 156(2):183 -- 209, 2008.

\bibitem{PavlovicD:CQStruct}
Bob Coecke, \'{E}ric Oliver~Paquette, and {Dusko Pavlovic}.
\newblock Classical and quantum structuralism.
\newblock In Simon Gay and Ian Mackie, editors, {\em Semantical Techniques in
  Quantum Computation}, pages 29--69. Cambridge University Press, 2009.

\bibitem{CoeckeB-Kissinger:book}
Bob Coecke and Aleks Kissinger.
\newblock {\em Picturing Quantum Processes: {A} First Course in Quantum Theory
  and Diagrammatic Reasoning}.
\newblock Cambridge University Press, 2017.

\bibitem{ConwayJH:ONAG}
John~H. Conway.
\newblock {\em On Numbers and Games}.
\newblock Number~6 in London Mathematical Society Monographs. Academic Press,
  1976.

\bibitem{Conway-Guy:numbers}
John~H. Conway and Richard Guy.
\newblock {\em The Book of Numbers}.
\newblock Springer, 2012.

\bibitem{Coquand-Huet:constructions}
Thierry Coquand and G{\'e}rard~P. Huet.
\newblock The calculus of constructions.
\newblock {\em Inf. Comput.}, 76:95--120, 1988.

\bibitem{CurryH:functionality}
Haskell~B. Curry.
\newblock Functionality in combinatory logic.
\newblock {\em Proceedings of the National Academy of Sciences}, 20:584--590,
  1934.

\bibitem{CantorG:dauben}
Joseph~W. Dauben.
\newblock {\em Georg Cantor: His Mathematics and Philosophy of the Infinite}.
\newblock Princeton University Press, 2020.

\bibitem{DavisM:undecidable}
Martin Davis, editor.
\newblock {\em {The Undecidable : Basic papers on undecidable propositions,
  unsolvable problems, and computable functions}}.
\newblock Raven Press, Helwett, N.Y., 1965.

\bibitem{DedekindR:zahlen}
Richard Dedekind.
\newblock {\em Essays on the Theory of Numbers}.
\newblock The Religion of Science Library. Open Court Publishing Company, 1901.
\newblock reprinted by Dover in 1963.

\bibitem{Descartes:spirits}
Dennis Des~Chene.
\newblock {\em Spirits and Clocks: Machine and Organism in Descartes}.
\newblock Cornell University Press, 2018.

\bibitem{DijkstraE:guarded}
Edsger~W. Dijkstra.
\newblock Guarded commands, nondeterminacy and formal derivation of programs.
\newblock {\em Communications of the ACM}, 18(8):453--457, 1975.

\bibitem{Doner-Tarski}
John Doner and Alfred Tarski.
\newblock An extended arithmetic of ordinal numbers.
\newblock {\em Fundamenta Mathematicae}, 65(1):95--127, 1969.

\bibitem{Eckmann-Hilton:fund}
Beno Eckmann and Peter Hilton.
\newblock Structure maps in group theory.
\newblock {\em Fundamenta Mathematicae}, 50(2):207--221, 1961.

\bibitem{Eilenberg-MacLane:natural}
Samuel Eilenberg and Saunders MacLane.
\newblock General theory of natural equivalences.
\newblock {\em Transactions of the American Mathematical Society},
  58(2):231--294, 1945.

\bibitem{Eilenberg-Steenrod}
Samuel Eilenberg and Norman Steenrod.
\newblock {\em Foundations of Algebraic Topology}.
\newblock Princeton University Press, 1952.

\bibitem{ErshovY:enum-siberian}
Yu.L. Ershov.
\newblock Enumeration of families of general recursive functions.
\newblock {\em Siberian Mathematical Journal}, 8(5):771--778, 1967.

\bibitem{ErshovY:enum-book}
Yu.L. Ershov.
\newblock {\em Theory of Enumerations}.
\newblock Mathematical Logic and Foundations of Mathematics. Nauka, Moscow,
  1977.
\newblock (in Russian).

\bibitem{FefermanS:pca}
Solomon Feferman.
\newblock A language and axioms for explicit mathematics.
\newblock In {\em Algebra and logic}, pages 87--139. Springer, 1975.

\bibitem{FeynmanR:character}
Richard Feynman.
\newblock {\em The Character of Physical Law}.
\newblock The MIT Press. MIT Press, 1967.

\bibitem{PavlovicD:Cantor-categories}
Peter Freyd and Dusko Pavlovic.
\newblock Conversation on the categories mailing list 12--13 fabruary 1994.
\newblock https://www.mta.ca/~cat-dist/archive/.

\bibitem{Futamura}
Yoshihiko Futamura.
\newblock Partial evaluation of computation process\&mdash;anapproach to a
  compiler-compiler.
\newblock {\em Higher Order Symbol. Comput.}, 12(4):381--391, December 1999.

\bibitem{GandyR:confluence}
Robin Gandy.
\newblock The confluence of ideas in 1936.
\newblock In Rolf Herken, editor, {\em A half-century survey on The Universal
  Turing Machine}, pages 55--111. Springer, 1988.

\bibitem{compendium:2003}
G.~Gierz, K.H. Hofmann, K.~Keimel, J.D. Lawson, M.~Mislove, and D.S. Scott.
\newblock {\em Continuous Lattices and Domains}.
\newblock EBSCO ebook academic collection. Cambridge University Press, 2003.

\bibitem{GirardJY:F}
Jean-Yves Girard.
\newblock The system f of variable types, fifteen years later.
\newblock {\em Theoretical computer science}, 45:159--192, 1986.

\bibitem{GirardJY:prot}
Jean-Yves Girard, Yves Lafont, and Paul Taylor.
\newblock {\em Proofs and Types}.
\newblock Number~7 in Cambridge Tracts in Theoretical Computer Science.
  Cambridge University Press, 1989.

\bibitem{GoedelK:ueber}
Kurt G{\"o}del.
\newblock {\"U}ber formal unentscheidbare {S}{\"a}tze der {P}rincipia
  {M}athematica und verwandter {S}ysteme {I}.
\newblock {\em Monatshefte f{\"u}r Mathematik und Physik}, 38:173--198, 1931.
\newblock English translations, ``On Formally Undecidable Propositions of
  `{P}rincipia {M}athematica' and Related Systems'' published by Oliver and
  Boyd, 1962 and Dover, 1992; also in \cite{vanHeijenoortJ}, pp.~596--616; also
  in \cite{DavisM:undecidable}, pages 5--38.

\bibitem{GoedelK:bicentennial}
Kurt G{\"o}del.
\newblock Remarks before the princeton bicentennial conference of prob- lems in
  mathematics.
\newblock In Martin Davis, editor, {\em The Undecidable : Basic papers on
  undecidable propositions, unsolvable problems, and computable functions},
  pages 84--87. Raven Press, Helwett, N.Y., 1965.

\bibitem{GrassmannH:new}
Hermann Grassmann.
\newblock {\em A New Branch of Mathematics: The \emph{"Ausdehnungslehre"} of
  1844 and Other Works}.
\newblock Open Court, 1995.

\bibitem{GrassmannH:extension}
Hermann Grassmann.
\newblock {\em Extension Theory}, volume~19 of {\em History of Mathematics}.
\newblock American Mathematical Soc., 2000.

\bibitem{GrothendieckA:tohoku}
Alexandre Grothendieck.
\newblock Sur quelques points d'alg{\`e}bre homologique.
\newblock {\em Tohoku Mathematical Journal, Second Series}, 9(2):119--183,
  1957.

\bibitem{Grzegorczyk1953}
Andrzej Grzegorczyk.
\newblock {\em Some classes of recursive functions}, volume~4 of {\em Rosprawy
  Matematyzne}.
\newblock Instytut Matematyczny Polskiej Akademi Nauk, 1953.

\bibitem{GunterC:book}
Carl Gunter.
\newblock {\em Semantics of Programming Languages: Structures and Techniques}.
\newblock Foundations of Computing. MIT Press, 1992.

\bibitem{GurevichY:east-west}
Yuri Gurevich.
\newblock Logic activities in {Europe}.
\newblock {\em ACM SIGACT News}, 25(2):11--24, June 1994.

\bibitem{HayashiS:semifunctors}
Susumu Hayashi.
\newblock Adjunction of semifunctors: Categorical structures in nonextensional
  lambda calculus.
\newblock {\em Theor. Comput. Sci.}, 41:95--104, 1985.

\bibitem{vanHeijenoortJ}
Jean~van Heijenoort, editor.
\newblock {\em From {F}rege to {G}{\"o}del: a Source Book in Mathematical
  Logic, 1879--1931}.
\newblock Harvard University Press, 1967.
\newblock Reprinted 1971, 1976.

\bibitem{HeytingA:algebra}
Arend Heyting.
\newblock Die formalen {Regeln} der intuitionistischen {Logik}. {I}, {II},
  {III}.
\newblock {\em Sitzungsber. Preu{{\ss}}. Akad. Wiss., Phys.-Math. Kl.},
  1930:42--56, 57--71, 158--169, 1930.

\bibitem{HilbertD:unendliche}
David Hilbert.
\newblock {\"{U}ber das Unendliche}.
\newblock {\em Mathematische Annalen}, 95:161--190, 1926.
\newblock English translation in \cite{vanHeijenoortJ}, pp.~367--392.

\bibitem{HinkisA:cantor-bernstein}
Arie Hinkis.
\newblock {\em {Proofs of the Cantor-Bernstein Theorem: A Mathematical
  Excursion}}, volume~45 of {\em Science Networks. Historical Studies}.
\newblock Springer, 2013.

\bibitem{Hinze-Marsden}
Ralph Hinze and Dan Marsden.
\newblock {\it The Art of Category Theory Part {I}: Introducing String
  Diagrams}, 2019.
\newblock Completed book, under review.

\bibitem{Hoare:hints}
C.A.R. Hoare.
\newblock Hints on the design of a programming language for real-time command
  and control.
\newblock In J.P. Spencer, editor, {\em Real-time Software: International State
  of the Art Report}, pages 685--99. Infotech International, 1976.

\bibitem{HofstadterD:GEB}
Douglas~R. Hofstadter.
\newblock {\em G{\"o}del, Escher, Bach}.
\newblock Basic Books. Basic Books, 1979.

\bibitem{book:Hopcroft-Ullman-Motwani}
John~E. Hopcroft, Rajeev Motwani, and Jeffrey~D. Ullman.
\newblock {\em Introduction to Automata Theory, Languages, and Computation}.
\newblock Addison-Wesley, Boston, MA, USA, 3rd edition edition, 2006.

\bibitem{HotzG:string}
G{\"u}nter Hotz.
\newblock {Eine Algebraisierung des Syntheseproblems von Schaltkreisen I--II}.
\newblock {\em Elektronische Informationsverarbeitung und Kybernetik},
  1--2(3):185--205, 209--231, 1965.

\bibitem{HowardW:curry}
William~A. Howard.
\newblock The fomul\ae-as-types notion of construction.
\newblock In J.P. Seldin and J.R. Hindley, editors, {\em To H.B.~Curry: Essays
  on combinatory logic, lambda calculus and formalism}, volume~2, pages
  479--490. Oxford University Press, 1980.

\bibitem{HudakP:history}
Paul Hudak.
\newblock Conception, evolution, and application of functional programming
  languages.
\newblock {\em ACM Comput. Surv.}, 21(3):359--411, sep 1989.

\bibitem{HylandJME:efft}
Martin Hyland.
\newblock The effective topos.
\newblock In Anne~Sjerp Troelstra and Dirk van Dalen, editors, {\em
  L.~E.~J.~Brouwer Centenary Symposium}, number 110 in Studies in Logic and the
  Foundations of Mathematics, pages 165--216. North-Holland, 1982.

\bibitem{JacobsB:book-coalg}
Bart Jacobs.
\newblock {\em Introduction to Coalgebra: Towards Mathematics of States and
  Observation}, volume~59 of {\em Cambridge Tracts in Theoretical Computer
  Science}.
\newblock Cambridge University Press, 2016.

\bibitem{JohnstoneP:elephant}
Peter~T. Johnstone.
\newblock {\em Sketches of an Elephant: A Topos Theory Compendium}.
\newblock Oxford Logic Guides. Clarendon Press, 2002.

\bibitem{JonesN:book-computability}
Neil~D. Jones.
\newblock {\em {Computability and Complexity: From a Programming Perspective}}.
\newblock Foundations of Computing. MIT Press, 1997.

\bibitem{Joyal-Street:geometry}
Andr\'{e} Joyal and Ross Street.
\newblock The geometry of tensor calculus {I}.
\newblock {\em Adv. in Math.}, 88:55--113, 1991.

\bibitem{Joyal-Street:geometry-2}
Andr\'{e} Joyal and Ross Street.
\newblock The geometry of tensor calculus {II}, 1995.
\newblock Draft available from Ross Street's website as item 53.

\bibitem{KanD:adj}
Daniel~M. Kan.
\newblock Adjoint functors.
\newblock {\em Transactions of the American Mathematical Society},
  87(2):294--329, 1958.

\bibitem{KellyGM:book-enriched}
Gregory~Maxwell Kelly.
\newblock {\em Basic Concepts of Enriched Category Theory}.
\newblock Number~64 in London Mathematical Society Lecture Note Series.
  Cambridge University Press, 1982.
\newblock Reprinted in Theory and Applications of Categories, No. 10 (2005) pp.
  1-136.

\bibitem{KleeneS:general-bulletin}
Stephen~C. Kleene.
\newblock {General recursive functions of natural numbers}.
\newblock {\em Bull. Amer. Math. Soc.}, 41, 1935.

\bibitem{KleeneS:general}
Stephen~C. Kleene.
\newblock {General recursive functions of natural numbers}.
\newblock {\em Math. Ann.}, 112:727--742, 1936.

\bibitem{KleeneS:lambda}
Stephen~C. Kleene.
\newblock {$\lambda$-definability and recursiveness.}
\newblock {\em {Duke Math. J.}}, 2:340--353, 1936.

\bibitem{KleeneS:ordinal-fund}
Stephen~C. Kleene.
\newblock On notation for ordinal numbers.
\newblock {\em Journal of Symbolic Logic}, 3(4):150--155, 1938.

\bibitem{KleeneS:realiz}
Stephen~C. Kleene.
\newblock On the interpretation of intuitionistic number theory.
\newblock {\em Journal of Symbolic Logic}, 10(4):109--124, 1945.

\bibitem{KleeneS:metamathematics}
Stephen~C. Kleene.
\newblock {\em Introduction to Meta-Mathematics}.
\newblock North-Holland Publ. Co., Amsterdam, 1952.

\bibitem{Koymans}
Christiaan~P.J. Koymans.
\newblock Models of the lambda calculus.
\newblock {\em Inf. Control.}, 52(3):306--332, 1982.

\bibitem{KreiselG:constructivity}
Georg Kreisel.
\newblock Interpretation of analysis by means of constructive functionals of
  finite types.
\newblock In Arend Heyting, editor, {\em Constructivity in Mathematics.
  Procedings of the Colloquium Held at Amsterdam 1957}, Studies in logic and
  the foundations of mathematics, pages 101--128. North-Holland, 1959.

\bibitem{Lambek-Scott:book}
Joachim Lambek and Philip Scott.
\newblock {\em Introduction to Higher Order Categorical Logic}.
\newblock Number~7 in Cambridge Studies in Advanced Mathematics. Cambridge
  University Press, 1986.

\bibitem{Landauer}
Rolf Landauer.
\newblock Irreversibility and heat generation in the computing process.
\newblock {\em IBM journal of research and development}, 5(3):183--191, 1961.

\bibitem{LawvereFW:funsat}
F.~William Lawvere.
\newblock Functorial semantics of algebraic theories.
\newblock {\em Proceedings of the National Academy of Sciences of the United
  States of America}, 50(1):869--872, 1963.

\bibitem{LawvereFW:Foundation}
F.~William Lawvere.
\newblock The category of categories as a foundation for mathematics.
\newblock In {\em Proceedings of the Conference on Categorical Algebra held in
  La Jolla}, pages 1--20. Springer, 1966.

\bibitem{LawvereFW:dialectica}
F.~William Lawvere.
\newblock Adjointness in foundations.
\newblock {\em Dialectica}, 23:281--296, 1969.
\newblock reprint in Theory and Applications of Categories, No. 16, 2006,
  pp.1--16.

\bibitem{LawvereFW:compreh}
F.~William Lawvere.
\newblock Equality in hyperdoctrines and the comprehension schema as an adjoint
  functor.
\newblock In Alex Heller, editor, {\em Applications of Categorical Algebra},
  number~17 in Proceedings of Symposia in Pure Mathematics, pages 1--14.
  American Mathematical Society, 1970.

\bibitem{LawvereFW:metric}
F.~William Lawvere.
\newblock Metric spaces, generalised logic, and closed categories.
\newblock In {\em Rendiconti del Seminario Matematico e Fisico di Milano},
  volume~43. Tipografia Fusi, Pavia, 1973.
\newblock Reprinted in \emph{Theory and Applications of Categories}, No. 1,
  2002, pp.~1--37.

\bibitem{LevinL:forbidden}
Leonid~A. Levin.
\newblock Forbidden information.
\newblock {\em J. ACM}, 60(2), may 2013.

\bibitem{Vitanyi-Li:book}
Ming Li and Paul M.~B. Vit{\'a}nyi.
\newblock {\em An introduction to Kolmogorov complexity and its applications
  (2. ed.)}.
\newblock Graduate texts in computer science. Springer, 1997.

\bibitem{LongleyJ:thesis}
John~R. Longley.
\newblock {\em Realizability toposes and language semantics}.
\newblock PhD thesis, University of Edinburgh, 1995.
\newblock available at www.lfcs.inf.ed.ac.uk/reports/95/ECS-LFCS-95-332/.

\bibitem{Longo-Moggi:enumerations}
Giuseppe Longo and Eugenio Moggi.
\newblock Cartesian closed categories of enumerations for effective type
  structures.
\newblock In Plotkin~G. Kahn~G., MacQueen~D.B., editor, {\em Semantics of Data
  Types (SDT)}, number 173 in Lecture Notes in Computer Science, pages
  235--255. Springer-Verlag, 1984.

\bibitem{MacLaneS:homology}
Saunders Mac\, Lane.
\newblock {\em Homology}.
\newblock {Grundlehren der mathematischen Wissenschaften}. Springer-Verlag,
  1963.

\bibitem{MacLaneS:CWM}
Saunders Mac\, Lane.
\newblock {\em Categories for the Working Mathematician}.
\newblock Number~5 in Graduate Texts in Mathematics. Springer-Verlag, 1971.

\bibitem{Olkin:book}
Albert~W. Marshall and Ingram Olkin.
\newblock {\em Inequalities: Theory of Majorization and Its Applications},
  volume 143 of {\em Mathematics in Science and Engineering}.
\newblock Academic Press, 1979.

\bibitem{Martin-LoefP:inttt}
Per Martin-L{\"o}f.
\newblock {\em Intuitionistic Type Theory}.
\newblock Bibliopolis, Naples, 1984.

\bibitem{MoggiE:monad}
Eugenio Moggi.
\newblock Notions of computation and monads.
\newblock {\em Information and Computation}, 93(1):55 -- 92, 1991.
\newblock Selections from 1989 IEEE Symposium on Logic in Computer Science.

\bibitem{MoschovakisY:Kleene}
Yiannis~N. Moschovakis.
\newblock {Kleene's Amazing Second Recursion Theorem}.
\newblock {\em Bulletin of Symbolic Logic}, 16(2):189--239, 2010.

\bibitem{MulryP:efft}
Philip~S. Mulry.
\newblock Generalized banach-mazur functionals in the topos of recursive sets.
\newblock {\em J. of Pure and Applied algebra}, 26(1):71--83, 1982.

\bibitem{MyhillJ:creative}
John Myhill.
\newblock Creative sets.
\newblock {\em Mathematical Logic Quarterly}, 1(2):97--108, 1955.

\bibitem{Automath}
Ron~P. Nederpelt, J.~Herman Geuvers, and Roel~C. de~Vrijer.
\newblock {\em Selected Papers on Automath}, volume 133 of {\em Studies in
  Logic and the Foundations of Mathematics}.
\newblock Elsevier Science, 1994.

\bibitem{NeumannJ:self-reproducing}
John~von Neumann and A.W. Burks.
\newblock {\em Theory of Self-Reproducing Automata}.
\newblock University of Illinois Press, 1984.

\bibitem{Cantor-Dedekind:Briefwechsel}
Emmy Noether and Jean Cavaill{\`e}s, editors.
\newblock {\em {Briefwechsel Cantor-Dedekind}}, volume 518 of {\em Actualit\'es
  Scientifiques et Industrielles}.
\newblock Hermann \& Cie., Paris, 1937.

\bibitem{Martin-Loef:programming}
Bent Nordstr{\"o}m, Kent Petersson, and Jan~M. Smith.
\newblock {\em Programming in Martin-L{\"o}f's Type Theory: An Introduction}.
\newblock International series of monographs on computer science. Clarendon
  Press, 1990.

\bibitem{Odifreddi:one}
Piergiorgio Odifreddi.
\newblock {\em {Classical Recursion Theory : The Theory of Functions and Sets
  of Natural Numbers}}, volume 125 of {\em Studies in logic and the foundations
  of mathematics}.
\newblock North-Holland, Amsterdam, New-York, Oxford, Tokyo, 1989.

\bibitem{OostenJ:book}
Jaap~van Oosten.
\newblock {\em Realizability: An Introduction to its Categorical Side}, volume
  152 of {\em Studies in Logic and the Foundations of Mathematics}.
\newblock Elsevier Science, 2008.

\bibitem{DiPaola-Heller:dominical}
Robert A.~Di Paola and Alex Heller.
\newblock Dominical categories: Recursion theory without elements.
\newblock {\em The Journal of Symbolic Logic}, 52(3):594--635, 1987.

\bibitem{PapaC:CX}
Christos~H. Papadimitriou.
\newblock {\em Computational Complexity}.
\newblock Addison-Wesley, 1994.

\bibitem{PareR:absolute}
Robert Par\'e.
\newblock On absolute colimits.
\newblock {\em J. Alg.}, 19:80--95, 1971.

\bibitem{PavlovicD:Como}
Dusko Pavlovic.
\newblock Categorical interpolation: descent and the {Beck-Chevalley} condition
  without direct images.
\newblock In A.~Carboni et~al., editor, {\em Category Theory, Como 1990},
  volume 1488 of {\em Lecture Notes in Mathematics}, pages 306--326. Springer
  Verlag, 1991.

\bibitem{PavlovicD:constructions}
Dusko Pavlovic.
\newblock Constructions and predicates.
\newblock In D.~Pitt et~al., editor, {\em Category Theory and Computer Science
  '91}, volume 530 of {\em Lecture Notes in Computer Science}, pages 173--197.
  Springer Verlag, 1991.

\bibitem{PavlovicD:mapsII}
Dusko Pavlovic.
\newblock Maps {II}: Chasing diagrams in categorical proof theory.
\newblock {\em J. of the IGPL}, 4(2):1--36, 1996.

\bibitem{PavlovicD:Qabs12}
Dusko Pavlovic.
\newblock Geometry of abstraction in quantum computation.
\newblock {\em Proceedings of Symposia in Applied Mathematics}, 71:233--267,
  2012.
\newblock arxiv.org:1006.1010.

\bibitem{PavlovicD:IC12}
Dusko Pavlovic.
\newblock Monoidal computer {I}: {Basic computability by string diagrams}.
\newblock {\em Information and Computation}, 226:94--116, 2013.
\newblock arxiv:1208.5205.

\bibitem{PavlovicD:spider}
Dusko Pavlovic.
\newblock {Lambek pregroups are Frobenius spiders in preorders}.
\newblock {\em Compositionality}, 4(1):1--21, 2022.
\newblock arxiv:2105.03038.

\bibitem{PavlovicD:ASE01}
Dusko Pavlovic and Douglas~R. Smith.
\newblock Composition and refinement of behavioral specifications.
\newblock In {\em Automated Software Engineering 2001. The Sixteenth
  International Conference on Automated Software Engineering}. IEEE, 2001.

\bibitem{PavlovicD:MonCom3}
Dusko Pavlovic and Muzamil Yahia.
\newblock {Monoidal computer {III}: {A coalgebraic view of computability and
  complexity}}.
\newblock In C.~C{\^{\i}}rstea, editor, {\em Coalgebraic Methods in Computer
  Science ({CMCS}) 2018 --- Selected Papers}, volume 11202 of {\em Lecture
  Notes in Computer Science}, pages 167--189. Springer, 2018.
\newblock arxiv:1704.04882.

\bibitem{PenroseR:negdim}
Roger Penrose.
\newblock Applications of negative dimensional tensors.
\newblock In D.~J.~A. Welsh, editor, {\em Combinatorial mathematics and its
  applications. Proceedings of a conference held at the Mathematical Institute,
  Oxford, 7--10 July, 1969}, pages 221--244. Academic Press, 1971.

\bibitem{Rozsa}
Rozsa P\'eter.
\newblock {\em Recursive Functions}.
\newblock Academic Press, 1967.

\bibitem{PopperK:logic}
Karl~R. Popper.
\newblock {\em The Logic of Scientific Discovery}.
\newblock Basic Books, 1954.

\bibitem{PostE:creative}
Emil~L. Post.
\newblock {Recursively enumerable sets of positive integers and their decision
  problems}.
\newblock {\em {Bull. Amer. math. Soc.}}, 50:284--316, 1944.

\bibitem{PoundstoneW:recursive}
William Poundstone.
\newblock {\em The Recursive Universe: Cosmic Complexity and the Limits of
  Scientific Knowledge}.
\newblock Dover Books on Science. Dover Publications, 2013.

\bibitem{RiceHG:rice-thm}
H.~Gordon Rice.
\newblock Classes of recursively enumerable sets and their decision problems.
\newblock {\em Transactions of the American Mathematical society},
  74(2):358--366, 1953.

\bibitem{RiehlE:CT}
Emily Riehl.
\newblock {\em Category Theory in Context}.
\newblock Aurora: Dover Modern Math Originals. Dover Publications, 2017.

\bibitem{Hartley:isomorphism}
Hartley Rogers, Jr.
\newblock G\"odel numberings of partial recursive functions.
\newblock {\em The Journal of Symbolic Logic}, 23(3):331--341, 1958.

\bibitem{RogersH:book}
Hartley Rogers, Jr.
\newblock {\em Theory of recursive functions and effective computability}.
\newblock MIT Press, Cambridge, MA, USA, 1987.

\bibitem{RussellB:paradox}
Bertrand Russell.
\newblock Letter to {Frege}, 1902.
\newblock In \cite{vanHeijenoortJ}.

\bibitem{RussellB:types}
Bertrand Russell.
\newblock Mathematical logic based on the theory of types.
\newblock {\em American Journal of Mathematics}, 30:222--262, 1908.
\newblock Reprinted in \cite{vanHeijenoortJ}, pages 150--182.

\bibitem{Russell-Whitehead:principia}
Bertrand Russell and Alfred~North Whitehead.
\newblock {\em Principia Mathematica}.
\newblock Cambridge University Press, 1910--13.

\bibitem{SchonfinkelM:bausteine}
Moses Sch{\"o}nfinkel.
\newblock {{\"U}ber die Bausteine der mathematischen Logik}.
\newblock {\em Mathematische Annalen}, 92(3):305--316, 1924.

\bibitem{ScottDS:dattl}
Dana Scott.
\newblock Data types as lattices.
\newblock {\em SIAM Journal on Computing}, 5:522--587, 1976.

\bibitem{ScottD:outline}
Dana~S. Scott.
\newblock Outline of mathematical theory of computation.
\newblock In {\em Proceedings of the 4th Annual Princeton Conf. on Information
  Sciences and Systems}, pages 169--176. Princeton University Press, 1970.

\bibitem{ScottD:continuous}
Dana~S. Scott.
\newblock Continuous lattices.
\newblock In F.W. Lawvere, editor, {\em Toposes, Algebraic Geometry and Logic},
  volume~2, pages 403--450. Springer, 1980.

\bibitem{ScottD:relating}
Dana~S. Scott.
\newblock Relating theories of the lambda calculus.
\newblock In J.P. Seldin and J.R. Hindley, editors, {\em To H.B.~Curry: Essays
  on combinatory logic, lambda calculus and formalism}, volume~2, pages
  403--450. Academic Press, 1980.

\bibitem{SeiferasJ:blum}
Joel~I. Seiferas.
\newblock Machine-independent complexity theory.
\newblock In J.~{van Leeuven}, editor, {\em Algorithms and Complexity},
  volume~A of {\em Handbook of Theoretical Computer Science}, pages 163--186.
  Elsevier, Amsterdam, 1990.

\bibitem{ShallitJ:second-automata}
Jeffrey Shallit.
\newblock {\em A second course in formal languages and automata theory}.
\newblock Cambridge University Press, 2008.

\bibitem{Shannon-Weaver}
Claude~E. Shannon and Warren Weaver.
\newblock {\em The Mathematical Theory of Communication}.
\newblock The Mathematical Theory of Communication. University of Illinois
  Press, 1962.

\bibitem{SiegW:cantor-bernstein}
Wilfried Sieg.
\newblock {The Cantor-Bernstein theorem: how many proofs?}
\newblock {\em Philosophical Transactions of the Royal Society A},
  377(2140):20180031, 2019.

\bibitem{SimmonsH:CT}
Harold Simmons.
\newblock {\em An Introduction to Category Theory}.
\newblock Cambridge University Press, 2011.

\bibitem{SipserM:book}
Michael Sipser.
\newblock {\em Introduction to the Theory of Computation}.
\newblock Cengage Learning, 2013.

\bibitem{SmorynskiC:self-reference}
Craig Smorynski.
\newblock {\em Self-Reference and Modal Logic}.
\newblock Universitext. Springer New York, 2012.

\bibitem{SmullyanR:recursion}
Raymond~M. Smullyan.
\newblock {\em Recursion Theory for Metamathematics}.
\newblock Oxford Logic Guides. Oxford University Press, 1993.

\bibitem{SmullyanR:diagonalization}
Raymond~M. Smullyan.
\newblock {\em Diagonalization and Self-reference}.
\newblock Logic Guides Series. Clarendon Press, 1994.

\bibitem{SmullyanR:what-name}
Raymond~M. Smullyan.
\newblock {\em What is the Name of this Book?}
\newblock Dover Recreational Math Series. Dover Publications, 2011.

\bibitem{SmullyanR:mock}
Raymond~M. Smullyan.
\newblock {\em To Mock a Mocking Bird}.
\newblock Knopf Doubleday, 2012.

\bibitem{smyth-plotkin}
M.B. Smyth and G.D. Plotkin.
\newblock The category-theoretic solution of recursive domain equations.
\newblock {\em SIAM Journal on Computing}, 11(4):761--783, 1982.

\bibitem{SmythM:topology}
Michael~B. Smyth.
\newblock Topology.
\newblock In S.~Abramsky and D.~Gabbay, editors, {\em Handbook of Logic in
  Computer Science (Vol. 1). Background: Mathematical Structures}, pages
  641--761. Oxford University Press, Inc., USA, 1993.

\bibitem{SoareR:RE}
Robert~I. Soare.
\newblock {\em Recursively Enumerable Sets and Degrees: A Study of Computable
  Functions and Computably Generated Sets}.
\newblock Perspectives in Mathematical Logic. Springer, 1999.

\bibitem{SoareR:CTA}
Robert~I. Soare.
\newblock {\em Turing Computability: Theory and Applications}.
\newblock Theory and Applications of Computability. Springer, 2016.

\bibitem{Viggo}
Viggo Stoltenberg-Hansen, Ingrid Lindstr{\"o}m, and Edward~R. Griffor.
\newblock {\em Mathematical theory of domains}.
\newblock Number~22 in Cambridge Tracts in Theoretical Computer Science.
  Cambridge University Press, 1994.

\bibitem{StreetR:kelly-penrose}
Ross Street.
\newblock Posting on categories mailing list of 5 may, 2017.

\bibitem{TarskiA:truth}
Alfred Tarski.
\newblock {Der Wahrheitsbegriff in den formalisierten Sprachen}.
\newblock {\em Studia Philosophica}, 1:261 -- 405, 1935.

\bibitem{TegmarkM:universe}
Max Tegmark.
\newblock {\em Our Mathematical Universe: My Quest for the Ultimate Nature of
  Reality}.
\newblock Knopf Doubleday Publishing Group, 2015.

\bibitem{TroelstraA:metamath}
Anne~S. Troelstra.
\newblock {\em Metamathematical Investigation of Intuitionistic Arithmetic and
  Analysis}, volume 344 of {\em Lecture Notes in Mathematics}.
\newblock Springer, 1973.
\newblock With contributions by C.A. Smory{\'n}ski, J.I. Zucker and W.A.
  Howard.

\bibitem{Troelstra-vanDalen:book}
Anne~S. Troelstra and Dirk van Dalen.
\newblock {\em Constructivism in Mathematics. An Introduction}, volume 121,123
  of {\em Studies in Logic and Foundations of Mathematics}.
\newblock Elsevier Science, 1988.

\bibitem{Turchin:supercompiler}
Valentin~F. Turchin.
\newblock The concept of a supercompiler.
\newblock {\em ACM Trans. Program. Lang. Syst.}, 8(3):292--325, June 1986.

\bibitem{TuringA:morphogenesis-book}
Alan Turing.
\newblock {\em Morphogenesis}.
\newblock Collected Works of A.M. Turing. Elsevier Science, 1992.
\newblock edited by Saunders, P.T.

\bibitem{TuringA:Entscheidung}
Alan~M. Turing.
\newblock On computable numbers, with an application to the
  {Entscheidungsproblem}.
\newblock {\em Proceedings of the London Mathematical Society. Second Series},
  42:230--265, 1936.
\newblock Reprinted in \cite{DavisM:undecidable}.

\bibitem{TuringA:thesis}
Alan~M. Turing.
\newblock Systems of logic based on ordinals.
\newblock {\em Proc.~of the London Mathematical Society. Second Series},
  45:161--228, 1939.

\bibitem{LeeuwenJ:handbook-A}
Jan van Leeuwen, editor.
\newblock {\em Handbook of Theoretical Computer Science, Volume {A:} Algorithms
  and Complexity}.
\newblock Elsevier and {MIT} Press, 1990.

\bibitem{LeeuwenJ:handbook-B}
Jan van Leeuwen, editor.
\newblock {\em Handbook of Theoretical Computer Science, Volume {B:} Formal
  Models and Semantics}.
\newblock Elsevier and {MIT} Press, 1990.

\bibitem{VardiM:CTA}
Moshe~Y. Vardi.
\newblock Why doesn't {ACM} have a {SIG} for {Theoretical Computer Science}?
\newblock {\em Commun. ACM}, 58(8):5, jul 2015.

\bibitem{NeumannJ:numbers}
Johann von Neumann.
\newblock {Zur Einf{\"u}hrung der transfiniten Zahlen}.
\newblock {\em Acta litt. Acad. Sc. Szeged}, X(1):199--208, 1923.
\newblock English translation, ``On the introduction of transfinite numbers''
  in \cite{vanHeijenoortJ}, pages 393--413.

\bibitem{WadlerP:curry-howard}
Philip Wadler.
\newblock Propositions as types.
\newblock {\em Commun. ACM}, 58(12):75--84, November 2015.

\bibitem{WinskelG:book}
Glynn Winskel.
\newblock {\em The Formal Semantics of Programming Languages: An Introduction}.
\newblock Foundations of Computing. MIT Press, 1994.

\bibitem{YanofskyN:outer}
Noson~S. Yanofsky.
\newblock {\em The Outer Limits of Reason: What Science, Mathematics, and Logic
  Cannot Tell Us}.
\newblock MIT Press, 2016.

\end{thebibliography}

\clearpage
\addcontentsline{toc}{chapter}{Index}
\printindex

\begin{figure}[!ht]
\begin{center}
\includegraphics[height=20cm
]{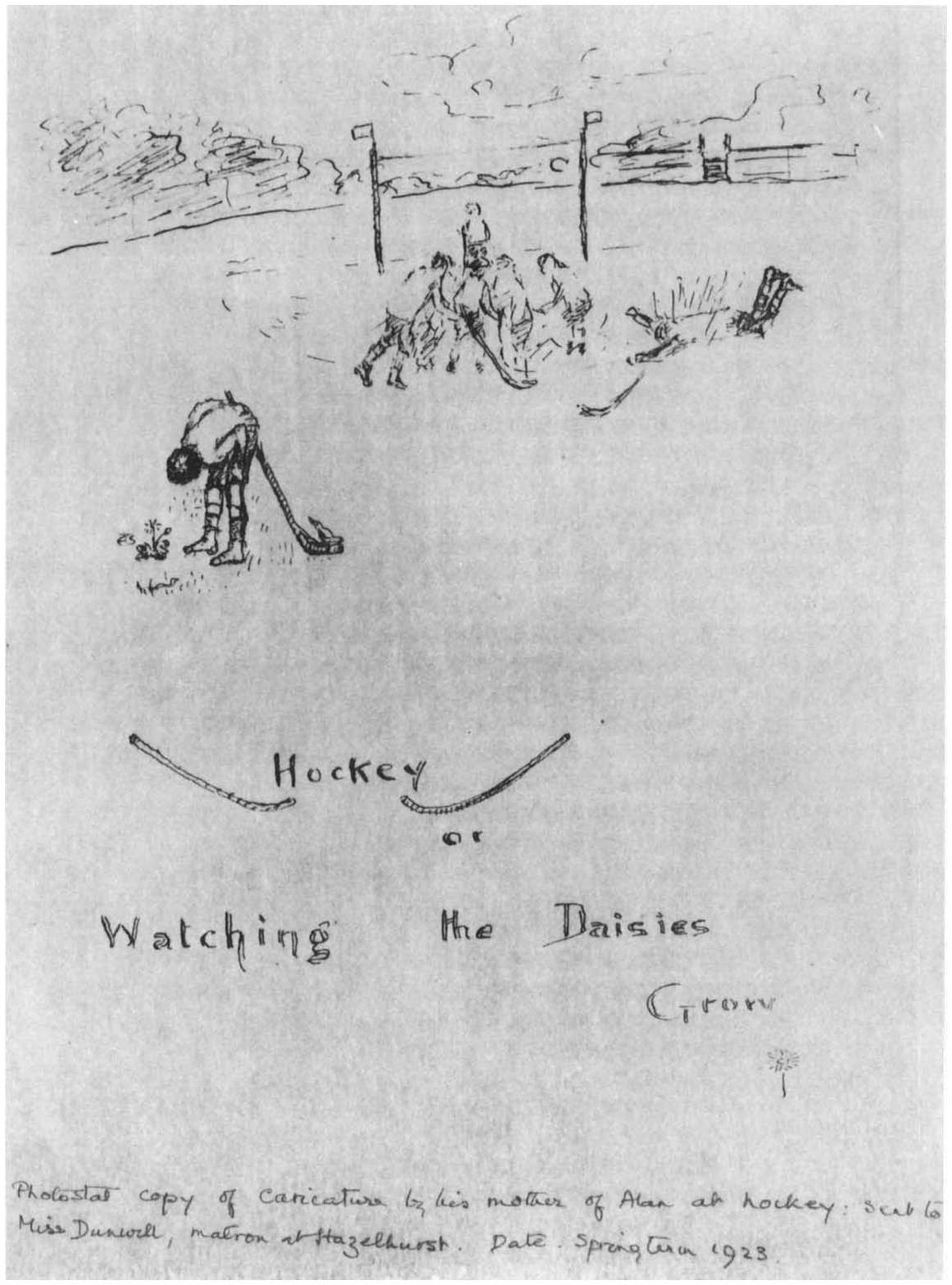}
\vspace{-2.5\baselineskip}
\caption{11-year old Alan Turing as seen by his mother}
\label{Fig:turing-daisy}
\end{center}
\end{figure}

\end{document}